\documentclass[11pt,a4paper]{article}
    \usepackage[format=hang]{caption}
    \usepackage{epsfig,amssymb,amsmath,graphicx,subcaption,verbatim,hyperref,ulem,bm,float,tensor,tikz,extarrows}
    \usepackage{blkarray}
    \usepackage{xcolor}
    \usepackage{color}
    \usepackage{makecell}%linebreak in a table cell
    \usepackage{tikz}
    \usetikzlibrary{decorations.markings}
    \usetikzlibrary{calc,quotes}
\usetikzlibrary{arrows.meta}
\usepackage{pgfmath}
\newcommand \mathtikz[1] {\quad \vcenter{\hbox{\tikz{#1}}} \quad}
\newcommand \mathtikzS[2] {
    \mathtikz{
        \begin{scope}[scale=#1]
            #2
        \end{scope}
    }
}
\def \triLen {0.5}
\def \diskRad {0.1}

\def \squareLen {0.2}

\newcommand \tri[3] {
\begin{scope} [xshift=#1, yshift=#2, rotate=#3-22.5] 
	\coordinate (A) at (0, 0);
	\coordinate (B) at (\triLen, 0);
	\coordinate (C) at (0.707*\triLen, 0.707*\triLen);
	\fill (A) -- (B) -- (C) -- cycle;
\end{scope}
}

\newcommand \triL[4] {
\begin{scope} [xshift=#1, yshift=#2, rotate=#3-22.5, scale=#4] 
	\coordinate (A) at (0, 0);
	\coordinate (B) at (\triLen, 0);
	\coordinate (C) at (0.707*\triLen, 0.707*\triLen);
	\fill (A) -- (B) -- (C) -- cycle;
\end{scope}
}

\newcommand \disk[3] {
\begin{scope}[xshift=#1, yshift=#2]
	\fill [fill=#3] (0,0) circle [radius=\diskRad];
\end{scope}
}

\newcommand \mysquare[3]{
	\begin{scope} [xshift=#1, yshift=#2]
		\coordinate (A) at (\squareLen/2,\squareLen/2);
		\coordinate (B) at (-\squareLen/2,\squareLen/2);
		\coordinate (C) at (-\squareLen/2,-\squareLen/2);
		\coordinate (D) at (\squareLen/2,-\squareLen/2);
		\draw (A) -- (B) -- (C) -- (D) -- cycle;
		\fill [fill=#3] (A) -- (B) -- (C) -- (D) -- cycle;
	\end{scope}
}

\newcommand\vertexI[3]{ %the same as above picture, with integer coordinates%
\begin{scope}[xshift=#1,yshift=#2]
\draw [thick] (-1*#3,-1*#3) -- (0,0);
\draw [thick] (1*#3,-1*#3) -- (0,0);
\draw [thick] (0,#3) -- (0,0);
\end{scope}
}
\newcommand\vertexIL[6]{ % = vertexI plus labels of legs
\begin{scope}[xshift=#1,yshift=#2]
    \draw [thick] (-1*#3,-1*#3) -- (0,0);
    \draw [thick] (1*#3,-1*#3) -- (0,0);
    \draw [thick] (0,#3) -- (0,0);
    \node[below] at (-#3,-#3) {#4};
    \node[below] at (#3,-#3) {#5};
    \node[above] at (0,#3) {#6};
\end{scope}
}
\newcommand\vertexILC[7]{ % = vertexI plus labels of legs
\begin{scope}[xshift=#1,yshift=#2]
    \draw [thick,#7] (-1*#3,-1*#3) -- (0,0);
    \draw [thick,#7] (1*#3,-1*#3) -- (0,0);
    \draw [thick,#7] (0,#3) -- (0,0);
    \node[below] at (-#3,-#3) {#4};
    \node[below] at (#3,-#3) {#5};
    \node[above] at (0,#3) {#6};
\end{scope}
}
\newcommand\vertexIC[4]{ %the same as above, the 4th parameter is the color%
\begin{scope}[xshift=#1,yshift=#2]
    \draw [thick, #4] (-1*#3,-1*#3) -- (0,0);
    \draw [thick, #4] (1*#3,-1*#3) -- (0,0);
    \draw [thick, #4] (0,#3) -- (0,0);
\end{scope}
}
\newcommand\antivertexI[3]{ %
\begin{scope}[xshift=#1,yshift=#2]
\draw [thick] (-1*#3,1*#3) -- (0,0);
\draw [thick] (1*#3,1*#3) -- (0,0);
\draw [thick] (0,-#3) -- (0,0);
\end{scope}
}
\newcommand\antivertexIL[6]{ % = antivertexI plus labels of legs
\begin{scope}[xshift=#1,yshift=#2]
    \draw [thick] (-1*#3,1*#3) -- (0,0);
    \draw [thick] (1*#3,1*#3) -- (0,0);
    \draw [thick] (0,-#3) -- (0,0);
    \node[above] at (-#3,#3) {#4};
    \node[above] at (#3,#3) {#5};
    \node[below] at (0,-#3) {#6};
\end{scope}
}

\newcommand\antivertexILC[7]{ % the last argument determines the color of vertex
\begin{scope}[xshift=#1,yshift=#2]
    \draw [thick,#7] (-1*#3,1*#3) -- (0,0);
    \draw [thick,#7] (1*#3,1*#3) -- (0,0);
    \draw [thick,#7] (0,-#3) -- (0,0);
    \node[above] at (-#3,#3) {#4};
    \node[above] at (#3,#3) {#5};
    \node[below] at (0,-#3) {#6};
\end{scope}
}

\newcommand\vertextI[4]{ %例图中的第二，三行%
\begin{scope}[xshift=#1,yshift=#2]
\draw [thick] (-1*#3*#4,-1*#3) -- (0,0);
\draw [thick] (1*#3*#4,-1*#3) -- (0,0);
\draw [thick] (1*#3*#4,1*#3) -- (0,0);
\end{scope}
}

\newcommand\crossingI[3]{ %crossing diagram with integer coordinates
\begin{scope}[xshift=#1,yshift=#2]
\draw [thick] (-1*#3,-1*#3) -- (1*#3,1*#3);
\draw [thick] (-1*#3,1*#3) -- (-0.15,0.15);
\draw [thick] (1*#3,-1*#3) -- (0.15,-0.15);
\end{scope}
}
\newcommand\anticrossingI[3]{ %crossing diagram with integer coordinates
\begin{scope}[xshift=#1,yshift=#2]
\draw [thick] (-1*#3,1*#3) -- (1*#3,-1*#3);
\draw [thick] (-1*#3,-1*#3) -- (-0.15,-0.15);
\draw [thick] (1*#3,1*#3) -- (0.15,0.15);
\end{scope}
}

\def \bulkCol {green}
\def \algCol {black}
\def \modCol {magenta}
\DeclareMathOperator{\Hom}{Hom}
\DeclareMathOperator{\id}{id}
\DeclareMathOperator{\Ind}{Ind}
\DeclareMathOperator{\Rep}{Rep}
\DeclareMathOperator{\Vect}{Vec}
\DeclareMathOperator{\Diag}{Diag}

\renewcommand{\triLen}{0.3}

    \usepackage{amsthm}
        \usepackage{multirow}
    \usepackage{longtable}
    \usepackage{physics}

    \definecolor{BrickRed}{RGB}{203, 65, 84}
    \definecolor{Violet}{HTML}{EE82EE}
    \definecolor{OliveGreen}{HTML}{556B2F}
    \definecolor{RoyalBlue}{RGB}{25,41,88}
    
    \hypersetup{colorlinks=true, linkcolor=BrickRed, citecolor=Violet, filecolor=OliveGreen, urlcolor=RoyalBlue, filebordercolor={.8 .8 1}, urlbordercolor={.8 .8 0}}
    \newcommand{\be}{\begin{equation}}
    \newcommand{\ee}{\end{equation}}

\newcommand{\mA}{\mathcal{A}}

\newcommand{\Z}{\mathbb{Z}}

\usepackage{pdfpages}
\usepackage{array} % for \newcolumntype
\newcolumntype{C}{>{$}c<{$}} %math-mode of "c" column type
    \def\ket#1{\left|#1\right>}

    \newcommand{\news}{\setcounter{equation}{0}}
    \def\bea{\begin{eqnarray}}
    \def\eea{\end{eqnarray}}
    \numberwithin{equation}{section}
    \makeatletter
    \renewcommand*\env@matrix[1][\arraystretch]{
      \edef\arraystretch{#1}
      \hskip -\arraycolsep
      \let\@ifnextchar\new@ifnextchar
      \array{*\c@MaxMatrixCols c}}
    \makeatother
    \usepackage{color}
    
\begin{document}

\title{\vskip -65pt
    \vskip 60pt
        {\bf {\large A (Dummy's) Guide to Working with Gapped Boundaries via (Fermion) Condensation }\\[20pt]}}
\author{ Jiaqi Lou, Ce Shen\footnote{Lou and Shen are co- first authors of the manuscript.}, Chaoyi Chen, Ling-Yan Hung\\[20pt]
    $^1$State Key Laboratory of Surface Physics, \\
    Fudan University, \\
    200433 Shanghai, China\\
    $^2$Shanghai Qi Zhi Institute, \\
    41st Floor, AI Tower, No. 701 Yunjin Road,  \\
    Xuhui District, Shanghai, 200232, China\\
    $^3$Department of Physics and Center for Field Theory and Particle Physics, \\
    Fudan University, \\
    200433 Shanghai, China\\
    $^4$Institute for Nanoelectronic devices and Quantum computing, \\
    Fudan University, \\
    200433 Shanghai , China\\
}

\date{\today}
\maketitle
\vskip 20pt

\begin{abstract}
We study gapped boundaries characterized by ``fermionic condensates'' in 2+1 d topological order.  Mathematically, each of these condensates can be described by  a super commutative Frobenius algebra.
We systematically obtain the species of excitations at the gapped boundary/ junctions, and study their endomorphisms (ability to trap a Majorana fermion) and fusion rules, and generalized the defect Verlinde formula to a twisted version. We illustrate these results with explicit examples. We also connect these results with topological defects in super modular invariant CFTs. 
To render our discussion self-contained, we provide a pedagogical review of relevant mathematical results, so that physicists without prior experience in tensor category should be able to pick them up  and apply them readily. 

\end{abstract}

\vfill

%PACS:
%\newpage
\newpage
\tableofcontents
\newpage

\section{Introduction}\news

There are many works that study gapped boundaries in 2+1 d bosonic topological orders that are
characterized by anyon condensation \cite{bais_broken_2002, bais_condensate-induced_2009, bais_theory_2009, hung_ground_2015, lan_gapped_2015, hung_generalized_2015, kapustin_topological_2011, davydov_witt_2010, kitaev_models_2012, Barkeshli:2013yta, barkeshli_classification_2013, 
 Kong:2013aya, fuchs2013bicategories, levin_protected_2013, 2018ARCMP...9..307B,cong_topological_2016}. 
 More recently, it is realized that some gapped boundaries of these bosonic orders  can
be characterized by ``fermion" condensations -- physically, it corresponds to emergent fermions pairing up with local
free fermions which subsequently condense at the boundary \cite{Aasen:2017ubm,Bhardwaj:2016clt, Lan_2016, Wan:2016php,Bhardwaj:2016clt}. The boundary thus necessarily becomes 
sensitive to the spin structure. 
Mathematically, these gapped boundaries can be characterized by (super) Frobenius algebra in the tensor category describing
the topological order concerned. 
\footnote{Let us emphasize here that gapped interfaces are also characterized by anyon condensation. However, every gapped interface can be understood in terms of a gapped boundary by the folding trick. The main difference between a gapped interface and a gapped boundary is that across the interface there are still non-trivial bulk excitations, while the phase is trivial across a gapped boundary. The discussion here, to avoid clutter, addresses directly the gapped boundaries. The discussion however can be easily turned around into a discussion of gapped interface.  }

The most signatory set of physical properties of a gapped boundary includes the collection of topological excitations and defects it supports,
their quantum dimensions and fusion properties which controls ground state degeneracies of the system in an open manifold.
As alluded to above, these gapped boundaries can be understood in terms of (super) Frobenius algebra. 
The physics of the gapped boundaries should be encoded in the mathematics, which in principle could be extracted systematically. 
This is indeed the case particularly for bosonic gapped boundaries -- except that the techniques are dispersed in the physics and mathematics literature,
the latter of which is often shrouded in a language completely foreign to physicists, and that the formal principles laid out may not be readily 
converted into a practical computation. A practical way of computing fusion rules of defects localized at junctions between different gapped boundaries have been elucidated in \cite{Shen_2019}. In the case of fermion condensation which receives attention only more recently \cite{Aasen:2017ubm,Bhardwaj:2016clt, Lan_2016, Wan:2016php}, a systematic study including non-Abelian fermion condensation and junctions remain largely an open problem. 

We propose that a super-commutative separable special Frobenius algebra in the bulk topological order is responsible for characterizing its fermionic boundary conditions -- to our knowledge this is the first systematic use of the ``super-commutative'' version of the Frobenius algebra to describe the fermionic boundaries and through which to work out their properties.
 
We elucidate properties of defects in a gapped boundary or junctions characterized by  fermion condensations.  
This includes obtaining the full collection of topological defects, identifying their endomorphisms, and also computing their fusion rules, that can be summarized by a (super) defect Verlinde formula. We extended the results of \cite{Aasen:2017ubm,Bhardwaj:2016clt, Lan_2016, Wan:2016php} to include non-Abelian condensates, and also the study of junctions between (fermionic) gapped boundaries. We also develop new ways to compute the half-linking numbers. 

To understand these results, it is most convenient if the reader is familiar with the computational tools available in braided tensor categories and their algebra objects. 
We hope to convey the power of computations using category theory -- while some of the  examples discussed in the current paper can be obtained using their explicit realization in field theory or lattice models, such as the boundaries and junctions of the toric code order -- category theory is really an elaborate and generalized group theory that allows one to work out the basic features of these gapped phase and their boundaries in a clean way without getting bogged down by extra details pertaining to a specific realization of the topological order that are in fact not universal to the order. 
Category theory is a powerful tool that keeps track of the combinatoric data assuming only that there are conserved (topological) charges, and that they can fuse in an associative way -- which are clearly model independent features of a topological order. 

Relevant mathematical results are mostly scattered in many different places, which maybe a major hindrance to entering the subject.
We therefore collect the most relevant tools to make the paper self-contained. We give up some mathematical rigor to make the language more readily accessible to working physicists with minimal prior experience in tensor category theories -- i.e. we give a collection of mathematical definitions we deem immediately relevant in doing computations at least in the current context that makes heavy use of algebra in categories, and explain how these mathematical results can be used in explicit computations illustrated in examples. This feature hopefully fills the gap in most of the mathematical literature that is dense on definition and theorems, but scarce in making connections with explicit computations. 

A formal and proper introduction to the subject can be found in numerous places in the literature. Of particular use are \cite{Fuchs_2002, Fuchs_2004, kirillov}, and references therein. 
The paper is organized as follows. 
In section 2 we first give a brief review of (super)  braided tensor categories. We review also algebra objects in a category, and their representations. These results are then extended to include super commutative algebra. The computation of half-linking number and the fusion rules of modules and bi-modules, in addition to their endomorphisms, are discussed. 

In section 3, we illustrate the results obtained in section 2 in explicit examples, namely the toric code model and the $D(S_3)$ quantum double, where we explicitly obtain the Frobenius algebra and the bi-modules that describe  fermionic boundaries. We also demonstrate how their fusion rules are computed.

In section 4, we describe the connection of the current results to supersymmetric CFT's and their topological defects. We also discuss the twisted version of the Verlinde formula that produces the difference in fermion parity even and odd channels in the fusion of primaries.

In section 5, we conclude with some miscellaneous facts about fermion condensation, and various open problems to be addressed in the future. 

There are several lengthy computations that we have relegated into the appendix. In particular, the computation of the 6j symbol describing associativity of fusion of the boundary excitations, are explained and illustrated in detail there. We also include more sophisticated examples of fermion condensates in the $SU(2)_{10}$ and $D(D_4)$, which involve condensates of multiple fermions. In particular, in the latter example some of these (super) modular invariants do not appear to correspond to a condensate that preserve fermion parity. Whether these examples have physical meanings should be explored in greater details in the future. 
We present in appendix B the counting of Majorana modes localized at junction using the Abelian Chern-Simons description of the toric code order, and compute entanglement entropy on a cylinder with different fermionic/bosonic boundary conditions. Some topological data of the $D(S_3)$ quantum double is reviewed in appendix C.

\section{A Physicist's skeletal manual to tensor category and gapped boundaries -- review and generalizations to fermionic boundaries} \label{sec:intro_cat}

Tensor category covers a huge class of mathematical structures. 
To the author, the framework has a structure not unlike an onion where extra structures can be 
included layer by layer, adding to the complexity of the situation. 

As far as 2+1 d topological order is concerned, the categories that are of interest are
(braided, or in fact modular) fusion categories. \footnote{Let us emphasize here that we are considering the mathematical abstraction of the topological order itself. We note that
the construction of explicit lattice models of topological order -- such as the Kitaev models and Levin-Wen models, require data of a fusion category as input data. That is often called ``input category'' as opposed to the resultant topological order which is called the ``output category''. Using this language, we are describing the ``output'' category only in the current paper.  }

In the following, we will collect the most important results that will actually be used in the rest of the paper. 
Rather than listing all the algebraic equations in one full swoop as if all the properties are supposed to appear together from the beginning, 
we are presenting these results in a way to emphasize that many of the properties are in fact independent. Each add-on property is an extra mathematical structure to the
construct, and each such addition has to be made consistent with all the other qualifiers already included, 
very often leading to extra consistency constraints, which is the origin of the many algebraic equations characterizing a certain tensor category.

\subsection{The basics of Braided fusion tensor category} \label{sec:introcat}
Let us summarize the most basic concepts below. \footnote{Materials here can be found reviewed for example in \cite{Bonderson_2008,Barkeshli_2019} and references therein.  Our emphasis on explicit basis construction for morphisms can be attributed to \cite{Fuchs_2002}.  }

\vspace{1cm}
{\bf \underline{Simple Objects}}\\

The most basic structure is the collection of objects.  
Simple objects describe different species of point excitations in a 2+1 d topological order. 
Physically interesting theories are semi-simple categories, where every object can be decomposed as direct sums of simple objects, which form
a basis of elementary particles.
\be
a = \oplus_i m_{ai} c_i, \qquad a, c_i \in C, \qquad m_{ai} \in \mathbb{Z}_{\ge 0}.
\ee
The multiplicity $m_{ai}$ should be non-negative integers. 

\vspace{0.5cm}
{\bf \underline{Morphism}}\\

Morphisms, or homomorphisms, often denoted Hom($a,b$) are maps taking $a$ to $b$.  Morphisms reveal structures of the objects. 
In a bosonic theory, simple objects, describing point particles having {\it no internal structure} has a 1-dimensional ``endomorphism'' space. 
i.e. there is a unique map mapping a simple object to itself ({\it endomorphism}). That map is just the identity map.  Graphically it is often represented as a straight line.  \\
Between two simple objects $a$ and $b$ there is no map between them, unless $a=b$. 
For example if we have 
\be
a = \oplus_i m_{a i } c_i, \qquad b= \oplus_i m_{b i}c_i,
\ee
where $c_i $ are the simple objects in $C$, 
\be
\textrm{dim} [\textrm{Hom}(a,b)] = \sum_i m_{ai} m_{bi}.
\ee

We can construct a basis for these morphisms from the composite object $a$ to $c_i$. This is illustrated in (\ref{eq:hombasis}), where the basis index $\alpha$ runs from 1 to $m_{ai}$. 
\begin{figure}[h]
\begin{equation} \label{eq:hombasis}
    \mathtikzS{0.7}{
    \coordinate["$a$" right] (A) at (0,1cm);
    \coordinate (O) at (0,-0.2cm);
    \node[right] at (O) {$\alpha$};
    \coordinate["$c_i$" right] (B) at (0,-1.5cm);
    \draw [thick] (A) -- (O);
    \draw [thick] (O) -- (B);
    \tri{0}{0}{270};
    }
    \in\quad \Hom(c_i,a), \quad \alpha\in \{1,\dots,m_{ai}\}
\end{equation}
\end{figure}

Let us note that in all these pictures here and in the rest of the paper, they have an orientation.   One could think of them as the likes of Feynman diagrams, that each describing a process, and the orientation here is chosen (unless otherwise specified) such that ``time'' is flowing from the bottom to the top. If one flips a diagram upside down, that is equivalent to taking conjugate, where individual anyon species should be replaced by its dual, and any coefficients should be replaced by its complex conjugate.

States in a Hilbert space are constructed from basis of morphisms.  Particularly when we attempt to count the number of states in a topological order with a given number of anyons, the Hilbert space is basically the space of morphisms that map the collection of anyons to appropriate objects -- e.g. to the trivial anyons if the system is in a closed manifold with no boundaries.  These ``maps of collection of anyons to another anyon" are part of an extra mathematical structure -- namely fusion, that is discussed below. 
One important technical issue is a phase ambiguity in constructing a basis for morphisms. 
Rescaling a given basis in (\ref{eq:hombasis}) by $\zeta^\alpha(a,c_i)$ defines an equally good basis.

\begin{figure}[h]
\begin{equation} \label{eq:hombasis_rescale}
    \mathtikzS{0.7}{
    \coordinate["$a$" right] (A) at (0,1cm);
    \coordinate (O) at (0,-0.2cm);
    \node[right] at (O) {$\alpha$};
    \coordinate["$c_i$" right] (B) at (0,-1.5cm);
    \draw [thick] (A) -- (O);
    \draw [thick] (O) -- (B);
    \tri{0}{0}{270};
    }
   \to \zeta^a_{c_i}(\alpha)   
   \mathtikzS{0.7}{
    \coordinate["$a$" right] (A) at (0,1cm);
    \coordinate (O) at (0,-0.2cm);
    \node[right] at (O) {$\alpha$};
    \coordinate["$c_i$" right] (B) at (0,-1.5cm);
    \draw [thick] (A) -- (O);
    \draw [thick] (O) -- (B);
    \tri{0}{0}{270};
    }
\end{equation}
\end{figure}

\vspace{0.5cm}
{\bf \underline{Fusion}}\\

Anyons obey a commutative and associative fusion algebra:
\begin{equation}
    a\otimes b = \sum_{c} N_{ab}^c c,
    \label{eq:bulk_fusion}
\end{equation}
where $N_{ab}^c=N_{ba}^c$ is a non-negative integer specifying the number of different ways in which anyons $a$ and $b$ can fuse to $c$.
%#done: unnormalized state in Hilbert space, associativity, F symbols, pentagon
A special object $\mathbf{1}$ called the trivial object(vacuum) fuses trivially with all other objects: $\mathbf{1}\otimes a=a$.\footnote{In the appendix and in some literature $\mathbf{0}$ is also used to label the trivial object.}
The building block of the states of anyonic Hilbert space is the fusion basis represented diagrammatically by a vertex:
%#done: picture here
\begin{equation} \label{eq:fusion_basis}
    \ket{a,b;c,\mu} \quad = \quad \left( \frac{d_c}{d_a d_b} \right)^{1/4} \quad
    \mathtikzS{0.5}{
        \vertexIC{0}{0}{1cm}{black};
        \node[right] at (0,0) {$\mu$};
        \node[above] at (0,1cm) {$c$};
        \node[below] at (1cm,-1cm) {$b$};
        \node[below] at (-1cm,-1cm) {$a$};
    }
\end{equation}
where $\mu=1,\dots,N_{ab}^c$. The number $d_i$ is the quantum dimension of anyon $i$, which will be briefly reviewed later in this section. The collection of all fusion trees with the same input/output legs spans a subspace of the Hilbert space, namely the fusion space $V^{ab}_c$.  Note that in the construction of explicit basis of these linear maps $V^{ab}_c$, or equivalently the definition of the states (\ref{eq:fusion_basis}), is ambiguous up to a phase $\xi^{ab}_c$, which is the same kind of rescaling of morphism basis as we have seen in (\ref{eq:hombasis_rescale}). 

Larger fusion bases are constructed from the building blocks by taking tensor product of the building blocks in an appropriate order.

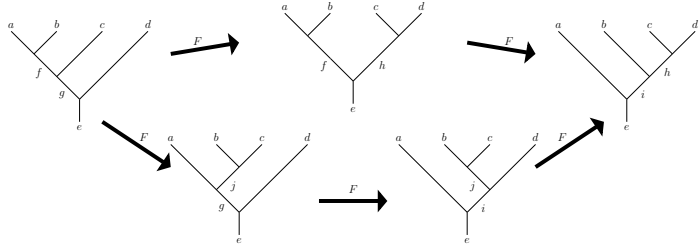
\begin{figure}
    \centering
    \scalebox{0.3}{
\begin{tikzpicture}
    %left
    \begin{scope}[xshift=0,yshift=0]
        %outer skeleton
        \draw[thick] (0,0) -- (3cm,3cm);
        \draw[thick] (0,0) -- (-3cm,3cm);
        \draw[thick] (0,0) -- (0,-1cm);
        %inner skeleton
        \draw[thick] (-2cm,2cm) -- (-1cm,3cm);
        \draw[thick] (-1cm,1cm) -- (1cm,3cm);
        % input/output legs
        \node[above] at (-3cm,3cm) {\Large $a$};
        \node[above] at (-1cm,3cm) {\Large $b$};
        \node[above] at (1cm,3cm) {\Large $c$};
        \node[above] at (3cm,3cm) {\Large $d$};
        \node[below] at (0,-1cm) {\Large $e$};
        % inner legs
        \node[below left] at (-0.5cm,0.5cm) {\Large $g$};
        \node[below left] at (-1.5cm,1.5cm) {\Large $f$};
    \end{scope}
    %%%%%%%%%%%%%%%%%%%%%%%%%%%%%%%%%%%%%%%%%%%%%%%%%%
    %upper
    \begin{scope}[xshift=12cm,yshift=0.8cm]
        \vertexI{0}{0}{-1cm};
        \vertextI{-2cm}{2cm}{-1cm}{-1};
        \vertextI{2cm}{2cm}{-1cm}{1};
        % input/output legs
        \node[above] at (-3cm,3cm) {\Large $a$};
        \node[above] at (-1cm,3cm) {\Large $b$};
        \node[above] at (1cm,3cm) {\Large $c$};
        \node[above] at (3cm,3cm) {\Large $d$};
        \node[below] at (0,-1cm) {\Large $e$};
        % inner legs
        \node[below left] at (-1cm,1cm) {\Large $f$};
        \node[below right] at (1cm,1cm) {\Large $h$};
    \end{scope}
    %%%%%%%%%%%%%%%%%%%%%%%%%%%%%%%%%%%%%%%%%%%%%%%%%%
    %right
    \begin{scope}[xshift=24cm,yshift=0]
         % x -> -x mirroring of the first diagram
        \begin{scope}[xscale=-1]
         %outer skeleton
         \draw[thick] (0,0) -- (3cm,3cm);
         \draw[thick] (0,0) -- (-3cm,3cm);
         \draw[thick] (0,0) -- (0,-1cm);
         %inner skeleton
         \draw[thick] (-2cm,2cm) -- (-1cm,3cm);
         \draw[thick] (-1cm,1cm) -- (1cm,3cm);
        \end{scope}
         % input/output legs
        \node[above] at (-3cm,3cm) {\Large $a$};
        \node[above] at (-1cm,3cm) {\Large $b$};
        \node[above] at (1cm,3cm) {\Large $c$};
        \node[above] at (3cm,3cm) {\Large $d$};
        \node[below] at (0,-1cm) {\Large $e$};
        % inner legs
        \node[below right] at (1.5cm,1.5cm) {\Large $h$};
        \node[below right] at (0.5cm,0.5cm) {\Large $i$};
    \end{scope}
    %%%%%%%%%%%%%%%%%%%%%%%%%%%%%%%%%%%%%%%%%%%%%%%%%%
    %lower left
    \begin{scope}[xshift=7cm,yshift=-5cm]
        %outer skeleton
        \draw[thick] (0,0) -- (3cm,3cm);
        \draw[thick] (0,0) -- (-3cm,3cm);
        \draw[thick] (0,0) -- (0,-1cm);
        %inner skeleton
        \vertextI{0}{2cm}{-1cm}{1}
         % input/output legs
         \node[above] at (-3cm,3cm) {\Large $a$};
         \node[above] at (-1cm,3cm) {\Large $b$};
         \node[above] at (1cm,3cm) {\Large $c$};
         \node[above] at (3cm,3cm) {\Large $d$};
         \node[below] at (0,-1cm) {\Large $e$};
         % inner legs
         \node[below left] at (-0.5cm,0.5cm) {\Large $g$};
         \node[below right] at (-0.5cm,1.5cm) {\Large $j$};
    \end{scope}
    %%%%%%%%%%%%%%%%%%%%%%%%%%%%%%%%%%%%%%%%%%%%%%%%%%
    %lower right
    \begin{scope}[xshift=17cm,yshift=-5cm]
        \begin{scope}[xscale=-1]
         %outer skeleton
         \draw[thick] (0,0) -- (3cm,3cm);
         \draw[thick] (0,0) -- (-3cm,3cm);
         \draw[thick] (0,0) -- (0,-1cm);
         %inner skeleton
         \vertextI{0}{2cm}{-1cm}{1}
        \end{scope}
        % input/output legs
        \node[above] at (-3cm,3cm) {\Large $a$};
        \node[above] at (-1cm,3cm) {\Large $b$};
        \node[above] at (1cm,3cm) {\Large $c$};
        \node[above] at (3cm,3cm) {\Large $d$};
        \node[below] at (0,-1cm) {\Large $e$};
        % inner legs
        \node[below right] at (0.5cm,0.5cm) {\Large $i$};
        \node[below left] at (0.5cm,1.5cm) {\Large $j$};
    \end{scope}
    %%%%%%%%%%%%%%%%%%%%%%%%%%%%%%%%%%%%%%%%%%%%%%%%%%
    %%%%%%%%%%%%%%%%%%%%%%%%%%%%%%%%%%%%%%%%%%%%%%%%%%
    %%%%%%%%%%%%%%%%%%%%%%%%%%%%%%%%%%%%%%%%%%%%%%%%%%
    %% the F arrows
    %%%%%%%%%%%%%%%%%%%%%%%%%%%%%%%%%%%%%%%%%%%%%%%%%%
    % upper left
    \begin{scope}[xshift=4cm,yshift=2cm]
        \draw[-{Triangle[width=28pt,length=12pt]}, line width=5pt] (0, 0) -- (3cm, 0.5cm);
        \node[above left] at (1.5cm,0.25cm) {\Large $F$};
    \end{scope}
    % upper right
    \begin{scope}[xshift=20cm,yshift=2cm]
        \draw[-{Triangle[width=28pt,length=12pt]}, line width=5pt] (-3cm, 0.5cm) -- (0,0);
        \node[above right] at (-1.5cm,0.25cm) {\Large $F$};
    \end{scope}
    % lower
    \begin{scope}[xshift=12cm,yshift=-4.5cm]
        \draw[-{Triangle[width=28pt,length=12pt]}, line width=5pt] (-1.5cm, 0) -- (1.5cm,0);
        \node[above] at (0,0.2cm) {\Large $F$};
    \end{scope}
    %lower left
    \begin{scope}[xshift=1cm,yshift=-1cm]
        \draw[-{Triangle[width=28pt,length=12pt]}, line width=5pt] (0, 0) -- (3cm,-2cm);
        \node[above right] at (1.5cm,-1cm) {\Large $F$};
    \end{scope}
    %lower right
    \begin{scope}[xshift=23cm,yshift=-1cm]
        \draw[-{Triangle[width=28pt,length=12pt]}, line width=5pt] (-3cm, -2cm) -- (0,0);
        \node[above left] at (-1.5cm,-1cm) {\Large $F$};
    \end{scope}
\end{tikzpicture}
}
    \caption{Pentagon equation w.r.t. the fusion space $(V^{abcd}_e)^*$.}
    \label{fig:pentagon}
\end{figure}

Associativity of anyon fusion is given by $(a\otimes b)\otimes c = a\otimes(b\otimes c)$.
It follows that the corresponding fusion space $V^{abc}_d$ has two sets of basis with respect to the fusion order, and the basis transformation in this fusion space is captured by the $F$-symbols\footnote{For simplicity we have suppressed the vertex multiplicity label $\mu$, which can be easily restored in case of non-trivial fusion multiplicity.} as shown in (\ref{eq:Fmove}).
%#done: F move figure
\begin{equation}
    \mathtikzS{0.4}{
        \coordinate["$i$" above] (A) at (-3cm,3cm);
        \coordinate["$j$" above] (B) at (-1cm,3cm);
        \coordinate["$k$" above] (C) at (1cm,3cm);
        \coordinate["$l$" below right] (O) at (0,0);
        \coordinate (D) at (-2cm,2cm);
        \coordinate (E) at (-1cm,1cm);
        \coordinate["$m$" below left] (M) at (-1.5cm,1.5cm);
        \draw[thick] (O) -- (A);
        \draw[thick] (D) -- (B);
        \draw[thick] (E) -- (C);
    }
    =\quad\sum_n\quad (F_l^{ijk})_{mn}^*
    \mathtikzS{0.4}{
        \coordinate["$i$" above] (A) at (-3cm,3cm);
        \coordinate["$j$" above] (B) at (-1cm,3cm);
        \coordinate["$k$" above] (C) at (1cm,3cm);
        \coordinate["$l$" below right] (O) at (0,0);
        \coordinate (E) at (-1cm,1cm);
        \coordinate (F) at (0,2cm);
        \coordinate["$n$" below right] (N) at (-0.5cm,1.5cm);
        \draw[thick] (O) -- (A);
        \draw[thick] (C) -- (E);
        \draw[thick] (B) -- (F);
    }
    \label{eq:Fmove}
\end{equation}
The $F$-symbols are not independent, they're related by the coherent condition known as the pentagon equation, as shown in figure \ref{fig:pentagon}. Pentagon equations are sufficient to solve for all $F$-symbols in an anyonic model.

Note that because of the phase ambiguity mentioned above the $F$-symbols are not invariant under these rescalings. It allows one to fix some of the components of $F$.

\vspace{1cm}
{\bf \underline{Quantum dimension}}\\

A category is called \textit{pivotal} if every simple object $a$ has a unique dual $a^*$. Given any simple object $a$, the dual of $a$ is a simple object $a^*$ satisfying 
\begin{equation}
a\otimes a^* = \mathbf{1} + \cdots,
\end{equation}
where $\mathbf{1}$ is the trivial object (vacuum). Diagrammatically, any line labeled by $a^*$ is equivalent to a line labeled by $a$ but with the direction reversed. The pivotal structure is essential in the definition of quantum dimension. The quantum dimension of a simple object is defined as the quantum trace (the pivotal trace) of an identity operator.
%#done: qd figure
    \begin{equation}
        d_a = \Tr(\id_a) = 
        \mathtikzS{0.6}{
            \draw[thick] (0,0) circle [radius=1cm];
            \node[right] at (1cm,0) {$a$};
        }
        \label{eq:qd_def}
    \end{equation}
The diagram in (\ref{eq:qd_def}) is direction-irrelevant, so we can freely replace $a$ by $a^*$ and therefore $d_a=d_{a^*}$. The quantum dimension of the trivial object $\mathbf{1}$ is $d_{\mathbf{1}}=1$ in any anyonic model. Quantum dimension is conserved under anyon fusion, following the fusion (\ref{eq:bulk_fusion}) we have
\begin{equation}
    d_a d_b = \sum_c N_{ab}^c d_c.
\end{equation}
The notion of quantum dimension can be generalized to an arbitrary object in the category $C$, in particular
\begin{equation}
    \dim(A)=\sum_a m_a d_a,\quad \text{for any object } A=\sum_a m_a a \in C.
\end{equation}
The total quantum dimension of category $C$ is defined as
\begin{equation}
    D_C=\sqrt{\sum_a d_a^2},
\end{equation}.

\begin{figure}[H]
    \centering
    \scalebox{0.3}{
\begin{tikzpicture}
    %left
    \begin{scope}[xshift=0,yshift=0]
        % lower outer skeleton
        \draw[thick] (0,0) -- (2cm,2cm);
        \draw[thick] (0,0) -- (-2cm,2cm);
        \draw[thick] (0,0) -- (0,-1cm);
        % crossing parts
        \crossingI{-1cm}{3cm}{1cm};
        \crossingI{0}{4cm}{1cm};
        % other parts
        \draw[thick] (-1cm,1cm) -- (0,2cm);
        \draw[thick] (2cm,2cm) -- (1cm,3cm);
        \draw[thick] (-2cm,4cm) -- (-3cm,5cm) -- (-3cm,6cm);
        \draw[thick] (-1cm,5cm) -- (-1cm,6cm);
        \draw[thick] (1cm,5cm) -- (1cm,6cm);
        % input/output legs
        \node[above] at (-3cm,6cm) {\Large $a$};
        \node[above] at (-1cm,6cm) {\Large $b$};
        \node[above] at (1cm,6cm) {\Large $c$};
        \node[below] at (0cm,-1cm) {\Large $d$};
        % inner leg
        \node[below left] at (-0.5cm,0.5cm) {\Large $e$};
    \end{scope}
    %upper left
    \begin{scope}[xshift=8cm,yshift=4cm]
        % lower outer skeleton
        \draw[thick] (0,0) -- (1cm,1cm);
        \draw[thick] (0,0) -- (-3cm,3cm);
        \draw[thick] (0,0) -- (0,-1cm);
        % crossing part
        \crossingI{0}{2cm}{1cm};
        % other parts
        \draw[thick] (-3cm,3cm) -- (-3cm,5cm);
        \draw[thick] (-1cm,3cm) -- (-1cm,5cm);
        \draw[thick] (1cm,3cm) -- (1cm,5cm);
        % input/output legs
        \node[above] at (-3cm,5cm) {\Large $a$};
        \node[above] at (-1cm,5cm) {\Large $b$};
        \node[above] at (1cm,5cm) {\Large $c$};
        \node[below] at (0cm,-1cm) {\Large $d$};
        % inner leg
        \node[below left] at (-0.5cm,0.5cm) {\Large $e$};
    \end{scope}
    %upper right
    \begin{scope}[xshift=18cm,yshift=4cm]
        % outer skeleton
        \draw[thick] (0,0) -- (2cm,2cm);
        \draw[thick] (0,0) -- (-2cm,2cm);
        \draw[thick] (0,0) -- (0,-1cm);
        % crossing part
        \crossingI{1cm}{3cm}{1cm};
        % other parts
        \draw[thick] (1cm,1cm) -- (0,2cm);
        \draw[thick] (-2cm,2cm) -- (-2cm,5cm);
        \draw[thick] (0,4cm) -- (0,5cm);
        \draw[thick] (2cm,4cm) -- (2cm,5cm);
        % input/output leg labels
        \node[above] at (-2cm,5cm) {\Large $a$};
        \node[above] at (0cm,5cm) {\Large $b$};
        \node[above] at (2cm,5cm) {\Large $c$};
        \node[below] at (0cm,-1cm) {\Large $d$};
        % inner leg
        \node[below right] at (0.5cm,0.5cm) {\Large $f$};
    \end{scope}
    %right
    \begin{scope}[xshift=26cm,yshift=0]
        % outer skeleton
        \draw[thick] (0,0) -- (2cm,2cm);
        \draw[thick] (0,0) -- (-2cm,2cm);
        \draw[thick] (0,0) -- (0,-1cm);
        % other parts
        \draw[thick] (-2cm,2cm) -- (-2cm,4cm);
        \draw[thick] (1cm,1cm) -- (0,2cm) -- (0,4cm);
        \draw[thick] (2cm,2cm) -- (2cm,4cm);
         % input/output legs
         \node[above] at (-2cm,4cm) {\Large $a$};
         \node[above] at (0cm,4cm) {\Large $b$};
         \node[above] at (2cm,4cm) {\Large $c$};
         \node[below] at (0cm,-1cm) {\Large $d$};
         % inner leg
        \node[below right] at (0.5cm,0.5cm) {\Large $f$};
    \end{scope}
    %lower left
    \begin{scope}[xshift=8cm,yshift=-5cm]
        % lower outer skeleton
        \draw[thick] (0,0) -- (2cm,2cm);
        \draw[thick] (0,0) -- (-2cm,2cm);
        \draw[thick] (0,0) -- (0,-1cm);
        % crossing parts
        \crossingI{-1cm}{3cm}{1cm};
        \crossingI{0}{4cm}{1cm};
        % other parts
        \draw[thick] (1cm,1cm) -- (0,2cm);
        \draw[thick] (2cm,2cm) -- (1cm,3cm);
        \draw[thick] (-2cm,4cm) -- (-3cm,5cm) -- (-3cm,6cm);
        \draw[thick] (-1cm,5cm) -- (-1cm,6cm);
        \draw[thick] (1cm,5cm) -- (1cm,6cm);
         % input/output legs
         \node[above] at (-3cm,6cm) {\Large $a$};
         \node[above] at (-1cm,6cm) {\Large $b$};
         \node[above] at (1cm,6cm) {\Large $c$};
         \node[below] at (0cm,-1cm) {\Large $d$};
         % inner leg
        \node[below right] at (0.5cm,0.5cm) {\Large $g$};
    \end{scope}
    % lower right
    \begin{scope}[xshift=18cm,yshift=-5cm]
         % outer skeleton
         \draw[thick] (0,0) -- (2cm,2cm);
         \draw[thick] (0,0) -- (-2cm,2cm);
         \draw[thick] (0,0) -- (0,-1cm);
         % other parts
         \draw[thick] (-2cm,2cm) -- (-2cm,6cm);
         \draw[thick] (-1cm,1cm) -- (0,2cm) -- (0,6cm);
         \draw[thick] (2cm,2cm) -- (2cm,6cm);
          % input/output legs
          \node[above] at (-2cm,6cm) {\Large $a$};
          \node[above] at (0cm,6cm) {\Large $b$};
          \node[above] at (2cm,6cm) {\Large $c$};
          \node[below] at (0cm,-1cm) {\Large $d$};
          % inner leg
        \node[below left] at (-0.5cm,0.5cm) {\Large $g$};
    \end{scope}
     %%%%%%%%%%%%%%%%%%%%%%%%%%%%%%%%%%%%%%%%%%%%%%%%%%
    %%%%%%%%%%%%%%%%%%%%%%%%%%%%%%%%%%%%%%%%%%%%%%%%%%
    %%%%%%%%%%%%%%%%%%%%%%%%%%%%%%%%%%%%%%%%%%%%%%%%%%
    %% the F arrows and R arrows
    %%%%%%%%%%%%%%%%%%%%%%%%%%%%%%%%%%%%%%%%%%%%%%%%%%
    % upper left
    \begin{scope}[xshift=2cm,yshift=4cm]
        \draw[-{Triangle[width=28pt,length=12pt]}, line width=5pt] (0, 0) -- (3cm, 2cm);
        \node[above left] at (1.5cm,1cm) {\Large $R$};
        % input/output legs
    \end{scope}
    % upper right
    \begin{scope}[xshift=24cm,yshift=5cm]
        \draw[-{Triangle[width=28pt,length=12pt]}, line width=5pt] (-3cm, 2cm) -- (0,0);
        \node[above right] at (-1.5cm,1cm) {\Large $R$};
    \end{scope}
    % upper
    \begin{scope}[xshift=13cm,yshift=7cm]
        \draw[-{Triangle[width=28pt,length=12pt]}, line width=5pt] (-1.5cm, 0) -- (1.5cm, 0);
        \node[above] at (0,0.2cm) {\Large $F$};
    \end{scope}
    % lower
    \begin{scope}[xshift=13cm,yshift=-4cm]
        \draw[-{Triangle[width=28pt,length=12pt]}, line width=5pt] (-1.5cm, 0) -- (1.5cm, 0);
        \node[above] at (0,0.2cm) {\Large $R$};
    \end{scope}
    % lower left
    \begin{scope}[xshift=1cm,yshift=-1cm]
        \draw[-{Triangle[width=28pt,length=12pt]}, line width=5pt] (0, 0) -- (3cm, -2cm);
        \node[above right] at (1.5cm,-1cm) {\Large $F$};
    \end{scope}
    % lower right
    \begin{scope}[xshift=24cm,yshift=-1cm]
        \draw[-{Triangle[width=28pt,length=12pt]}, line width=5pt] (-3cm, -2cm) -- (0,0);
        \node[above left] at (-1.5cm,-1cm) {\Large $F$};
    \end{scope}
\end{tikzpicture}
}
    \caption{Hexagon equation w.r.t. the fusion space $(V^{abc}_d)^*$.}
    \label{fig:hexagon}
\end{figure}
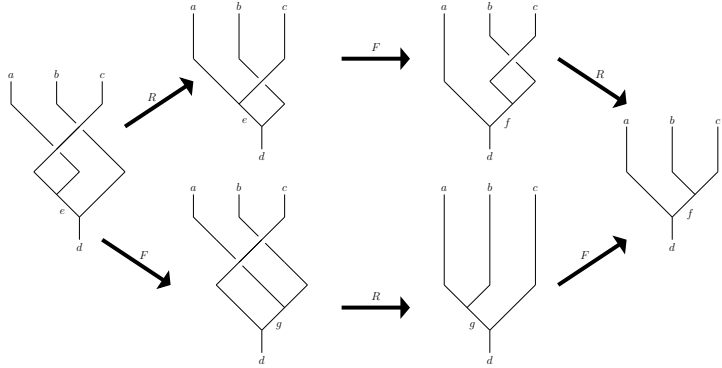

\vspace{1cm}
{\bf \underline{Braiding and Twist (spin) }}\\
Using the fusion tree basis, the braiding exchange operator can be represented by R symbols shown below:
%#done: R symbols figure
    \begin{equation}
        \mathtikzS{0.35}{
            \vertexI{0}{0}{1cm};
            \crossingI{0cm}{-2cm}{1cm};
            \node[below] at (-1cm,-3cm) {$a$};
            \node[below] at (1cm,-3cm) {$b$};
            \node[above] at (0,1cm) {$c$};
        }
        =\quad R^{ab}_c
        \mathtikzS{0.5}{
            \vertexI{0}{0cm}{1cm};
            \node[below] at (-1cm,-1cm) {$a$};
            \node[below] at (1cm,-1cm) {$b$};
            \node[above] at (0,1cm) {$c$};
        }.
    \end{equation}
In the special case where $b=a^*$ and $c=\mathbf{1}$, the R symbol is reduced to the \textit{spin} of an anyon.
%#done: twist figure
    \begin{equation}
        \mathtikzS{0.5}{
            \anticrossingI{0}{0}{0.5cm};
            \draw[thick] (-0.5cm,-0.5cm) -- (-1cm,0) -- (-0.5cm,0.5cm);
            \draw[thick] (0.5cm,0.5cm) -- (0.5cm,1.5cm);
            \draw[thick] (0.5cm,-0.5cm) -- (0.5cm,-1.5cm);
            \node[right] at (0.5cm,-1.5cm) {$a$};
        }
        = \quad \theta_a
        \mathtikzS{0.5}{
            \draw[thick] (0,-1.5cm) -- (0,1.5cm);
            \node[right] at (0,-1.5cm) {$a$};
        }.
    \end{equation}
Taking the trace of the above relation we know from (\ref{eq:qd_def}) that the spin can be expressed as
\begin{equation}
    \theta_a = \frac{1}{d_a} 
    \mathtikzS{0.8}{
        \anticrossingI{0}{0}{0.5cm};
        \draw[thick] (-0.5cm,-0.5cm) -- (-1cm,0) -- (-0.5cm,0.5cm);
        \draw[thick] (0.5cm,-0.5cm) -- (1cm,0) -- (0.5cm,0.5cm);
        \node[below] at (0.5cm,-0.5cm) {$a$};
    }.
\end{equation}
Each anyon has a definite spin, bosons are spin=$1$ particles while fermions are spin=$-1$ particles. 
We require that braiding and fusion commute. Diagrammatically this means we can freely move lines across a vertex.
%#done: line slide across vertex figure
\begin{equation}
    \mathtikzS{0.5}{
        \vertexI{0}{0}{-0.5cm};
        \draw[thick] (-0.5cm,0.5cm) -- (-0.5cm,2.5cm);
        \draw[thick] (0.5cm,0.5cm) -- (0.5cm,2.5cm);
        \draw[thick] (0,-0.5cm) -- (0,-1cm);
        \draw[thick] (-1cm,-1cm) -- (-1cm,0.5cm) -- (-0.6cm,0.9cm);
        \draw[thick] (-0.4cm,1.1cm) -- (0.4cm,1.9cm);
        \draw[thick] (0.6cm,2.1cm) -- (1cm,2.5cm);
    }
    =
    \mathtikzS{0.5}{
        \vertexI{0}{0}{-0.5cm};
        \draw[thick] (-0.5cm,0.5cm) -- (-0.5cm,2.5cm);
        \draw[thick] (0.5cm,0.5cm) -- (0.5cm,2.5cm);
        \draw[thick] (0,-0.5cm) -- (0,-1cm);
        \draw[thick] (-0.5cm,-1cm) -- (-0.1cm,-0.6cm);
        \draw[thick] (0.1cm,-0.4cm) -- (1cm,0.5cm) -- (1cm,2.5cm);
    }
\end{equation}
$R$-symbols and $F$-symbols are not independent. 
In order for braiding to be compatible with fusion, it is found that some coherent condition must be satisfied by the $F$-symbols and $R$-symbols, which may be expressed diagrammatically as the Hexagon equation shown in figure \ref{fig:hexagon}. Given the solution of $F$-symbols, hexagon equations are sufficient to solve for all $R$-symbols.

\vspace{1cm}
{\bf \underline{A note on Super-fusion category}} 
A useful place for the discussion of super-fusion category can be found in \cite{Gu_2015, Aasen:2017ubm}.
The above discussion applies to a generic fusion category. In the presence of fermion condensation which
we will be interested below, the resultant gapped boundary would carry extra structure connected to the $\mathbb{Z}_2$ fermion
parity. To accommodate that structure, we need to upgrade the notion of a fusion category to a super-fusion category. There are many definitions
of super-categories.
At the level of the objects, it may involve a decomposition of the objects to a direct sum of even and odd parity objects \footnote{See for example \cite{Brundan_2017} and also some of the references that appear in \cite{Aasen:2017ubm} }.
\be \label{eq:decomposeC}
C= C_0 \oplus C_1.
\ee
However, this definition is quite restrictive.  In the gapped boundaries that are considered, and more so when it comes to defects localized between boundaries, such
a decomposition is not very clear. Therefore, we will adopt the discussion in  \cite{Gu_2015, Aasen:2017ubm} -- which keeps track of fermion parity of morphisms, and not discuss  a decomposition of the objects themselves as in (\ref{eq:decomposeC}).

Then the most distinctive characteristic of a super-fusion category is the appearance of fermion parity odd morphisms. 
First, the allowed endo-morphism space of simple objects would be enlarged. 
Simple objects could potentially carry a ``fermion parity odd" map to itself, in addition to the usual identity map which carries even fermion parity. 
i.e. In this case,  dim $[Hom(a,a) =2]$.  Pictorially, the two ``basis maps'' to itself  are represented as in figure below. 
Simple objects having a two dimensional endomorphism are referred to as  {\bf q-type objects} in the literature \cite{Aasen:2017ubm}. 
    \begin{equation} 
        \mathtikzS{0.7}{
        \coordinate["$a$" right] (A) at (0,1cm);
        \coordinate (O) at (0,-0.2cm);
        \node[right] at (O) {$\alpha$};
        \coordinate["$a$" right] (B) at (0,-1.5cm);
        \draw [thick] (A) -- (O);
        \draw [thick] (O) -- (B);
        \tri{0}{0}{270};
        }
        \in\quad \Hom(a,a), \quad \alpha\in\{1,2\}
    \end{equation}

Second, fusion, being morphisms from $C\otimes C \to C$, could also acquire both parity even and odd channels. They are illustrated in (\ref{eq:even_odd_fuse}). Odd channels are often represented with an extra red dot on the vertex.
    \begin{equation} \label{eq:even_odd_fuse}
        \mathtikzS{0.6}{
            \vertexIL{0}{0}{1cm}{$a$}{$b$}{$c$};
        }
        \quad \text{and} \quad
        \mathtikzS{0.6}{
            \vertexIL{0}{0}{1cm}{$a$}{$b$}{$c$};
            \fill [fill=red] (0,0) circle [radius=0.1cm];
        }
    \end{equation}

There is an ambiguity in the definition of the fusion coefficients. Consider the following fusion process:
\be \label{eq:superfuse}
a\otimes b = \otimes \Delta_{ab}^c c
\ee
The fusion is an element in Hom$(a\otimes b, c)$. If $c$ is a q-type object that has a two dimensional endomorphism space, the fusion map could
be concatenated by a non-trivial endomorphism in $c$ and remains an element in Hom$(a\otimes b, c)$. In other words, while defining $\Delta_{ab}$, we have implicitly
made a choice in discarding possible endomorphism in $c$. This does not appear natural. Therefore, it is proposed in \cite{Aasen:2017ubm} to enlarge the fusion space to 
\be \label{eq:superfuse}
V_{ab}^c = \Delta_{ab}^c \otimes \textrm{End}(c).
\ee

Under this definition, it would mean that the fermionic ``dots'' can be freely moved from the vertex to the connecting anyon lines if they are q-type objects \cite{Aasen:2017ubm}. 

Since fusion spaces are $\mathbb{Z}_2$ graded, and that the old channels essentially carry a Majorana mode which leads to sign changes under swapping of labels \cite{Gu_2015, Aasen:2017ubm}, the pentagon equation has to be upgraded to keep track of these labellings and signs. This leads to the super-pentagon equation illustrated in (\ref{fig:super_pent}). 

\begin{figure}[H]
    \centering
    \be \label{fig:super_pent}
    \scalebox{0.3}{
\begin{tikzpicture}
    %%%%%%%%%%%%%%%%%%%%%%%%%%%%%%%%%%%%%%%%%%%%%%%%%%
    %%%%%%%%%%%%%%%%%%%%%%%%%%%%%%%%%%%%%%%%%%%%%%%%%%
    %%%%%%%%%%%%%%%%%%%%%%%%%%%%%%%%%%%%%%%%%%%%%%%%%%
    %% the diagrams
    %%%%%%%%%%%%%%%%%%%%%%%%%%%%%%%%%%%%%%%%%%%%%%%%%%
    %left
    \begin{scope}[xshift=0,yshift=0]
        %outer skeleton
        \draw[thick] (0,0) -- (3cm,3cm);
        \draw[thick] (0,0) -- (-3cm,3cm);
        \draw[thick] (0,0) -- (0,-1cm);
        %inner skeleton
        \draw[thick] (-2cm,2cm) -- (-1cm,3cm);
        \draw[thick] (-1cm,1cm) -- (1cm,3cm);
        % input/output legs
        \node[above] at (-3cm,3cm) {\Large $a$};
        \node[above] at (-1cm,3cm) {\Large $b$};
        \node[above] at (1cm,3cm) {\Large $c$};
        \node[above] at (3cm,3cm) {\Large $d$};
        \node[below] at (0,-1cm) {\Large $e$};
        % inner legs
        \node[below left] at (-0.5cm,0.5cm) {\Large $g$};
        \node[below left] at (-1.5cm,1.5cm) {\Large $f$};
    \end{scope}
    %%%%%%%%%%%%%%%%%%%%%%%%%%%%%%%%%%%%%%%%%%%%%%%%%%
    %upper left
    \begin{scope}[xshift=8cm,yshift=2cm]
        \vertexI{0}{0}{-1cm};
        \vertextI{-2cm}{2cm}{-1cm}{-1};
        \vertextI{2cm}{2cm}{-1cm}{1};
        \node[left] at (-2cm,2cm) {\Large $\mu$};
        \node[left] at (2cm,2cm) {\Large $\nu$};
        % input/output legs
        \node[above] at (-3cm,3cm) {\Large $a$};
        \node[above] at (-1cm,3cm) {\Large $b$};
        \node[above] at (1cm,3cm) {\Large $c$};
        \node[above] at (3cm,3cm) {\Large $d$};
        \node[below] at (0,-1cm) {\Large $e$};
        % inner legs
        \node[below left] at (-1cm,1cm) {\Large $f$};
        \node[below right] at (1cm,1cm) {\Large $h$};
        % vertices
        \disk{-2cm}{2cm}{black};
        \disk{2cm}{2cm}{black};
    \end{scope}
     %%%%%%%%%%%%%%%%%%%%%%%%%%%%%%%%%%%%%%%%%%%%%%%%%%
    %upper right
    \begin{scope}[xshift=16cm,yshift=2cm]
        \vertexI{0}{0}{-1cm};
        \vertextI{-2cm}{2cm}{-1cm}{-1};
        \vertextI{2cm}{2cm}{-1cm}{1};
        \node[left] at (-2cm,2cm) {\Large $\nu$};
        \node[left] at (2cm,2cm) {\Large $\mu$};
        % input/output legs
        \node[above] at (-3cm,3cm) {\Large $a$};
        \node[above] at (-1cm,3cm) {\Large $b$};
        \node[above] at (1cm,3cm) {\Large $c$};
        \node[above] at (3cm,3cm) {\Large $d$};
        \node[below] at (0,-1cm) {\Large $e$};
        % inner legs
        \node[below left] at (-1cm,1cm) {\Large $f$};
        \node[below right] at (1cm,1cm) {\Large $h$};
        % vertices
        \disk{-2cm}{2cm}{black};
        \disk{2cm}{2cm}{black};
    \end{scope}
    %%%%%%%%%%%%%%%%%%%%%%%%%%%%%%%%%%%%%%%%%%%%%%%%%%
    %right
    \begin{scope}[xshift=24cm,yshift=0]
         % x -> -x mirroring of the first diagram
        \begin{scope}[xscale=-1]
         %outer skeleton
         \draw[thick] (0,0) -- (3cm,3cm);
         \draw[thick] (0,0) -- (-3cm,3cm);
         \draw[thick] (0,0) -- (0,-1cm);
         %inner skeleton
         \draw[thick] (-2cm,2cm) -- (-1cm,3cm);
         \draw[thick] (-1cm,1cm) -- (1cm,3cm);
        \end{scope}
         % input/output legs
        \node[above] at (-3cm,3cm) {\Large $a$};
        \node[above] at (-1cm,3cm) {\Large $b$};
        \node[above] at (1cm,3cm) {\Large $c$};
        \node[above] at (3cm,3cm) {\Large $d$};
        \node[below] at (0,-1cm) {\Large $e$};
        % inner legs
        \node[below right] at (1.5cm,1.5cm) {\Large $h$};
        \node[below right] at (0.5cm,0.5cm) {\Large $i$};
    \end{scope}
    %%%%%%%%%%%%%%%%%%%%%%%%%%%%%%%%%%%%%%%%%%%%%%%%%%
    %lower left
    \begin{scope}[xshift=8cm,yshift=-4.5cm]
        %outer skeleton
        \draw[thick] (0,0) -- (3cm,3cm);
        \draw[thick] (0,0) -- (-3cm,3cm);
        \draw[thick] (0,0) -- (0,-1cm);
        %inner skeleton
        \vertextI{0}{2cm}{-1cm}{1}
         % input/output legs
         \node[above] at (-3cm,3cm) {\Large $a$};
         \node[above] at (-1cm,3cm) {\Large $b$};
         \node[above] at (1cm,3cm) {\Large $c$};
         \node[above] at (3cm,3cm) {\Large $d$};
         \node[below] at (0,-1cm) {\Large $e$};
         % inner legs
         \node[below left] at (-0.5cm,0.5cm) {\Large $g$};
         \node[below right] at (-0.5cm,1.5cm) {\Large $j$};
    \end{scope}
    %%%%%%%%%%%%%%%%%%%%%%%%%%%%%%%%%%%%%%%%%%%%%%%%%%
    %lower right
    \begin{scope}[xshift=16cm,yshift=-4.5cm]
        \begin{scope}[xscale=-1]
         %outer skeleton
         \draw[thick] (0,0) -- (3cm,3cm);
         \draw[thick] (0,0) -- (-3cm,3cm);
         \draw[thick] (0,0) -- (0,-1cm);
         %inner skeleton
         \vertextI{0}{2cm}{-1cm}{1}
        \end{scope}
        % input/output legs
        \node[above] at (-3cm,3cm) {\Large $a$};
        \node[above] at (-1cm,3cm) {\Large $b$};
        \node[above] at (1cm,3cm) {\Large $c$};
        \node[above] at (3cm,3cm) {\Large $d$};
        \node[below] at (0,-1cm) {\Large $e$};
        % inner legs
        \node[below right] at (0.5cm,0.5cm) {\Large $i$};
        \node[below left] at (0.5cm,1.5cm) {\Large $j$};
    \end{scope}
    %%%%%%%%%%%%%%%%%%%%%%%%%%%%%%%%%%%%%%%%%%%%%%%%%%
    %%%%%%%%%%%%%%%%%%%%%%%%%%%%%%%%%%%%%%%%%%%%%%%%%%
    %%%%%%%%%%%%%%%%%%%%%%%%%%%%%%%%%%%%%%%%%%%%%%%%%%
    %% the F arrows
    %%%%%%%%%%%%%%%%%%%%%%%%%%%%%%%%%%%%%%%%%%%%%%%%%%
    % upper
    \begin{scope}[xshift=12cm,yshift=3cm]
        \draw[purple, -{Triangle[width=28pt,length=12pt]}, line width=5pt] (-1.5cm, 0) -- (1.5cm,0);
        \node[above] at (0,0.2cm) {\Large $K$};
    \end{scope}
    % upper left
    \begin{scope}[xshift=3cm,yshift=1cm]
        \draw[-{Triangle[width=28pt,length=12pt]}, line width=5pt] (0, 0) -- (3cm, 1cm);
        \node[above left] at (1.5cm,0.5cm) {\Large $F$};
    \end{scope}
    % upper right
    \begin{scope}[xshift=21cm,yshift=1cm]
        \draw[-{Triangle[width=28pt,length=12pt]}, line width=5pt] (-3cm, 1cm) -- (0,0);
        \node[above right] at (-1.5cm,0.5cm) {\Large $F$};
    \end{scope}
    % lower
    \begin{scope}[xshift=12cm,yshift=-5cm]
        \draw[-{Triangle[width=28pt,length=12pt]}, line width=5pt] (-1.5cm, 0) -- (1.5cm,0);
        \node[above] at (0,0.2cm) {\Large $F$};
    \end{scope}
    %lower left
    \begin{scope}[xshift=2cm,yshift=-2cm]
        \draw[-{Triangle[width=28pt,length=12pt]}, line width=5pt] (0, 0) -- (3cm,-2cm);
        \node[above right] at (1.5cm,-1cm) {\Large $F$};
    \end{scope}
    %lower right
    \begin{scope}[xshift=22cm,yshift=-2cm]
        \draw[-{Triangle[width=28pt,length=12pt]}, line width=5pt] (-3cm, -2cm) -- (0,0);
        \node[above left] at (-1.5cm,-1cm) {\Large $F$};
    \end{scope}
\end{tikzpicture}
}
    \ee
\end{figure}

Note that the operation $K$ refers to the exchange of the labels of the vertices. Depending on their fermion parity, (i.e. if both vertices carry odd parity), there is a sign change there. 

\subsection{Gapped boundaries and (super)- Frobenius Algebra in tensor categories} \label{sec:frobeniusalgebra}

It is well known that each bosonic gapped boundary of a non-chiral bosonic topological order $C$ in 2+1 dimensions is characterized by
a (commutative separable symmetric) Frobenius algebra in $C$  \cite{Kong:2013aya, kirillov}. \footnote{It is pointed out to us that the relevant mathematics first appeared in \cite{B_ckenhauer_1999,B_ckenhauer_2000,bckenhauer2000longorehren}.} Physically the algebra encodes the condensation of a collection of bosons at the boundary. 
 While there are already ample hints elsewhere such as \cite{Aasen:2017ubm, Wan:2016php, Bhardwaj:2016clt}, that attempts to obtain the collection of excitations in the condensed child theory, the discussion does not provide a systematic framework to compute the excitations in the child theory, not to mention excitations localized between different boundaries. Here, we propose that a fermionic gapped boundary is also encoded in a separable symmetric Frobenius algebra, except where ``commutativity'' is relaxed to ``super-commutativity'', to accommodate condensation of fermionic anyons. i.e. To reiterate, the relevant mathematical structure is a super-commutative separable symmetric Frobenius algebra. 
 
 Each of these labels will be discussed below.  
 
We collect some necessary facts about Frobenius algebra  and anyon condensation below.  In this section, we rely heavily on \cite{kirillov}, and particularly \cite{Fuchs_2002, Fuchs_2004}, which have developed many useful tools and proved numerous identities related to Frobenius algebra and their modules, that assist us significantly in our quest -- the reason being that a super-commutative Frobenius algebra is a Frobenius algebra after all. 
Applications of these tools to understand super Frobenius algebra and their q-type modules/bi-modules are some of the main goals of the current paper. 

\vspace{1cm}
{\bf \underline{Algebra and co-algebra}}

An {\bf algebra} in the category $C$ is a collection $\mathcal{A} $ of simple objects equipped with a product $\mu$ and unit $\iota_\mathcal{A}$. This collection is expressible as \cite{kirillov, Fuchs_2002} 
\be
\mathcal{A} = \oplus_i W_{ i 1} c_i, \qquad W_{ i 1} \in \mathbb{Z}_{\ge 0}, \qquad c_i \in C. 
\ee
This collection of anyons when equipped with the appropriate set of structures that we will discuss below, would be identified with the set of anyons that condense at the gapped boundary.
The product $\mu$ maps $\mathcal{A} \times \mathcal{A} \to \mathcal{A}$. It is trivially associated (\ref{fig:protrivialass}). Further we have already constructed all the basis of maps (or homomorphisms) from $C\times C \to C$ in the previous section. Therefore this product $\mu$ must be expressible in terms of the basis constructed out of the simple objects. 
      \be
      \centering
      \begin{tikzpicture}\label{fig:protrivialass}
      \draw [black, ultra thick] (0,0.5) to [out=90, in=90] (1,0.5);
      \draw [black, ultra thick] (0,0) -- (0,0.5);
      \draw [black, ultra thick] (1,0) -- (1,0.5);
      \draw [black, ultra thick] (0.5,1.5) to [out=90, in=90] (2,1.5);
      \draw [black, ultra thick] (0.5,0.8) -- (0.5,1.5);
      \draw [black, ultra thick] (2,0) -- (2,1.5);
      \draw [black, ultra thick] (1.25,1.92) -- (1.25,2.5);

      \draw [black, thick] (2.325,0.9) -- (2.675,0.9);
      \draw [black, thick] (2.325,1) -- (2.675,1);

      \draw [black, ultra thick] (4,0.5) to [out=90, in=90] (5,0.5);
      \draw [black, ultra thick] (4,0) -- (4,0.5);
      \draw [black, ultra thick] (5,0) -- (5,0.5);
      \draw [black, ultra thick] (3,1.5) to [out=90, in=90] (4.5,1.5);
      \draw [black, ultra thick] (4.5,0.8) -- (4.5,1.5);
      \draw [black, ultra thick] (3,0) -- (3,1.5);
      \draw [black, ultra thick] (3.75,1.92) -- (3.75,2.5);

      \node at (0.3,0) {$\mathcal{A}$}
      node at (1.3,0) {$\mathcal{A}$}
      node at (2.3,0) {$\mathcal{A}$} 
      node at (3.3,0) {$\mathcal{A}$} 
      node at (4.3,0) {$\mathcal{A}$} 
      node at (5.3,0) {$\mathcal{A}$} 
      node at (0.8,1) {$\mathcal{A}$} 
      node at (4.8,1) {$\mathcal{A}$} 
      node at (1.55,2.2) {$\mathcal{A}$} 
      node at (4.05,2.2) {$\mathcal{A}$}; 
      \end{tikzpicture}
      \ee\\
This is illustrated in (\ref{eq:Aproduct}).  The $\zeta$ in (\ref{eq:Aproduct}) labels the fusion channel $i\otimes j \rightarrow k$ in the bulk.

%#todo: make labels consistent
\begin{figure}[h]
    \centering
    \begin{equation} \label{eq:Aproduct}
        \mathtikzS{0.8}{
            \vertexI{0}{0}{1.5cm};
            \node[right] at (0,0) {$\mu$};
            \node[above] at (0,1.5cm) {$\mA$};
            \node[below] at (-1.5cm,-1.5cm) {$\mA$};
            \node[below] at (1.5cm,-1.5cm) {$\mA$};
        }
        =\quad\sum_{i,j,k\in \mA}~\sum_{\alpha,\beta,\gamma,\zeta} \quad\mu_{(i\alpha)(j\beta)}^{(k\gamma);\zeta}
        \mathtikzS{0.8}{
            \vertexI{0}{0}{1.5cm};
            \node[above] at (0,1.5cm) {$\mA$};
            \node[below] at (-1.5cm,-1.5cm) {$\mA$};
            \node[below] at (1.5cm,-1.5cm) {$\mA$};
            \vertexIC{0}{0}{1cm}{\bulkCol};
            \tri{-1cm}{-1cm}{45};
            \tri{0}{1.1cm}{270};
            \tri{1cm}{-1cm}{135};
            \node[right] at (0,0.5cm) {$k$};
            \node[above] at (-0.65cm,-0.65cm) {$i$};
            \node[above] at (0.65cm,-0.65cm) {$j$};
            %junc label
            \node[right] at (0,1cm) {$\gamma$};
            \node[left] at (-0.8cm,-0.8cm) {$\bar{\alpha}$};
            \node[right] at (0.8cm,-0.8cm) {$\bar{\beta}$};
            % fusion channe;
            \filldraw [fill=yellow] (0,0) circle [radius=0.05cm];
            \node[right] at (0,0) {$\zeta$};
        }
    \end{equation}
\end{figure}

The multiplicity $W_{i 1}$ determines the dimension of the maps (homomorphisms) from $\mathcal{A}$ to the simple object $c_i$. We
therefore introduce a label $\alpha$ as we did in  (\ref{eq:hombasis}). 
Defining the product $\mu$ is equivalent to solving for the coefficients defining the linear combination of basis maps -- note that they are subjected to the same phase ambiguity as discussed in (\ref{eq:hombasis_rescale}). 

\begin{figure}[h]
    \centering
    \begin{equation}  \label{fig:condensation_mapA}
        \mathtikzS{0.7}{
            \coordinate["$\mathcal{A}$" right] (A) at (0,1cm);
            \coordinate (O) at (0,-0.2cm);
            \node[right] at (O) {$\alpha$};
            \coordinate["$i$" right] (B) at (0,-1.5cm);
            \draw [thick] (A) -- (O);
            \draw [thick, \bulkCol] (O) -- (B);
            \tri{0}{0}{270};
        }
        \quad \alpha\in\{ 1,2,\dots,W_{i1} \}
    \end{equation}
  
\end{figure}

%#done: review unit
A {\bf unit} $\iota_\mathcal{A}$ is a morphism from the vacuum $\mathbf{1}$ to $\mathcal{A}$. This morphism is in fact an embedding $\iota_\mathcal{A}:\mathbf{1}\hookrightarrow\mathcal{A}$, so in any mathematical expression the unit can be simply understood as the vacuum object $\mathbf{1}$ despite its nature of morphism. A more accurate yet pedagogical understanding of the unit is to think of it as ``taking out the vacuum $\mathbf{1}$ from $\mathcal{A}$".  Or alternatively, when we do arithmetic, any number $x = 1. x$. The unit map just means we can freely multiply any number by unity.  

The vacuum fuses trivially with all objects in the category, translating this back to a morphism we see immediately that the unit has to satisfy the morphism equality $\mu\circ(\iota_\mathcal{A}\otimes \id_{\mathcal{A}})=\id_{\mathcal{A}}$. Here $\id_{\mathcal{A}}$ is the identity map in $\mathcal{A}$. This equality is illustrated in (\ref{eq:unit}).
\begin{equation}
    \mathtikzS{0.7}{
        \draw[thick] (0,-1cm) -- (0,1cm);
        \draw[thick] (-0.5cm, -0.5cm) -- (0,0);
        \fill (-0.5cm,-0.5cm) circle [radius=0.05cm];
        \node[below] at (0,-1cm) {$\mathcal{A}$};
        \node[left] at (-0.5cm,-0.5cm) {$\iota_\mathcal{A}$};
    }
    =
    \mathtikzS{0.7}{
        \draw[thick] (0,-1cm) -- (0,1cm);
        \node[below] at (0,-1cm) {$\mathcal{A}$};
    }
    \label{eq:unit}
\end{equation}
Each algebra in a category has a unique unit, which means the vacuum appears exactly once in the algebra $\mathcal{A}$.

A {\bf co-algebra} $ \mathcal{A}$ is a collection of simple objects $\mathcal{A}$ equipped with a co-product $\Delta$, which maps $\mathcal{A} \to \mathcal{A} \times \mathcal{A}$, and a counit $\varepsilon_\mathcal{A}$.
While this co-product operation may look mysterious, we have a very familiar example in physics. Consider for example two electrons with spins $S_1$ and  $S_2$ respectively. The action of spatial rotations on these spins are effected through the total angular momentum operator $\hat S = \hat S_1 + \hat S_2$.  This is an example where we are ``splitting'' an $SU(2)$ group element into the products of two $SU(2)$ group elements, expressed as a sum of operators in the corresponding Lie algebra. 
Like in the case of the product $\mu$, the map $\Delta$ can be expressed in terms of the basis maps constructed in $C$. This is illustrated in the following. 
\begin{equation}
        \mathtikzS{0.8}{
            \vertexIC{0}{0}{-1.5cm}{black};
            \node[right] at (0,0) {$\Delta$};
            \node[below] at (0,-1.5cm) {$\mA$};
            \node[above] at (1.5cm,1.5cm) {$\mA$};
            \node[above] at (-1.5cm,1.5cm) {$\mA$};
        }
        =\quad\sum_{i,j,k\in\mA}~\sum_{\alpha,\beta,\gamma,\xi}\quad \Delta_{(k\gamma);\xi}^{(i\alpha)(j\beta)}
        \mathtikzS{0.8}{
            \vertexIC{0}{0}{-1.5cm}{black};
            \node[below] at (0,-1.5cm) {$\mA$};
            \node[above] at (1.5cm,1.5cm) {$\mA$};
            \node[above] at (-1.5cm,1.5cm) {$\mA$};
            \vertexIC{0}{0}{-1cm}{\bulkCol};
            \tri{1cm}{1cm}{225};
            \tri{-1cm}{1cm}{315};
            \tri{0cm}{-1.1cm}{90};
            \node[right] at (0.5cm,0.5cm) {$j$};
            \node[right] at (-0.5cm,0.5cm) {$i$};
            \node[right] at (0,-0.5cm) {$k$};
            \node[right] at (0.8cm,0.8cm) {$\beta$};
            \node[left] at (-0.8cm,0.8cm) {$\alpha$};
            \node[right] at (0,-1cm) {$\bar{\gamma}$};
            \filldraw [fill=yellow] (0,0) circle [radius=0.05cm];
            \node[right] at (0,0) {$\xi$};
        }
    \end{equation}\\
Similar to the product, the co-product $\Delta$ is also trivially associated (\ref{fig:coprotrivialass}).
 \be
      \centering
      \begin{tikzpicture}\label{fig:coprotrivialass}
      \draw [black, ultra thick] (0,-0.5) to [out=-90, in=-90] (1,-0.5);
      \draw [black, ultra thick] (0,0) -- (0,-0.5);
      \draw [black, ultra thick] (1,0) -- (1,-0.5);
      \draw [black, ultra thick] (0.5,-1.5) to [out=-90, in=-90] (2,-1.5);
      \draw [black, ultra thick] (0.5,-0.8) -- (0.5,-1.5);
      \draw [black, ultra thick] (2,0) -- (2,-1.5);
      \draw [black, ultra thick] (1.25,-1.92) -- (1.25,-2.5);

      \draw [black, thick] (2.325,-0.9) -- (2.675,-0.9);
      \draw [black, thick] (2.325,-1) -- (2.675,-1);

      \draw [black, ultra thick] (4,-0.5) to [out=-90, in=-90] (5,-0.5);
      \draw [black, ultra thick] (4,0) -- (4,-0.5);
      \draw [black, ultra thick] (5,0) -- (5,-0.5);
      \draw [black, ultra thick] (3,-1.5) to [out=-90, in=-90] (4.5,-1.5);
      \draw [black, ultra thick] (4.5,-0.8) -- (4.5,-1.5);
      \draw [black, ultra thick] (3,0) -- (3,-1.5);
      \draw [black, ultra thick] (3.75,-1.92) -- (3.75,-2.5);

      \node at (0.3,0) {$\mathcal{A}$}
      node at (1.3,0) {$\mathcal{A}$}
      node at (2.3,0) {$\mathcal{A}$} 
      node at (3.3,0) {$\mathcal{A}$} 
      node at (4.3,0) {$\mathcal{A}$} 
      node at (5.3,0) {$\mathcal{A}$} 
      node at (0.8,-1) {$\mathcal{A}$} 
      node at (4.8,-1) {$\mathcal{A}$} 
      node at (1.55,-2.2) {$\mathcal{A}$} 
      node at (4.05,-2.2) {$\mathcal{A}$}; 
      \end{tikzpicture}
      \ee\\
A {\bf counit} $\varepsilon_\mathcal{A}$ is the same morphism as the unit, but with the direction reversed, namely a morphism from $\mathcal{A}$ to the vacuum $\mathbf{1}$. Like the unit, the counit satisfies a similar morphism equality: $(\varepsilon_\mathcal{A} \otimes \id_{\mathcal{A}})\circ\Delta = \id_{\mathcal{A}}$. The equality is illustrated by 
\begin{equation}
    \mathtikzS{0.7}{
        \draw[thick] (0,-1cm) -- (0,1cm);
        \draw[thick] (-0.5cm, 0.5cm) -- (0,0);
        \fill (-0.5cm,0.5cm) circle [radius=0.05cm];
        \node[below] at (0,-1cm) {$\mathcal{A}$};
        \node[left] at (-0.5cm,0.5cm) {$\varepsilon_\mathcal{A}$};
    }
    =
    \mathtikzS{0.7}{
        \draw[thick] (0,-1cm) -- (0,1cm);
        \node[below] at (0,-1cm) {$\mathcal{A}$};
    },
\end{equation}
which is a direction reversed version of (\ref{eq:unit}).

A {\bf Frobenius algebra} $\mathcal{A}$ is both an algebra and co-algebra. It is equipped with a product and a co-product at the same time. By now, it should be familiar that every time an extra structure is introduced, we have to determine how the new structure and all the previous structures already in place should fit together. Its properties are conveniently summarised in the following picture. 
\begin{equation}
    \mathtikzS{0.6}{
        \vertexIC{0}{0}{-1cm}{black};
        \node[left] at (0,0) {$\Delta$};
        \vertexI{1cm}{1cm}{1cm};
        \node[right] at (1cm,1cm) {$\mu$};
        %extra long legs
        \draw[thick] (2cm,0cm) -- (2cm,-1cm);
        \draw[thick] (-1cm,1cm) -- (-1cm,2cm);
        \node[below] at (0,-1cm) {$\mathcal{A}$};
        \node[below] at (2,-1cm) {$\mathcal{A}$};
        \node[above] at (-1cm,2cm) {$\mathcal{A}$};
        \node[above] at (1cm,2cm) {$\mathcal{A}$};
    }
    =
    \mathtikzS{0.6}{
        \vertexI{0}{0}{1cm};
        \antivertexI{0}{1cm}{1cm};
        \node[left] at (0,0) {$\mu$};
        \node[left] at (0,1cm) {$\Delta$};
        \node[below] at (-1cm,-1cm) {$\mathcal{A}$};
        \node[below] at (1,-1cm) {$\mathcal{A}$};
        \node[above] at (-1cm,2cm) {$\mathcal{A}$};
        \node[above] at (1cm,2cm) {$\mathcal{A}$};
    }
    =
    \mathtikzS{0.6}{
        \vertexI{0}{0}{1cm};
        \antivertexI{1cm}{-1cm}{1cm}
        \draw[thick] (-1cm,-1cm) -- (-1cm,-2cm);
        \draw[thick] (2cm,0cm) -- (2cm,1cm);
        %\draw[thick] (0,1cm) -- (1cm,0);
        \node[left] at (0,0cm) {$\mu$};
        \node[right] at (1cm,-1cm) {$\Delta$};
        \node[below] at (-1cm,-2cm) {$\mathcal{A}$};
        \node[below] at (1,-2cm) {$\mathcal{A}$};
        \node[above] at (0cm,1cm) {$\mathcal{A}$};
        \node[above] at (2cm,1cm) {$\mathcal{A}$};
    }
\end{equation}
We have also included the conditions for the algebra being ``separable" and symmetric.

%#done: separable algebra
A Frobenius algebra $\mathcal{A}$ is called {\bf separable} if there exists a map $e:\mathcal{A}\rightarrow \mathcal{A}\otimes\mathcal{A}$ such that $\mu\circ e = \id_{\mathcal{A}}$. A Frobenius algebra $\mathcal{A}$ is called {\bf special}(a.k.a. strongly-separable) if the product $\mu$ is the inverse of the coproduct $\Delta$ and the unit $\iota_{\mathcal{A}}$ is the inverse of the counit $\varepsilon_{\mathcal{A}}$, namely\footnote{This condition is sometimes called \textit{normalized special}.}
\begin{equation} \label{eq:separability}
    \mathtikzS{0.5}{
        \antivertexI{0}{-1cm}{1cm};
        \vertexI{0}{1cm}{1cm};
        \node[below] at (0,-2cm) {$\mathcal{A}$};
        \node[above] at (0,2cm) {$\mathcal{A}$};
        \node[left] at (0,-1cm) {$\Delta$};
        \node[left] at (0,1cm) {$\mu$};
    }
    =
    \mathtikzS{0.5}{
        \draw[thick] (0,-2cm) -- (0,2cm);
        \node[below] at (0,-2cm) {$\mathcal{A}$};
        %\node[above] at (0,2cm) {$\mathcal{A}$};
    }
    \qquad \text{and} \qquad
    \mathtikzS{0.7}{
        \draw[thick] (0,-1cm) -- (0,1cm);
        \fill (0,1cm) circle [radius=0.1cm];
        \fill (0,-1cm) circle [radius=0.1cm];
        \node[left] at (0,-1cm) {$\iota_{\mathcal{A}}$};
        \node[left] at (0,1cm) {$\varepsilon_{\mathcal{A}}$};
    }
    =\quad\dim\mathcal{A}
    %\mathtikz{
    %    \draw[thick] (0,-1cm) -- (0,1cm);
    %    \node[right] at (0,-1cm) {$\mathbf{1}$};
    %}
\end{equation}
The separability of $\mathcal{A}$ allows a well-defined notion of simple objects in the representation category $\Rep \mathcal{A}$, and of non-simple objects as direct sums of simple objects.

%#done: symmetric algebra
A Frobenius algebra $\mathcal{A}$ is called {\bf symmetric} if the product $\mu$ and the counit $\varepsilon_{\mathcal{A}}$ satisfy\footnote{This condition looks slightly different from (3.33) of \cite{Fuchs_2002}, there the authors have made implicit the composition of the unit $\iota_{\mA}$ and the coproduct $\Delta$.}
\begin{equation}
    \mathtikzS{0.4}{
        \vertexI{0}{0}{1cm};
        \draw[thick] (-1cm,-1cm) -- (-1cm,-3cm);
        \draw[thick] (1cm,-1cm) -- (2cm,0) -- (2cm,2cm);
        \draw[thick] (1cm,-1cm) -- (1cm,-2cm);
        \fill (1cm,-2cm) circle [radius=0.1cm];
        \node[below] at (1cm,-2cm) {$\iota_{\mA}$};
        \node[right] at (1cm,-1cm) {$\Delta$};
        \fill (0,1cm) circle [radius=0.1cm];
        \node[left] at (0,1cm) {$\varepsilon_{\mathcal{A}}$};
        \node[below] at (-1cm,-3cm) {$\mathcal{A}$};
        \node[above] at (2cm,2cm) {$\mathcal{A}$};
        \node[left] at (0,0) {$\mu$};
    }
    =
    \mathtikzS{0.4}{
        \vertexI{0}{0}{1cm};
        \draw[thick] (1cm,-1cm) -- (1cm,-3cm);
        \draw[thick] (-1cm,-1cm) -- (-2cm,0) -- (-2cm,2cm);
        \draw[thick] (-1cm,-1cm) -- (-1cm,-2cm);
        \fill (-1cm,-2cm) circle [radius=0.1cm];
        \node[below] at (-1cm,-2cm) {$\iota_{\mA}$};
        \node[right] at (-1cm,-1cm) {$\Delta$};
        \fill (0,1cm) circle [radius=0.1cm];
        \node[left] at (0,1cm) {$\varepsilon_{\mathcal{A}}$};
        \node[below] at (1cm,-3cm) {$\mathcal{A}$};
        \node[above] at (-2cm,2cm) {$\mathcal{A}$};
        \node[left] at (0,0) {$\mu$};
    }.
\end{equation}

A {\bf commutative algebra} is one where $ \mu \circ R_{\mathcal{A}, \mathcal{A}} = \mu$. This is illustrated in the following figure. 
The collection of anyons $\mathcal{A}$ can condense physically if they are mutually local, and that they are bosonic. It turns out that the above condition is sufficient to imply both. 
    \begin{equation}
        \mathtikzS{0.6}{
            \vertexI{0}{0}{0.7cm};
            \crossingI{0cm}{-1.4cm}{0.7cm};
            \node[below] at (0.7cm,-2.1cm) {$\mathcal{A}$};
            \node[below] at (-0.7cm,-2.1cm) {$\mathcal{A}$};
            \node[above] at (0,0.7cm) {$\mathcal{A}$};
        }
        =
        \mathtikzS{0.6}{
            \vertexI{0}{0}{1cm};
            \node[above] at (0,1cm) {$\mathcal{A}$};
            \node[below] at (-1cm,-1cm) {$\mathcal{A}$};
            \node[below] at (1cm,-1cm) {$\mathcal{A}$};
        }.
    \end{equation}

We have explained all the necessary qualifiers of the algebra object that describe a condensate. 
The condensate should ultimately behave like the vacuum, or trivial anyon in the condensed phase, where intuitively it could be freely created or annihilated without causing any changes to the states. These mathematical structures introduced above are therefore physical requirements of the condensed anyons.

To accommodate also fermionic anyons condensing, one has to relax the commutative condition to ``super-commutativity".

A {\bf super algebra} is one which is graded by the $\mathbb{Z}_2$ (fermion parity) symmetry. Therefore, the algebra $\mathcal{A}$ acquires a decomposition \cite{Creutzig:2017anl}
\be
\mathcal{A} = \mathcal{A}_0 \oplus \mathcal{A}_1.
\ee
The fermion parity $\sigma(i)$ of an anyon $i$ belonging to $\mathcal{A}_{p}$  is $(-1)^p$, for $p\in{0,1}$.

This decomposition allows us to define {\bf super-commutativity}, which is given by \footnote{There are again many mathematics papers discussing super algebra. However, many of the structures are probably not suited for the purpose here. We will restrict to a bare minimum of structures which are actually used in the current paper.  Our discussion of super-commutativity is inspired by \cite{Creutzig:2017anl}. However, we have somewhat modified it to make it compatible with non-Abelian fermions taking part in $\mathcal{A}$, which is beyond \cite{Creutzig:2017anl} or \cite{Aasen:2017ubm}. }
\be  \label{eq:super_braid}
\mu \circ R_{c_i, c_j} (-1)^{\sigma(c_i) . \sigma(c_j)} = \mu, \qquad c_i, c_j \in \mathcal{A},
\ee
where $R_{c_i,c_j}$ is the half-braid of the modular tensor category $C$. The mathematical definition has a simple physical interpretation -- this is precisely
to bind the fermionic anyon with a ``free fermion'' so that the pair together behaves like a boson and condenses. The idea of pairing is also discussed in \cite{Aasen:2017ubm}. Here we made it explicit that this is the physical realization of super-commutativity in the mathematics literature.

The dimension $D_{\mathcal{A}}$ of the algebra is defined as
\be
D_{\mathcal{A}} = \sum_{i}W_{i1} d_i, \qquad i \in \mathcal{A}.
\ee

A {\bf (super)-Lagrangian algebra} is a (super)-commutative Frobenius algebra satisfying
\be
D_{\mathcal{A}} = D_C
\ee. 

This is sufficient and necessary condition for the bosonic algebra to recover a modular invariant, 
and apparently a sufficient condition to recover a super-modular category. As we will demonstrate in the example of $D(D_4)$ in the appendix \ref{sec:dd4}, there are examples where a super-modular invariant with positive integer coefficients that
does not admit an interpretation as a fermion condensate. 
In the following, we will study the representations (modules) of the (super)-algebra. It is the modules that form a super-fusion category defined in the sense in section \ref{sec:introcat}. These modules are the boundary excitations. Note that the braided structure is not preserved in this fusion category describing boundary excitations. 

%[[[ {\bf Can we prove that $D_{A_0} = D_{A_1}$? }]]

\subsection{Defects via construction of modules}

Having introduced the concept of an algebra in a category $C$ which plays the role of our condensate at the gapped boundary, we would like to obtain the collection of 
allowed defects, or excitations, at the boundary. 

When an anyon in the bulk phase approaches the gapped boundary, it would generally become an excitation at the boundary. 
However, since a bunch of anyons are ``condensed'' at the boundary, which can be freely created and annihilated there, it means that the fusion product between a bulk anyon and a condensed anyon might no longer be distinguishable at the boundary. In other words, the bulk anyons form "multiplets" under fusion with the condensed anyons. The corresponding mathematical jargon would be that the boundary excitations are modules (or representations) of the condensate algebra $\mathcal{A}$. We note that for a (super)-commutative Frobenius algebra, these left/right modules form a fusion category.  In physical terms -- excitations at the gapped boundary have well defined fusion rules. 

In this subsection, we would summarise how to recover these modules.

\subsubsection{Left (Right) Modules}
Each module $M$ of  $\mathcal{A}$ in $C$ is also a collection of anyons in $C$, i.e.,
\be \label{eq:Mdecompose}
M = \oplus_i W_{i M} c_i.
\ee
Again, there are maps from $M \to C$,  as well as its dual map, from $ C \to M$, which are illustrated in  (\ref{fig:condensation_map}).
\renewcommand{\triLen}{0.5}
\begin{figure}[htbp]
    \centering
   \be \label{fig:condensation_map}
    \begin{tikzpicture}[scale=0.7]
    \draw [green, thick] (0,2) -- (0,0);
    \draw [magenta, thick] (0,0) -- (0,-2);
    \tri{0}{-0.25cm}{90};
    \node at (0.3,2) {$i$};
    \node at (0.4,-2) {$M$};
    \node at (0.4,0) {$\bar\alpha$};
    \node at (-1,0) {$b_M^{(i\alpha)}:=$};
    \begin{scope}[xshift=4cm,yshift=0cm]{
    \draw [green, thick] (0,-2) -- (0,0);
    \draw [magenta, thick] (0,0) -- (0,2);
    \tri{0}{0.25cm}{270};
    \node at (0.3,-2) {$i$};
    \node at (0.4,2) {$M$};
    \node at (0.4,0) {$\alpha$};
    \node at (-1,0) {$b_{(i\alpha)}^M:=$};
    }
    \end{scope}
    \begin{scope}[xshift=10cm,yshift=-1cm]{
    \scalebox{0.7}[0.7]{
    \draw [magenta, thick] (0,-2) -- (0,0);
    \draw [green, thick] (0,0) -- (0,2);
    \draw [magenta, thick] (0,2) -- (0,4);
    \tri{0}{-0.25cm}{90}
    \tri{0}{2.25cm}{270}
    \node at (0.3,1) {\Large{$i$}}
    node at (0.4,-2) {\Large{$M$}}
    node at (0.4,4) {\Large{$M$}}
    node at (-1,1) {\Large{$\sum_{i,\alpha}$}}
    node at (1.5,1) {\Large{$=$}}
    node at (0.4,2) {\Large{$\alpha$}}
    node at (0.4,0) {\Large{$\bar{\alpha}$}};
    \draw [magenta, thick] (3,-2) -- (3,4);
    \node at (3.5,4) {\Large{$M$}};
    }
    }
    \end{scope}
    \begin{scope}[xshift=17cm,yshift=-1cm]{
    \scalebox{0.7}[0.7]{
    \draw [green, thick] (0,-2) -- (0,0);
    \draw [magenta, thick] (0,0) -- (0,2);
    \draw [green, thick] (0,2) -- (0,4);
    \tri{0}{0.25cm}{270}
    \tri{0}{1.75cm}{90}
    \node at (-0.6,1) {\Large{$M$}}
    node at (0.4,4) {\LARGE{$i$}}
    node at (0.4,-2) {\LARGE{$i$}}
    node at (1.5,1) {\LARGE{$=$}}
    node at (0.4,0) {\Large{$\beta$}}
    node at (0.4,2) {\Large{$\bar{\alpha}$}};
    \draw [green, thick] (4,-2) -- (4,4);
    \node at (2.8,1) {\LARGE{$\delta_{ij}\delta_{\alpha\beta}$}}
    node at (4.4,4) {\LARGE{$i$}};
    }
    }
    \end{scope}
    \node at (0.8,-0.2) {,}
    node at (4.8,-0.2) {,}
    node at (10,0) {and};
    \end{tikzpicture}
    \ee
\end{figure}

As a ``representation" of the algebra $\mathcal{A}$, the modules admit an action by $\mathcal{A}$ onto it, i.e., there is a linear map $\rho^M_\mathcal{A} : \mathcal{A} \times M \to M$.

Since these anyons have non-trivial mutual braiding, we should specify whether $\mathcal{A}$ is acting on the left or on the right of $M$ at the gapped boundary. Here we will assume that the action is on the left, making $M$ a ``left-module''.  In the case of commutative and super-commutative algebra $\mathcal{A}$, the right action can be generated from the left action, simply by composing the product with an R-crossing \cite{Fuchs_2002}. In the following, unless otherwise specified, we will first explicitly discuss left actions, which would automatically apply to right actions. 

 Again, these (left or right) actions are linear maps which can be expressed in terms of the basis of morphisms $C \times C \to C$ (fusion) we have constructed in the previous section. 
The map $\rho^M_\mathcal{A}$ can thus be explicitly expressed in terms of these basis, as illustrated in (\ref{eq:reps}). 
%\begin{figure}[htbp]
%    \centering
%    \be \label{eq:reps}
%    \begin{tikzpicture}
%    \scalebox{0.7}[0.7]{
%    \draw [thick, magenta] (-4*0.75,-4*0.75) -- (-4*0.75,4*0.75);
%    \draw [thick] (-7*0.75,-3*0.75) -- (-4*0.75,0*0.75);
%    \node at (-5*0.75,0) {$\rho$}
%    node at (-7.8*0.75,-3*0.75) {$A$}
%    node at (-3.2*0.75,4*0.75) {$M$}
%    node at (-3.2*0.75,-4*0.75) {$M$}
%    node at (-1.9*0.75,0) {\LARGE{$=$}}
%    node at (0.5*0.75,0) {$\sum_{a,i,j,\alpha,\gamma,\beta}$}
%    node at (2.5*0.75,0) {$\sum_{\delta}$}
%    node at (4.2*0.75,0) {$\rho^{M(i\gamma);\delta}_{(a\alpha)(j\beta)}$};
%    \draw [thick, magenta] (8*0.75,-4*0.75) -- (8*0.75,4*0.75);
%    \draw [thick] (5*0.75,-3*0.75) -- (8*0.75,0);
%    \draw [green, thick] (8*0.75,-1.5*0.75) -- (8*0.75,-2.4*0.75);
%    \draw [green, thick] (8*0.75,-2*0.75) -- (8*0.75,2*0.75);
%    \draw [green, thick] (6.5*0.75,-1.5*0.75) -- (8*0.75,0);
%    \tri{6.2*0.75cm}{-1.8*0.75cm}{45}
%    \tri{8*0.75cm}{2*0.75cm}{270}
%    \tri{8*0.75cm}{-2.4*0.75cm}{90}
%    \node at (5.8*0.75,-1.2*0.75) {$\bar{\alpha}$}
%    node at (8.8*0.75,-2.2*0.75) {$\bar{\beta}$}
%    node at (8.8*0.75,1.7*0.75) {$\gamma$}
%    node at (4.2*0.75,-3*0.75) {$A$}
%    node at (8.8*0.75,4*0.75) {$M$}
%    node at (8.8*0.75,-4*0.75) {$M$}
%    node at (7*0.75,-0.3*0.75) {$a$}
%    node at (8.6*0.75,0.7*0.75) {$i$}
%    node at (8.6*0.75,-0.9*0.75) {$j$};
%    \draw [fill=yellow] (8*0.75,0) circle [radius=0.1];
%    \node at (8.4*0.75,-0.2) {$\delta$};
%    }
%    \end{tikzpicture}
%    \ee
%\end{figure}

\begin{equation}
    \label{eq:reps}
    \mathtikzS{0.7}{
        \draw[thick, \modCol] (0,-2cm) -- (0,2cm);
        \draw[thick] (-2cm,-2cm) -- (0,0);
        \node[below] at (0,-2cm) {$M$};
        \node[above] at (0,2cm) {$M$};
        \node[below] at (-2cm,-2cm) {$\mA$};
        \node[left] at (0,0) {$\rho$};
    }
    \quad=\quad
    \sum_{a,i,j,\alpha,\gamma,\beta} \sum_{\delta} \quad \rho^{M(i\gamma);\delta}_{(a\alpha)(j\beta)}
    \mathtikzS{0.7}{
        \draw[thick, \modCol] (0,-2cm) -- (0,2cm);
        \draw[thick] (-2cm,-2cm) -- (0,0);
        \node[below] at (0,-2cm) {$M$};
        \node[above] at (0,2cm) {$M$};
        \node[below] at (-2cm,-2cm) {$\mA$};
        % middle
        \draw[thick,\bulkCol] (0,-1.3cm) -- (0,1.3cm);
        \draw[thick,\bulkCol] (-1.3cm,-1.3cm) -- (0,0);
        \node[right] at (0,0.5cm) {$i$};
        \node[right] at (0,-0.5cm) {$j$};
        \node[left] at (-0.5cm,-0.5cm) {$a$};
        \filldraw [fill=yellow] (0,0) circle [radius=0.08cm];
        \triL{0}{1.3cm}{270}{0.5};
        \triL{0}{-1.3cm}{90}{0.5};
        \triL{-1.3cm}{-1.3cm}{45}{0.5};
        % labels
        \node[right] at (0,0) {$\delta$};
        \node[right] at (0,-1.2cm) {$\bar{\beta}$};
        \node[left] at (-1.1cm,-1.1cm) {$\bar{\alpha}$};
        \node[right] at (0,1.2cm) {$\gamma$};
    }
\end{equation}

As a representation of the algebra $\mathcal{A}$, $\rho^M_{\mathcal{A}}$ must satisfy (\ref{eq:associativity_module}). This is nothing but a generalization of our familiar property of a group representation, in which
\be \label{eq:reps_eq}
\rho^M(gh)_{ab} = \sum_{c} \rho^M (g)_{ac} \rho^M(h)_{cb},
\ee
where $\rho^M(g)_{ab}$ is the representation matrix of the group element $g\in G$. 
Besides, it is well known that the irreducible representations of a group satisfy an orthogonality relation
\be
\frac{1}{|G|}\sum_{g\in G}(\rho^M(g)^*)^b_a(\rho^{M'}(g))^d_c=\frac{1}{dim(M)}\delta^d_a\delta^b_c\delta^{M,M'}
\ee
where $M$ and $M'$ are two irreducible representations of the group $G$. A similar orthogonality relation is satisfied by the simple modules $M$, $M'$ of a special Frobenius algebra $\mA$, as illustrated in (\ref{eq:orthogonality}) 
\cite{Fuchs_2002}.
\begin{figure}[htbp]
\be\label{eq:orthogonality}
\centering
\begin{tikzpicture}[scale=0.8]
\begin{scope}[xshift=11cm,yshift=0.5cm]{
\draw [thick, green] (-3,-2) -- (-3,3.5);
\node at (-6.5,0.5) {$=\frac{dim(i)}{dim(M)}\delta^{(j,\beta)}_{(j',\beta')}\delta^{(i,\alpha)}_{(i',\alpha')}\delta_{M,M'}$};
\node at (-2.8,3.5) {\tiny{$j$}};
}
\end{scope}
\draw [thick, green] (0.3,2.5) -- (0.3,4);
\node at (0.5,4) {\tiny{$j$}};
\node at (0.5,2.85) {\tiny{$\bar\beta$}};
\draw [thick, magenta] (0.3,3) -- (0.3,2);
\draw [thick, green] (0.3,2) -- (0.3,1);
\scalebox{0.5}[0.5]{\tri{0.6cm}{4.3cm}{270}}
\scalebox{0.5}[0.5]{\tri{0.6cm}{5.7cm}{90}}
\node at (0.2,1.6) {\tiny{$i$}};
\node at (0.5,2.15) {\tiny{$\alpha$}};
\draw [thick] (0.3,2.5) to [out=180,in=90] (-0.3,2);
\draw [thick] (-0.3,2) to [out=-90,in=60] (-1,0.75);
\node at (0,2.7) {\tiny{$\rho^{M}_A$}};
\draw [thick, green] (0.3,1) -- (0.3,0.5);
\node at (0.5,0.9) {\tiny{$i'$}};
\node at (0.5,0.35) {\tiny{$\bar\alpha'$}};
\draw [thick, magenta] (0.3,0.5) -- (0.3,-0.5);
\draw [thick, green] (0.3,-0.5) -- (0.3,-1.5);
\scalebox{0.5}[0.5]{\tri{0.6cm}{-0.7cm}{270}}
\scalebox{0.5}[0.5]{\tri{0.6cm}{0.7cm}{90}}
\node at (0.1,-1.5) {\tiny{$j'$}};
\node at (0.5,-0.35) {\tiny{$\beta'$}};
\draw [thick] (0,-0.3) -- (-0.5,-0.8);
\draw [thick] (-1,-0.3) -- (-0.5,-0.8);
\draw [thick] (-0.5,-1.3) -- (-0.5,-0.8);
\draw [fill] (-0.5,-1.3) circle [radius=0.05];
\draw [thick] (0,-0.3) to [out=45,in=180] (0.3,0);
\draw [thick] (-1,-0.3) to [out=135,in=-120] (-1,0.75);
\node at (-0.7,-0.2) {$\mathcal{A}$};
\node at (0,0.2) {\tiny{$\rho^{M'}_A$}};
\end{tikzpicture}
\ee
 \end{figure}
 In fact after the left actions $\rho^M$ are expand in terms of basis, they form a basis of $Hom(\mathcal{A}\otimes j,k)$\cite{Fuchs_2002}. 
 Hence for a morphism $\phi \in$ Hom$(\mathcal{A}\otimes j,k)$, it can be expressed in terms of linear combinations of the left actions on the modules, schematically as
\be
\phi=\sum_{M} \lambda_{M,\{\alpha\}} \rho^{M, \{\alpha\}}.
\ee
Here $\{\alpha\}$ are labels of basis of $\rho^M$ when expanded explicitly as maps in $C$. It will be made explicit in the following. 

To extract the coefficients $\lambda$ we can use the orthogonality relation (\ref{eq:orthogonality}) which then gives  (\ref{eq:extract_lambda}) \cite{Fuchs_2002}. 

\begin{figure}[htbp]
\be \label{eq:extract_lambda}
\centering
\begin{tikzpicture}
\begin{scope}[xshift=0.75cm,yshift=0.5cm]{
\draw [thick, green] (-4,-2) -- (-4,2);
\node at (-4.7,0) {$\lambda_{M,\alpha}^\beta$};
\node at (-3,0) {$=\frac{dim(M)}{dim(i)}$};
\node at (-3.8,2) {\tiny{$j$}};
}
\end{scope}
\draw [thick, green] (0,2.5) -- (0,1.5);
\draw [thick, green, fill=yellow] (-0.5,1.5)--(0.5,1.5)--(0.5,1)--(-0.5,1)--(-0.5,1.5);
\node at (0.2,2.5) {\tiny{$j$}};
\node at (0,1.25) {$\phi$};
\draw [thick, green] (0.3,1) -- (0.3,0.5);
\draw [thick] (-0.3,1) -- (-0.3,0.75);
\node at (0.5,0.75) {\tiny{$i$}};
\node at (0.5,0.35) {\tiny{$\bar\alpha$}};
\draw [thick, magenta] (0.3,0.5) -- (0.3,-0.5);
\draw [thick, green] (0.3,-0.5) -- (0.3,-1.5);
\scalebox{0.5}[0.5]{\tri{0.6cm}{-0.7cm}{270}}
\scalebox{0.5}[0.5]{\tri{0.6cm}{0.7cm}{90}}
\node at (0.5,-1.5) {\tiny{$j$}};
\node at (0.5,-0.35) {\tiny{$\beta$}};
\draw [thick] (0,-0.3) -- (-0.5,-0.8);
\draw [thick] (-1,-0.3) -- (-0.5,-0.8);
\draw [thick] (-0.5,-1.3) -- (-0.5,-0.8);
\draw [fill] (-0.5,-1.3) circle [radius=0.05];
\draw [thick] (0,-0.3) to [out=45,in=180] (0.3,0);
\draw [thick] (-1,-0.3) to [out=135,in=-90] (-0.3,0.75);
\node at (-0.7,-0.2) {$\mathcal{A}$};
\node at (0.1,0.2) {\tiny{$\rho^{M}_\mathcal{A}$}};
\end{tikzpicture}
\ee
\end{figure}
Note that the basis abstractly denoted $\{\alpha\}$ shows up above actually correspond to labels of the basis map projecting the modules to anyons in $C$ in (\ref{eq:extract_lambda}). i.e. $\{\alpha\} \to \alpha, \beta$.  
This identity is very useful. We note that (super)-commutativity of the  Frobenius algebra $\mathcal{A}$ allows us to work simply with left modules. It also ensures that the resultant collection of boundary excitations (modules) form a (super) fusion category (i.e. the structure of fusion is well defined.) \cite{kirillov, Fuchs_2002} . To a physicist -- it means it makes sense to look for edge excitations which are always expressible as a linear combination of some basic excitations (the simple/irreducible representations) of the algebra, and that these excitations have well defined fusion rules. 

The 6j-symbols responsible for associativity of the fusion of the boundary excitations can be computed systematically, as soon as the precise multiplication $\mu$  of the condensate algebra $\mathcal{A}$ and the left/right action of the algebra on its modules are solved. Since this is relatively tedious and lengthy, we would relegate the computation to the appendix. 

In practice,  $W_{c_i M}$ is crucial data to work out $\rho^M_{\mathcal{A}}$ using equation (\ref{eq:associativity_module}). One important handle towards solving for $W$ is via the inspection of induced modules. 

Modules can be ``induced'' by fusion with the condensate $\mathcal{A}$. The product $\mu$ would automatically supply the correct structure to produce a left (right) action satifying (\ref{eq:associativity_module}) described above.  This is illustrated in (\ref{eq:ind_mod}) \cite{kirillov, Fuchs_2002}. 
\begin{figure}[htbp]
\be  \label{eq:associativity_module}
    \centering
    \begin{tikzpicture}
    \node at (-0.5,0) {\LARGE{$=$}};
    \draw [thick] (-3,-0.4) -- (-2,0.6);
    \draw [thick] (-3,-1.4) -- (-2,-0.4);
    \draw [thick] (-0.3,-1.4) -- (1.7,0.6);
    \draw [thick] (0.7,-0.4) -- (0.7,-1.4);
    \draw [thick, magenta] (1.7,1.6) -- (1.7,-1.4);
    \draw [thick, magenta] (-2,1.6) -- (-2,-1.4);
    \node at (-3.2,-0.4) {$\mathcal{A}$}
    node at (-3.2,-1.4) {$\mathcal{A}$}
    node at (-0.5,-1.4) {$\mathcal{A}$}
    node at (0.5,-1.4) {$\mathcal{A}$}
    node at (0.5,-0.2) {$\mathcal{A}$}
    node at (-1.6,1.6) {$M$}
    node at (2.1,1.6) {$M$};
    \end{tikzpicture}
    \ee
\end{figure}		
\begin{equation}
    \label{eq:ind_mod}
    \mathtikzS{1}{
        \draw[thick, purple] (0,-1cm) -- (0,1cm);
        \draw[thick] (-1cm,-1cm) -- (0,0);
        \node[right] at (0,0) {$\rho_{\Ind_{\mA}(c_i)}$};
        \node[below] at (0,-1cm) {$\mA\otimes c_i$};
        \node[above] at (0,1cm) {$\mA\otimes c_i$};
        \node[below] at (-1cm,-1cm) {$\mA$};
    }
    =
    \mathtikzS{1}{
        \vertexI{0}{0}{1cm};
        \node[left] at (0,0) {$\mu$};
        \node[below] at (-1cm,-1cm) {$\mA$};
        \node[below] at (1cm,-1cm) {$\mA$};
        \node[above] at (0,1cm) {$\mA$};
        \begin{scope}[xshift=2cm]
            \draw[thick, \bulkCol] (0,-1cm) -- (0,1cm);
            \node[below] at (0,-1cm) {$c_i$};
            \node[above] at (0,1cm) {$c_i$};
        \end{scope}
    }
\end{equation}
These induced representations following from $\mathcal{A} \otimes c_i$, denoted Ind${}_{\mathcal{A}}(c_i)$ are generally reducible. They can be expressed in terms of the simple (irreducible) modules as \cite{Fuchs_2002}
\be
\textrm{ Ind${}_{\mathcal{A}}(c_i)$} = \sum_{x} \lambda_x  \rho^{M_x}_\mathcal{A}
\ee

These parameters $\lambda$ can be solved using the identity illustrated in (\ref{eq:extract_lambda}). This identity allows very efficient computation of the modules -- particularly when the induced module is itself simple.  It is possible to generate all the simple modules through constructing induced modules. The deduction of the $W$-matrix is greatly faciliated by the identities relating quantum dimensions discussed in \ref{sec:quantumdim} below.

{\bf  \underline{Endomorphisms}}

As we have emphasized in multiple instances, one novel ingredient in a fermionic gapped system (describable by a super-fusion category)  is that some of the modules have non-trivial {\it endomorphism} -- i.e. the q-type objects we have referred to earlier. We need to identify which of the modules we obtained correspond to q-type objects. 
 This is discussed in \cite{Aasen:2017ubm} in the context of Abelian fermion condensation. There, it is observed that anyons that are ``fixed points'' under fusion with the condensing fermion are q-type objects when considered as modules (or boundary excitations) of the condensate algebra $\mathcal{A}$. 
 i.e. fixed point anyons satisfy 
 \be \label{eq:fixedpoint}
 \psi \otimes  a = a,
 \ee
 where $\psi \in \mathcal{A}$, and $d_{\psi} =1$. 
 
 In the case of non-Abelian condensation and where the defects are localized at the junctions, the above condition (\ref{eq:fixedpoint}) is not well defined. 
 In these cases, endomorphisms of the modules can be deduced by applying identities discussed in \cite{Fuchs_2002} and also solving for the modules explicitly. 
 These methods are to be reviewed and extended in the next two subsections. 
 The identities applicable to junction defects will be discussed separately in section \ref{sec:junctions}.
 A necessary signature of non-trivial endomorphism is that the {\it same} module acquires two independent left (right) action.  
 The situation can easily be confused with the case when a single anyon is {\it splitted} into two boundary excitations (i.e. two simple modules).  This situation has been discussed for example in \cite{Eliens:2013epa}. 
 These two situations are distinguished precisely using identities relating quantum dimensions in the parent and the condensed theory that we will discuss below. 

\subsubsection{Quantum dimension and endomorphism-- some useful identities}
\label{sec:quantumdim}
This section explains a novel application of various useful identities proved in \cite{Fuchs_2002, Fuchs_2004}.
These identitis connect the quantum dimensions of the defect and that of the anyons composing it .
Since they are very useful and powerful, we would like to reproduce some of them here. 
To simplify notations, let us follow \cite{Fuchs_2002, Fuchs_2004} and denote
\be
\textrm{dim [Hom$(a,b)$]} = \langle a, b\rangle.  
\ee

A module $M$ as a collection of anyons in $C$ has a quantum dimension in $C$ given by
\be
\textrm{dim}_C(M) = \sum_i W_{i M} d_{c_i},
\ee
where now we can write 
\be
W_{iM } = \langle c_i, M\rangle_C. 
\ee
The subscript $C$ serves to remind us that this is counting the dimension of homeomorphisms (or maps) from the point of view of $C$. 

The dimension of a module $M$ as an object in the representation category of $\mathcal{A}$ can be defined by the quantum trace in $\mathcal{A}$. 
This is illustrated in the following. 
%#done: add this quantum dim figure
\begin{equation}
    \frac{1}{\dim\mathcal{A}}
    \mathtikzS{0.5}{
        \draw[thick,\modCol] (-0.4cm,1.6cm) -- (0,2cm) -- (1cm,1cm) -- (1cm,-1cm) -- (0,-2cm) -- (-1cm,-1cm) -- (-1cm,1cm) -- (-0.6cm,1.4cm);
        \draw[thick, \algCol] (-1cm,0) -- (-2cm,-1cm) -- (-2cm,-2cm);
        \draw[thick, \algCol] (1cm,0) -- (0.5cm,-0.5cm) -- (-0.5cm,0.5cm) -- (-0.5cm,2.5cm);
        \node[below] at (-2cm,-2cm) {$\mathcal{A}$};
        \node[above] at (-0.5cm,2.5cm) {$\mathcal{A}$};
        \node[left] at (-1cm,1cm) {$M$};
    }
    =\quad
    \dim_{\mathcal{A}} (M)
    \mathtikzS{0.5}{
        \draw[thick, \algCol] (0,-2cm) -- (0,2cm);
        \node[below] at (0,-2cm) {$\mathcal{A}$};
    }
\end{equation}
This gives
\be \label{eq:qd_defect}
{\textrm{dim} }_\mathcal{A}(M) = \frac{{\textrm{dim}}_C (M) \times \langle M,M\rangle_{\mathcal{A}}  }{{\textrm{dim}} \mathcal{A}}.
\ee
The shorthand $\langle M,M\rangle_\mathcal{A}$ denotes the dimension of endomorphism of $M$ as a ``simple'' object in the representation category of $\mathcal{A}$. i.e. 
This is equal to 2 for a q-type representation, and 1 otherwise.  This is a generalization of the result in \cite{kirillov, Fuchs_2002}, allowing for the left (right) action coming in two independent copies for q-type excitations.

There is also a very useful theorem.

{The {\bf Reciprocity theorem}. It states that 
\be
\langle c_i, M\rangle_C = \langle \textrm{Ind${}_\mathcal{A}(c_i)$}, M\rangle_\mathcal{A}, \qquad  \langle M, c_i \rangle_C = \langle  M, \textrm{Ind${}_\mathcal{A}(c_i)$},\rangle_\mathcal{A}.
\ee
In words, it says the dimension of space of maps between $M$ and $c_i$ in $C$ is the same as  that of maps between $M$ and the induced module of $c_i$ when they are treated as objects in the representation category of $\mathcal{A}$, or in other words, as boundary excitations.  The is proved in \cite{Fuchs_2002}.

Two useful relations follow from the above theorem. They are given by
\be \label{eq:indM}
{\textrm{Ind}}_\mathcal{A}(c_i) \cong \oplus_x   W_{i M_x} M_x,
\ee
and 
\be \label{eq:indMdim}
\textrm{dim}(\mathcal{A}) d_{c_i} = \sum_x  \langle M_x ,M_x\rangle_{\mathcal{A}} \textrm{dim}_C (M_x) W_{i M_x} = \sum_{x,j}  \langle M_x ,M_x\rangle_{\mathcal{A}} W_{i M_x} W_{j M_x} d_{c_j}
\ee

Equation (\ref{eq:indMdim}) is a new result. 
It is a generalization of Corollary 4.14 in \cite{Fuchs_2002} allowing for q-type objects. The generalization follows from the fact that a q-type object
carries two independent left action which should be implicitly summed over in $x$.  
Since the two sets of left action belong to the same module $M_x$ for $M_x$ a q-type object, we replace the sum by the dimension of the endomorphism of
$M_x$.   Physically, this has a very simple interpretation, which applies to both fermionic and bosonic condensation. It can be re-written as
\be
d_{c_i}= \sum_x W_{i M_x}   \textrm{dim}_\mathcal{A}(M_x).
\ee
This means that quantum dimension is ``conserved'' as a bulk anyon is ``decomposed'' into boundary excitations in (\ref{eq:indM}).

Moreover, (\ref{eq:indM}) together with (\ref{eq:Mdecompose}) imply that 
\be  \label{eq:getW}
\mathcal{A} \otimes c_i = \oplus_j  \sum_x \, \langle M_x, M_x\rangle_{\mathcal{A}} W_{i M_x} W_{j M_x} c_j.
\ee

Equation (\ref{eq:qd_defect}, \ref{eq:indMdim}, \ref{eq:getW}) are powerful handles to determining $W$, and also the endomorphism -- specifically to distinguish it from the situation in which an anyon ``splits'' at the boundary, actually participating in two distinct representations. Specifically, when two independent solutions of an irreducible representation can be solved involving the same collection of anyons, the dimension of the endomorphism must be consistent with a quantum dimension of the resultant excitation that is greater than 1. 
This will be illustrated in the example section with explicit solutions. 

\subsubsection{Fermion parity and spin structures}

We note that there are two other new ingredients in working with fermion condensation. 
Here, we extended the techniques in \cite{Fuchs_2002, cong_topological_2016} to accommodate these new ingredients.  
 
Firstly, one would expect to work out $\sigma_{c_i}^M$, which is the fermion parity assignment to the anyon $c_i$ in the representation $M$.
 In a non-Abelian theory, it is possible that $c_i$ participates in multiple modules. 
 Rather than assigning a fermion parity to individual anyons, it is more appropriate to determine fermion parity of the ``condensation channel''.
 i.e. Among the $W_{i x}$ different ways a bulk anyon is mapped to the boundary excitation $M_x$, some of the maps have even parity and others, odd parity. 
 
 One can work out the fermion parity of these condensation channels systematically starting from the parity assignment of the condensate algebra $\mathcal{A}$. 
 This follows from a twist of the relation (\ref{eq:getW}).
 
\be \label{eq:getOmega}
\tilde{\mathcal{A}} \otimes c_i = \oplus_j  \sum_x \,  \Omega_{i x} \Omega_{j x} c_j,
\ee
where $\Omega_{cx}$ gives the difference between the number of even and odd participation channels for $c$ in $x$, and
\be
\tilde{\mathcal{A}} \equiv \oplus_i W_{i1} \exp(2\pi i h_{c_i}) c_i.
\ee
i.e. there is a minus sign for every fermion in the condensate.

For $x=1$ we actually have
\be \label{eq:Omega0}
\Omega_{i1} = W_{i1} e^{2\pi i h_{c_i}}.
\ee

Since $\Omega_{ix}$ is the difference between the number of even and odd ``condensation'' channels of $c_i$ in $x$, one can see that for $x$ a q-type excitation, $\Omega_{ix} =0.$ This has an impact on the derivation of the ``twisted Verlinde formula'' to be discussed below. 
In the case where all $W_{i1} <2$, $\Omega_{i1}$ can be directly treated as the fermion parity $\sigma_i$ of the anyon $c_i$ in defining the super commutative algebra (\ref{eq:super_braid}).
For cases where some $W_{i1}\ge 2$, on first sight we might have to assign multiple parities to the same anyon participating in the algebra $\mathcal{A}$. However, we suspect this could never happen -- i.e. a fermionic anyon could never enter a super Frobenius algebra $\mathcal{A}$ twice having $W_{i1} >2$, since supercommutativity should be violated.

We note that since (\ref{eq:getOmega}) shows up quadratically on the r.h.s., there is a sign ambiguity for $\Omega_{cx}$ for $x\neq 0$.
Practically in the examples we work with, we make a specific choice. We are not aware of a canonical choice at present.

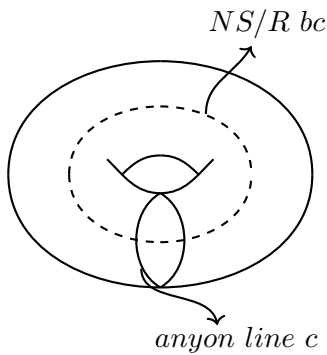
\begin{figure}[h]
    \centering
    \begin{tikzpicture}
    \draw [thick] (-2,0) to [out=90,in=180] (0,1.5);
    \draw [thick] (0,1.5) to [out=0,in=90] (2,0);
    \draw [thick] (-2,0) to [out=270,in=180] (0,-1.5);
    \draw [thick] (0,-1.5) to [out=0,in=270] (2,0);
    \draw [thick] (-0.5,0) to [out=45,in=180] (0,0.25);
    \draw [thick] (0,0.25) to [out=0,in=135] (0.5,0);
    \draw [thick] (-0.7,0.2) to [out=-45,in=180] (0,-0.25);
    \draw [thick] (0,-0.25) to [out=0,in=-135] (0.7,0.2);
    \draw [thick, dashed] (-1.2,0) to [out=90,in=180] (0,0.9);
    \draw [thick, dashed] (0,0.9) to [out=0,in=90] (1.2,0);
    \draw [thick, dashed] (-1.2,0) to [out=270,in=180] (0,-0.9);
    \draw [thick, dashed] (0,-0.9) to [out=0,in=270] (1.2,0);
    \draw [thick] (0,-0.25) to [out=-150,in=150] (0,-1.5);
    \draw [thick] (0,-1.5) to [out=30,in=-30] (0,-0.25);
    \draw [thick, ->] (0.6,0.8) to [out=90,in=240] (1.2,1.7);
    \draw [thick, ->] (-0.25,-1.25) to [out=-90,in=100] (0.75,-2);
    \node at (1.4,2) {$NS/R\ bc$}
    node at (1,-2.2) {$anyon\ line\ c$};
    \end{tikzpicture}
    \caption{A defect wrapping a cycle on the torus, while generating either periodic or anti-periodic boundary conditions for 
    free fermions in the other cycle, determining the spin-structure on the torus.}
    \label{fig:spin_structure}
\end{figure}	
Secondly, in the presence of fermions which are sensitive to spin-structures, there are  anyons that are responsible for Ramond (or anti-periodic) boundary conditions for the free fermions -- or in other words, they fit with the Ramond type spin-structure when they are inserted in a closed manifold with non-contractible cycle. This is illustrated in figure \ref{fig:spin_structure}.  Under a fermion condensation, the boundary excitations can either be responsible for the Neveu- Schwarz (NS) type spin structure or Ramond (R) type in a non-trivial cycle.  
This can be checked by checking the monodromy matrix:
\be
M_{\psi c_i}^{c_j} = -\frac{\theta_{c_j}}{\theta_{c_i}}
\ee
where $\psi \in \mathcal{A}$ has fermionic self-statistics,  with $\theta_{\psi} = -1$, and both $c_{i,j}$ belong to the same boundary excitation, with $W_{ix}$  and $W_{jx}$ non-vanishing for some $x$, and that their $\Omega_{ix}$ and $\Omega_{jx}$ comes with opposite signs.
The boundary excitation $x$ is R type if the monodromy matrix defined above evaluates to -1 for all $i,j$ in $x$, and NS type in the case of +1. 
This generalizes the discussion in \cite{Wan:2016php,Aasen:2017ubm, Lan_2016}  to accommodate non-Abelian fermionic condensates. 

We note however, that in a condensate involving bosonic anyons not generated by fusion of two fermions, the confined anyons do not necessarily have a well defined spin structure -- since they are non-local wrt the bosonic components of the new vacuum made up of the condensed anyons.

\subsubsection{Fusion rules} \label{sec:b_fusion}

Physically, when we observe a cluster of excitations from sufficiently far away i.e. at a distance large compared to the separation between them, then the cluster would effectively appear as some point excitation. The fusion maps in the bulk is part of the data that defines the topological theory. The fusion between boundary excitations however are ``derived'' properties that can be worked out from the choice of the condensed anyons $\mathcal{A}$ and the bulk fusion rules. 

Mathematically, we have contended that boundary excitations are representations (or modules) of the condensed algebra $\mathcal{A}$. Therefore, the physical concept of fusion simply correspond to fusion of representations. 
Diagrammatically, when we have a pair of representations, we should be able to define a new left (and  right) action on the combined system. Directly analogous to the situation in combining spins where we take $\hat S = \hat S_1 +  \hat S_2$ -- as already explained when we introduce the co-product -- the new left action on the combined system should require the use of the co-product. 
This is illustrated diagrammatically in the middle figure in (\ref{eq:fuse_modules}) for fusion of left modules. 
An extra intermediate $\mathcal{A}$ line connecting $M_1$ and $M_2$ is introduced. As illustrated in (\ref{eq:fuse_modules}), it implements automatically 
\be \label{eq:modA}
%(\mathcal{A} \circ M_1) \otimes M_2  \sim M_1 \otimes (\mathcal{A}\circ M_2). 
\rho_\mathcal{A}^{M_1}\otimes id_{M_2}\sim (id_{M_1}\otimes\rho_\mathcal{A}^{M_2})\circ R_{A,M_1}
\ee
The intermediate A-line indeed acts a projector. i.e. If one puts in two parallel A-lines between the modules $M_1, M_2$ as in (\ref{eq:modA}), using the fact that the algebra $\mathcal{A}$ is Frobenius and separable (\ref{eq:separability}) and that $M_{1,2}$ both satisfy (\ref{eq:reps_eq}), one can show that it can be reduced to having only one line. This would be left as a simple exercise for the readers.

\begin{figure}[h]
    \centering
    \be  \label{eq:fuse_modules}
    \begin{tikzpicture}
    \begin{scope}[xshift=0,yshift=0]{
    \draw [thick, magenta] (0,0.1) -- (0,2);
    \draw [thick, magenta] (0,-1) -- (0,-0.1);
    \draw [thick, magenta] (1.5,-1) -- (1.5,2);
    \draw [thick] (0,1) to [out=225,in=180] (0,0);
    \draw [thick] (0,0) to [out=0,in=230] (1.5,1);
    \draw [thick] (-1,0.5) -- (0,1.5);
    \node at (0.4,2) {$M_1$}
    node at (1.9,2) {$M_2$}
    node at (-0.5,0.2) {$\mathcal{A}$}
    node at (-0.5,1.5) {$\mathcal{A}$};
    }
    \end{scope}
    \begin{scope}[xshift=4cm,yshift=0]{
    \draw [thick, magenta] (0,0.1) -- (0,2);
    \draw [thick, magenta] (0,-1) -- (0,-0.1);
    \draw [thick, magenta] (1.5,-1) -- (1.5,2);
    \draw [thick] (0,1) to [out=225,in=180] (0,0);
    \draw [thick] (0,0) to [out=0,in=230] (1.5,1);
    \draw [thick] (-1,-1) -- (0,0);
    \node at (0.4,2) {$M_1$}
    node at (1.9,2) {$M_2$}
    node at (-1.3,0.5) {\Large{$=$}};
    }
    \end{scope}
    \begin{scope}[xshift=8cm,yshift=0]{
    \draw [thick, magenta] (0,0.1) -- (0,2);
    \draw [thick, magenta] (0,-0.7) -- (0,-0.1);
    \draw [thick, magenta] (0,-1) -- (0,-0.9);
    \draw [thick, magenta] (1.5,-1) -- (1.5,2);
    \draw [thick] (0,1) to [out=225,in=180] (0,0);
    \draw [thick] (0,0) to [out=0,in=230] (1.5,1);
    \draw [thick] (-0.2,-1) -- (1.5,0.7);
    \node at (0.4,2) {$M_1$}
    node at (1.9,2) {$M_2$}
    node at (-1.3,0.5) {\Large{$=$}};
    }
    \end{scope}
    \end{tikzpicture}
    \ee
\end{figure}	

To summarise, the fusion map of modules of $\mathcal{A}$ is defined such that one mods out the relation (\ref{eq:modA}). This fusion map is denoted $\otimes_{\mathcal{A}}$, and practically implemented using the projector involving the A-line introduced in (\ref{eq:fuse_modules}).

Note however, the module resulting from the fusion is not irreducible. It is important information to recover the decomposition of the fusion 
product in terms of irreducible representations. 
This has been considered in \cite{Fuchs_2002}. The decomposition coefficients can be computed using (\ref{eq:extract_lambda}). 

To recover only the fusion coefficients however, we can make use of the identity \cite{kirillov} (which has been generalized
here to accommodate for non-trivial endomorphisms :

\begin{align}  \label{eq:evenfuse}
\mathcal{A} \otimes c_i\otimes c_j &= \oplus_l  \sum_{x,k} \langle M_x, M_x\rangle_\mathcal{A} \, W_{i M_x} W_{k M_x} N_{k j}^{l}  \,c_l= \oplus_{l} \sum_{k,x}\langle M_x, M_x\rangle_\mathcal{A} \, N_{i j}^{k} W_{k M_x}W_{l M_x} c_l   \nonumber \\
&=  \oplus_l \sum_{x,y,z} W_{i M_x} W_{j M_y} n_{xy}^z W_{l M_z} c_l
\end{align}
where $n_{xy}^z$ are the fusion coefficients counting the total number of fusion channels mapping $M_x\times M_y$ to $M_z$ in the boundary, as defined in (\ref{eq:superfuse}) for a super-fusion category. i.e.
\be
n_{xy}^z = \textrm{dim[ Hom}_{\mathcal{A}}(M_x\otimes M_y, M_z)]
\ee
As reviewed already in the previous section, the fusion channels in the condensed phase describable by a super fusion category also come with  even or
odd fermion parities. The above equation should thus be refined. We found a twisted version of the above relation, using also (\ref{eq:getOmega})

\begin{align} \label{eq:tfusionM}
\tilde{\mathcal{A}} \otimes c_i\otimes c_j  & =    \oplus_l  \sum_{x,k} \langle M_x, M_x\rangle_\mathcal{A}^{\delta} \, \Omega_{k x} \Omega_{i x} N_{k j}^{l}  \,c_l \nonumber \\
&=  \oplus_l \sum_{x,y,z} \, \Omega_{ix} \Omega_{jy} \tilde{n}_{xy}^z \Omega_{l z} \, c_l.
\end{align}
The twisted fusion coefficient  $\tilde{n}_{xy}^z$ is the difference between the number of even and odd fusion channels taking $x\otimes y$ to $z$. 
Here, $\langle M_x, M_x\rangle_{\mathcal{A}}^\delta$ denotes the difference between the number of even and odd endomorphisms of $M_x$. For example, a q-type object with one even and one odd endomorphic maps satisfies $\langle M_x, M_x\rangle_{\mathcal{A}}^\delta=0$. As discussed previously,  the dimension of endomorphism space for simple objects is either 1 (non q-type) or 2 (q-type) in a super-fusion category. 
 Also,  $\Omega_{ix}$ vanishes for $x$ a q-type object. Therefore, the sum over $x$ in (\ref{eq:tfusionM}) might as well be restricted to non-q-type excitations, to give
\begin{align} \label{eq:tfusionM2}
\tilde{\mathcal{A}} \otimes c_i\otimes c_j  & =    \oplus_l  \sum_{x \neq \textrm{q-type},k} \, \Omega_{k x} \Omega_{i x} N_{k j}^{l}  \,c_l \nonumber \\
&=  \oplus_l \sum_{x,y,z \neq \textrm{q-type}} \, \Omega_{ix} \Omega_{jy} \tilde{n}_{xy}^z \Omega_{l z} \, c_l.
\end{align}
Now this totally parallels (\ref{eq:evenfuse}).

\subsubsection{(twisted) Defect Verlinde formula} \label{sec:twistedDVF}

In \cite{Shen_2019}, we described a formula relating the fusion coefficients of the boundary excitations with the ``half-link'' between the boundary excitations and the condensed anyons.
Here we would like to generalize it to the case accommodating fermion condensation, and also to express the ``half-link'' in terms of a trace of the different linear maps whose basis we have constructed explicitly in the previous section. 

First, let us obtain the pair of twisted defect Verlinde formula for a given gapped boundary characterized by $\mathcal{A}$.

This can be derived using (\ref{eq:evenfuse}), which takes an identical form in fermionic condensates as in bosonic ones discussed in \cite{Shen_2019}
\be \label{eq:defect_verlinde_e}
n_{xy}^{z} = \sum_{i}\frac{\langle M_z, M_z\rangle_\mathcal{A} \,V_{x i}V_{y i}V^{-1}_{i z}}{S_{1i}}, \qquad  V^{-1}_{i x} = 
\sum_{k}  \bar{S}_{i k} W_{k x}
\ee
The matrix $V$ is invertible -- the first index $i$ runs over only $c_i \in \mathcal{A}$ and the index $x$ enumerates the boundary excitations. As we argued in \cite{hung_ground_2015, Shen_2019}, the number of anyons in $\mathcal{A}$ is always equal to the number of boundary excitations, so that $V$ is a square matrix. 

As observed in \cite{Shen_2019}, the matrix $V$ is related to the ``half-linking'' number as follows:
\be \label{eq:Vgamma}
\frac{\gamma_{xi}}{\gamma_{1i}} = V_{ix}^{-1}, \qquad \gamma_{1i} = \sqrt{S_{1i}}.
\ee

Here, we would also like to express the half-linking number  in terms of the basic defining properties of the condensate algebra $\mathcal{A}$ and the modules, in the incarnation of a  quantum trace that is illustrated in (\ref{eq:gamma_pic}). 
\begin{figure}[h]
    \centering
    \be \label{eq:gamma_pic}
    \begin{tikzpicture}[scale=0.8]
    \draw [thick, green] (0,1) -- (0,0.1);
    \draw [thick, green] (0,-0.1) -- (0,-2);
    \draw [thick, green] (-0.5,1) -- (-0.5,0.5);
    \draw [thick, green] (-0.5,-1.5) -- (-0.5,-2);
    \draw [thick, green] (-0.5,0.5) to [out=270, in=180] (0.1,0);
    \draw [thick, green] (0.1,0) to [out=0, in=0] (0.1,-1);
    \draw [thick, green] (-0.1,-1) to [out=180, in=90] (-0.5,-1.5);
    \draw [thick, green] (0,1) to [out=90, in=90] (1.5,1);
    \draw [thick, green] (-0.5,1) to [out=90, in=90] (2.5,1);
    \draw [thick, green] (0,-2) to [out=270, in=270] (1.5,-2);
    \draw [thick, green] (-0.5,-2) to [out=270, in=270] (2.5,-2);
    \draw [thick, green] (1.5,1) -- (1.5,-2);
    \draw [thick, green] (2.5,1) -- (2.5,-2);
    \scalebox{0.5}[0.5]{\tri{3cm}{0}{90}}
    \scalebox{0.5}[0.5]{\tri{5cm}{0}{90}}
    \scalebox{0.5}[0.5]{\tri{3cm}{-2cm}{270}}
    \scalebox{0.5}[0.5]{\tri{5cm}{-2cm}{270}}
    \draw [thick] (2.5,0) -- (2.5,-1);
    \draw [thick, magenta] (1.5,0) -- (1.5,-1);
    \draw [thick] (1.5,-1/3) to [out=-45, in=225] (2.5,-1/3);
    \draw [thick] (2,-0.55) -- (2,-0.8);
    \draw [thick, fill] (2,-0.85) circle [radius=0.05];
    \node at (2.7,-0.5) {$\mathcal{A}$}
    node at (1.3,-0.5) {$x$}
    node at (2,-0.3) {$\mathcal{A}$}
    node at (2.7,0.5) {$c$}
    node at (1.3,0.5) {$i$}
    node at (-2.3,-0.5) {\Large{$\gamma_{xc} = \mathcal{N}^x_c \sum_i$}};
    \end{tikzpicture}
    \ee
\end{figure}	

We observe that the normalization constant takes the following form
\be \label{eq:NAB}
\mathcal{N}^x_c = \frac{1}{ \sqrt{D_{bulk}d_c} }
%\mathcal{N}^x_c = \frac{\sqrt{p D_{bulk}}}{d_c}, \qquad p=1 (\textrm{where $x$ a left/right module}),\qquad p=2 (\textrm{where $x$ a bimodule})
\ee

In a super fusion category, fusion channels can acquire even or odd fermion parities. The defect Verlinde formula given above relates the total number of fusion channels with the half-linking numbers. 

There is an independent equation that relates the difference between the number of even and odd fusion channels to a ``twisted'' half-linking number. This can be derived using (\ref{eq:tfusionM}) using very similar techniques as the derivation of 
(\ref{eq:defect_verlinde_e}).  We first define the matrix $v^{-1}$
\be  \label{eq:littlev}
v^{-1}_{ix} = \sum_j \bar{S}_{ij} \Omega_{jx}.
\ee
This is the analogue of the $V$ matrix defined in (\ref{eq:defect_verlinde_e}). As already noted earlier when $\Omega_{jx} $ was defined, 
$\Omega_{jx} =0$ for $x$ a q-type excitation. Therefore, in the matrix $v_{ix}$, $x$ only runs over the gapped excitations
that are not q-type. The other index $j$ now runs over the anyons $c_j$ belonging to the gapped excitation $x_f$ responsible for generating the fermion parity. i.e. There is a special gapped boundary excitation $x_f$ such that the monodromy with the condensate produces a +1 on all the bosonic condensed anyons, and a -1 on all the fermionic ones.  -- This will be further discussed in section \ref{sec:CFT} below, where this special excitation can be readily worked out by a simple modular transformation in the bulk using (\ref{eq:work_xf}).  
Surprisingly, $v$ is also a square matrix -- there is always an equal number of anyons in the special boundary excitation  $x_f$ as there are non-q-type boundary excitations! 
i.e. Once restricting $x$ to non-q-type objects, equation (\ref{eq:tfusionM2})  and (\ref{eq:evenfuse}) take the same form.

Thus we obtain the following {\it twisted Verlinde formula} simply by replacing $V$ by $v$ in (\ref{eq:defect_verlinde_e}), which gives
\be \label{eq:twisted_VLF}
\tilde{n}_{xy}^z = \sum_{i\in x_f}\frac{v_{x i}v_{y i}v^{-1}_{i z}}{S_{1i}}, \qquad x,y,z \neq \, \textrm{q-type}.
\ee
The sum over $i$ runs over anyons participating in $x_f$. We note that this is not imposed by hand -- but simply follows from the property of $v$.  This is one of the main results of this paper. 

\subsection{Defects at junctions and bimodules} \label{sec:junctions}

In the previous section, we have focussed on excitations in a given gapped boundary where $\mathcal{A}$ condensed. 
Here, we would like to extend the discussion to  excitations localized at the junction of two different gapped boundaries characterized by condensate algebra $\mathcal{A}$ and $\mathcal{B}$ should correspond to irreducible {\bf left-right bi-modules} of $\mathcal{A}$ and $\mathcal{B}$.  Each bi-module is again a collection of anyons in $C$. The left and right action of $\mathcal{A}$ and $\mathcal{B}$ respectively should commute. This is illustrated in (\ref{eq:bimodule}). 
\begin{figure}[h]
    \centering
    \be \label{eq:bimodule}
    \begin{tikzpicture}
    \draw [thick, magenta] (-2,-1.5) -- (-2,2);
    \draw [thick, magenta] (2,-1.5) -- (2,2);
    \draw [thick] (-2,0.5) -- (-4,-1.5);
    \draw [thick] (-2,-0.5) -- (-1,-1.5);
    \draw [thick] (2,0.5) -- (4,-1.5);
    \draw [thick] (2,-0.5) -- (1,-1.5);
    \node at (0,0) {\Large{$=$}}
    node at (-3,0) {$\mathcal{A}$}
    node at (-1.4,2) {$M^{\mathcal{A}|\mathcal{B}}$}
    node at (-1,-1) {$\mathcal{B}$}
    node at (1,-1) {$\mathcal{A}$}
    node at (2.6,2) {$M^{\mathcal{A}|\mathcal{B}}$}
    node at (3,0) {$\mathcal{B}$};
    \end{tikzpicture}
    \ee
\end{figure}

It is shown that the bimodules together also form a semi-simple fusion category \cite{Fuchs_2002}. Exactly analogous to the case of
left (right) modules, one can generate induced modules from any anyon $c_i \in C$, by sandwiching $c_i$ by $\mathcal{A}$ and $\mathcal{B}$ on the left and right of $c_i$ respectively. i.e Repeating (\ref{eq:ind_mod}) with a copy of $\mathcal{B}$ on the right as well. 

The induced bimodule obtained are reducible (not simple), and thus can be decomposed in terms of simple ones. 
By inspecting the fusion $\mathcal{A} \otimes c_i \otimes \mathcal{B}$ that generates the induced bi-module, it is possible to isolate all the independent simple (irreducible) modules, and recover the W-matrix. Without the simple conservation formula for quantum dimensions as in (\ref{eq:indMdim}) and also the analogue of (\ref{eq:evenfuse}), it is not apparent if there is a simple formula for the W-matrix. Moreover,  as in the case of left (right) modules, one has to work out the endomorphism of a given module.

An identity particularly useful for the purpose is the following \cite{Fuchs_2004} : 
\be
\textrm{Hom}_{\mathcal{A}|\mathcal{B}}(\textrm{Ind}_{\mathcal{A}|\mathcal{B}}(c_i), \textrm{Ind}_{\mathcal{A}|\mathcal{B}}(c_j)) \cong \textrm{Hom}(c_i, \mathcal{A} \otimes c_j \otimes \mathcal{B})
\ee
where here we are using $\cong$ loosely to mean the two sides are isomorphic. 
This also implies
\be \label{eq:double_reciprocity2}
\textrm{End}_{\mathcal{A}|\mathcal{B}} (\textrm{Ind}_{\mathcal{A}|\mathcal{B}} (c_i)) 
\cong \textrm{Hom}(c_i, \mathcal{A} \otimes c_i \otimes \mathcal{B}).
\ee
As we will see in the example of $D(S_3)$ in section \ref{sec:bfjunction_take2}, the formula assists us in determining non-trivial endomorphisms in a junction excitation.

Exactly as in the case of left (right) modules, one can work out the endomorphism of a given module.

There is a new complication in the presence of fermionic condensates. As we have discussed, free fermions have been introduced into the system to enrich the theory into a spin-TQFT, and the condensate algebra a super-Frobenius algebra. 
In a spin TQFT, it is possible to introduce localized Majorana modes. Therefore, every excitation at the junctions would become $\mathbb{Z}_{>0}$ graded -- the non-negative integer ``grading'' keeping track of the number of Majorana modes that have been added to the spot. This has been observed in \cite{barkeshli_classification_2013,Barkeshli:2013yta} where gapped boundaries of Abelian spin TQFT were discussed. For each extra Majorana mode that is added, the quantum dimension of the defect would be raised by a factor of $\sqrt{2}$. We note that when we add a pair of Majorana mode to the same spot, they could pair up as a Dirac fermion mode and be gapped out by a local Hamiltonian. Therefore, the grading is not topologically robust, and could be reduced to a $\mathbb{Z}_2$ structure. 

In the current paper where we focus on bosonic bulk topological orders, it is observed that fusion of defects between bosonic junctions with defects at bosonic-fermionic junctions could generate different flavours (or grading) of the excitations.

\subsubsection{Fusion rules and the Defect Verlinde formula} \label{sec:fusion_junc}

Fusion of bi-modules (or excitations localized at junctions) follows a similar playbook as the fusion of the modules.
For $\mathcal{A}, \mathcal{B}, \mathcal{C} \subset C$, the fusion map is given by
\be
M^{\mathcal{A}|\mathcal{B}} \otimes_{\mathcal{B}} M^{\mathcal{B}|\mathcal{C}} = M^{\mathcal{A}|\mathcal{C}}.
\ee
Practically, $\otimes_{\mathcal{B}}$, which we have already discussed while defining the fusion map for left (right) modules, can be implemented by inserting a $\mathcal{B}$ line. This is illustrated in (\ref{eq:fuse_bimod}).
\begin{figure}[h]
    \centering
    \be \label{eq:fuse_bimod}
    \begin{tikzpicture}
    \draw [thick, magenta] (0,-1) -- (0,2);
    \draw [thick, magenta] (2,-1) -- (2,2);
    \draw [thick] (0,1) to [out=-45,in=180] (1,0);
    \draw [thick] (1,0) to [out=0,in=230] (2,1);
    \draw [thick] (-1,-0.5) -- (0,0.5);
    \draw [thick] (2,0.5) -- (3,-0.5);
    \node at (0.6,2) {$M^{\mathcal{A}|\mathcal{B}}$}
    node at (2.6,2) {$M^{\mathcal{B}|\mathcal{C}}$}
    node at (-0.5,0.5) {$\mathcal{A}$}
    node at (1,0.5) {$\mathcal{B}$}
    node at (2.5,0.5) {$\mathcal{C}$};
    \end{tikzpicture}
    \ee
\end{figure}

Again, we can decompose the resultant $\mathcal{A}|\mathcal{C}$ bimodule in terms of irreducible (or simple) $\mathcal{A}|\mathcal{C}$ modules.
This can be done by using equation (\ref{eq:orthogonality}, \ref{eq:extract_lambda}) again. 
We note that these equations work equally well for a bimodule -- we simply need to view the bimodule as the left-module of the algebra $\mathcal{A} \otimes \mathcal{C}^{\textrm{rev}}$ -- where the super-script {\it rev} refers to folding $\mathcal{C}$. Practically however since the left and right action of $\mathcal{A}$ and $\mathcal{C}$ respectively commute, one is basically including in (\ref{eq:orthogonality})
an extra $\mathcal{C}$ loop also on the right. 

In actual applications, one is often only interested in working out the fusion coefficients. As such, we might hope to adopt a strategy similar to  (\ref{eq:evenfuse}).  
However, for bi-modules, we do not know of a simple analogue.
For Abelian bulk topological order however, we can work it out as follows. For
simplicity, we will assume that  $\mathcal{A} \cap \mathcal{B} = 1$. i.e. The trivial anyon in $C$ is the only anyon in the intersection between the two condensates. In this case, every induced bimodule Ind${}_{\mathcal{A}|\mathcal{B}}(c_i)$ is also simple. The strategy to work out the fusion coefficients is that the fusion operation $\otimes_{\mathcal{A}}$ can be implemented by modding out redundant copies of the condensates as the induced bimodules fuse. i.e.
\begin{align} \label{eq:bimodule_fuse}
&\textrm{Ind}_{\mathcal{A}|\mathcal{B}}(c_i) \otimes_{\mathcal{B}} \textrm{Ind}_{\mathcal{B}|\mathcal{C}}(c_j)
= \mathcal{A} \otimes c_i \otimes \mathcal{B} \otimes_{\mathcal{B}} \mathcal{B} \otimes c_j \otimes \mathcal{C} \nonumber \\
&= \mathcal{A} \otimes c_i \otimes \mathcal{B} \otimes c_j \otimes \mathcal{C} \nonumber \\
& = \oplus_k     (\sum_{l,m} N_{il}^m N_{mj}^k W_{l1}^{\mathcal{B}}) \,  \mathcal{A} \otimes c_k \otimes \mathcal{C} 
\end{align}
We have included extra super-scripts over the W-matrix to distinguish the data of different condensates. 
The above is clearly a $\mathcal{A}|\mathcal{C}$ bimodule which is then decomposed into simple bimodules. 

\begin{itemize}
\item Extra trapped Majorana modes at junctions

The above rules apply to bosonic condensates $\mathcal{A}, \mathcal{B}, \mathcal{C}$. There is a caveat when it comes to 
algebra involving fermionic condensate as well.  As already mentioned in the previous sub-section, junctions between boundaries can host Majorana zero modes in the presence of free fermions, so that every  simple bimodules that we work out based on seeking representations of the algebra $\mathcal{A},\mathcal{B}$ comes in an infinite number of versions -- differing by the number of extra Majorana zero modes that they host. When considering fusion of junctions, different versions
are often being generated, even if we start with a {\it canonical} choice obtained via the induced modules $\mathcal{A} \otimes c_i\otimes \mathcal{B}$.

To accommodate this complication, we note the following. The fermionic anyons that condensed had in fact formed Cooper pairs with free fermions introduced at the gapped boundary. Therefore it is more proper to write the condensate algebra as
\be \label{eq:Ap}
\mathcal{A}' = \oplus_i W_{i 1} c_i \otimes \psi_0^{\sigma(c_i)}%, \qquad \sigma_i = \frac{1+ \theta_i}{2}.
\ee

i.e. A free fermion is denoted $\psi_0$ which pairs up with the condensed fermion. 

An extra majorana mode trapped at a junction would correspond to an extra copy of $ \otimes (1 \oplus \psi_0)$ introduced there. This is an appropriate way of  keeping track of these Majorana modes, since they can absorb or release a Dirac fermion, and so behaves as a genuine fermion condensate at a point -- and whose only module would be the ``condensate'' itself : $\chi \equiv 1\oplus \psi_0$. When we fuse two such modes (which are modules of $\chi$) we should get
\be \label{eq:fusemajorana}
(1\oplus \psi_0)_{\textrm{0 d}} \otimes_{\chi}  (1\oplus \psi_0)_{\textrm{0 d}} = 1\oplus \psi_0.
\ee
On the rhs, it is understood that the fermion is no longer localized at a 0-dimensional junction.
This then correctly recovers the fusion rule of Majorana modes -- that produces the direct sum of the trivial state and a single fermion state.  When considering general fusion of junctions with possible extra Majorana modes, it is key to keep track of copies of $\chi$. We will illustrate this technicality in the examples section \ref{sec:bfjunction} and \ref{sec:bfjunction_take2}, which is crucial towards keeping track of the quantum dimensions of junction excitations.

\end{itemize}

If $\mathcal{C} = \mathcal{A}$, the resultant gapped boundary excitations should be further reduced from an $\mathcal{A}|\mathcal{A}$
bimodule to a left (right) module (recall that the left and right modules can be generated from each other since we are considering (super)- commutative Frobenius algebra).

Then we have
\begin{align}
&\textrm{Ind}_{\mathcal{A}|\mathcal{B}}(c_i) \otimes_{\mathcal{B}} \textrm{Ind}_{\mathcal{B}|\mathcal{A}}(c_j)
= \mathcal{A} \otimes c_i \otimes \mathcal{B} \otimes_{\mathcal{B}} \mathcal{B} \otimes c_j \otimes \mathcal{A} \nonumber \\
&= \mathcal{A} \otimes c_i \otimes \mathcal{B} \otimes c_j \otimes \mathcal{A} \nonumber \\
& \xlongleftrightarrow{\textrm{reducing to left-modules}} \oplus_k     (\sum_{l,m} N_{il}^m N_{mj}^k W_{l1}^{\mathcal{B}}) \,  \mathcal{A} \otimes c_k  \nonumber \\
& = \oplus_x    \sum_{l,m,k} N_{il}^m N_{mj}^k \langle M^\mathcal{A}_x, M^\mathcal{A}_x\rangle_{\mathcal{A}}W_{l1}^{\mathcal{B}}  W^{\mathcal{A}}_{kx}  M^\mathcal{A}_x,
\end{align}
 where we have made use of (\ref{eq:getW}, \ref{eq:Mdecompose}) in the last equality.

In \cite{Shen_2019}, we obtained a defect Verlinde formula describing the fusion of bi-modules.  
In the presence of both bosonic and fermionic condensates, one can write down a defect Verlinde formula too.
Given the extra complication of Majorana fermion modes, we will have to fix the ambiguity in the defect Verlinde formula too. 
The half-linking number across a junction  expressed as a quantum trace is illustrated in (\ref{eq:gamma_pic}). The junction excitation involved here is the ``canonical'' choice obtainable from the induced modules, and our defect Verlinde formula would describe the fusion of these canonical junction excitations. 

The defect Verlinde formula that describes the canonical fusion coefficients as defined above takes exactly the same form as in \cite{Shen_2019}.
For completeness, we reproduce it here
\be  \label{eq:defect_verlinde_junction}
n_{x y}^z
 = \sum_c \sum_{\alpha_{\mathcal{A}}, \beta_{\mathcal{B}},\beta'_{\mathcal{B}}, \sigma_{\mathcal{C}}}\langle M_z,M_z\rangle_{A|C}~\gamma_{xc_{\alpha_{\mathcal{A}}, \beta_{\mathcal{B}}}}^{(\mathcal{A}|\mathcal{B})} (M^{\mathcal{B}}_c)^{-1}_{\beta_{\mathcal{B}} \beta'_{\mathcal{B}}} \gamma^{(\mathcal{B}|\mathcal{C})}_{yc_{\beta'_{\mathcal{B}}, \sigma_{\mathcal{C}}}} (\gamma^{(\mathcal{A}|\mathcal{C})})^{-1}_{c_{\sigma_{\mathcal{C}}, \alpha_{\mathcal{A}}}z},
\ee

where 
\be
(M^\mathcal{B}_{c})_{\alpha, \beta} =  \gamma^\mathcal{B}_{1c_{\alpha, \beta}},
\ee
and the inverse of $M$ is taken by treating it as a matrix with indices $\alpha,\beta$, while the inverse of $\gamma$ is taken wrt the indices $\{c_{\alpha,\beta}, z\}$ i.e. the number of $z$ indices is equal to $c_{\alpha,\beta}$. 
We have taken extra pains to include subscripts for the $\alpha, \beta$ to indicate the precise condensate these condensation channels are related to. 
It should be clear that $x$ lives at the junction between $\mathcal{A}$ and $\mathcal{B}$, and $y$ between $\mathcal{B}$ and $\mathcal{C}$, and finally $z$ between $\mathcal{A}$ and $\mathcal{C}$.

\begin{figure}[h]
\centering
\be\label{eq:gamma_pic_2}
\begin{tikzpicture}
\draw [thick, green] (-0.45,0.5) -- (0.7,0.5);
\draw [thick] (0.7,0.5) to [out=0,in=120] (1.5,-0.5);
\draw [thick] (0.5,-0.5) -- (1.7,-0.5);
\draw [thick, green] (-0.55,0.5) -- (-0.7,0.5);
\draw [thick] (-0.7,0.5) to [out=180,in=60] (-1.5,-0.5);
\draw [thick] (-0.5,-0.5) -- (-1.7,-0.5);
\draw [thick, magenta] (-0.5,0) -- (-0.5,-1);
\draw [thick, green] (-0.5,0) -- (-0.5,1);
\draw [thick, green] (0.5,0) -- (0.5,0.45);
\draw [thick, green] (0.5,1) -- (0.5,0.55);
\draw [thick, magenta] (0.5,0) -- (0.5,-1);
\draw [thick, green] (-0.5,1) to [out=90,in=90] (0.5,1);
\draw [thick, green] (-0.5,-1) to [out=270,in=270] (0.5,-1);
\draw [thick, fill] (1.75,-0.5) circle [radius=0.05];
\draw [thick, fill] (-1.75,-0.5) circle [radius=0.05];
\scalebox{0.4}[0.4]{\tri{-1*1.25cm}{0}{90}}
\scalebox{0.4}[0.4]{\tri{1*1.25cm}{0}{90}}
\scalebox{0.4}[0.4]{\tri{-1*1.25cm}{-1.6*1.25cm}{270}}
\scalebox{0.4}[0.4]{\tri{1*1.25cm}{-1.6*1.25cm}{270}}
\scalebox{0.4}[0.4]{\tri{-1.7*1.25cm}{1*1.25cm}{0}}
\scalebox{0.4}[0.4]{\tri{1.7*1.25cm}{1*1.25cm}{180}}
\node at (0,1.5) {$i$}
node at (0,-1.5) {$j$}
node at (-1.6,0) {$\mathcal{A}$}
node at (-1,-0.8) {$\mathcal{A}$}
node at (1,-0.8) {$\mathcal{B}$}
node at (1.6,0) {$\mathcal{B}$}
node at (-0.7,-0.2) {$x$}
node at (0.7,-0.2) {$x$}
node at (0,0.7) {$c$}
node at (-5,0) {\Large{$\gamma_{xc_{\alpha_\mathcal{A},\beta_\mathcal{B}}}^{(\mathcal{A}|\mathcal{B})}=\mathcal{N}^x_{c_{\alpha_\mathcal{A},\beta_\mathcal{B}}}\sum_{i,j}$}};
\node at (-0.8,0.8) {$\alpha_\mathcal{A}$};
\node at (0.8,0.8) {$\beta_\mathcal{B}$};
\end{tikzpicture}
\ee
\end{figure}	
Similarly to (\ref{eq:gamma_pic}), for half-linking numbers considering the fusion of bi-modules we propose (\ref{eq:gamma_pic_2}) with normalization constant given by
\be\label{eq:NAB_2}
\mathcal{N}^x_{c_{\alpha_\mathcal{A},\beta_\mathcal{B}}}=\frac{1}{\sqrt{2D_{bulk}d_c}}
\ee

\subsection{Note: the M3J and M6J symbols and VLCs}

To assist our readers in the sea of literature, here we would like to comment on the relationship between the condensate algebra and some of the linear maps introduced elsewhere. 

{\bf \underline{M-symbols}}
The notion of M-symbols were introduced in \cite{cong_topological_2016}. 
The idea is that the gapped boundary is an interface where bulk anyons could end on it. 
As one consider multiple bulk anyons ending on the boundary, it is possible to consider changing the order of fusion of the bulk anyons before they end on the boundary. These different processes should be related by linear maps, which are given by the M3J and M6J symbols. This is illustrated in (\ref{eq:M3JM6J}). 
\begin{figure}[htbp]
\centering
\be\label{eq:M3JM6J}
\begin{tikzpicture}
\draw [thick] (1.25,4) -- (1.25,7);
\draw [thick] (2,4) -- (2,7);
\draw [ultra thick] (0.75,5.5) -- (2.5,5.5);
\node at (1.05,4) {$x$}
node at (1.05,7) {$a$}
node at (1.8,4) {$y$}
node at (1.8,7) {$b$}
node at (1.45,5.3) {$\nu$}
node at (2.2,5.3) {$\lambda$};
\begin{scope}[xshift=1cm,yshift=-2cm]{
\draw [ultra thick] (5,7.5) -- (7,7.5);
\draw [thick] (5.5,6.5) to [out=90, in=90] (6.5,6.5);
\draw [thick] (5.5,6) -- (5.5,6.5);
\draw [thick] (6.5,6) -- (6.5,6.5);
\draw [thick] (6,6.8) -- (6,8.2);
\draw [thick] (5.5,8.5) -- (5.5,9);
\draw [thick] (6.5,8.5) -- (6.5,9);
\draw [thick] (5.5,8.5) to [out=270, in=270] (6.5,8.5);
\node at (6.2,7.3) {$\psi$}
node at (5.3,6) {$x$}
node at (6.3,6) {$y$}
node at (5.3,9) {$a$}
node at (6.3,9) {$b$}
node at (5.8,7) {$z$}
node at (5.8,7.9) {$c$}
node at (3.3,7.5) {$=\sum_{c,z}[M^{ab;z}_{c;xy}]^{\mu\nu}_{\psi}$};
}
\end{scope}
\begin{scope}[xshift=0cm,yshift=2.25cm]{
\draw [thick] (1.25,5.5) -- (1.25,7);
\draw [thick] (2,5.5) -- (2,7);
\draw [ultra thick] (0.75,5.5) -- (2.5,5.5);
\node at (1.05,7) {$a$}
node at (1.8,7) {$b$}
node at (1.25,5.3) {$\nu$}
node at (2,5.3) {$\lambda$};
\begin{scope}[xshift=0cm,yshift=-2cm]{
\draw [ultra thick] (5,7.5) -- (7,7.5);
\draw [thick] (6,7.5) -- (6,8.2);
\draw [thick] (5.5,8.5) -- (5.5,9);
\draw [thick] (6.5,8.5) -- (6.5,9);
\draw [thick] (5.5,8.5) to [out=270, in=270] (6.5,8.5);
\node at (6,7.3) {$\psi$}
node at (5.3,9) {$a$}
node at (6.3,9) {$b$}
node at (5.8,7.9) {$c$}
node at (3.8,8.25) {$=\sum_{c}[M^{ab}_{c}]^{\mu\nu}_{\psi}$};
}
\end{scope}
}
\end{scope}
\end{tikzpicture}
\ee
\end{figure}
 As expected, the M-symbols are directly related to the defining data of the condensate algebra $\mathcal{A}$ and its modules (up to appropriate normalizations). 
This is illustrated in  (\ref{eq:Msymbols}). 
\begin{figure}[htbp]
\centering
\be\label{eq:Msymbols}
\begin{tikzpicture}[scale=0.5]
%\draw [ultra thick] (-1.5,2*0.5) -- (1.5,2*0.5);
%\draw [ultra thick] (-1.5,-2*0.5) -- (1.5,-2*0.5);
\draw [thick, green] (-0.8,5*0.5) -- (-0.8,2*0.5);
\draw [thick, green] (-0.8,-2*0.5) -- (-0.8,-5*0.5);
\draw [thick, magenta] (-0.8,-2*0.5) -- (-0.8,-0.1);
\draw [thick, magenta] (-0.8,0.1) -- (-0.8,2*0.5);
\draw [thick, green] (0.8,5*0.5) -- (0.8,2*0.5);
\draw [thick, green] (0.8,-2*0.5) -- (0.8,-5*0.5);
\draw [thick, magenta] (0.8,-2*0.5) -- (0.8,2*0.5);
\draw [thick] (-0.8,1*0.5) to [out=225,in=180] (-0.8,0);
\draw [thick] (-0.8,0) to [out=0,in=230] (0.8,1*0.5);
\draw [thick] (-1.8,-1*0.5) -- (-0.8,0);
\draw [fill] (-1.8,-1*0.5) circle [radius=0.08];
\node at (-0.5,0.4*0.5) {\tiny{$M_i$}}
node at (1.1,0.4*0.5) {\tiny{$M_j$}}
node at (-1.5,-0.2*0.5) {\tiny{$\mathcal{A}$}}
node at (-0.5,2.2*0.5) {\tiny{$\bar\mu$}}
node at (1.1,2.2*0.5) {\tiny{$\bar\nu$}}
node at (-0.6,5*0.5) {\tiny{$a$}}
node at (1,5*0.5) {\tiny{$b$}}
node at (-0.5,-2.3*0.5) {\tiny{$\mu'$}}
node at (1.1,-2.3*0.5) {\tiny{$\nu'$}}
node at (-0.6,-5*0.5) {\tiny{$a'$}}
node at (1,-5*0.5) {\tiny{$b'$}};
\scalebox{0.5}[0.5]{\tri{-1.6cm}{4cm*0.5}{90}}
\scalebox{0.5}[0.5]{\tri{-1.6cm}{-4cm*0.5}{270}}
\scalebox{0.5}[0.5]{\tri{1.6cm}{4cm*0.5}{90}}
\scalebox{0.5}[0.5]{\tri{1.6cm}{-4cm*0.5}{270}}
\begin{scope}[xshift=9.5cm,yshift=0cm]{
%\draw [ultra thick] (-1.5,2*0.5) -- (1.5,2*0.5);
%\draw [ultra thick] (-1.5,-2*0.5) -- (1.5,-2*0.5);
\draw [thick, green] (-0.8,5*0.5) -- (0,4.2*0.5);
\draw [thick, green] (0.8,5*0.5) -- (0,4.2*0.5);
\draw [thick, green] (0,-4.2*0.5) -- (-0.8,-5*0.5);
\draw [thick, green] (0,-4.2*0.5) -- (0.8,-5*0.5);
\draw [thick, magenta] (-0.8,-2*0.5) -- (-0.8,-0.1);
\draw [thick, magenta] (-0.8,0.1) -- (-0.8,2*0.5);
\draw [thick, green] (0,2.8*0.5) -- (0.8,2*0.5);
\draw [thick, green] (0,2.8*0.5) -- (-0.8,2*0.5);
\draw [thick, green] (0.8,-2*0.5) -- (0,-2.8*0.5);
\draw [thick, green] (-0.8,-2*0.5) -- (0,-2.8*0.5);
\draw [thick, green] (0,2.8*0.5) -- (0,4.2*0.5);
\draw [thick, green] (0,-4.2*0.5) -- (0,-2.8*0.5);
\draw [thick, magenta] (0.8,-2*0.5) -- (0.8,2*0.5);
\draw [thick] (-0.8,1*0.5) to [out=225,in=180] (-0.8,0);
\draw [thick] (-0.8,0) to [out=0,in=230] (0.8,1*0.5);
\draw [thick] (-1.8,-1*0.5) -- (-0.8,0);
\draw [fill] (-1.8,-1*0.5) circle [radius=0.08];
\node at (-4.8,0) {\tiny{$=\sum_{c,c'}[F^{ab}_{ab}]_{0c}[F^{a'b'}_{a'b'}]_{0c'}$}};
\node at (-0.5,0.4*0.5) {\tiny{$M_i$}}
node at (1.1,0.4*0.5) {\tiny{$M_j$}}
node at (-1.5,-0.2*0.5) {\tiny{$\mathcal{A}$}}
node at (-1,1.7*0.5) {\tiny{$\bar\mu$}}
node at (1.1,1.7*0.5) {\tiny{$\bar\nu$}}
node at (-1,5*0.5) {\tiny{$a$}}
node at (1,5*0.5) {\tiny{$b$}}
node at (-0.4,2.8*0.5) {\tiny{$a$}}
node at (0.4,2.8*0.5) {\tiny{$b$}}
node at (0.2,3.5*0.5) {\tiny{$c$}}
node at (-1,-1.7*0.5) {\tiny{$\mu'$}}
node at (1.1,-1.7*0.5) {\tiny{$\nu'$}}
node at (-1,-5*0.5) {\tiny{$a'$}}
node at (1,-5*0.5) {\tiny{$b'$}}
node at (-0.4,-2.9*0.5) {\tiny{$a'$}}
node at (0.4,-2.9*0.5) {\tiny{$b'$}}
node at (-0.3,-3.5*0.5) {\tiny{$c'$}};
\scalebox{0.5}[0.5]{\tri{-1.6cm+9.5cm}{4cm*0.5}{90-45-20}}
\scalebox{0.5}[0.5]{\tri{-1.6cm+9.5cm}{-4cm*0.5}{270+45+20}}
\scalebox{0.5}[0.5]{\tri{1.6cm+9.5cm}{4cm*0.5}{90+45+20}}
\scalebox{0.5}[0.5]{\tri{1.6cm+9.5cm}{-4cm*0.5}{270-45-20}}
}
\end{scope}
\begin{scope}[xshift=24cm,yshift=0cm]{
%\draw [ultra thick] (-1.5,2*0.5) -- (1.5,2*0.5);
%\draw [ultra thick] (-1.5,-2*0.5) -- (1.5,-2*0.5);
\draw [thick, green] (-0.8,5*0.5) -- (0,4.2*0.5);
\draw [thick, green] (0.8,5*0.5) -- (0,4.2*0.5);
\draw [thick, green] (0,-4.2*0.5) -- (-0.8,-5*0.5);
\draw [thick, green] (0,-4.2*0.5) -- (0.8,-5*0.5);
\draw [thick, green] (0,2*0.5) -- (0,4.2*0.5);
\draw [thick, green] (0,-4.2*0.5) -- (0,-2*0.5);
\draw [thick, magenta] (0,-2*0.5) -- (0,2*0.5);
\draw [thick] (-1,-1*0.5) -- (0,0);
\draw [fill] (-1,-1*0.5) circle [radius=0.08];
\node at (0.4,0) {\tiny{$M_k$}}
node at (0.3,1.7*0.5) {\tiny{$\bar\alpha$}}
node at (-1,5*0.5) {\tiny{$a$}}
node at (1,5*0.5) {\tiny{$b$}}
node at (0.2,3.5*0.5) {\tiny{$c$}}
node at (0.3,-1.7*0.5) {\tiny{$\alpha'$}}
node at (-1,-5*0.5) {\tiny{$a'$}}
node at (1,-5*0.5) {\tiny{$b'$}}
node at (0.3,-3.5*0.5) {\tiny{$c'$}}
node at (-0.7,-0.2*0.5) {\tiny{$\mathcal{A}$}};
\node at (-6.5,0) {\tiny{$=\sum_{c,c'}[F^{ab}_{ab}]_{0c}[F^{a'b'}_{a'b'}]_{0c'}\lambda^{[M_i,M_j,M_k](c\alpha)(a\mu)(b\nu)}_{(c'\alpha')(a'\mu')(b'\nu')}$}};
\scalebox{0.5}[0.5]{\tri{24cm}{4cm*0.5}{90}}
\scalebox{0.5}[0.5]{\tri{24cm}{-4cm*0.5}{270}}
}
\end{scope}
\begin{scope}[xshift=0cm,yshift=-6cm]{
\draw [ultra thick] (-1.5,2*0.5) -- (1.5,2*0.5);
\draw [ultra thick] (-1.5,-2*0.5) -- (1.5,-2*0.5);
\draw [thick, green] (-0.8,5*0.5) -- (-0.8,2*0.5);
\draw [thick, green] (-0.8,-2*0.5) -- (-0.8,-5*0.5);
\draw [thick, magenta] (-0.8,-2*0.5) -- (-0.8,2*0.5);
\draw [thick, green] (0.8,5*0.5) -- (0.8,2*0.5);
\draw [thick, green] (0.8,-2*0.5) -- (0.8,-5*0.5);
\draw [thick, magenta] (0.8,-2*0.5) -- (0.8,2*0.5);
\node at (-0.5,0) {\tiny{$i$}}
node at (1.1,0) {\tiny{$j$}}
node at (-0.5,2.4*0.5) {\tiny{$\mu$}}
node at (1.1,2.4*0.5) {\tiny{$\nu$}}
node at (-0.6,5*0.5) {\tiny{$a$}}
node at (1,5*0.5) {\tiny{$b$}}
node at (-0.5,-2.5*0.5) {\tiny{$\mu'$}}
node at (1.1,-2.5*0.5) {\tiny{$\nu'$}}
node at (-0.6,-5*0.5) {\tiny{$a'$}}
node at (1,-5*0.5) {\tiny{$b'$}};
\begin{scope}[xshift=11.5cm,yshift=0cm]{
\draw [ultra thick] (-1.5,2*0.5) -- (1.5,2*0.5);
\draw [ultra thick] (-1.5,-2*0.5) -- (1.5,-2*0.5);
\draw [thick, green] (-0.8,5*0.5) -- (0,4.2*0.5);
\draw [thick, green] (0.8,5*0.5) -- (0,4.2*0.5);
\draw [thick, green] (0,-4.2*0.5) -- (-0.8,-5*0.5);
\draw [thick, green] (0,-4.2*0.5) -- (0.8,-5*0.5);
\draw [thick, green] (0,2*0.5) -- (0,4.2*0.5);
\draw [thick, green] (0,-4.2*0.5) -- (0,-2*0.5);
\draw [thick, magenta] (0,2*0.5) -- (0,0.8*0.5);
\draw [thick, magenta] (0,-2*0.5) -- (0,-0.8*0.5);
\draw [thick, magenta] (-0.8,0) -- (0,0.8*0.5);
\draw [thick, magenta] (0.8,0) -- (0,0.8*0.5);
\draw [thick, magenta] (-0.8,0) -- (0,-0.8*0.5);
\draw [thick, magenta] (0.8,0) -- (0,-0.8*0.5);
\node at (-5.5,0) {\tiny{$=\sum_{c,c'}[M^{ab;k}_{c;ij}]^{\mu\nu}_{\alpha}\left\{[M^{a'b';k'}_{c';i'j'}]^{\mu'\nu'}_{\alpha'}\right\}^*$}};
\node at (0.4,-1.4*0.5) {\tiny{$k$}}
node at (0.4,1.4*0.5) {\tiny{$k$}}
node at (-0.5,0) {\tiny{$i$}}
node at (1.1,0) {\tiny{$j$}}
node at (0.3,2.5*0.5) {\tiny{$\alpha$}}
node at (-1,5*0.5) {\tiny{$a$}}
node at (1,5*0.5) {\tiny{$b$}}
node at (0.2,3.5*0.5) {\tiny{$c$}}
node at (0.3,-2.5*0.5) {\tiny{$\alpha'$}}
node at (-1,-5*0.5) {\tiny{$a'$}}
node at (1,-5*0.5) {\tiny{$b'$}}
node at (0.3,-3.5*0.5) {\tiny{$c'$}};
}
\end{scope}
\begin{scope}[xshift=25cm,yshift=0cm]{
\draw [ultra thick] (-1.5,2*0.5) -- (1.5,2*0.5);
\draw [ultra thick] (-1.5,-2*0.5) -- (1.5,-2*0.5);
\draw [thick, green] (-0.8,5*0.5) -- (0,4.2*0.5);
\draw [thick, green] (0.8,5*0.5) -- (0,4.2*0.5);
\draw [thick, green] (0,-4.2*0.5) -- (-0.8,-5*0.5);
\draw [thick, green] (0,-4.2*0.5) -- (0.8,-5*0.5);
\draw [thick, green] (0,2*0.5) -- (0,4.2*0.5);
\draw [thick, green] (0,-4.2*0.5) -- (0,-2*0.5);
\draw [thick, magenta] (0,-2*0.5) -- (0,2*0.5);
\node at (0.4,0) {\tiny{$k$}}
node at (0.3,1.5*0.5) {\tiny{$\alpha$}}
node at (-1,5*0.5) {\tiny{$a$}}
node at (1,5*0.5) {\tiny{$b$}}
node at (0.2,3.5*0.5) {\tiny{$c$}}
node at (0.3,-1.5*0.5) {\tiny{$\alpha'$}}
node at (-1,-5*0.5) {\tiny{$a'$}}
node at (1,-5*0.5) {\tiny{$b'$}}
node at (0.3,-3.5*0.5) {\tiny{$c'$}};
\node at (-6,0) {\tiny{$=\sum_{c,c'}[M^{ab;k}_{c;ij}]^{\mu\nu}_{\alpha}\left\{[M^{a'b';k'}_{c';i'j'}]^{\mu'\nu'}_{\alpha'}\right\}^*\sqrt{\frac{d_id_j}{d_k}}$}};
}
\end{scope}
}
\end{scope}
\end{tikzpicture}
\ee
\end{figure}

It is not very convenient to solve for $M$ using the above relation.  It is often easier to solve for $M$ directly based on its consistency conditions that is the analogue of the pentagon equation. 
Therefore, we generalize the consistency condition for purely bosonic condensate in \cite{cong_topological_2016} to accommodate fermionic condensates, 
where the M6J symbols would satisfy a twisted form of such a consistency  identity.  The major difference is to recognize that the junction at which a bulk anyon enters the gapped boundary is precisely described by the condensation maps that we have defined in (\ref{fig:condensation_map}) and (\ref{fig:condensation_mapA}). They come in fermion parity even and odd versions in the presence of a fermionic condensate, and one has to keep track of the ordering of these junctions, similar to the derivation of the super-pentagon identity. 

\begin{figure}[htbp]
\centering
\be\label{eq:M_super_pentagon}
\begin{tikzpicture}
\draw [thick] (2.2,0.5) to [out=90, in=90] (3.2,0.5);
\draw [thick] (2.2,0) -- (2.2,0.5);
\draw [thick] (3.2,0) -- (3.2,0.5);
\draw [thick] (2.7,0.8) -- (2.7,2.2);
\draw [thick] (3.7,0) -- (3.7,3);
\draw [ultra thick] (1.7,1.5) -- (4.2,1.5);
\draw [thick] (2.2,2.5) -- (2.2,3);
\draw [thick] (3.2,2.5) -- (3.2,3);
\draw [thick] (2.2,2.5) to [out=270, in=270] (3.2,2.5);
\node at (2,3) {$a$}
node at (3,3) {$b$}
node at (3.5,3) {$c$}
node at (2,0) {$x$}
node at (3,0) {$y$}
node at (3.5,0) {$z$}
node at (2.5,1) {$w$}
node at (2.5,2) {$e$}
node at (2.9,1.3) {$\sigma$}
node at (3.9,1.3) {$\lambda$}
node at (2.7, 0.6) {$\omega$};

\draw [fill,red] (2.7,1.5) circle [radius=0.05];
\draw [fill,red] (3.7,1.5) circle [radius=0.05];
\draw [fill,red] (2.7,0.8) circle [radius=0.05];

\draw [thick] (7.8,0.3) to [out=90, in=90] (8.8,0.3);
\draw [thick] (7.8,-0.2) -- (7.8,0.3);
\draw [thick] (8.8,-0.2) -- (8.8,0.3);
\draw [thick] (8.3,0.8) to [out=90, in=90] (9.3,0.8);
\draw [thick] (9.3,-0.2) -- (9.3,0.8);
\draw [thick] (8.3,0.6) -- (8.3,0.8);
\draw [thick] (8.8,1.1) -- (8.8,1.9);
\draw [ultra thick] (7.3,1.5) -- (9.8,1.5);
\draw [thick] (8.3,2.2) to [out=270, in=270] (9.3,2.2);
\draw [thick] (7.8,2.7) to [out=270, in=270] (8.8,2.7);
\draw [thick] (7.8,2.7) -- (7.8,3.2);
\draw [thick] (8.8,2.7) -- (8.8,3.2);
\draw [thick] (9.3,2.2) -- (9.3,3.2);
\draw [thick] (8.3,2.2) -- (8.3,2.4);
\node at (7.6,-0.2) {$x$}
node at (8.6,-0.2) {$y$}
node at (9.1,-0.2) {$z$}
node at (8.3,0.3) {$\omega$}
node at (8.8,0.8) {$\zeta$}
node at (8.1,0.8) {$w$}
node at (8.6,1.2) {$v$}
node at (9,1.4) {$\phi$}
node at (8.6,1.8) {$d$}
node at (8.1,2.2) {$e$}
node at (7.6,3.2) {$a$}
node at (8.6,3.2) {$b$}
node at (9.1,3.2) {$c$};
\draw [fill,red] (8.8,1.5) circle [radius=0.05];
\draw [fill,red] (8.8,1.1) circle [radius=0.05];
\draw [fill,red] (8.3,0.6) circle [radius=0.05];

\draw [thick] (0.5,4) -- (0.5,7);
\draw [thick] (1.25,4) -- (1.25,7);
\draw [thick] (2,4) -- (2,7);
\draw [ultra thick] (0,5.5) -- (2.5,5.5);
\node at (0.3,4) {$x$}
node at (0.3,7) {$a$}
node at (1.05,4) {$y$}
node at (1.05,7) {$b$}
node at (1.8,4) {$z$}
node at (1.8,7) {$c$}
node at (0.7,5.3) {$\mu$}
node at (1.45,5.3) {$\nu$}
node at (2.2,5.3) {$\lambda$};
\draw [fill,red] (0.5,5.5) circle [radius=0.05];
\draw [fill,red] (1.25,5.5) circle [radius=0.05];
\draw [fill,red] (2,5.5) circle [radius=0.05];

\draw [ultra thick] (4.5,7.5) -- (7,7.5);
\draw [thick] (5,6) -- (5,9);
\draw [thick] (5.5,6.5) to [out=90, in=90] (6.5,6.5);
\draw [thick] (5.5,6) -- (5.5,6.5);
\draw [thick] (6.5,6) -- (6.5,6.5);
\draw [thick] (6,6.8) -- (6,8.2);
\draw [thick] (5.5,8.5) -- (5.5,9);
\draw [thick] (6.5,8.5) -- (6.5,9);
\draw [thick] (5.5,8.5) to [out=270, in=270] (6.5,8.5);
\node at (5.2,7.3) {$\mu$}
node at (6.2,7.3) {$\psi$}
node at (4.8,6) {$x$}
node at (4.8,9) {$a$}
node at (5.3,6) {$y$}
node at (6.3,6) {$z$}
node at (5.3,9) {$b$}
node at (6.3,9) {$c$}
node at (5.8,7) {$u$}
node at (5.8,7.9) {$f$}
node at (6,6.5) {$\kappa$};
\draw [fill,red] (5,7.5) circle [radius=0.05];
\draw [fill,red] (6,7.5) circle [radius=0.05];
\draw [fill,red] (6,6.8) circle [radius=0.05];

\draw [ultra thick] (9,5.5) -- (11.5,5.5);
\draw [thick] (10,4.3) to [out=90, in=90] (11,4.3);
\draw (10,3.8) -- (10,4.3);
\draw (11,3.8) -- (11,4.3);
\draw (10.5,4.6) -- (10.5,4.8);
\draw [thick] (9.5,4.8) to [out=90, in=90] (10.5,4.8);
\draw (9.5,3.8) -- (9.5,4.8);
\draw (10,5.1) -- (10,5.9);
\draw [thick] (9.5,6.2) to [out=270, in=270] (10.5,6.2);
\draw (9.5,6.2) -- (9.5,7.2);
\draw (10.5,6.2) -- (10.5,6.4);
\draw [thick] (10,6.7) to [out=270, in=270] (11,6.7);
\draw (10,6.7) -- (10,7.2);
\draw (11,6.7) -- (11,7.2);
\node at (10.2,5.4) {$\phi$}
node at (9.3,3.8) {$x$}
node at (9.8,3.8) {$y$}
node at (10.8,3.8) {$z$}
node at (9.3,7.2) {$a$}
node at (9.8,7.2) {$b$}
node at (10.8,7.2) {$c$}
node at (9.8,5.2) {$v$}
node at (9.8,5.8) {$d$}
node at (10.3,4.7) {$u$}
node at (10.3,6.2) {f}
node at (10.5,4.4) {$\kappa$}
node at (10,4.9) {$\rho$};
\draw [fill,red] (10,5.5) circle [radius=0.05];
\draw [fill,red] (10,5.1) circle [radius=0.05];
\draw [fill,red] (10.5,4.6) circle [radius=0.05];

\draw [thick,->] (4.6,1.5) -- (6.9,1.5);
\draw [thick,->] (10.2,1.5) to [out=0,in=270] (11,3.5);
\draw [thick,->] (0.5,3.5) to [out=270,in=180] (1.3,1.5);
\draw [thick,->] (2.5,6.5) to [out=60,in=180] (4.2,7.5);
\draw [thick,->] (7.3,7.5) to [out=0,in=120] (9,6.5);

\node at (5.7,1.9) {$[M^{ec;v}_{d;wz}]^{\sigma\lambda}_{\phi\zeta}$}
node at (0,1.3) {$[M^{ab;w}_{e;xy}]^{\mu\nu}_{\sigma\omega}$}
node at (2.8,7.8) {$[M^{bc;u}_{f;yz}]^{\nu\lambda}_{\psi\kappa}$}
node at (8.7,7.8) {$[M^{af;v}_{d;xu}]^{\mu\psi}_{\phi\rho}$}
node at (12,1.3) {$F^{abc}_{d;ef}([F^{xyz}_{v;w u}]^{\omega\zeta}_{\kappa\rho})^\dagger$};
\end{tikzpicture}
\ee
\end{figure}
Similarly to bosonic condensation the M6J symbol for fermion condensation carries several groups of indices. First, there are labels of bulk anyons  $a,b,c$. Second, there are boundary excitations $x,y,z$ they condense to and third, condensation channel labels $\mu,\nu,\lambda$.  In addition to those, it has an extra index to label the fusion channels of the boundary excitations. Then we introduce $s^a_x(\mu),s^{xy}_w(\omega)$ to denote the parity of the condensation channel and fusion channel respectively, 0 for even and 1 for odd. To avoid dependence of fermionic wave functions on the ordering of odd channels , we introduce one ``Majorana number''  $\theta_{\underline{x}}$ on each vertex $\underline{x}$ of condensation and also each fusion vertex of the boundary excitations, which are denoted by red dots in the above diagram. 
These Majorana numbers satisfy

\begin{align*}
&\theta^2_{\underline{x}}=1\\
&\theta_{\underline{x}}\theta_{\underline{y}}=-\theta_{\underline{y}}\theta_{\underline{x}}\\
&\theta_{\underline{x}}^\dagger=\theta_{\underline{x}}
\end{align*}

In [zheng-cheng gu, zhenghan wang and xiao-gang wen, 1010.1517] they introduce 6j symbols that carry the Majorana numbers along. They are defined as
\be
[\mathcal{F}^{xyz}_{v;wu}]^{\omega\zeta}_{\kappa\rho}=\theta^{s^{xy}_w(\omega)}_{\underline{\omega}}\theta^{s^{wz}_v(\zeta)}_{\underline{\zeta}}\theta^{s^{xu}_v(y)}_{\underline{y}}\theta^{s^{yz}_u(\kappa)}_{\underline{\kappa}}[F^{xyz}_{v;wu}]^{\omega\zeta}_{\kappa\rho}
\ee

Similarly, one could define a new M-tensor carrying the Majorana numbers:

\be
[\mathcal{M}^{bc;u}_{f;yz}]^{\nu\lambda}_{\psi\kappa}=(\theta^{s^b_y(\nu)}_{\underline{\nu}}\theta^{s^c_z(\lambda)}_{\underline{\lambda}}\theta^{s^f_u(\psi)}_{\underline{\psi}}\theta^{s^{yz}_u(\kappa)}_{\underline{\kappa}})^\dagger[M^{bc;u}_{f;yz}]^{\nu\lambda}_{\psi\kappa}
\ee

Therefore the pentagon identity for M6J symbols with Majorana numbers are given by

\be
\sum_{e,\epsilon,\sigma\omega,\zeta}\mathcal{F}^{abc}_{d;ef}([\mathcal{F}^{xyz}_{v;wu}]^{\omega\zeta}_{\kappa\rho})^\dagger[\mathcal{M}^{ec;v}_{d;wz}]^{\sigma\lambda}_{\phi\zeta}[\mathcal{M}^{ab;w}_{e;xy}]^{\mu\nu}_{\sigma\omega}\simeq\sum_\psi[\mathcal{M}^{af;v}_{d;xu}]^{\mu\psi}_{\phi\rho}[\mathcal{M}^{bc;u}_{f;yz}]^{\nu\lambda}_{\psi\kappa}
\ee

The order of M and F tensors now matters because they carry with them Majorana numbers.  
Finally by removing the Majorana numbers we arrive at the fermionic pentagon identity for M6J symbols

\be
\sum_{e,w,\sigma,\omega,\zeta}[M^{ab;w}_{e;xy}]^{\mu\nu}_{\sigma\omega}[M^{ec;v}_{d;wz}]^{\sigma\lambda}_{\phi\zeta}F^{abc}_{d;ef}([F^{xyz}_{v;wu}]^{\omega\zeta}_{\kappa\rho})^\dagger=(-1)^{s^a_x(\mu)s^{yz}_u(\kappa)}\sum_\psi[M^{bc;u}_{f;yz}]^{\nu\lambda}_{\psi\kappa}[M^{af;v}_{d;xu}]^{\mu\psi}_{\phi\rho}.
\ee

\vspace{1cm}
{\bf \underline{Vertex Lifting Coefficients  (VLC)}}

VLC's were introduced in \cite{Eliens:2013epa}. They are linear maps that map fusion basis in the bulk theory to the boundary fusion basis.
This is illustrated in (\ref{eq:defineVLC}).
\begin{equation} \label{eq:defineVLC}
    \mathtikzS{0.4}{
        \vertexILC{0}{0}{2cm}{$X$}{$Y$}{$Z$}{\modCol};
    }
    = \quad\sum_{i,j,k} \quad
    \begin{bmatrix}
        X & Y & Z \\
        i & j & k
    \end{bmatrix}\
    \mathtikzS{0.4}{
        \vertexILC{0}{0}{2cm}{$i$}{$j$}{$k$}{\bulkCol};
    }
\end{equation}
Note that the notation introduced there is applicable for all $W_{i x} \in \{0,1\}$. 

These VLC's can be separated into three classes. 

First, there are the ``vacuum vertex'' where three vacuum lines in the boundary theory meet. These vertices are precisely defining the product and co-product of the condensed algebra $\mathcal{A}$.

Then, there are vertices where the boundary vacuum line meets a boundary excitation, leaving it ``invariant''. These vertices are precisely defining the left (right) action of the algebra on the module corresponding to the boundary excitation. 

Finally, there are three boundary excitations meeting at a vertex, defining fusion maps in the condensed theory.  Fusion of modules are defined in (\ref{eq:fuse_modules}). These can be decomposed in terms of irreducible (simple) modules using (\ref{eq:extract_lambda}) which we have discussed. 
These decomposition coefficients can be related to the VLC's defining the fusion map as illustrated in (\ref{eq:lambda_VLC}) by mapping them to the parent theory, and compare fusion basis by basis. 
This connection with \cite{Eliens:2013epa} is valid when we restrict to the situation where  $W_{i x} \in \{0,1\}$, and so the channel labels do not feature here.  \footnote{We have not worked very hard to ensure that the normalisation implied in the relation shown is identical to the normalisation taken in \cite{Eliens:2013epa}. But to work it out is straightforward, and beside the point.}

\begin{figure}[htbp]
\centering
\be \label{eq:lambda_VLC}
\begin{tikzpicture}[scale=0.7]
\scalebox{0.7}[0.7]{
\draw [thick, magenta] (0,0.1) -- (0,2);
\draw [thick, magenta] (0,-1) -- (0,-0.4);
\draw [thick, magenta] (0,-0.2) -- (0,-0.1);
\draw [thick, magenta] (2,-1) -- (2,2);
\draw [thick, green] (1,3) -- (0,2);
\draw [thick, green] (1,3) -- (2,2);
\draw [thick, green] (0,-1) -- (1,-2);
\draw [thick, green] (2,-1) -- (1,-2);
\draw [thick, green] (1,4) -- (1,3);
\draw [thick, green] (1,-3) -- (1,-2);
\node at (0.4,1.5) {$M_1$}
node at (2.4,1.5) {$M_2$}
node at (1.2,4) {$i$}
node at (1.2,-3) {$i'$}
node at (0.5,2.8) {$j$}
node at (1.5,2.8) {$k$}
node at (0.5,-1.9) {$j'$}
node at (1.5,-1.9) {$k'$};
\scalebox{0.5}[0.5]{\tri{0}{4cm}{45}}
\scalebox{0.5}[0.5]{\tri{4cm}{4cm}{45+90}}
\scalebox{0.5}[0.5]{\tri{4cm}{-2cm}{45-180}}
\scalebox{0.5}[0.5]{\tri{0}{-2cm}{45-90}}
\draw [thick] (0,1) to [out=225,in=180] (0,0);
\draw [thick] (0,0) to [out=0,in=230] (2,1);
\draw [thick] (0.3,0) -- (-0.7,-1);
\draw [fill] (-0.7,-1) circle [radius=0.08]; 
\node at (-0.5,-0.3) {$\mathcal{A}$};
\begin{scope}[xshift=8cm,yshift=0cm]{
\node at (0.2,4) {$i$}
node at (0.2,-3) {$i'$}
node at (-3,0.5) {$\tiny{=\sum_{M_3}\lambda^{[M_1,M_2,M_3]ijk}_{i'j'k'}}$}
node at (-0.5,-0.3) {$\mathcal{A}$}
node at (0.4,0.5) {$M_3$};
\draw [thick,green] (0,4) -- (0,-3);
\draw [thick] (-1,-0.5) -- (0,0.5);
\draw [fill] (-1,-0.5) circle [radius=0.08];
\draw [thick, magenta] (0,-1) -- (0,2);
\scalebox{0.5}[0.5]{\tri{8cm}{3.8cm}{90}}
\scalebox{0.5}[0.5]{\tri{8cm}{-1.8cm}{270}}
}
\end{scope}
\begin{scope}[xshift=21cm,yshift=0cm]{
\node at (-6,0.5) {$\tiny{=\sum_{M_3,a}\rho^{M_1j}_{aj'}\rho^{M_2k}_{ak'}\begin{bmatrix}
 M_1&M_2&M_3\\ 
 j&k&i 
\end{bmatrix}
\begin{bmatrix}
 M_1&M_2&M_3\\ 
 j'&k'&i'
\end{bmatrix}^*}$};
\draw [thick, green] (0,0.1) -- (0,1.5);
\draw [thick, green] (0,-0.6) -- (0,-0.1);
\draw [thick, green] (2,-0.6) -- (2,1.5);
\draw [thick, green] (1,2.5) -- (0,1.5);
\draw [thick, green] (1,2.5) -- (2,1.5);
\draw [thick, green] (0,-0.6) -- (1,-1.6);
\draw [thick, green] (2,-0.6) -- (1,-1.6);
\draw [thick, green] (1,3.5) -- (1,2.5);
\draw [thick, green] (1,-2.6) -- (1,-1.6);
\draw [thick, green] (1,3.5) -- (1,4);
\draw [thick, green] (1,-2.6) -- (1,-3);
\node at (1.2,4) {$i$}
node at (1.2,-3) {$i'$}
node at (0,2) {$j$}
node at (2,2) {$k$}
node at (0,-1) {$j'$}
node at (2,-1) {$k'$};
\draw [thick, green] (0,1) to [out=225,in=180] (0,0);
\draw [thick, green] (0,0) to [out=0,in=230] (2,1);
\node at (0.5,-0.2) {$a$};
}
\end{scope}
}
\end{tikzpicture}
\ee
\end{figure}	

Note that the right most diagram in (\ref{eq:lambda_VLC}) is nothing but the expansion of the following diagram (\ref{fig:bais_bubble}) in basis form. 

\begin{figure}[htbp]
\centering
\be \label{fig:bais_bubble}
\begin{tikzpicture}[scale=0.6]
\draw [thick, magenta] (0,0.1) -- (0,1.5);
\draw [thick, magenta] (0,-0.6) -- (0,-0.1);
\draw [thick, magenta] (2,-0.6) -- (2,1.5);
\draw [thick, magenta] (1,2.5) -- (0,1.5);
\draw [thick, magenta] (1,2.5) -- (2,1.5);
\draw [thick, magenta] (0,-0.6) -- (1,-1.6);
\draw [thick, magenta] (2,-0.6) -- (1,-1.6);
\draw [thick, magenta] (1,3.5) -- (1,2.5);
\draw [thick, magenta] (1,-2.6) -- (1,-1.6);
\node at (0.4,0.5) {\tiny{$M_1$}}
node at (2.4,0.5) {\tiny{$M_2$}}
node at (1.4,3) {\tiny{$M_3$}}
node at (1.4,-2) {\tiny{$M_3$}};
\draw [thick] (0,1) to [out=225,in=180] (0,0);
\draw [thick] (0,0) to [out=0,in=230] (2,1);
\node at (-0.5,-0.3) {\tiny{$\mA$}};
\node at (-1.5,0.5) {$\tiny{\sum_{M_3}}$};
\end{tikzpicture}
\ee
\end{figure}	

A lesson learned here is that the defining property of the condensed theory is basically the product of the algebra $\mathcal{A}$, and its left (right) modules, from which everything else derives. The precise mathematical formulation also allows extension to cases where $W_{ix}>1 $ rather seamlessly.

\section{Illustrating with examples}

Having developed the formal computational tools based on (super)-Frobenius algebra and their modules, here we would like to illustrate these tools in explicit examples, and in the process, understanding interesting features of boundaries.

\subsection{Beginner's level -- the Toric code} \label{sec:toric}

The toric code is the paradigmatic example of a bosonic topological order in 2+1 dimensions. We are going to see that most of the important physics of gapped boundaries and junctions can already be understood here. 

It has four excitations, $\{1,e,m,f\}$. 
As it is well known, there are {\it three} kinds of gapped boundaries for the toric code topological order in 2+1 dimensions. 
Among these boundaries, two of them are conventional ones obtained from condensing bosons. Specifically one is termed the electric boundary where the electric charges condense (i.e. $\mathcal{A}_e = 1 \oplus e$), and the other, termed the magnetic boundary where the magnetic charges condense (i.e. $\mathcal{A}_m = 1\oplus m$). 

Here we would like to discuss in detail the third type of gapped boundary following from condensing the $e-m$ bound state which is a fermion.   This has been mentioned before in \cite{Bhardwaj:2016clt}. 
We will also study junctions between these boundaries. 

  \subsubsection{The fermion condensate} \label{sec:toricf_bc}
      
      The fermionic Frobenius algebra is given by
      \be
      \mathcal{A}_f = 1 \oplus f.
      \ee
      
      The boundary excitations are characterized by 
      \be
      X_f = e\oplus m.
      \ee
     We can summarize this data in terms of the $W$ and $\Omega$ matrices :
     \begin{eqnarray}
     W=
     \begin{blockarray}{ccccc}
     1 & e & m & f\\
     \begin{block}{(cccc)l}
     1 & 0 & 0 & 1 & ~\mathcal{A}_f\\
     0 & 1 & 1 & 0 & ~X_f\\
     \end{block}
     \end{blockarray}
     ~~~~\&~~~~\Omega=
     \begin{blockarray}{ccccc}
     1 & e & m & f\\
     \begin{block}{(cccc)l}
     1 & 0 & 0 & -1 & ~\mathcal{A}_f\\
     0 & 1 & -1 & 0 & ~X_f\\
     \end{block}
     \end{blockarray}
     \end{eqnarray}

The fusion rules can be obtained using (\ref{eq:evenfuse}). 
   
     \begin{longtable}{l|ll}
     $\otimes$ & $\mathcal{A}_f$ & $X_f$\\\hline
     $\mathcal{A}_f$ & $\mathcal{A}_f$ & $X_f$\\
     $X_f$ & $X_f$  & $\mathcal{A}_f$\\
     \end{longtable}    
      One should also check that $X_f$ is a non-q-type object with trivial endomorphism. Further, one can check that $X_f$ is responsible for generating the fermion parity. The number of objects it contains as a module of $\mathcal{A}_f$ is equals 2. This is the same as the total number of non-q-type defects in the gapped boundary -- which is also 2 (where the "trivial defect" has to be included.)   This is a confirmation of the claim made after  (\ref{eq:littlev}).\\   
      The 6j symbols of the condensed phase can be read off following the discussion in the previous section. They are given by $F^{\mathcal{A}_f\mathcal{A}_f\mathcal{A}_f}_{\mathcal{A}_f;\mathcal{A}_f\mathcal{A}_f}=F^{X_f\mathcal{A}_f\mathcal{A}_f}_{X_f;X_f\mathcal{A}_f}=F^{\mathcal{A}_fX_f\mathcal{A}_f}_{X_f;X_fX_f}=F^{\mathcal{A}_f\mathcal{A}_fX_f}_{X_f;\mathcal{A}_fX_f}=F^{X_fX_f\mathcal{A}_f}_{\mathcal{A}_f;\mathcal{A}_fX_f}=F^{X_f\mathcal{A}_fX_f}_{0;X_fX_f}=F^{\mathcal{A}_fX_fX_f}_{\mathcal{A}_f;X_f\mathcal{A}_f}=F^{X_fX_fX_f}_{X_f;\mathcal{A}_f\mathcal{A}_f}=1$

      \subsubsection{The Bosonic-Fermionic junctions} \label{sec:bfjunction}
      
    As alluded to in the previous subsection, the toric code model admits two bosonic gapped boundaries that correspond to the electric $\mathcal{A}_e$ and magnetic $\mathcal{A}_m$ condensates. i.e.
      \be
      \mathcal{A}_e = 1 \oplus e, \qquad \mathcal{A}_m = 1 \oplus m. 
      \ee
For completeness, let us recall also that in each of these bosonic boundaries, there is one non-trivial excitation. 
Let us denote the one in the electric boundary by $X_e$ and that in the magnetic boundary by $X_m$. They are
given by 
\be
X_e = m\oplus f,  \qquad X_m = e\oplus f.
\ee
One can readily check using (\ref{eq:evenfuse}) that they satisfy a $\mathbb{Z}_2$ fusion rules. 

We would like to consider junctions between these bosonic boundaries with the fermionic boundary introduced in the previous sub section. 

First, we consider the $e-f$ junction. Results of the $m-f$ would follow in a completely analogous manner. 
By considering the ``induced'' bimodule $\mathcal{A}_e \otimes c_i \otimes \mathcal{A}_f$, we find that there is only one excitation  $X_{ef}$ localized at the $e-f$ junction. i.e. The four different anyons $c_i \in \{1,e,m,f\}$ in the toric code model would generate exactly the same bimodule. i.e.
\be
X_{ef} = 1\oplus e \oplus m \oplus f, \qquad \textrm{i.e. } W^{\mathcal{A}_e| \mathcal{A}_f}_{c_i X_{ef}} = 1, \forall i
\ee
We note that $W^{\mathcal{A}_e| \mathcal{A}_f}_{c_i X_{ef}} = W^{\mathcal{A}_f| \mathcal{A}_e}_{c_i X_{fe}}$. 
 Since this is an Abelian model, one can work out the fusions readily using (\ref{eq:bimodule_fuse}).  
 The fusion rules are given by
\be
\begin{aligned} \label{eq:ef_fuse}
&X_{ef}\otimes_{\mathcal{A}_f} X_{fe}= \mathcal{A}_e \oplus X_e\\
&X_{fe}\otimes_{\mathcal{A}_f} X_{ef}=  \mathcal{A}_f \oplus X_f
\end{aligned}
\ee
Where $X_e$ and $X_f$ are non-trivial excitations of $e$ and $f$ boundaries respectively. From these fusion rules we can conclude that the quantum dimension of this excitation is $\sqrt{2}$.\\

We note that (\ref{eq:ef_fuse}) takes the same form as that in the fusion of defects localized at the $e-m$ junction. There, 
one could also readily check that
\be
\begin{aligned} \label{eq:em_fuse}
&X_{em}\otimes X_{me}= \mathcal{A}_e \oplus X_e,\\
&X_{me}\otimes X_{em}=  \mathcal{A}_m \oplus X_m,  \qquad X_{em} = X_{me} = 1\oplus e \oplus m\oplus f.
\end{aligned}
\ee
Therefore it is known that $X_{em}$ also has quantum dimension $ \sqrt{2}$ \cite{Barkeshli:2013yta, barkeshli_classification_2013}.

Recall in section \ref{sec:fusion_junc} that there is generically a subtlety regarding Majorana modes, and that in the computation above one should make the replacement  
\be \label{eq:AftoAfp}
\mathcal{A}_f \to \mathcal{A}_f' \equiv 1 \oplus f \otimes \psi_0.
\ee 
This does not affect the conclusion in (\ref{eq:ef_fuse}), or any of the fusion rules in a single gapped boundary -- all it does is to tag an odd fusion channel by an explicit factor of $\psi_0$. 

It however makes a crucial difference below as we are going to see.

Consider the fusion of excitations in different types of junctions. Specifically, this is illustrated in figure \ref{fig:ef-fm}.
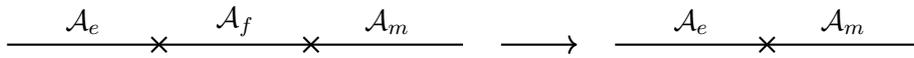
\begin{figure}[h]
\centering
\begin{tikzpicture}
\draw [thick] (0,0) -- (6,0);
\draw [thick] (1.9,-0.1) -- (2.1,0.1);
\draw [thick] (1.9,0.1) -- (2.1,-0.1);
\draw [thick] (3.9,-0.1) -- (4.1,0.1);
\draw [thick] (3.9,0.1) -- (4.1,-0.1);
\node at (1,0.3) {$\mathcal{A}_e$};
\node at (3,0.3) {$\mathcal{A}_f$};
\node at (5,0.3) {$\mathcal{A}_m$};
\draw [thick] (8,0) -- (12,0);
\draw [thick] (9.9,-0.1) -- (10.1,0.1);
\draw [thick] (9.9,0.1) -- (10.1,-0.1);
\draw [thick,->] (6.5,0) -- (7.5,0);
\node at (9,0.3) {$\mathcal{A}_e$};
\node at (11,0.3) {$\mathcal{A}_m$};
\end{tikzpicture}
\caption{Fusion of $ef$ and $fm$ junctions. }
\label{fig:ef-fm}
\end{figure}\\

We might expect the following fusion rule
\be \label{eq:fuse_effm1}
X_{ef}\otimes X_{fm}=\#\cdot X_{em}
\ee
where $\#$ should be some positive integer. But one could readily see that this is not possible if quantum dimensions are conserved
in the process of fusion -- which it should -- to ensure that the counting of ground state degeneracy a robust topological number.
This is because using the methods above, we claimed that all the defects should have quantum dimension $\sqrt{2}$, so that $\#$
could not possibly be an integer. 

Now we reconsider (\ref{eq:fuse_effm1}) by introducing the free fermions $\psi_0$. The fusion described in (\ref{eq:fuse_effm1}) can now be computed as follows:
\begin{align} \label{eq:fuse_effm2}
&[(1\oplus e )\otimes (1\oplus f \otimes \psi_0)] \otimes_{1 \oplus f\psi_0} [(1 \oplus f \otimes \psi_0) \otimes (1+ m)]  \nonumber \\
&= (1\oplus e \oplus m \oplus f)  \otimes (1\oplus f\otimes \psi_0) \nonumber  \\
&= (1\oplus e \oplus m \oplus f) \otimes (1\oplus \psi_0).
\end{align}

This shows that we obtain at the $e-m$ junction $X_{em}$ and also a Majorana mode $(1 \oplus
\psi_0)$ -- here this has to be interpreted as such since it is localized at the $e-m$ junction!
This may appear somewhat mysterious. To elucidate the physics, we demonstrate it using two different methods.

First, let us study lattice model of the toric code and also explicit constructions of its gapped boundaries. 
It is convenient to describe these boundaries using the Wen-Plaquette version of the toric code topological order\cite{Wen:2003yv}, as had been thoroughly discussed in \cite{Yu:2012eu}.

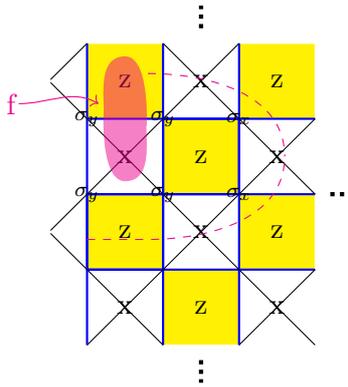
\begin{figure}[h]
\centering
\begin{tikzpicture}
\draw [white, fill=yellow] (1,-1) -- (1,0) -- (2,0) -- (2,-1) -- (1,-1);
\draw [white, fill=yellow] (0,2) -- (0,3) -- (1,3) -- (1,2) -- (0,2);
\foreach \y in {0,2}
\draw [white, fill=yellow] (2,\y) -- (2,\y+1) -- (3,\y+1) -- (3,\y) -- (2,\y);
\foreach \x in {0,1}
\foreach \y in {0}
\draw [blue, thick, fill=yellow] (\x,\y+\x) -- (\x,1+\y+\x) -- (\x+1,1+\y+\x) -- (\x+1,\y+\x) -- (\x,\y+\x);
\foreach \x in {0,1,2}
\draw [blue, thick] (\x,-1) -- (\x,3);
\foreach \y in {0,1,2}
\draw [blue, thick] (0,\y) -- (3,\y);
\foreach \y in {-0.5,1.5}
\node at (0.5,\y) {x} 
node at (1.5,\y) {z}
node at (2.5,\y) {x} 
node at (0.5,\y+1) {z} 
node at (1.5,\y+1) {x}
node at (2.5,\y+1) {z};
\node at (0,2) {\scriptsize$\sigma_y$}
node at (1,2) {\scriptsize$\sigma_y$}
node at (2,2) {\scriptsize$\sigma_x$}
node at (2,1) {\scriptsize$\sigma_x$}
node at (1,1) {\scriptsize$\sigma_y$}
node at (0,1) {\scriptsize$\sigma_y$};
\draw (-0.48,0.52) -- (2,3);
\draw (0,-1) -- (3,2);
\draw (2,-1) -- (3,0);
\draw (1,3) -- (3,1);
\draw (-0.48,2.48) -- (3,-1);
\draw (-0.48,0.48) -- (1,-1);
\draw (-0.48,2.52) -- (0,3);
\draw [ultra thick, magenta, fill=magenta, opacity=0.5] (0.5,1.2) to [out=180, in=90] (0.25,2) to [out=90, in=180] (0.5,2.8) to [out=0, in=270] (0.75,2) to [out=270, in=0] (0.5,1.2);
\node [magenta] at (-1,2.2) {f};
\draw [->, magenta] (-0.9,2.2) to [out=0,in=150] (0.15,2.2);
\draw [dashed, magenta] (0.8,2.6) to [out=0, in= 90] (2.6,1.5);
\draw [dashed, magenta] (0,0.4) -- (0.8,0.4) to [out=0, in= 270] (2.6,1.5);
\draw [dotted, ultra thick] (1.5,3.2) -- (1.5,3.5);
\draw [dotted, ultra thick] (1.5,-1.2) -- (1.5,-1.5);
\draw [dotted, ultra thick] (3.2,1) -- (3.5,1);
\end{tikzpicture}
\caption{Both the Kitaev model \cite{kitaev_fault-tolerant_2003} and the Wen Plaquette model \cite{Wen:2003yv} realize the toric code topological order.  They are illustrated in the same picture here. The black lines denote the lattice of Kitaev's toric code model and the blue lines denote the lattice of the Wen plaquette model \cite{Wen:2003yv}.
Note that in the former, the spin-degrees of freedom lives on the links, whereas in the latter, they live on the vertices. Therefore, where the black lines intersect the blue lattice lives a spin 1/2 degree of freedom. The plaquettes are divided into two sets, the $Z$ and $X$ plaquettes. The Hamiltonian acts in a way depending on this division, as reviewed briefly in (\ref{eq:Wen_H}).}
\label{fig:WenPlaq_Kitaev}
\end{figure}

For completeness, the Hamiltonian (viewed from the perspective of the Wen Plaquette model) is reproduced here
\be
\begin{aligned} \label{eq:Wen_H}
&H=-\sum_{i\in Z\ plaquette}\hat{Z_i}-\sum_{i\in X\ plaquette}\hat{X_i}\\
&\hat{X_i}=\prod_{e\in sites\ around\ plauette\ X_i}\sigma_e^x\,, & \, \hat{Z_i}=\prod_{e \in sites\ around\ plauette\ Z_i}\sigma_e^z
\end{aligned}
\ee
where $\sigma's$ are Pauli matrices. By acting a $\sigma^z$ operator on a vertex, a pair of $e$ can be created in two adjacent $X$ plaquettes. Similarly by acting a $\sigma^x$ on a vertex, a pair of $m$ will be created in two adjacent $Z$ plaquettes.\\ 
The fermion gapped boundary appears as a ``smooth'' boundary in the blue lattice. 
A ``smooth'' fermionic boundary on the Wen plaquette model was discussed in \cite{Yu:2012eu}. To visualize the boundary modes, it is most convenient to fermionize the boundary spin degrees of freedom, and turn it into a set of Majorana modes $\{c_i\}$, one at each boundary vertex $i$. 
As a check,  a fermion string operator can be applied at the boundary as shown in the figure,  showing that individual $f$ anyon can be created or destroyed there -- justifying the claim that $f$ condenses at the boundary  \cite{Yu:2012eu,Bhardwaj:2016clt}.

The boundary is gapless if translation invariance is preserved \cite{Yu:2012eu}. There are multiple ways to gap it. One way, discussed in \cite{Bhardwaj:2016clt}, is to introduce an extra set of Majorana modes $\{\gamma_i\}$, one at each vertex at the boundary. Another possibility is simply to give up translation invariance, and introduce a boundary Hamiltonian that pairs neighbouring Majorana modes. For our purpose, this suffices to illustrate the fusion rules of junctions discussed above. 

From the perspective of the Kitaev lattice, the fermionic boundary looks like a zig-zag rugged edge. 
On the other hand, it is well known that  the {\it rough} and {\it smooth} boundaries in the Kitaev lattice  correspond to gapped boundaries characterized by the electric condensate and the magnetic condensate respectively \cite{kitaev_models_2012, beigi_quantum_2011}. They in turn show up as a rugged surface in the Wen Plaquette lattice. The $e$ bounday only consists of $Z$ plaquettes. Therefore the boundary Hamitonian only includes $\hat{Z}$ operators, which commute with $\sigma^z$, the creation operator of the electric charge. When an electric charge approaches the boundary it will disappear, while a magnetic vortex will be stuck on it and becomes an excitation on the boundary. Note that on the boundary there are only three sites around the plaquette. 
The $m$ boundary works similarly -- one simply replaces $\hat{Z}$ by $\hat{X}$, and $\sigma_z$ by $\sigma_x$.

Now we are ready to study $e-f$, $m-f$ and $e-m$ junctions on the lattice model. 

First, as a warm up,  consider the most familiar situation of an $e-m$ junction. This is illustrated in figure \ref{fig:emlattice} below. One can see that an odd number of Majorana mode must be trapped between the boundaries. (i.e. 1 extra Majorana mode is left at the junction in the figure.) This is the well known conclusion that we have re-derived based on bi-modules in (\ref{eq:em_fuse}).\\
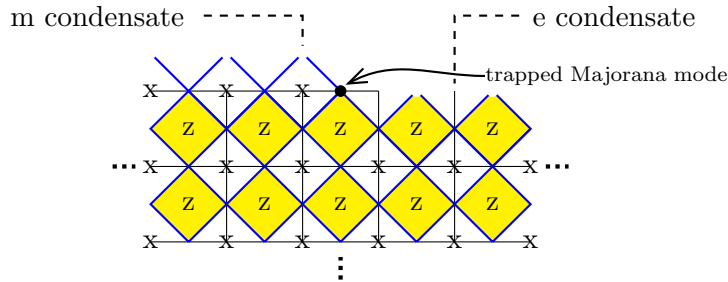
\begin{figure}[h]
\centering
\begin{tikzpicture}
\draw (0,0) -- (5,0);
\draw (0,1) -- (5,1);
\draw (0,2) -- (3,2);
\draw (1,0) -- (1,2);
\draw (2,0) -- (2,2);
\draw (3,0) -- (3,2);
\draw (4,0) -- (4,2);

\draw [blue, thick, fill=yellow] (0.5,0) -- (1,0.5) -- (0.5,1) -- (0,0.5) -- (0.5,0);
\draw [blue, thick, fill=yellow] (1.5,0) -- (2,0.5) -- (1.5,1) -- (1,0.5) -- (1.5,0);
\draw [blue, thick, fill=yellow] (2.5,0) -- (3,0.5) -- (2.5,1) -- (2,0.5) -- (2.5,0);
\draw [blue, thick, fill=yellow] (3.5,0) -- (4,0.5) -- (3.5,1) -- (3,0.5) -- (3.5,0);
\draw [blue, thick, fill=yellow] (4.5,0) -- (5,0.5) -- (4.5,1) -- (4,0.5) -- (4.5,0);

\draw [blue, thick, fill=yellow] (0.5,1) -- (1,1.5) -- (0.5,2) -- (0,1.5) -- (0.5,1);
\draw [blue, thick, fill=yellow] (1.5,1) -- (2,1.5) -- (1.5,2) -- (1,1.5) -- (1.5,1);
\draw [blue, thick, fill=yellow] (2.5,1) -- (3,1.5) -- (2.5,2) -- (2,1.5) -- (2.5,1);
\draw [blue, thick, fill=yellow] (3.5,1) -- (4,1.5) -- (3.5,2) -- (3,1.5) -- (3.5,1);
\draw [blue, thick, fill=yellow] (4.5,1) -- (5,1.5) -- (4.5,2) -- (4,1.5) -- (4.5,1);

\draw [blue, thick] (3,1.5) -- (3.5,1) -- (4,1.5) -- (4.5,1) -- (5,1.5);
\draw [blue, thick] (2,1.5) -- (2.5,2) -- (2,2.5) -- (1.5,2) -- (2,1.5);
\draw [blue, thick] (1,1.5) -- (1.5,2) -- (1,2.5) -- (0.5,2) -- (1,1.5);
\draw [blue, thick] (0,2.5) -- (0.5,2);

\foreach \x in {0.5,1.5,2.5,3.5,4.5}
\foreach \y in {0.5,1.5}
\node at (\x,\y) {z};

\foreach \x in {0,1,2,3,4,5}
\foreach \y in {0,1}
\node at (\x,\y) {x};
\node at (0,2) {x};
\node at (1,2) {x};
\node at (2,2) {x};

\foreach \x in {0,1,2}
\draw[white, fill=white] (\x,2.5) circle [radius=0.07];
\draw[white, fill=white] (4.5,2) circle [radius=0.07];
\draw[white, fill=white] (3.5,2) circle [radius=0.07];

\draw[fill] (2.5,2) circle [radius=0.07];
\node at (6,2.2) {\scriptsize trapped Majorana mode};
\draw [thick] (4.4,2.2) to [out=180,in=30] (2.6,2.1);
\draw [thick] (2.8,2.3) -- (2.6,2.1) -- (2.86, 2.12);
\node at (-0.7,3) {m condensate};
\draw [thick,dashed] (0.7,3) -- (2,3) --(2,2.6);
\node at (6.1,3) {e condensate};
\draw [thick,dashed] (4.85,3) -- (4,3) -- (4,2.1);

\draw [dotted, ultra thick] (-0.5,1) -- (-0.2,1);
\draw [dotted, ultra thick] (2.5,-0.5) -- (2.5,-0.2);
\draw [dotted, ultra thick] (5.2,1) -- (5.5,1);
\end{tikzpicture}
\caption{An illustration of the $e-m$ junction on the lattice. The black lines denote the lattice of Kitaev's toric code model and the blue lines denote the lattice of the Wen plaquette model.}
\label{fig:emlattice}
\end{figure}\\

Now we can look into the problem we met in the previous subsection figure \ref{fig:ef-fm}. 
We notice that indeed there is an ambiguity in the result!
As illustrated in figure \ref{fig:efmjunction}, whether an unpaired Majorana mode is trapped in the $e-f$ junction depends on how we choose to gap the fermion boundary pairing up Majorana modes. There is thus an ambiguity of $\sqrt{2}$ in the quantum dimension of the defect in the $e-f$ or $m-f$ junction. Nonetheless, at the end of the day, there is an odd number of Majorana modes shared between the $e-f$ and $f-m$ junctions.
If we put such $m-f-e$ boundaries on a circle, we will find a total of two Majorana modes, one shared between the $e-f$ and $f-m$ junctions, and another located at the $e-m$ junction. \\
\begin{figure}[h]
\centering
\begin{tikzpicture}
\draw [blue, thick, fill=yellow] (0,0) rectangle (0.5,0.5);
\draw [blue, thick, fill=white] (0.5,0) rectangle (1,0.5);
\draw [blue, thick, fill=yellow] (0.5,0.5) rectangle (1,1);
\draw [blue, thick, fill=white] (1,0.5) rectangle (1.5,1);
\draw [blue, thick, fill=yellow] (1,1) rectangle (1.5,1.5);
\draw [blue, thick, fill=yellow] (1,0) rectangle (1.5,0.5);

\draw [blue, thick, fill=white] (1.5,0) rectangle (2,0.5);
\draw [blue, thick, fill=white] (2,0.5) rectangle (2.5,1);
\draw [blue, thick, fill=white] (1.5,1) rectangle (2,1.5);
\draw [blue, thick, fill=yellow] (2,0) rectangle (2.5,0.5);
\draw [blue, thick, fill=yellow] (1.5,0.5) rectangle (2,1);
\draw [blue, thick, fill=yellow] (2,1) rectangle (2.5,1.5);

\draw [blue, thick, fill=white] (2.5,1) rectangle (3,1.5);
\draw [blue, thick, fill=white] (2.5,0) rectangle (3,0.5);
\draw [blue, thick, fill=yellow] (2.5,0.5) rectangle (3,1);
\draw [blue, thick, fill=white] (3,0.5) rectangle (3.5,1);
\draw [blue, thick, fill=yellow] (3,0) rectangle (3.5,0.5);
\draw [blue, thick, fill=white] (3.5,0) rectangle (4,0.5);

\draw [blue, thick, fill=yellow] (5,0) rectangle (5.5,0.5);
\draw [blue, thick, fill=white] (5.5,0) rectangle (6,0.5);
\draw [blue, thick, fill=yellow] (5.5,0.5) rectangle (6,1);
\draw [blue, thick, fill=white] (6,0.5) rectangle (6.5,1);
\draw [blue, thick, fill=yellow] (6,1) rectangle (6.5,1.5);
\draw [blue, thick, fill=yellow] (6,0) rectangle (6.5,0.5);

\draw [blue, thick, fill=white] (6.5,0) rectangle (7,0.5);
\draw [blue, thick, fill=white] (7,0.5) rectangle (7.5,1);
\draw [blue, thick, fill=white] (6.5,1) rectangle (7,1.5);
\draw [blue, thick, fill=yellow] (7,0) rectangle (7.5,0.5);
\draw [blue, thick, fill=yellow] (6.5,0.5) rectangle (7,1);
\draw [blue, thick, fill=yellow] (7,1) rectangle (7.5,1.5);

\draw [blue, thick, fill=white] (7.5,1) rectangle (8,1.5);
\draw [blue, thick, fill=white] (7.5,0) rectangle (8,0.5);
\draw [blue, thick, fill=yellow] (7.5,0.5) rectangle (8,1);
\draw [blue, thick, fill=white] (8,0.5) rectangle (8.5,1);
\draw [blue, thick, fill=yellow] (8,0) rectangle (8.5,0.5);
\draw [blue, thick, fill=white] (8.5,0) rectangle (9,0.5);

\draw [dotted, ultra thick] (2,-0.2) -- (2,-0.5);
\draw [dotted, ultra thick] (7,-0.2) -- (7,-0.5);

\draw[fill] (1.5,1.5) circle [radius=0.06];
\draw[fill] (7.5,1.5) circle [radius=0.06];
\draw[fill] (2.5,1.5) circle [radius=0.06];
\draw[fill] (6.5,1.5) circle [radius=0.06];
\draw[fill] (2,1.5) circle [radius=0.06];
\draw[fill] (7,1.5) circle [radius=0.06];
\draw (1.75,1.5) to [out=90, in=90] (2.75,1.5) to [out=270, in=270] (1.75,1.5);
\draw (6.25,1.5) to [out=90, in=90] (7.25,1.5) to [out=270, in=270] (6.25,1.5);

\node at (0.25,1) {\scriptsize m}
node at (5.25,1) {\scriptsize m}
node at (3.75,1) {\scriptsize e}
node at (8.75,1) {\scriptsize e}
node at (2,2) {\scriptsize f}
node at (7,2) {\scriptsize f}
node at (3.25,1.75) {\scriptsize unit cell}
node at (5.75,1.75) {\scriptsize unit cell};

\end{tikzpicture}
\caption{There is an odd number of Majorana modes shared between the $e-f$ and $f-m$ junction.}
\label{fig:efmjunction}
\end{figure}
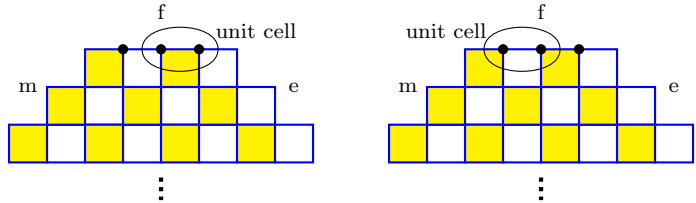\\

As already noted in section \ref{sec:fusion_junc}, in the presence of a fermionic condensate, the quantum dimension of junctions could acquire a $\sqrt{2}$ factor ambiguity, corresponding to adding/subtracting a Majorana mode. Now the fusion rules worked out using methods of bi-modules in (\ref{eq:fuse_effm2}) can be understood as a {\it canonical} choice, where we beef up the junctions by inserting an extra Majorana mode at one of the two $e-f$ or $f-m$ junctions that originally lacks a Majorana mode, so that the two junctions become symmetric, and each carry a quantum dimension of $\sqrt{2}$. Of course, the resultant fusion product would carry two Majorana modes, instead of one that is always expected to be trapped at the $e-m$ junction. 
It is not possible to add only a single Majorana mode in a physical state.  
On a disk, one would have to add a Majorana 
at one of the $e-f$ and $f-m$ junctions, and another at the $e-m$ junction.

The same results can also be understood from the perspective of Abelian Chern-Simons theory. 

This will be relegated to the appendix.

\subsubsection{The bimodules and computing the half-linking number} 

In the previous sections, we have obtained some ``coarse-grained'' data regarding the bimodules. Here we would like to provide details of some ``fine-grained'' data of these boundaries -- namely the actual Frobenius algebra characterizing the boundary, and the left/right action of the bi-modules, to illustrate the general principles laid out in earlier sections. 

For concreteness, let us focus on the $e-f$ junction. 
To begin with, we need to solve for two Frobenius algebra $\mathcal{A}_e$ and $\mathcal{A}_f$. Using the conditions discussed in section \ref{sec:frobeniusalgebra} and also the 6j-symbols of the toric code topological order, we obtain (\ref{eq:algebra-Am}, \ref{eq:algebra-Af}).

\begin{figure}[htbp]
\be
\label{eq:algebra-Am}
    \centering
    \begin{tikzpicture}[scale=0.6]
      \begin{scope}[xshift=0cm,yshift=0cm]{
            \node at (-1.2,0) {$\sqrt{2}$};
            \vertexI{0}{0}{1.5cm};
            \node[above] at (0,1.5cm) {\scriptsize{$\mathcal{A}_e$}};
            \node[below] at (-1.5cm,-1.5cm) {\scriptsize{$\mathcal{A}_e$}};
            \node[below] at (1.5cm,-1.5cm) {\scriptsize{$\mathcal{A}_e$}};
        }
        \end{scope}
        \begin{scope}[xshift=4cm,yshift=0cm]{
            \node at (-2,0) {$=$};
            \vertexI{0}{0}{1.5cm};
            \node[above] at (0,1.5cm) {\scriptsize{$\mathcal{A}_e$}};
            \node[below] at (-1.5cm,-1.5cm) {\scriptsize{$\mathcal{A}_e$}};
            \node[below] at (1.5cm,-1.5cm) {\scriptsize{$\mathcal{A}_e$}};
            \vertexIC{0}{0}{1cm}{\bulkCol};
            \scalebox{0.5}[0.5]{\tri{-2cm+4cm}{-2cm}{45}};
            \scalebox{0.5}[0.5]{\tri{0cm+4cm}{2.2cm}{270}};
            \scalebox{0.5}[0.5]{\tri{2cm+4cm}{-2cm}{135}};
            \node[right] at (0,0.5cm) {\scriptsize{$1$}};
            \node[above] at (-0.5cm,-0.5cm) {\scriptsize{$1$}};
            \node[above] at (0.5cm,-0.5cm) {\scriptsize{$1$}};
        }
        \end{scope}
        \begin{scope}[xshift=8cm,yshift=0cm]{
            \node at (-2,0) {$+$};
            \vertexI{0}{0}{1.5cm};
            \node[above] at (0,1.5cm) {\scriptsize{$\mathcal{A}_e$}};
            \node[below] at (-1.5cm,-1.5cm) {\scriptsize{$\mathcal{A}_e$}};
            \node[below] at (1.5cm,-1.5cm) {\scriptsize{$\mathcal{A}_e$}};
            \vertexIC{0}{0}{1cm}{\bulkCol};
            \scalebox{0.5}[0.5]{\tri{-2cm+8cm}{-2cm}{45}};
            \scalebox{0.5}[0.5]{\tri{0cm+8cm}{2.2cm}{270}};
            \scalebox{0.5}[0.5]{\tri{2cm+8cm}{-2cm}{135}};
            \node[right] at (0,0.5cm) {\scriptsize{$1$}};
            \node[above] at (-0.5cm,-0.5cm) {\scriptsize{$e$}};
            \node[above] at (0.5cm,-0.5cm) {\scriptsize{$e$}};
        }
        \end{scope}
        \begin{scope}[xshift=12cm,yshift=0cm]{
            \node at (-2,0) {$+$};
            \vertexI{0}{0}{1.5cm};
            \node[above] at (0,1.5cm) {\scriptsize{$\mathcal{A}_e$}};
            \node[below] at (-1.5cm,-1.5cm) {\scriptsize{$\mathcal{A}_e$}};
            \node[below] at (1.5cm,-1.5cm) {\scriptsize{$\mathcal{A}_e$}};
            \vertexIC{0}{0}{1cm}{\bulkCol};
            \scalebox{0.5}[0.5]{\tri{-2cm+12cm}{-2cm}{45}};
            \scalebox{0.5}[0.5]{\tri{0cm+12cm}{2.2cm}{270}};
            \scalebox{0.5}[0.5]{\tri{2cm+12cm}{-2cm}{135}};
            \node[right] at (0,0.5cm) {\scriptsize{$m$}};
            \node[above] at (-0.5cm,-0.5cm) {\scriptsize{$e$}};
            \node[above] at (0.5cm,-0.5cm) {\scriptsize{$1$}};
        }
        \end{scope}
        \begin{scope}[xshift=16cm,yshift=0cm]{
            \node at (-2,0) {$+$};
            \vertexI{0}{0}{1.5cm};
            \node[above] at (0,1.5cm) {\scriptsize{$\mathcal{A}_e$}};
            \node[below] at (-1.5cm,-1.5cm) {\scriptsize{$\mathcal{A}_e$}};
            \node[below] at (1.5cm,-1.5cm) {\scriptsize{$\mathcal{A}_e$}};
            \vertexIC{0}{0}{1cm}{\bulkCol};
            \scalebox{0.5}[0.5]{\tri{-2cm+16cm}{-2cm}{45}};
            \scalebox{0.5}[0.5]{\tri{0cm+16cm}{2.2cm}{270}};
            \scalebox{0.5}[0.5]{\tri{2cm+16cm}{-2cm}{135}};
            \node[right] at (0,0.5cm) {\scriptsize{$m$}};
            \node[above] at (-0.5cm,-0.5cm) {\scriptsize{$1$}};
            \node[above] at (0.5cm,-0.5cm) {\scriptsize{$e$}};
        }
        \end{scope}
    \end{tikzpicture}
    \ee
\end{figure}

\begin{figure}[htbp]
\be
\label{eq:algebra-Af}
    \centering
    \begin{tikzpicture}[scale=0.6]
      \begin{scope}[xshift=0cm,yshift=0cm]{
            \node at (-1.2,0) {$\sqrt{2}$};
            \vertexI{0}{0}{1.5cm};
            \node[above] at (0,1.5cm) {\scriptsize{$\mathcal{A}_f$}};
            \node[below] at (-1.5cm,-1.5cm) {\scriptsize{$\mathcal{A}_f$}};
            \node[below] at (1.5cm,-1.5cm) {\scriptsize{$\mathcal{A}_f$}};
        }
        \end{scope}
        \begin{scope}[xshift=4cm,yshift=0cm]{
            \node at (-2,0) {$=$};
            \vertexI{0}{0}{1.5cm};
            \node[above] at (0,1.5cm) {\scriptsize{$\mathcal{A}_f$}};
            \node[below] at (-1.5cm,-1.5cm) {\scriptsize{$\mathcal{A}_f$}};
            \node[below] at (1.5cm,-1.5cm) {\scriptsize{$\mathcal{A}_f$}};
            \vertexIC{0}{0}{1cm}{\bulkCol};
            \scalebox{0.5}[0.5]{\tri{-2cm+4cm}{-2cm}{45}};
            \scalebox{0.5}[0.5]{\tri{0cm+4cm}{2.2cm}{270}};
            \scalebox{0.5}[0.5]{\tri{2cm+4cm}{-2cm}{135}};
            \node[right] at (0,0.5cm) {\scriptsize{$1$}};
            \node[above] at (-0.5cm,-0.5cm) {\scriptsize{$1$}};
            \node[above] at (0.5cm,-0.5cm) {\scriptsize{$1$}};
        }
        \end{scope}
        \begin{scope}[xshift=8cm,yshift=0cm]{
            \node at (-2,0) {$+$};
            \vertexI{0}{0}{1.5cm};
            \node[above] at (0,1.5cm) {\scriptsize{$\mathcal{A}_f$}};
            \node[below] at (-1.5cm,-1.5cm) {\scriptsize{$\mathcal{A}_f$}};
            \node[below] at (1.5cm,-1.5cm) {\scriptsize{$\mathcal{A}_f$}};
            \vertexIC{0}{0}{1cm}{\bulkCol};
            \scalebox{0.5}[0.5]{\tri{-2cm+8cm}{-2cm}{45}};
            \scalebox{0.5}[0.5]{\tri{0cm+8cm}{2.2cm}{270}};
            \scalebox{0.5}[0.5]{\tri{2cm+8cm}{-2cm}{135}};
            \node[right] at (0,0.5cm) {\scriptsize{$1$}};
            \node[above] at (-0.5cm,-0.5cm) {\scriptsize{$f$}};
            \node[above] at (0.5cm,-0.5cm) {\scriptsize{$f$}};
        }
        \end{scope}
        \begin{scope}[xshift=12cm,yshift=0cm]{
            \node at (-2,0) {$+$};
            \vertexI{0}{0}{1.5cm};
            \node[above] at (0,1.5cm) {\scriptsize{$\mathcal{A}_f$}};
            \node[below] at (-1.5cm,-1.5cm) {\scriptsize{$\mathcal{A}_f$}};
            \node[below] at (1.5cm,-1.5cm) {\scriptsize{$\mathcal{A}_f$}};
            \vertexIC{0}{0}{1cm}{\bulkCol};
            \scalebox{0.5}[0.5]{\tri{-2cm+12cm}{-2cm}{45}};
            \scalebox{0.5}[0.5]{\tri{0cm+12cm}{2.2cm}{270}};
            \scalebox{0.5}[0.5]{\tri{2cm+12cm}{-2cm}{135}};
            \node[right] at (0,0.5cm) {\scriptsize{$f$}};
            \node[above] at (-0.5cm,-0.5cm) {\scriptsize{$f$}};
            \node[above] at (0.5cm,-0.5cm) {\scriptsize{$1$}};
        }
        \end{scope}
        \begin{scope}[xshift=16cm,yshift=0cm]{
            \node at (-2,0) {$+$};
            \vertexI{0}{0}{1.5cm};
            \node[above] at (0,1.5cm) {\scriptsize{$\mathcal{A}_f$}};
            \node[below] at (-1.5cm,-1.5cm) {\scriptsize{$\mathcal{A}_f$}};
            \node[below] at (1.5cm,-1.5cm) {\scriptsize{$\mathcal{A}_f$}};
            \vertexIC{0}{0}{1cm}{\bulkCol};
            \scalebox{0.5}[0.5]{\tri{-2cm+16cm}{-2cm}{45}};
            \scalebox{0.5}[0.5]{\tri{0cm+16cm}{2.2cm}{270}};
            \scalebox{0.5}[0.5]{\tri{2cm+16cm}{-2cm}{135}};
            \node[right] at (0,0.5cm) {\scriptsize{$f$}};
            \node[above] at (-0.5cm,-0.5cm) {\scriptsize{$1$}};
            \node[above] at (0.5cm,-0.5cm) {\scriptsize{$f$}};
        }
        \end{scope}
    \end{tikzpicture}
    \ee
\end{figure}

Then we would like to obtain the unique simple bimodule $X_{ef}$ already discussed in (\ref{sec:bfjunction}).
Note that here we have used the freedom to rescale discussed in (\ref{eq:hombasis_rescale}) and introduce  $\zeta^{\mathcal{A}_e}_1$, $\zeta^{\mathcal{A}_f}_1$, $\zeta^{\mathcal{A}_e}_e$ and $\zeta^{\mathcal{A}_f}_f$ to set all the coefficients in (\ref{eq:algebra-Am}, \ref{eq:algebra-Af}) to 1. Then we would like to obtain the unique simple bimodule $X_{ef}$ already discussed in (\ref{sec:bfjunction}).
A bi-module is separately a left module of $\mA_e$ and right module of $\mA_f$. Therefore the left-right actions must separately satisfy (\ref{eq:reps_eq}) and its right-action counterpart. But as a bi-module, it must satisfy commutativity between the left and right action as illustrated in  (\ref{eq:bimodule}) too. These results in the bimodule are illustrated in (\ref{eq:bimodule-M1-1}) and (\ref{eq:bimodule-M1-1}).

\begin{figure}[htbp]
    \centering
    \be\label{eq:bimodule-M1-1}
    \begin{tikzpicture}[scale=0.3]
    \begin{scope}[xshift=0cm,yshift=0cm]{
    \node at (-7,0) {$\sqrt{2}$};
    \draw [thick, magenta] (-4,-4) -- (-4,4);
    \draw [thick] (-7,-3) -- (-4,0);
    \node at (-7.8,-3) {\footnotesize{$\mathcal{A}_e$}}
    node at (-3.2,4) {\footnotesize{$X_{ef}$}}
    node at (-3.2,-4) {\footnotesize{$X_{ef}$}};
    }
    \end{scope}
    \begin{scope}[xshift=-5cm,yshift=0cm]{
    \node at (4.5,0) {\Large{$=$}};
    \draw [thick, magenta] (8,-4) -- (8,4);
    \draw [thick] (5,-3) -- (8,0);
    \draw [green, thick] (8,-1.5) -- (8,-2.4);
    \draw [green, thick] (8,-2) -- (8,2);
    \draw [green, thick] (6.5,-1.5) -- (8,0);
    \tri{6.2cm}{-1.8cm}{45}
    \tri{8cm}{2cm}{270}
    \tri{8cm}{-2.4cm}{90}
    \node at (4.2,-3) {\footnotesize{$\mathcal{A}_e$}}
    node at (8.8,4) {\footnotesize{$X_{ef}$}}
    node at (8.8,-4) {\footnotesize{$X_{ef}$}}
    node at (7,-0.3) {\footnotesize{$1$}}
    node at (8.6,0.7) {\footnotesize{$1$}}
    node at (8.6,-0.9) {\footnotesize{$1$}};
    }
    \end{scope}
    \begin{scope}[xshift=2cm,yshift=0cm]{
    \node at (4.5,0) {\Large{$+$}};
    \draw [thick, magenta] (8,-4) -- (8,4);
    \draw [thick] (5,-3) -- (8,0);
    \draw [green, thick] (8,-1.5) -- (8,-2.4);
    \draw [green, thick] (8,-2) -- (8,2);
    \draw [green, thick] (6.5,-1.5) -- (8,0);
    \tri{6.2cm}{-1.8cm}{45}
    \tri{8cm}{2cm}{270}
    \tri{8cm}{-2.4cm}{90}
    \node at (4.2,-3) {\footnotesize{$\mathcal{A}_e$}}
    node at (8.8,4) {\footnotesize{$X_{ef}$}}
    node at (8.8,-4) {\footnotesize{$X_{ef}$}}
    node at (7,-0.3) {\footnotesize{$1$}}
    node at (8.6,0.7) {\footnotesize{$e$}}
    node at (8.6,-0.9) {\footnotesize{$e$}};
    }
    \end{scope}
    \begin{scope}[xshift=9cm,yshift=0cm]{
    \node at (4.5,0) {\Large{$+$}};
    \draw [thick, magenta] (8,-4) -- (8,4);
    \draw [thick] (5,-3) -- (8,0);
    \draw [green, thick] (8,-1.5) -- (8,-2.4);
    \draw [green, thick] (8,-2) -- (8,2);
    \draw [green, thick] (6.5,-1.5) -- (8,0);
    \tri{6.2cm}{-1.8cm}{45}
    \tri{8cm}{2cm}{270}
    \tri{8cm}{-2.4cm}{90}
    \node at (4.2,-3) {\footnotesize{$\mathcal{A}_e$}}
    node at (8.8,4) {\footnotesize{$X_{ef}$}}
    node at (8.8,-4) {\footnotesize{$X_{ef}$}}
    node at (7,-0.3) {\footnotesize{$1$}}
    node at (8.6,0.7) {\footnotesize{$e$}}
    node at (8.6,-0.9) {\footnotesize{$e$}};
    }
    \end{scope}
    \begin{scope}[xshift=16cm,yshift=0cm]{
    \node at (4.5,0) {\Large{$+$}};
    \draw [thick, magenta] (8,-4) -- (8,4);
    \draw [thick] (5,-3) -- (8,0);
    \draw [green, thick] (8,-1.5) -- (8,-2.4);
    \draw [green, thick] (8,-2) -- (8,2);
    \draw [green, thick] (6.5,-1.5) -- (8,0);
    \tri{6.2cm}{-1.8cm}{45}
    \tri{8cm}{2cm}{270}
    \tri{8cm}{-2.4cm}{90}
    \node at (4.2,-3) {\footnotesize{$\mathcal{A}_e$}}
    node at (8.8,4) {\footnotesize{$X_{ef}$}}
    node at (8.8,-4) {\footnotesize{$X_{ef}$}}
    node at (7,-0.3) {\footnotesize{$1$}}
    node at (8.6,0.7) {\footnotesize{$f$}}
    node at (8.6,-0.9) {\footnotesize{$f$}};
    }
    \end{scope}
    \begin{scope}[xshift=-8cm,yshift=-10cm]{
    \node at (4.5,0) {\Large{$+$}};
    \draw [thick, magenta] (8,-4) -- (8,4);
    \draw [thick] (5,-3) -- (8,0);
    \draw [green, thick] (8,-1.5) -- (8,-2.4);
    \draw [green, thick] (8,-2) -- (8,2);
    \draw [green, thick] (6.5,-1.5) -- (8,0);
    \tri{6.2cm}{-1.8cm}{45}
    \tri{8cm}{2cm}{270}
    \tri{8cm}{-2.4cm}{90}
    \node at (4.2,-3) {\footnotesize{$\mathcal{A}_e$}}
    node at (8.8,4) {\footnotesize{$X_{ef}$}}
    node at (8.8,-4) {\footnotesize{$X_{ef}$}}
    node at (7,-0.3) {\footnotesize{$m$}}
    node at (8.6,0.7) {\footnotesize{$1$}}
    node at (8.6,-0.9) {\footnotesize{$m$}};
    }
    \end{scope}
    \begin{scope}[xshift=-1cm,yshift=-10cm]{
    \node at (4.5,0) {\Large{$+$}};
    \draw [thick, magenta] (8,-4) -- (8,4);
    \draw [thick] (5,-3) -- (8,0);
    \draw [green, thick] (8,-1.5) -- (8,-2.4);
    \draw [green, thick] (8,-2) -- (8,2);
    \draw [green, thick] (6.5,-1.5) -- (8,0);
    \tri{6.2cm}{-1.8cm}{45}
    \tri{8cm}{2cm}{270}
    \tri{8cm}{-2.4cm}{90}
    \node at (4.2,-3) {\footnotesize{$\mathcal{A}_e$}}
    node at (8.8,4) {\footnotesize{$X_{ef}$}}
    node at (8.8,-4) {\footnotesize{$X_{ef}$}}
    node at (7,-0.3) {\footnotesize{$m$}}
    node at (8.6,0.7) {\footnotesize{$m$}}
    node at (8.6,-0.9) {\footnotesize{$1$}};
    }
    \end{scope}
    \begin{scope}[xshift=6cm,yshift=-10cm]{
    \node at (4.5,0) {\Large{$+$}};
    \draw [thick, magenta] (8,-4) -- (8,4);
    \draw [thick] (5,-3) -- (8,0);
    \draw [green, thick] (8,-1.5) -- (8,-2.4);
    \draw [green, thick] (8,-2) -- (8,2);
    \draw [green, thick] (6.5,-1.5) -- (8,0);
    \tri{6.2cm}{-1.8cm}{45}
    \tri{8cm}{2cm}{270}
    \tri{8cm}{-2.4cm}{90}
    \node at (4.2,-3) {\footnotesize{$\mathcal{A}_e$}}
    node at (8.8,4) {\footnotesize{$X_{ef}$}}
    node at (8.8,-4) {\footnotesize{$X_{ef}$}}
    node at (7,-0.3) {\footnotesize{$m$}}
    node at (8.6,0.7) {\footnotesize{$e$}}
    node at (8.6,-0.9) {\footnotesize{$f$}};
    }
    \end{scope}
    \begin{scope}[xshift=13cm,yshift=-10cm]{
    \node at (4.5,0) {\Large{$+$}};
    \draw [thick, magenta] (8,-4) -- (8,4);
    \draw [thick] (5,-3) -- (8,0);
    \draw [green, thick] (8,-1.5) -- (8,-2.4);
    \draw [green, thick] (8,-2) -- (8,2);
    \draw [green, thick] (6.5,-1.5) -- (8,0);
    \tri{6.2cm}{-1.8cm}{45}
    \tri{8cm}{2cm}{270}
    \tri{8cm}{-2.4cm}{90}
    \node at (4.2,-3) {\footnotesize{$\mathcal{A}_e$}}
    node at (8.8,4) {\footnotesize{$X_{ef}$}}
    node at (8.8,-4) {\footnotesize{$X_{ef}$}}
    node at (7,-0.3) {\footnotesize{$m$}}
    node at (8.6,0.7) {\footnotesize{$f$}}
    node at (8.6,-0.9) {\footnotesize{$e$}};
    }
    \end{scope}
    \end{tikzpicture}
    \ee
    \end{figure}	
    \begin{figure}[htbp]
    \centering
    \be\label{eq:bimodule-M1-2}
    \begin{tikzpicture}[scale=0.3]
    \begin{scope}[xshift=0cm,yshift=0cm]{
    \node at (-6,0) {$\sqrt{2}$};
    \draw [thick, magenta] (-4,-4) -- (-4,4);
    \draw [thick] (-1,-3) -- (-4,0);
    \node at (-0.2,-3) {\footnotesize{$\mathcal{A}_f$}}
    node at (-3.2,4) {\footnotesize{$X_{ef}$}}
    node at (-3.2,-4) {\footnotesize{$X_{ef}$}};
    }
    \end{scope}
    \begin{scope}[xshift=-5cm,yshift=0cm]{
    \node at (4.5,0) {\Large{$=$}};
    \draw [thick, magenta] (8,-4) -- (8,4);
    \draw [thick] (11,-3) -- (8,0);
    \draw [green, thick] (8,-1.5) -- (8,-2.4);
    \draw [green, thick] (8,-2) -- (8,2);
    \draw [green, thick] (9.5,-1.5) -- (8,0);
    \tri{9.8cm}{-1.8cm}{45+90}
    \tri{8cm}{2cm}{270}
    \tri{8cm}{-2.4cm}{90}
    \node at (11.8,-3) {\footnotesize{$\mathcal{A}_f$}}
    node at (8.8,4) {\footnotesize{$X_{ef}$}}
    node at (8.8,-4) {\footnotesize{$X_{ef}$}}
    node at (9,-0.3) {\footnotesize{$1$}}
    node at (7.4,0.7) {\footnotesize{$1$}}
    node at (7.4,-0.9) {\footnotesize{$1$}};
    }
    \end{scope}
    \begin{scope}[xshift=2cm,yshift=0cm]{
    \node at (4.5,0) {\Large{$+$}};
    \draw [thick, magenta] (8,-4) -- (8,4);
    \draw [thick] (11,-3) -- (8,0);
    \draw [green, thick] (8,-1.5) -- (8,-2.4);
    \draw [green, thick] (8,-2) -- (8,2);
    \draw [green, thick] (9.5,-1.5) -- (8,0);
    \tri{9.8cm}{-1.8cm}{45+90}
    \tri{8cm}{2cm}{270}
    \tri{8cm}{-2.4cm}{90}
    \node at (11.8,-3) {\footnotesize{$\mathcal{A}_f$}}
    node at (8.8,4) {\footnotesize{$X_{ef}$}}
    node at (8.8,-4) {\footnotesize{$X_{ef}$}}
    node at (9,-0.3) {\footnotesize{$1$}}
    node at (7.4,0.7) {\footnotesize{$e$}}
    node at (7.4,-0.9) {\footnotesize{$e$}};
    }
    \end{scope}
    \begin{scope}[xshift=9cm,yshift=0cm]{
    \node at (4.5,0) {\Large{$+$}};
    \draw [thick, magenta] (8,-4) -- (8,4);
    \draw [thick] (11,-3) -- (8,0);
    \draw [green, thick] (8,-1.5) -- (8,-2.4);
    \draw [green, thick] (8,-2) -- (8,2);
    \draw [green, thick] (9.5,-1.5) -- (8,0);
    \tri{9.8cm}{-1.8cm}{45+90}
    \tri{8cm}{2cm}{270}
    \tri{8cm}{-2.4cm}{90}
    \node at (11.8,-3) {\footnotesize{$\mathcal{A}_f$}}
    node at (8.8,4) {\footnotesize{$X_{ef}$}}
    node at (8.8,-4) {\footnotesize{$X_{ef}$}}
    node at (9,-0.3) {\footnotesize{$1$}}
    node at (7.4,0.7) {\footnotesize{$m$}}
    node at (7.4,-0.9) {\footnotesize{$m$}};
    }
    \end{scope}
    \begin{scope}[xshift=16cm,yshift=0cm]{
    \node at (4.5,0) {\Large{$+$}};
    \draw [thick, magenta] (8,-4) -- (8,4);
    \draw [thick] (11,-3) -- (8,0);
    \draw [green, thick] (8,-1.5) -- (8,-2.4);
    \draw [green, thick] (8,-2) -- (8,2);
    \draw [green, thick] (9.5,-1.5) -- (8,0);
    \tri{9.8cm}{-1.8cm}{45+90}
    \tri{8cm}{2cm}{270}
    \tri{8cm}{-2.4cm}{90}
    \node at (11.8,-3) {\footnotesize{$\mathcal{A}_f$}}
    node at (8.8,4) {\footnotesize{$X_{ef}$}}
    node at (8.8,-4) {\footnotesize{$X_{ef}$}}
    node at (9,-0.3) {\footnotesize{$1$}}
    node at (7.4,0.7) {\footnotesize{$f$}}
    node at (7.4,-0.9) {\footnotesize{$f$}};
    }
    \end{scope}
    \begin{scope}[xshift=-8cm,yshift=-10cm]{
    \node at (4.5,0) {\Large{$+$}};
    \draw [thick, magenta] (8,-4) -- (8,4);
    \draw [thick] (11,-3) -- (8,0);
    \draw [green, thick] (8,-1.5) -- (8,-2.4);
    \draw [green, thick] (8,-2) -- (8,2);
    \draw [green, thick] (9.5,-1.5) -- (8,0);
    \tri{9.8cm}{-1.8cm}{45+90}
    \tri{8cm}{2cm}{270}
    \tri{8cm}{-2.4cm}{90}
    \node at (11.8,-3) {\footnotesize{$\mathcal{A}_f$}}
    node at (8.8,4) {\footnotesize{$X_{ef}$}}
    node at (8.8,-4) {\footnotesize{$X_{ef}$}}
    node at (9,-0.3) {\footnotesize{$f$}}
    node at (7.4,0.7) {\footnotesize{$f$}}
    node at (7.4,-0.9) {\footnotesize{$1$}};
    }
    \end{scope}
    \begin{scope}[xshift=-1cm,yshift=-10cm]{
    \node at (4.5,0) {\Large{$+$}};
    \draw [thick, magenta] (8,-4) -- (8,4);
    \draw [thick] (11,-3) -- (8,0);
    \draw [green, thick] (8,-1.5) -- (8,-2.4);
    \draw [green, thick] (8,-2) -- (8,2);
    \draw [green, thick] (9.5,-1.5) -- (8,0);
    \tri{9.8cm}{-1.8cm}{45+90}
    \tri{8cm}{2cm}{270}
    \tri{8cm}{-2.4cm}{90}
    \node at (11.8,-3) {\footnotesize{$\mathcal{A}_f$}}
    node at (8.8,4) {\footnotesize{$X_{ef}$}}
    node at (8.8,-4) {\footnotesize{$X_{ef}$}}
    node at (9,-0.3) {\footnotesize{$f$}}
    node at (7.4,0.7) {\footnotesize{$1$}}
    node at (7.4,-0.9) {\footnotesize{$f$}};
    }
    \end{scope}
    \begin{scope}[xshift=6cm,yshift=-10cm]{
    \node at (4.5,0) {\Large{$+$}};
    \draw [thick, magenta] (8,-4) -- (8,4);
    \draw [thick] (11,-3) -- (8,0);
    \draw [green, thick] (8,-1.5) -- (8,-2.4);
    \draw [green, thick] (8,-2) -- (8,2);
    \draw [green, thick] (9.5,-1.5) -- (8,0);
    \tri{9.8cm}{-1.8cm}{45+90}
    \tri{8cm}{2cm}{270}
    \tri{8cm}{-2.4cm}{90}
    \node at (11.8,-3) {\footnotesize{$\mathcal{A}_f$}}
    node at (8.8,4) {\footnotesize{$X_{ef}$}}
    node at (8.8,-4) {\footnotesize{$X_{ef}$}}
    node at (9,-0.3) {\footnotesize{$f$}}
    node at (7.4,0.7) {\footnotesize{$e$}}
    node at (7.4,-0.9) {\footnotesize{$m$}};
    }
    \end{scope}
    \begin{scope}[xshift=13cm,yshift=-10cm]{
    \node at (4.5,0) {\Large{$+$}};
    \draw [thick, magenta] (8,-4) -- (8,4);
    \draw [thick] (11,-3) -- (8,0);
    \draw [green, thick] (8,-1.5) -- (8,-2.4);
    \draw [green, thick] (8,-2) -- (8,2);
    \draw [green, thick] (9.5,-1.5) -- (8,0);
    \tri{9.8cm}{-1.8cm}{45+90}
    \tri{8cm}{2cm}{270}
    \tri{8cm}{-2.4cm}{90}
    \node at (11.8,-3) {\footnotesize{$\mathcal{A}_f$}}
    node at (8.8,4) {\footnotesize{$X_{ef}$}}
    node at (8.8,-4) {\footnotesize{$X_{ef}$}}
    node at (9,-0.3) {\footnotesize{$f$}}
    node at (7.4,0.7) {\footnotesize{$m$}}
    node at (7.4,-0.9) {\footnotesize{$e$}};
    }
    \end{scope}
    \end{tikzpicture}
    \ee
\end{figure}	
%Substituting the algebra and the bimodules into (\ref{eq:gamma_pic}), we obtain

We still have enough phase rescaling freedom here ($\zeta^{X_{ef}}_e$, $\zeta^{X_{ef}}_m$ and $\zeta^{X_{ef}}_f$) to set all the coefficients to unity. Substituting the algebra and the bimodules into (\ref{eq:gamma_pic_2}), we obtain
\be 
\gamma_{X_{ef}\, 1}= \mathcal{N}_{ef} \sum_{i,j\in X_{ef}}\rho^{X_{ef} j}_{1i}\rho^{X_{ef}j}_{i1}(R^{1i}_jR^{i1}_j)^*\sqrt{d_id_jd_1}=\frac{\mathcal{N}_{ef}}{2}\sum_{i\in M_1}d_i=1.
\ee
 
The normalization $\mathcal{N}_{ef}$ is given by
 \be \label{eq:Nef}
 \mathcal{N}_{ef} = \frac{1}{\sqrt{2D_{Toric\ code}}},
 %\mathcal{N}_{ef} = \sqrt{2D_{Toric\ code}},
 \ee
 which recovers the fusion rules (\ref{eq:ef_fuse}), confirming (\ref{eq:NAB_2}).

\subsection{Intermediate level -- $D(S_3)$}

The quantum double model $D(S_3)$ is the paradigmatic example of non-Abelian topological orders that illustrate
non-trivial features that could arise. 

For completeness, we include the topological data of the bulk theory in the appendix, which sets the notations of the anyons that we will use below. 
The bosonic gapped boundaries of $D(S_3)$ have been studied in many places \cite{cong_topological_2016, Cong_2017, Shen_2019}. 
These condensates and the junctions between them are also summarized in the appendix. 

In addition to the well known bosonic gapped boundaries, there is also one fermionic gapped boundary. This is already noted in \cite{Wan:2016php}.
Let us study it in somewhat more detail below.

\subsubsection{The fermionic boundary}
The condensate is given by
\be \label{eq:finDs3}
\mathcal{A}_f = A \oplus C \oplus E.
\ee

This condensation is closely related to the fermionic boundary of the toric code. 
In this case, one can readily work out the $W$ matrix and $\Omega$ matrix using the methods in section \ref{sec:b_fusion}.
The fusion rules between defects are also readily obtainable. 
We will slightly delay the presentation of these results, by taking a somewhat longer route.

As discussed in \cite{Wan:2016php}, for a fermionic condensate that preserves fermion parity, one could consider splitting the condensation into two steps -- first condensing the bosons in $\mathcal{A}_f$, which should form a closed Frobenius sub-algebra, before condensing the fermion, which would be reduced to an Abelian anyon in the intermediate condensed phase. Applying this logic here, it would imply that one could first consider condensing 
\be \alpha_{AC} \equiv A \oplus C \subset \mathcal{A}_f.
\ee

%For completeness, let us present the Frobenius algebra $\alpha_{AC}$ in (\ref{eq:algebra-condenseC}). The virtue of the sequential condensation is that it allows one to work out $\mathcal{A}_f$ in (\ref{eq:finDs3}) as a {\it Lagrangian algebra} of 
For completeness, let us present the Frobenius algebra $\alpha_{AC}$ in (\ref{eq:algebra-condenseC}). Similarly to the case of the Toric code, here we have chosen the phase ambiguities ($\zeta^1_A$ and $\zeta^1_C$) such that all the coefficients including $A$ to be 1. The virtue of the sequential condensation is that it allows one to work out $\mathcal{A}_f$ in (\ref{eq:finDs3}) as a {\it Lagrangian algebra} of 
$D(S_3)$ by treating it as condensation  of simple modules of $\alpha_{AC}$. Un-packaging it into fusion basis in $D(S_3)$ is simple.

\begin{figure}[ht]
\be
\label{eq:algebra-condenseC}
    \centering
    \begin{tikzpicture}[scale=0.5]
      \begin{scope}[xshift=0cm,yshift=0cm]{
            \node at (-1.2,0) {$\sqrt{3}$};
            \vertexI{0}{0}{1.5cm};
            \node[above] at (0,1.5cm) {\scriptsize{$1$}};
            \node[below] at (-1.5cm,-1.5cm) {\scriptsize{$1$}};
            \node[below] at (1.5cm,-1.5cm) {\scriptsize{$1$}};
        }
        \end{scope}
        \begin{scope}[xshift=4.5cm,yshift=0cm]{
            \node at (-2.25,0) {$=$};
            \vertexI{0}{0}{1.5cm};
            \node[above] at (0,1.5cm) {\scriptsize{$1$}};
            \node[below] at (-1.5cm,-1.5cm) {\scriptsize{$1$}};
            \node[below] at (1.5cm,-1.5cm) {\scriptsize{$1$}};
            \vertexIC{0}{0}{1cm}{\bulkCol};
            \scalebox{0.5}[0.5]{\tri{-2cm+4.5cm}{-2cm}{45}};
            \scalebox{0.5}[0.5]{\tri{0cm+4.5cm}{2.2cm}{270}};
            \scalebox{0.5}[0.5]{\tri{2cm+4.5cm}{-2cm}{135}};
            \node[right] at (0,0.5cm) {\tiny{$A$}};
            \node[above] at (-0.5cm,-0.5cm) {\tiny{$A$}};
            \node[above] at (0.5cm,-0.5cm) {\tiny{$A$}};
        }
        \end{scope}
        \begin{scope}[xshift=9cm,yshift=0cm]{
            \node at (-2.25,0) {$+$};
            \vertexI{0}{0}{1.5cm};
            \node[above] at (0,1.5cm) {\scriptsize{$1$}};
            \node[below] at (-1.5cm,-1.5cm) {\scriptsize{$1$}};
            \node[below] at (1.5cm,-1.5cm) {\scriptsize{$1$}};
            \vertexIC{0}{0}{1cm}{\bulkCol};
            \scalebox{0.5}[0.5]{\tri{-2cm+9cm}{-2cm}{45}};
            \scalebox{0.5}[0.5]{\tri{0cm+9cm}{2.2cm}{270}};
            \scalebox{0.5}[0.5]{\tri{2cm+9cm}{-2cm}{135}};
            \node[right] at (0,0.5cm) {\tiny{$A$}};
            \node[above] at (-0.5cm,-0.5cm) {\tiny{$C$}};
            \node[above] at (0.5cm,-0.5cm) {\tiny{$C$}};
        }
        \end{scope}
        \begin{scope}[xshift=13.5cm,yshift=0cm]{
            \node at (-2.25,0) {$+$};
            \vertexI{0}{0}{1.5cm};
            \node[above] at (0,1.5cm) {\scriptsize{$1$}};
            \node[below] at (-1.5cm,-1.5cm) {\scriptsize{$1$}};
            \node[below] at (1.5cm,-1.5cm) {\scriptsize{$1$}};
            \vertexIC{0}{0}{1cm}{\bulkCol};
            \scalebox{0.5}[0.5]{\tri{-2cm+13.5cm}{-2cm}{45}};
            \scalebox{0.5}[0.5]{\tri{0cm+13.5cm}{2.2cm}{270}};
            \scalebox{0.5}[0.5]{\tri{2cm+13.5cm}{-2cm}{135}};
            \node[right] at (0,0.5cm) {\tiny{$C$}};
            \node[above] at (-0.5cm,-0.5cm) {\tiny{$A$}};
            \node[above] at (0.5cm,-0.5cm) {\tiny{$C$}};
        }
        \end{scope}
        \begin{scope}[xshift=18cm,yshift=0cm]{
            \node at (-2.25,0) {$+$};
            \vertexI{0}{0}{1.5cm};
            \node[above] at (0,1.5cm) {\scriptsize{$1$}};
            \node[below] at (-1.5cm,-1.5cm) {\scriptsize{$1$}};
            \node[below] at (1.5cm,-1.5cm) {\scriptsize{$1$}};
            \vertexIC{0}{0}{1cm}{\bulkCol};
            \scalebox{0.5}[0.5]{\tri{-2cm+18cm}{-2cm}{45}};
            \scalebox{0.5}[0.5]{\tri{0cm+18cm}{2.2cm}{270}};
            \scalebox{0.5}[0.5]{\tri{2cm+18cm}{-2cm}{135}};
            \node[right] at (0,0.5cm) {\tiny{$C$}};
            \node[above] at (-0.5cm,-0.5cm) {\tiny{$C$}};
            \node[above] at (0.5cm,-0.5cm) {\tiny{$A$}};
        }
        \end{scope}
        \begin{scope}[xshift=23cm,yshift=0cm]{
            \node at (-2,0) {$+\ \phi$};
            \vertexI{0}{0}{1.5cm};
            \node[above] at (0,1.5cm) {\scriptsize{$1$}};
            \node[below] at (-1.5cm,-1.5cm) {\scriptsize{$1$}};
            \node[below] at (1.5cm,-1.5cm) {\scriptsize{$1$}};
            \vertexIC{0}{0}{1cm}{\bulkCol};
            \scalebox{0.5}[0.5]{\tri{-2cm+23cm}{-2cm}{45}};
            \scalebox{0.5}[0.5]{\tri{0cm+23cm}{2.2cm}{270}};
            \scalebox{0.5}[0.5]{\tri{2cm+23cm}{-2cm}{135}};
            \node[right] at (0,0.5cm) {\tiny{$C$}};
            \node[above] at (-0.5cm,-0.5cm) {\tiny{$C$}};
            \node[above] at (0.5cm,-0.5cm) {\tiny{$C$}};
            \node at (4,0) {$\phi=2^{-\frac{1}{4}}$};
        }
        \end{scope}
    \end{tikzpicture}
    \ee
\end{figure}

One could work out the intermediate phase where $C$ is condensed. The methods discussed in section \ref{sec:b_fusion} continues to apply, even though $\alpha_{AC}$ is not a {\it Lagrangian} algebra that defines a gapped boundary. In this case, one finds that the condensed phase is described by a fusion category that contains the toric code category as a sub-category. It has been noted that the toric code order remains ``deconfined'' such that the braiding structure is preserved, in addition to sectors identified as ``confined defects''  that are non-local wrt to the condensate, and thus whose braided structure is lost \cite{kirillov}. 

Let us summarize the properties of the intermediate phase in the table below:
\be \label{tab:AC}
\begin{tabular}{|c|c|c|c|c|c|c|}
\hline 
sectors $x$ & 1 & $ e$ & $ m$ & $ f$ & $X$ & $Y$ \\
\hline 
$W^{\alpha_{AC}}_{ix}$ & $A\oplus C$ & $ B\oplus C$ & $D$ & $ E $ & $D\oplus E$ & $F\oplus G\oplus H$   \\
\hline  
\end{tabular}
\ee

One could also work out the precise left actions of the algebra $\alpha_{AC}$ on these modules. They are presented in (\ref{eq:modules-condenseC}).  
One observes that there are multiple solutions in each given module in (\ref{eq:modules-condenseC}). In this case however, they all correspond to a phase redundancy following from the choice of phase for the fusion basis discussed in (\ref{eq:hombasis_rescale}).  In other words, they do not lead to independent modules. This should be contrasted with a q-type object that we will study below, where a single module gives rise to two truly independent solutions.

\begin{figure}[htbp!]
    \centering
    \be
    \label{eq:modules-condenseC}
    \begin{tikzpicture}[scale=0.3]
    \begin{scope}[xshift=-12cm,yshift=0cm]{
    \begin{scope}[xshift=0cm,yshift=0cm]{
    \node at (-7,0) {$\sqrt{3}$};
    \draw [thick, magenta] (-4,-4) -- (-4,4);
    \draw [thick] (-7,-3) -- (-4,0);
    \node at (-7.8,-3) {\footnotesize{$1$}}
    node at (-3.2,4) {\footnotesize{$e$}}
    node at (-3.2,-4) {\footnotesize{$e$}};
    }
    \end{scope}
    \begin{scope}[xshift=-5cm,yshift=0cm]{
    \node at (4.5,0) {\Large{$=$}};
    \draw [thick, magenta] (8,-4) -- (8,4);
    \draw [thick] (5,-3) -- (8,0);
    \draw [green, thick] (8,-1.5) -- (8,-2.4);
    \draw [green, thick] (8,-2) -- (8,2);
    \draw [green, thick] (6.5,-1.5) -- (8,0);
    \tri{6.2cm}{-1.8cm}{45}
    \tri{8cm}{2cm}{270}
    \tri{8cm}{-2.4cm}{90}
    \node at (4.2,-3) {\footnotesize{$1$}}
    node at (8.8,4) {\footnotesize{$e$}}
    node at (8.8,-4) {\footnotesize{$e$}}
    node at (7,-0.3) {\footnotesize{$A$}}
    node at (8.6,0.7) {\footnotesize{$B$}}
    node at (8.6,-0.9) {\footnotesize{$B$}};
    }
    \end{scope}
    \begin{scope}[xshift=2cm,yshift=0cm]{
    \node at (4.5,0) {\Large{$+$}};
    \draw [thick, magenta] (8,-4) -- (8,4);
    \draw [thick] (5,-3) -- (8,0);
    \draw [green, thick] (8,-1.5) -- (8,-2.4);
    \draw [green, thick] (8,-2) -- (8,2);
    \draw [green, thick] (6.5,-1.5) -- (8,0);
    \tri{6.2cm}{-1.8cm}{45}
    \tri{8cm}{2cm}{270}
    \tri{8cm}{-2.4cm}{90}
    \node at (4.2,-3) {\footnotesize{$1$}}
    node at (8.8,4) {\footnotesize{$e$}}
    node at (8.8,-4) {\footnotesize{$e$}}
    node at (7,-0.3) {\footnotesize{$A$}}
    node at (8.6,0.7) {\footnotesize{$C$}}
    node at (8.6,-0.9) {\footnotesize{$C$}};
    }
    \end{scope}
    \begin{scope}[xshift=9cm,yshift=0cm]{
    \node at (4.5,0) {$+\ \theta$};
    \draw [thick, magenta] (8,-4) -- (8,4);
    \draw [thick] (5,-3) -- (8,0);
    \draw [green, thick] (8,-1.5) -- (8,-2.4);
    \draw [green, thick] (8,-2) -- (8,2);
    \draw [green, thick] (6.5,-1.5) -- (8,0);
    \tri{6.2cm}{-1.8cm}{45}
    \tri{8cm}{2cm}{270}
    \tri{8cm}{-2.4cm}{90}
    \node at (4.2,-3) {\footnotesize{$1$}}
    node at (8.8,4) {\footnotesize{$e$}}
    node at (8.8,-4) {\footnotesize{$e$}}
    node at (7,-0.3) {\footnotesize{$C$}}
    node at (8.6,0.7) {\footnotesize{$C$}}
    node at (8.6,-0.9) {\footnotesize{$B$}};
    }
    \end{scope}
    \begin{scope}[xshift=16cm,yshift=0cm]{
    \node at (4.5,0) {$-\ \theta$};
    \draw [thick, magenta] (8,-4) -- (8,4);
    \draw [thick] (5,-3) -- (8,0);
    \draw [green, thick] (8,-1.5) -- (8,-2.4);
    \draw [green, thick] (8,-2) -- (8,2);
    \draw [green, thick] (6.5,-1.5) -- (8,0);
    \tri{6.2cm}{-1.8cm}{45}
    \tri{8cm}{2cm}{270}
    \tri{8cm}{-2.4cm}{90}
    \node at (4.2,-3) {\footnotesize{$1$}}
    node at (8.8,4) {\footnotesize{$e$}}
    node at (8.8,-4) {\footnotesize{$e$}}
    node at (7,-0.3) {\footnotesize{$C$}}
    node at (8.6,0.7) {\footnotesize{$C$}}
    node at (8.6,-0.9) {\footnotesize{$B$}};
    }
    \end{scope}
    \begin{scope}[xshift=23cm,yshift=0cm]{
    \node at (4.5,0) {$-\ \phi$};
    \draw [thick, magenta] (8,-4) -- (8,4);
    \draw [thick] (5,-3) -- (8,0);
    \draw [green, thick] (8,-1.5) -- (8,-2.4);
    \draw [green, thick] (8,-2) -- (8,2);
    \draw [green, thick] (6.5,-1.5) -- (8,0);
    \tri{6.2cm}{-1.8cm}{45}
    \tri{8cm}{2cm}{270}
    \tri{8cm}{-2.4cm}{90}
    \node at (4.2,-3) {\footnotesize{$1$}}
    node at (8.8,4) {\footnotesize{$e$}}
    node at (8.8,-4) {\footnotesize{$e$}}
    node at (7,-0.3) {\footnotesize{$C$}}
    node at (8.6,0.7) {\footnotesize{$C$}}
    node at (8.6,-0.9) {\footnotesize{$C$}};
    \node at (14,0) {$\theta=\pm i$};
    }
    \end{scope}
    }
    \end{scope}
    
    \begin{scope}[xshift=0cm,yshift=-10cm]{
    \begin{scope}[xshift=-13cm,yshift=0cm]{
    \begin{scope}[xshift=0cm,yshift=0cm]{
    \node at (-7,0) {$\sqrt{3}$};
    \draw [thick, magenta] (-4,-4) -- (-4,4);
    \draw [thick] (-7,-3) -- (-4,0);
    \node at (-7.8,-3) {\footnotesize{$1$}}
    node at (-3.2,4) {\footnotesize{$m$}}
    node at (-3.2,-4) {\footnotesize{$m$}};
    }
    \end{scope}
    \begin{scope}[xshift=-5cm,yshift=0cm]{
    \node at (4.5,0) {\Large{$=$}};
    \draw [thick, magenta] (8,-4) -- (8,4);
    \draw [thick] (5,-3) -- (8,0);
    \draw [green, thick] (8,-1.5) -- (8,-2.4);
    \draw [green, thick] (8,-2) -- (8,2);
    \draw [green, thick] (6.5,-1.5) -- (8,0);
    \tri{6.2cm}{-1.8cm}{45}
    \tri{8cm}{2cm}{270}
    \tri{8cm}{-2.4cm}{90}
    \node at (4.2,-3) {\footnotesize{$1$}}
    node at (8.8,4) {\footnotesize{$m$}}
    node at (8.8,-4) {\footnotesize{$m$}}
    node at (7,-0.3) {\footnotesize{$A$}}
    node at (8.6,0.7) {\footnotesize{$D$}}
    node at (8.6,-0.9) {\footnotesize{$D$}};
    }
    \end{scope}
    \begin{scope}[xshift=3cm,yshift=0cm]{
    \node at (4.5,0) {$+\ \phi^{-1}$};
    \draw [thick, magenta] (8,-4) -- (8,4);
    \draw [thick] (5,-3) -- (8,0);
    \draw [green, thick] (8,-1.5) -- (8,-2.4);
    \draw [green, thick] (8,-2) -- (8,2);
    \draw [green, thick] (6.5,-1.5) -- (8,0);
    \tri{6.2cm}{-1.8cm}{45}
    \tri{8cm}{2cm}{270}
    \tri{8cm}{-2.4cm}{90}
    \node at (4.2,-3) {\footnotesize{$1$}}
    node at (8.8,4) {\footnotesize{$m$}}
    node at (8.8,-4) {\footnotesize{$m$}}
    node at (7,-0.3) {\footnotesize{$C$}}
    node at (8.6,0.7) {\footnotesize{$D$}}
    node at (8.6,-0.9) {\footnotesize{$D$}};
    }
    \end{scope}
    }
    \end{scope}
    \begin{scope}[xshift=13cm,yshift=0cm]{
    \begin{scope}[xshift=0cm,yshift=0cm]{
    \node at (-7,0) {$\sqrt{3}$};
    \draw [thick, magenta] (-4,-4) -- (-4,4);
    \draw [thick] (-7,-3) -- (-4,0);
    \node at (-7.8,-3) {\footnotesize{$1$}}
    node at (-3.2,4) {\footnotesize{$f$}}
    node at (-3.2,-4) {\footnotesize{$f$}};
    }
    \end{scope}
    \begin{scope}[xshift=-5cm,yshift=0cm]{
    \node at (4.5,0) {\Large{$=$}};
    \draw [thick, magenta] (8,-4) -- (8,4);
    \draw [thick] (5,-3) -- (8,0);
    \draw [green, thick] (8,-1.5) -- (8,-2.4);
    \draw [green, thick] (8,-2) -- (8,2);
    \draw [green, thick] (6.5,-1.5) -- (8,0);
    \tri{6.2cm}{-1.8cm}{45}
    \tri{8cm}{2cm}{270}
    \tri{8cm}{-2.4cm}{90}
    \node at (4.2,-3) {\footnotesize{$1$}}
    node at (8.8,4) {\footnotesize{$f$}}
    node at (8.8,-4) {\footnotesize{$f$}}
    node at (7,-0.3) {\footnotesize{$A$}}
    node at (8.6,0.7) {\footnotesize{$E$}}
    node at (8.6,-0.9) {\footnotesize{$E$}};
    }
    \end{scope}
    \begin{scope}[xshift=2cm,yshift=0cm]{
    \node at (4.5,0) {$-\ \phi^{-1}$};
    \draw [thick, magenta] (8,-4) -- (8,4);
    \draw [thick] (5,-3) -- (8,0);
    \draw [green, thick] (8,-1.5) -- (8,-2.4);
    \draw [green, thick] (8,-2) -- (8,2);
    \draw [green, thick] (6.5,-1.5) -- (8,0);
    \tri{6.2cm}{-1.8cm}{45}
    \tri{8cm}{2cm}{270}
    \tri{8cm}{-2.4cm}{90}
    \node at (4.2,-3) {\footnotesize{$1$}}
    node at (8.8,4) {\footnotesize{$f$}}
    node at (8.8,-4) {\footnotesize{$f$}}
    node at (7,-0.3) {\footnotesize{$C$}}
    node at (8.6,0.7) {\footnotesize{$E$}}
    node at (8.6,-0.9) {\footnotesize{$E$}};
    }
    \end{scope}
    }
    \end{scope}
    }
    \end{scope}
    
    \begin{scope}[xshift=-16cm,yshift=-25cm]{
    \scalebox{0.8}[0.8]{
    \begin{scope}[xshift=0cm,yshift=0cm]{
    \node at (-7,0) {$\sqrt{3}$};
    \draw [thick, magenta] (-4,-4) -- (-4,4);
    \draw [thick] (-7,-3) -- (-4,0);
    \node at (-7.8,-3) {\footnotesize{$1$}}
    node at (-3.2,4) {\footnotesize{$X$}}
    node at (-3.2,-4) {\footnotesize{$X$}};
    }
    \end{scope}
    \begin{scope}[xshift=-5cm,yshift=0cm]{
    \node at (4.5,0) {\Large{$=$}};
    \draw [thick, magenta] (8,-4) -- (8,4);
    \draw [thick] (5,-3) -- (8,0);
    \draw [green, thick] (8,-1.5) -- (8,-2.4);
    \draw [green, thick] (8,-2) -- (8,2);
    \draw [green, thick] (6.5,-1.5) -- (8,0);
    \tri{6.2cm}{-1.8cm}{45}
    \tri{8cm}{2cm}{270}
    \tri{8cm}{-2.4cm}{90}
    \node at (4.2,-3) {\footnotesize{$1$}}
    node at (8.8,4) {\footnotesize{$X$}}
    node at (8.8,-4) {\footnotesize{$X$}}
    node at (7,-0.3) {\footnotesize{$A$}}
    node at (8.6,0.7) {\footnotesize{$D$}}
    node at (8.6,-0.9) {\footnotesize{$D$}};
    }
    \end{scope}
    \begin{scope}[xshift=2cm,yshift=0cm]{
    \node at (4.5,0) {$+$};
    \draw [thick, magenta] (8,-4) -- (8,4);
    \draw [thick] (5,-3) -- (8,0);
    \draw [green, thick] (8,-1.5) -- (8,-2.4);
    \draw [green, thick] (8,-2) -- (8,2);
    \draw [green, thick] (6.5,-1.5) -- (8,0);
    \tri{6.2cm}{-1.8cm}{45}
    \tri{8cm}{2cm}{270}
    \tri{8cm}{-2.4cm}{90}
    \node at (4.2,-3) {\footnotesize{$1$}}
    node at (8.8,4) {\footnotesize{$X$}}
    node at (8.8,-4) {\footnotesize{$X$}}
    node at (7,-0.3) {\footnotesize{$A$}}
    node at (8.6,0.7) {\footnotesize{$E$}}
    node at (8.6,-0.9) {\footnotesize{$E$}};
    }
    \end{scope}
    \begin{scope}[xshift=9cm,yshift=0cm]{
    \node at (4.5,0) {$-\ \phi^3$};
    \draw [thick, magenta] (8,-4) -- (8,4);
    \draw [thick] (5,-3) -- (8,0);
    \draw [green, thick] (8,-1.5) -- (8,-2.4);
    \draw [green, thick] (8,-2) -- (8,2);
    \draw [green, thick] (6.5,-1.5) -- (8,0);
    \tri{6.2cm}{-1.8cm}{45}
    \tri{8cm}{2cm}{270}
    \tri{8cm}{-2.4cm}{90}
    \node at (4.2,-3) {\footnotesize{$1$}}
    node at (8.8,4) {\footnotesize{$X$}}
    node at (8.8,-4) {\footnotesize{$X$}}
    node at (7,-0.3) {\footnotesize{$C$}}
    node at (8.6,0.7) {\footnotesize{$D$}}
    node at (8.6,-0.9) {\footnotesize{$D$}};
    }
    \end{scope}
    \begin{scope}[xshift=16cm,yshift=0cm]{
    \node at (4.5,0) {$+\ \phi^3$};
    \draw [thick, magenta] (8,-4) -- (8,4);
    \draw [thick] (5,-3) -- (8,0);
    \draw [green, thick] (8,-1.5) -- (8,-2.4);
    \draw [green, thick] (8,-2) -- (8,2);
    \draw [green, thick] (6.5,-1.5) -- (8,0);
    \tri{6.2cm}{-1.8cm}{45}
    \tri{8cm}{2cm}{270}
    \tri{8cm}{-2.4cm}{90}
    \node at (4.2,-3) {\footnotesize{$1$}}
    node at (8.8,4) {\footnotesize{$X$}}
    node at (8.8,-4) {\footnotesize{$X$}}
    node at (7,-0.3) {\footnotesize{$C$}}
    node at (8.6,0.7) {\footnotesize{$E$}}
    node at (8.6,-0.9) {\footnotesize{$E$}};
    }
    \end{scope}
    \begin{scope}[xshift=26cm,yshift=0cm]{
    \node at (3,0) {$+\ \sqrt{3}\phi^3\alpha$};
    \draw [thick, magenta] (8,-4) -- (8,4);
    \draw [thick] (5,-3) -- (8,0);
    \draw [green, thick] (8,-1.5) -- (8,-2.4);
    \draw [green, thick] (8,-2) -- (8,2);
    \draw [green, thick] (6.5,-1.5) -- (8,0);
    \tri{6.2cm}{-1.8cm}{45}
    \tri{8cm}{2cm}{270}
    \tri{8cm}{-2.4cm}{90}
    \node at (4.2,-3) {\footnotesize{$1$}}
    node at (8.8,4) {\footnotesize{$X$}}
    node at (8.8,-4) {\footnotesize{$X$}}
    node at (7,-0.3) {\footnotesize{$C$}}
    node at (8.6,0.7) {\footnotesize{$E$}}
    node at (8.6,-0.9) {\footnotesize{$D$}};
    }
    \end{scope}
    \begin{scope}[xshift=36cm,yshift=0cm]{
    \node at (3,0) {$+\ i\sqrt{3}\phi^3\alpha$};
    \draw [thick, magenta] (8,-4) -- (8,4);
    \draw [thick] (5,-3) -- (8,0);
    \draw [green, thick] (8,-1.5) -- (8,-2.4);
    \draw [green, thick] (8,-2) -- (8,2);
    \draw [green, thick] (6.5,-1.5) -- (8,0);
    \tri{6.2cm}{-1.8cm}{45}
    \tri{8cm}{2cm}{270}
    \tri{8cm}{-2.4cm}{90}
    \node at (4.2,-3) {\footnotesize{$1$}}
    node at (8.8,4) {\footnotesize{$X$}}
    node at (8.8,-4) {\footnotesize{$X$}}
    node at (7,-0.3) {\footnotesize{$C$}}
    node at (8.6,0.7) {\footnotesize{$D$}}
    node at (8.6,-0.9) {\footnotesize{$E$}};
    \node at (13,0) {$\alpha=\pm e^{i\frac{3\pi}{4}}$};
    }
    \end{scope}
    }
    }
    \end{scope}

    \begin{scope}[xshift=-12cm,yshift=-30cm]{
    \begin{scope}[xshift=0cm,yshift=0cm]{
    \node at (-7,0) {$\sqrt{3}$};
    \draw [thick, magenta] (-4,-4) -- (-4,4);
    \draw [thick] (-7,-3) -- (-4,0);
    \node at (-7.8,-3) {\footnotesize{$1$}}
    node at (-3.2,4) {\footnotesize{$Y$}}
    node at (-3.2,-4) {\footnotesize{$Y$}};
    }
    \end{scope}
    \begin{scope}[xshift=-5cm,yshift=0cm]{
    \node at (4.5,0) {$=$};
    \draw [thick, magenta] (8,-4) -- (8,4);
    \draw [thick] (5,-3) -- (8,0);
    \draw [green, thick] (8,-1.5) -- (8,-2.4);
    \draw [green, thick] (8,-2) -- (8,2);
    \draw [green, thick] (6.5,-1.5) -- (8,0);
    \tri{6.2cm}{-1.8cm}{45}
    \tri{8cm}{2cm}{270}
    \tri{8cm}{-2.4cm}{90}
    \node at (4.2,-3) {\footnotesize{$1$}}
    node at (8.8,4) {\footnotesize{$Y$}}
    node at (8.8,-4) {\footnotesize{$Y$}}
    node at (7,-0.3) {\footnotesize{$A$}}
    node at (8.6,0.7) {\footnotesize{$F$}}
    node at (8.6,-0.9) {\footnotesize{$F$}};
    }
    \end{scope}
    \begin{scope}[xshift=2cm,yshift=0cm]{
    \node at (4.5,0) {$+$};
    \draw [thick, magenta] (8,-4) -- (8,4);
    \draw [thick] (5,-3) -- (8,0);
    \draw [green, thick] (8,-1.5) -- (8,-2.4);
    \draw [green, thick] (8,-2) -- (8,2);
    \draw [green, thick] (6.5,-1.5) -- (8,0);
    \tri{6.2cm}{-1.8cm}{45}
    \tri{8cm}{2cm}{270}
    \tri{8cm}{-2.4cm}{90}
    \node at (4.2,-3) {\footnotesize{$1$}}
    node at (8.8,4) {\footnotesize{$Y$}}
    node at (8.8,-4) {\footnotesize{$Y$}}
    node at (7,-0.3) {\footnotesize{$A$}}
    node at (8.6,0.7) {\footnotesize{$G$}}
    node at (8.6,-0.9) {\footnotesize{$G$}};
    }
    \end{scope}
    \begin{scope}[xshift=9cm,yshift=0cm]{
    \node at (4.5,0) {$+$};
    \draw [thick, magenta] (8,-4) -- (8,4);
    \draw [thick] (5,-3) -- (8,0);
    \draw [green, thick] (8,-1.5) -- (8,-2.4);
    \draw [green, thick] (8,-2) -- (8,2);
    \draw [green, thick] (6.5,-1.5) -- (8,0);
    \tri{6.2cm}{-1.8cm}{45}
    \tri{8cm}{2cm}{270}
    \tri{8cm}{-2.4cm}{90}
    \node at (4.2,-3) {\footnotesize{$1$}}
    node at (8.8,4) {\footnotesize{$Y$}}
    node at (8.8,-4) {\footnotesize{$Y$}}
    node at (7,-0.3) {\footnotesize{$A$}}
    node at (8.6,0.7) {\footnotesize{$H$}}
    node at (8.6,-0.9) {\footnotesize{$H$}};
    }
    \end{scope}
    \begin{scope}[xshift=16cm,yshift=0cm]{
    \node at (4.5,0) {$+\ \phi\beta$};
    \draw [thick, magenta] (8,-4) -- (8,4);
    \draw [thick] (5,-3) -- (8,0);
    \draw [green, thick] (8,-1.5) -- (8,-2.4);
    \draw [green, thick] (8,-2) -- (8,2);
    \draw [green, thick] (6.5,-1.5) -- (8,0);
    \tri{6.2cm}{-1.8cm}{45}
    \tri{8cm}{2cm}{270}
    \tri{8cm}{-2.4cm}{90}
    \node at (4.2,-3) {\footnotesize{$1$}}
    node at (8.8,4) {\footnotesize{$Y$}}
    node at (8.8,-4) {\footnotesize{$Y$}}
    node at (7,-0.3) {\footnotesize{$C$}}
    node at (8.6,0.7) {\footnotesize{$G$}}
    node at (8.6,-0.9) {\footnotesize{$F$}};
    }
    \end{scope}
    \begin{scope}[xshift=23cm,yshift=0cm]{
    \node at (4.5,0) {$+\ \phi\gamma$};
    \draw [thick, magenta] (8,-4) -- (8,4);
    \draw [thick] (5,-3) -- (8,0);
    \draw [green, thick] (8,-1.5) -- (8,-2.4);
    \draw [green, thick] (8,-2) -- (8,2);
    \draw [green, thick] (6.5,-1.5) -- (8,0);
    \tri{6.2cm}{-1.8cm}{45}
    \tri{8cm}{2cm}{270}
    \tri{8cm}{-2.4cm}{90}
    \node at (4.2,-3) {\footnotesize{$1$}}
    node at (8.8,4) {\footnotesize{$Y$}}
    node at (8.8,-4) {\footnotesize{$Y$}}
    node at (7,-0.3) {\footnotesize{$C$}}
    node at (8.6,0.7) {\footnotesize{$H$}}
    node at (8.6,-0.9) {\footnotesize{$F$}};
    }
    \end{scope}
    \begin{scope}[xshift=-7cm,yshift=-10cm]{
    \node at (3.5,0) {$+\ \phi\beta^{-1}$};
    \draw [thick, magenta] (8,-4) -- (8,4);
    \draw [thick] (5,-3) -- (8,0);
    \draw [green, thick] (8,-1.5) -- (8,-2.4);
    \draw [green, thick] (8,-2) -- (8,2);
    \draw [green, thick] (6.5,-1.5) -- (8,0);
    \tri{6.2cm}{-1.8cm}{45}
    \tri{8cm}{2cm}{270}
    \tri{8cm}{-2.4cm}{90}
    \node at (4.2,-3) {\footnotesize{$1$}}
    node at (8.8,4) {\footnotesize{$Y$}}
    node at (8.8,-4) {\footnotesize{$Y$}}
    node at (7,-0.3) {\footnotesize{$C$}}
    node at (8.6,0.7) {\footnotesize{$F$}}
    node at (8.6,-0.9) {\footnotesize{$G$}};
    }
    \end{scope}
    \begin{scope}[xshift=2cm,yshift=-10cm]{
    \node at (3.5,0) {$+\ \phi\gamma\beta^{-1}$};
    \draw [thick, magenta] (8,-4) -- (8,4);
    \draw [thick] (5,-3) -- (8,0);
    \draw [green, thick] (8,-1.5) -- (8,-2.4);
    \draw [green, thick] (8,-2) -- (8,2);
    \draw [green, thick] (6.5,-1.5) -- (8,0);
    \tri{6.2cm}{-1.8cm}{45}
    \tri{8cm}{2cm}{270}
    \tri{8cm}{-2.4cm}{90}
    \node at (4.2,-3) {\footnotesize{$1$}}
    node at (8.8,4) {\footnotesize{$Y$}}
    node at (8.8,-4) {\footnotesize{$Y$}}
    node at (7,-0.3) {\footnotesize{$C$}}
    node at (8.6,0.7) {\footnotesize{$H$}}
    node at (8.6,-0.9) {\footnotesize{$G$}};
    }
    \end{scope}
    \begin{scope}[xshift=10.5cm,yshift=-10cm]{
    \node at (3.5,0) {$+\ \phi\omega\gamma$};
    \draw [thick, magenta] (8,-4) -- (8,4);
    \draw [thick] (5,-3) -- (8,0);
    \draw [green, thick] (8,-1.5) -- (8,-2.4);
    \draw [green, thick] (8,-2) -- (8,2);
    \draw [green, thick] (6.5,-1.5) -- (8,0);
    \tri{6.2cm}{-1.8cm}{45}
    \tri{8cm}{2cm}{270}
    \tri{8cm}{-2.4cm}{90}
    \node at (4.2,-3) {\footnotesize{$1$}}
    node at (8.8,4) {\footnotesize{$Y$}}
    node at (8.8,-4) {\footnotesize{$Y$}}
    node at (7,-0.3) {\footnotesize{$C$}}
    node at (8.6,0.7) {\footnotesize{$F$}}
    node at (8.6,-0.9) {\footnotesize{$H$}};
    }
    \end{scope}
    \begin{scope}[xshift=19cm,yshift=-10cm]{
    \node at (3.5,0) {$+\ \phi\omega\beta\gamma$};
    \draw [thick, magenta] (8,-4) -- (8,4);
    \draw [thick] (5,-3) -- (8,0);
    \draw [green, thick] (8,-1.5) -- (8,-2.4);
    \draw [green, thick] (8,-2) -- (8,2);
    \draw [green, thick] (6.5,-1.5) -- (8,0);
    \tri{6.2cm}{-1.8cm}{45}
    \tri{8cm}{2cm}{270}
    \tri{8cm}{-2.4cm}{90}
    \node at (4.2,-3) {\footnotesize{$1$}}
    node at (8.8,4) {\footnotesize{$Y$}}
    node at (8.8,-4) {\footnotesize{$Y$}}
    node at (7,-0.3) {\footnotesize{$C$}}
    node at (8.6,0.7) {\footnotesize{$G$}}
    node at (8.6,-0.9) {\footnotesize{$H$}};
    \node at (13,1.5) {$\beta=\pm \omega^2$}
    node at (12.6,0) {$\gamma=\pm \omega$}
    node at (13,-1.5) {$\omega=e^{i\frac{2\pi}{3}}$};
    }
    \end{scope}
    }
    \end{scope}
    \end{tikzpicture}
    \ee
\end{figure}

\newpage

The fermionic gapped boundary is then generated by condensing $f$, exactly as in the toric code case. 
What is new here is that $X$ and $Y$ are non-Abelian defects, and they display interesting properties, particularly when we begin considering junctions between gapped boundaries. 

The fusion rules involving $X$ and $Y$ are summarized in the table below:
\be
\begin{tabular}{|c|c|c|}
\hline
$\otimes_{\alpha_{AC}}$ & $X$ & $Y$ \\
\hline
$e$ & $ X$ & $Y$\\
$m$ & $Y$ & $X$\\
$f$ & $Y$ & $X$\\
$X$ & $ 1\oplus e \oplus Y$      &   $m \oplus f  \oplus X$ \\
$Y$ &  $m \oplus f  \oplus X$    &   $ 1\oplus e \oplus Y$ \\
\hline
\end{tabular}
\ee

Finally, as we condense $f$, it is not hard to see that $X$ and $Y$ together form a module, while the toric code sub-category behaves in exactly the same way described in section \ref{sec:toricf_bc}.

Let us summarize the overall $W$ and $\Omega$ matrix: 
\be  \label{tab:Af}
\begin{tabular}{|c|c|c|c|} 
\hline
modules $x$ & 1 & $ X_{f}$ & $Z$ \\
\hline
$W^{\mathcal{A}_f}_{ix}$ & $A\oplus C\oplus E$ & $B\oplus C\oplus D$ & $D\oplus E \oplus F\oplus G\oplus H$ \\
\hline
$\Omega^{\mathcal{A}_f}_{ix}$  & $A \oplus C \oplus -E$ & $B\oplus C \oplus - D$ & $-D \oplus -E \oplus F \oplus G \oplus H$\\
\hline
\end{tabular}
\ee

The fusion rules between these defects are given by
\be \label{eq:fuseAf}
X_f \otimes_{\mathcal{A}_f} X_f = 1, \qquad X_f \otimes_{\mathcal{A}_f} Z = Z, \qquad Z \otimes_{\mathcal{A}_f} Z = 1 \oplus X_f \oplus Z.
\ee
These fusion rules satisfy the (twisted) defect Verlinde formula. 
It confirms that all the defects have trivial endomorphism. Now here, we again confirm the claim made after  (\ref{eq:littlev}), that the number of objects $N_f$ in the defect responsible for generating the fermion parity -- in this case it is $X_f = B\oplus C\oplus D$ containing 3 objects i.e. $N_f =3$ -- equals the total number of non-q-type defects in the gapped phase -- i.e. $\{1, X_f, Z\}$!

\subsubsection{A bosonic-fermionic junction -- Take 2} \label{sec:bfjunction_take2}

It is particularly interesting to revisit the bosonic-fermionic junction corresponding to juxtaposing the magnetic boundary and the fermionic
boundary in the toric code. There is new physics precisely because of the presence of non-Abelian confined defects. 

The magnetic boundary described in terms of a Frobenius algebra in $D(S_3)$ can be summarized by the following W-matrix:
\be
\begin{tabular}{|c|c|c|c|}
\hline
modules $x$ & 1 & $ X_{m}$ & $Z_m$ \\
\hline
$W^{\mathcal{A}_m}_{ix}$ & $A\oplus C\oplus D$ & $B\oplus C\oplus E$ & $D\oplus E \oplus F\oplus G\oplus H$ \\
\hline
\end{tabular}
\ee
Their fusion rules are identical to (\ref{eq:fuseAf}) by replacing $X_f \to X_m$ and $Z \to Z_m$.
One observes that the above table is equivalent to (\ref{tab:Af}) upon exchanging $D$ and $E$.
There is this curious situation that $Z_m$ as a defect in the magnetic boundary contains the same list of anyons as the $Z$ defect in the fermionic boundary.

Now we would like to work out the junction defects. Following the playbook in section \ref{sec:junctions}, one could identify two different bimodules by inspecting all the induced modules one by one. i.e. We inspect $\mathcal{A}_m\otimes c_i \otimes \mathcal{A}_f, \,\, \forall c_i \in D(S_3)$. The two junction modules are summarized as follows:
 
\be
X_{mf} = A\oplus C\oplus E \oplus B\oplus C \oplus D, \qquad Z_{mf} =  D\oplus E\oplus F \oplus G\oplus H.  
\ee
One can see from the table (\ref{tab:AC}), that $X_{mf}$ is the same $X_{mf}$ we have discussed previously in (\ref{eq:ef_fuse}).
i.e. The fusion of $X_{mf}$ is given by
\be \label{eq:xxmf}
X_{mf}\otimes_{\mathcal{A}_f} X_{fm} = \mathcal{A}_m \oplus  X_m,
\ee
and thus it should carry a quantum dimension of $\sqrt{2}$ -- up to the ambiguity of the addition of extra Majorana modes. 

The fusion of $Z_{mf}$ is slightly trickier. To understand it, it is necessary to see that it is actually a q-type object with non-trivial endomorphism. This is where studying the intermediate phase where $A\oplus C$ has already been condensed simplifies the problem
significantly. 

We can make use of the identity  (\ref{eq:double_reciprocity2}), applying it on the intermediate phase where $\alpha_{AC}$ has condensed. We notice that
\be
\mathcal{A}_m \otimes X \otimes \mathcal{A}_f = 2 (X\oplus Y) = 2 Z_{mf}.
\ee
The identity (\ref{eq:double_reciprocity2}) then implies 
\be
\textrm{End}_{\mathcal{A}_m|\mathcal{A}_f} (\textrm{Ind}_{\mathcal{A}_m|\mathcal{A}_f} (X)) \equiv \textrm{Hom}_{\mathcal{A}_m|\mathcal{A}_f} (\textrm{Ind}_{\mathcal{A}_m|\mathcal{A}_f} (X),\textrm{Ind}_{\mathcal{A}_m|\mathcal{A}_f} (X) ) = \textrm{Hom}(X, 2(X\oplus Y))
\ee
The space of maps from $X$ to $2(X\oplus Y)$ has to be 2 dimensional, for $X,Y$ {\it simple} objects. Therefore the endomorphism space of Ind$_{\mathcal{A}_m|\mathcal{A}_f}(X) = Z$ is also 2 dimensional. We note that one might entertain the possibility that  $Z_{mf}$ is {\it  not} a simple object (irreducible representation) -- that could also support a non-trivial space of endomorphism. However, the dimension should take the form of $\sum_x n_x^2 \langle \mathcal{M}_x, \mathcal{M}_x\rangle$, where $x$ runs through all the irreducible representations contained in $Z_{mf}$ and $n_x$ is the multiplicity of $M_x$ appearing in $Z$ and so $n_x \in \Z_{>0}$. 
Again for simple objects $\langle \mathcal{M}_x, \mathcal{M}_x\rangle$ can only be 1 for a non-q-type object, or 2 for a q-type object. Clearly, $\langle Z_{mf} , Z_{mf} \rangle_{\mathcal{A}_m|\mathcal{A}_f} =2$ is only compatible with $Z_{mf}$ being a simple q-type object.

There is another manifestation of the non-trivial endomorphism. Consider solving for the left-right action of $\mathcal{A}_m$ and $\mathcal{A}_f$ on $Z_{mf}$ using the methods described in (\ref{eq:reps_eq}) and (\ref{eq:bimodule}). One should be able to obtain 2 independent solutions, despite the fact that $Z_{mf}$ remains simple. These would form basis of the two generators of the endomorphism maps! For illustration purpose, we solve for them explicitly in the next subsection.

Finally, we are ready to recover the fusion rules of the junctions. Again to be careful with Majorana modes, we should upgrade the Frobenius algebra to include the free fermion explicitly, exactly as in (\ref{eq:Ap}, \ref{eq:AftoAfp}).

Using notations of the intermediate phase, we obtain the following fusion rules
\begin{align} 
&X_{mf} \otimes_{\mathcal{A}_f'} Z_{fm} =_{\textrm{reduction to left $\mathcal{A}_m$ module}} (1\oplus m) \otimes 1 \otimes (1\oplus  f\otimes \psi_0) X = (1\oplus \psi_0) \otimes Z_m ,  \label{eq:xzmf} \\
&Z_{mf}  \otimes_{\mathcal{A}_f'} Z_{fm} =_{\textrm{reduction to left $\mathcal{A}_m$ module}} (1\oplus m) \otimes X \otimes (1\oplus  f\otimes \psi_0) X  \nonumber \\
&= (1\oplus \psi_0) \otimes (\mathcal{A}_m \oplus X_m \oplus Z_m).  \label{eq:zzmf}
\end{align}

Note that we have made the reduction from a $\mathcal{A}_m|\mathcal{A}_m$ bimodule to a left (right) $\mathcal{A}_m$ module by implicitly modding out by a factor of $\mathcal{A}_m$. 

A warning here has to be flagged: the extra factor of $(1\oplus \psi_0)$ is {\bf not} a Majorana mode. We have actually confronted this situation in (\ref{eq:fusemajorana}), 
Recall that Majorana modes are localized at a point. Here, it is roaming free along the entire magnetic condensate boundary. 
It is only making explicit that there are two fusion channels, one with even and the other odd fermion parity. Had we kept $\psi_0$ explicit in (\ref{eq:evenfuse}) the odd channels would be tagged by a copy of $\psi_0$ too.  Now we see that $\psi_0$ is an important
book-keeping device -- if localized at a junction it accounts for quantum dimensions of $\sqrt{2}$. On the other hand if it roams free in the 1 dimensional boundary or in the bulk,  {\it they account for factors of 2 }. Their introduction allows one to keep track of quantum dimensions of defects clearly -- which are naturally conserved under fusion. 
Recall that $Z_{mf}$ is a q-type object, and in cases as such, it is expected that its fusion maps always carry an even number of even and odd channels, since they could be converted between each other by composing with an odd endomorphism. This has been briefly discussed in the introduction of super-category in section \ref{sec:intro_cat}.

The quantum dimension of $Z_{mf}$ is thus given by
\be
d_{Z_{mf}} = 2 \sqrt{2},
\ee
again with the ambiguity of adding Majorana modes at the junction on top of this ``canonical'' basis. 
This is rather amusing, since $Z_{mf}, Z_{m} $ and $Z$ all contain exactly the same list of anyons $D\oplus E \oplus F\oplus G \oplus H$!

The computation of the half-linking numbers and a check of the defect Verlinde formula will be discussed below. 

\subsubsection{Half-linking numbers and the defect Verlinde formula -- Take 2}

In the previous subsection we studied the bi-modules are argued that $Z_{mf}$ should carry non-trivial endomorphism. 
%One manifestation of the non-trivial endomorphism is the emergence of two independent solutions when one solves for the left-right action of the algebras on the module. To illustrate this point, we solve for the left-right actions explicitly below. 
One manifestation of the non-trivial endomorphism is the emergence of two independent solutions when one solves for the left-right action of the algebras on the module after fixing all the phase ambiguities. To illustrate this point, we solve for the left-right actions explicitly below. 
We note that the Frobenius algebra $\mathcal{A}_m = 1\oplus m$ is identical to (\ref{eq:algebra-Am}), replacing $e$ by $m$.  
\begin{figure}[h!]
    \centering
    \be
    \begin{tikzpicture}[scale=0.3]
    \begin{scope}[xshift=4cm,yshift=0cm]{
    \begin{scope}[xshift=0cm,yshift=0cm]{
    \node at (-7,0) {$\sqrt{2}$};
    \draw [thick, magenta] (-4,-4) -- (-4,4);
    \draw [thick] (-7,-3) -- (-4,0);
    \node at (-7.8,-3) {\footnotesize{$\mathcal{A}_m$}}
    node at (-3,4) {\footnotesize{$Z_{mf}$}}
    node at (-3,-4) {\footnotesize{$Z_{mf}$}};
    }
    \end{scope}
    \begin{scope}[xshift=-5cm,yshift=0cm]{
    \node at (4.5,0) {$=$};
    \draw [thick, magenta] (8,-4) -- (8,4);
    \draw [thick] (5,-3) -- (8,0);
    \draw [green, thick] (8,-1.5) -- (8,-2.4);
    \draw [green, thick] (8,-2) -- (8,2);
    \draw [green, thick] (6.5,-1.5) -- (8,0);
    \tri{6.2cm}{-1.8cm}{45}
    \tri{8cm}{2cm}{270}
    \tri{8cm}{-2.4cm}{90}
    \node at (4.2,-3) {\footnotesize{$\mathcal{A}_m$}}
    node at (9,4) {\footnotesize{$Z_{mf}$}}
    node at (9,-4) {\footnotesize{$Z_{mf}$}}
    node at (7,-0.3) {\footnotesize{$1$}}
    node at (8.6,0.7) {\footnotesize{$X$}}
    node at (8.6,-0.9) {\footnotesize{$X$}};
    }
    \end{scope}
    \begin{scope}[xshift=2cm,yshift=0cm]{
    \node at (4.5,0) {$+$};
    \draw [thick, magenta] (8,-4) -- (8,4);
    \draw [thick] (5,-3) -- (8,0);
    \draw [green, thick] (8,-1.5) -- (8,-2.4);
    \draw [green, thick] (8,-2) -- (8,2);
    \draw [green, thick] (6.5,-1.5) -- (8,0);
    \tri{6.2cm}{-1.8cm}{45}
    \tri{8cm}{2cm}{270}
    \tri{8cm}{-2.4cm}{90}
    \node at (4.2,-3) {\footnotesize{$\mathcal{A}_m$}}
    node at (9,4) {\footnotesize{$Z_{mf}$}}
    node at (9,-4) {\footnotesize{$Z_{mf}$}}
    node at (7,-0.3) {\footnotesize{$1$}}
    node at (8.6,0.7) {\footnotesize{$Y$}}
    node at (8.6,-0.9) {\footnotesize{$Y$}};
    }
    \end{scope}
    \begin{scope}[xshift=9cm,yshift=0cm]{
    \node at (4.5,0) {$+$};
    \draw [thick, magenta] (8,-4) -- (8,4);
    \draw [thick] (5,-3) -- (8,0);
    \draw [green, thick] (8,-1.5) -- (8,-2.4);
    \draw [green, thick] (8,-2) -- (8,2);
    \draw [green, thick] (6.5,-1.5) -- (8,0);
    \tri{6.2cm}{-1.8cm}{45}
    \tri{8cm}{2cm}{270}
    \tri{8cm}{-2.4cm}{90}
    \node at (4.2,-3) {\footnotesize{$\mathcal{A}_m$}}
    node at (9,4) {\footnotesize{$Z_{mf}$}}
    node at (9,-4) {\footnotesize{$Z_{mf}$}}
    node at (7,-0.3) {\footnotesize{$m$}}
    node at (8.6,0.7) {\footnotesize{$X$}}
    node at (8.6,-0.9) {\footnotesize{$Y$}};
    }
    \end{scope}
    \begin{scope}[xshift=16cm,yshift=0cm]{
    \node at (4.5,0) {$+$};
    \draw [thick, magenta] (8,-4) -- (8,4);
    \draw [thick] (5,-3) -- (8,0);
    \draw [green, thick] (8,-1.5) -- (8,-2.4);
    \draw [green, thick] (8,-2) -- (8,2);
    \draw [green, thick] (6.5,-1.5) -- (8,0);
    \tri{6.2cm}{-1.8cm}{45}
    \tri{8cm}{2cm}{270}
    \tri{8cm}{-2.4cm}{90}
    \node at (4.2,-3) {\footnotesize{$\mathcal{A}_m$}}
    node at (9,4) {\footnotesize{$Z_{mf}$}}
    node at (9,-4) {\footnotesize{$Z_{mf}$}}
    node at (7,-0.3) {\footnotesize{$m$}}
    node at (8.6,0.7) {\footnotesize{$Y$}}
    node at (8.6,-0.9) {\footnotesize{$X$}};
    }
    \end{scope}
     }
    \end{scope}
    \begin{scope}[xshift=0cm,yshift=-10cm]{
    \begin{scope}[xshift=0cm,yshift=0cm]{
    \node at (-6,0) {$\sqrt{2}$};
    \draw [thick, magenta] (-4,-4) -- (-4,4);
    \draw [thick] (-1,-3) -- (-4,0);
    \node at (-0.2,-3) {\footnotesize{$\mathcal{A}_f$}}
    node at (-3,4) {\footnotesize{$Z_{mf}$}}
    node at (-3,-4) {\footnotesize{$Z_{mf}$}};
    }
    \end{scope}
    \begin{scope}[xshift=-5cm,yshift=0cm]{
    \node at (4.5,0) {$=$};
    \draw [thick, magenta] (8,-4) -- (8,4);
    \draw [thick] (11,-3) -- (8,0);
    \draw [green, thick] (8,-1.5) -- (8,-2.4);
    \draw [green, thick] (8,-2) -- (8,2);
    \draw [green, thick] (9.5,-1.5) -- (8,0);
    \tri{9.8cm}{-1.8cm}{45+90}
    \tri{8cm}{2cm}{270}
    \tri{8cm}{-2.4cm}{90}
    \node at (11.8,-3) {\footnotesize{$\mathcal{A}_f$}}
    node at (9,4) {\footnotesize{$Z_{mf}$}}
    node at (9,-4) {\footnotesize{$Z_{mf}$}}
    node at (9,-0.3) {\footnotesize{$1$}}
    node at (7.4,0.7) {\footnotesize{$X$}}
    node at (7.4,-0.9) {\footnotesize{$X$}};
    }
    \end{scope}
    \begin{scope}[xshift=2cm,yshift=0cm]{
    \node at (4.5,0) {$+$};
    \draw [thick, magenta] (8,-4) -- (8,4);
    \draw [thick] (11,-3) -- (8,0);
    \draw [green, thick] (8,-1.5) -- (8,-2.4);
    \draw [green, thick] (8,-2) -- (8,2);
    \draw [green, thick] (9.5,-1.5) -- (8,0);
    \tri{9.8cm}{-1.8cm}{45+90}
    \tri{8cm}{2cm}{270}
    \tri{8cm}{-2.4cm}{90}
    \node at (11.8,-3) {\footnotesize{$\mathcal{A}_f$}}
    node at (9,4) {\footnotesize{$Z_{mf}$}}
    node at (9,-4) {\footnotesize{$Z_{mf}$}}
    node at (9,-0.3) {\footnotesize{$1$}}
    node at (7.4,0.7) {\footnotesize{$Y$}}
    node at (7.4,-0.9) {\footnotesize{$Y$}};
    }
    \end{scope}
    \begin{scope}[xshift=11cm,yshift=0cm]{
    \node at (4.5,0) {$+\ \psi\ $};
    \draw [thick, magenta] (8,-4) -- (8,4);
    \draw [thick] (11,-3) -- (8,0);
    \draw [green, thick] (8,-1.5) -- (8,-2.4);
    \draw [green, thick] (8,-2) -- (8,2);
    \draw [green, thick] (9.5,-1.5) -- (8,0);
    \tri{9.8cm}{-1.8cm}{45+90}
    \tri{8cm}{2cm}{270}
    \tri{8cm}{-2.4cm}{90}
    \node at (11.8,-3) {\footnotesize{$\mathcal{A}_f$}}
    node at (9,4) {\footnotesize{$Z_{mf}$}}
    node at (9,-4) {\footnotesize{$Z_{mf}$}}
    node at (9,-0.3) {\footnotesize{$f$}}
    node at (7.4,0.7) {\footnotesize{$X$}}
    node at (7.4,-0.9) {\footnotesize{$Y$}};
    }
    \end{scope}
    \begin{scope}[xshift=21cm,yshift=0cm]{
    \node at (4.5,0) {$+\ \psi^{-1}\ $};
    \draw [thick, magenta] (8,-4) -- (8,4);
    \draw [thick] (11,-3) -- (8,0);
    \draw [green, thick] (8,-1.5) -- (8,-2.4);
    \draw [green, thick] (8,-2) -- (8,2);
    \draw [green, thick] (9.5,-1.5) -- (8,0);
    \tri{9.8cm}{-1.8cm}{45+90}
    \tri{8cm}{2cm}{270}
    \tri{8cm}{-2.4cm}{90}
    \node at (11.8,-3) {\footnotesize{$\mathcal{A}_f$}}
    node at (9,4) {\footnotesize{$Z_{mf}$}}
    node at (9,-4) {\footnotesize{$Z_{mf}$}}
    node at (9,-0.3) {\footnotesize{$f$}}
    node at (7.4,0.7) {\footnotesize{$Y$}}
    node at (7.4,-0.9) {\footnotesize{$X$}};
    }
    \end{scope}
    \node at (37,0) {\large{$\psi=\pm i\omega$}};
    }
    \end{scope}
    \end{tikzpicture}
    \ee
\end{figure}

We can make use of the explicit form of the algebra and modules and compute the gamma matrix using (\ref{eq:gamma_pic_2}).
This gives
\begin{eqnarray} \label{eq:gammamf}
    \gamma^{(m|f)}=
    \begin{blockarray}{ccc}
        A & C \\
        \begin{block}{(cc)c}
          2 \mathcal{N}_A & 4  \mathcal{N}_C & ~X_{mf} \\
          4  \mathcal{N}_A & -4  \mathcal{N}_C & ~Z_{mf} \\
        \end{block}
        \end{blockarray}
\end{eqnarray}
We found that
\be \label{eq:NAB2}
\mathcal{N}_{c_i} = \frac{1 } {\sqrt{2D_{D(S_3)  } d_{c_i} }}, \qquad c_i \in   \mathcal{A}_m \cap \mathcal{A}_f.
%\mathcal{N}_{c_i} = \frac{\sqrt{2D_{D(S_3)  } } } {d_{c_i}}, \qquad c_i \in   \mathcal{A}_m \cap \mathcal{A}_f.
\ee
This expression again confirms (\ref{eq:NAB_2}). 
Substituting (\ref{eq:gammamf}, \ref{eq:NAB2}) into the defect Verlinder formula (\ref{eq:defect_verlinde_junction}), we recover the fusion rules (\ref{eq:xxmf}, \ref{eq:xzmf}, \ref{eq:zzmf}).  (Recall that the untwisted Verlinde formula gives the total number of fusion channels -- adding even and odd channels. Therefore the $1\oplus \psi_0$ factors in (\ref{eq:xzmf}, \ref{eq:zzmf}) should be interpreted as factors of 2. )
%Substituting (\ref{eq:gammamf}, \ref{eq:NAB2}) into the defect verlinder formula, we recover the fusion rules (\ref{eq:xxmf}, \ref{eq:xzmf}, \ref{eq:zzmf}).  (Recall that the untwisted Verlinde formula gives the total number of fusion channels -- adding even and odd channels. Therefore the $1\oplus \psi_0$ factors in (\ref{eq:xzmf}, \ref{eq:zzmf}) should be interpreted as factors of 2. )

\section{(super)-modular invariants and twisted characters}
\label{sec:CFT}

Bosonic gapped boundaries in a topological order are in 1-1 correspondence with modular invariants. For topological orders corresponding to the representation category $C$ of the tensor product of a chiral algebra and an anti-chiral algebra , each of these bosonic gapped boundaries correspond to a modular invariant CFT. 

In the case of fermionic gapped boundaries, each of them certainly defines a ``super''- modular invariant i.e. a modular invariant under $S$ and $T^2$ transformation on a torus \cite{levin_protected_2013}. The reverse is not true however, as we will describe interesting examples in the appendix. It is argued that invariance under $T^2$ and $S$ is the appropriate generalization of the concept of modular invariance for spin CFT \cite{levin_protected_2013}. \footnote{These ``super'' modular invariants should not be confused with modular invariants of supersymmetric CFT's discussed in the CFT literature. While the chiral symmetry algebra would contain a fermionic sector, the requirement of invariance under $T$ remains. }

Meanwhile, the (super) modular invariant essentially defines a Hilbert space $H^\mathcal{A}$. 
\be
H^{\mathcal{A}} = \oplus_i W^{\mathcal{A}}_{i 1}  H_i,
\ee
where 
\be
H_i = \mathcal{V}_{i}  \otimes \overline{\mathcal{V}_{\bar{i}}},
\ee
and $\mathcal{V}_i$ are the representations of the chiral algebra that defines the topological order introduced at the beginning of the section.

The excitations at the gapped boundary correspond to topological defects in the (super) modular invariant CFT. 

The defect operator takes the following form
\be
\hat X =\sum_{i \in H^\mathcal{A}} \frac{\gamma_{x \,i_{\alpha,\beta}}}{\sqrt{S_{0c}}} | c_i, \alpha \rangle \langle c_i, \beta |, 
\ee
where $|c_i\rangle$ is a short hand for the primary and also descendents in $H_i$. As a topological defect, the descendents are summed over in a way where the levels match in the bra and ket \cite{petkova_generalised_2001}. The indices $\alpha, \beta \in \{1, \cdots, W^{\mathcal{A}}_{i1}\}$.
The coefficients $\gamma_{x \,i_{\alpha,\beta}}$ correspond precisely to half-linking numbers between the condensed anyon $c_i$ and the boundary excitation $x$ in the topological theory. 

Taking the trace on a torus produces the twisted character $\chi_X(-1/\tau, - 1/\bar \tau)$,
\be \label{eq:chiX1}
\chi_X(\tilde q, \bar{\tilde q}) \equiv \textrm{tr} (\tilde q^{L_0 - c/24} \bar {\tilde q}^{\bar L_0 - \bar c/24} X) = \sum_{i, \alpha} \frac{\gamma_{x \,i_{\alpha,\alpha}}}{\sqrt{S_{1i}}}   \chi_i(\tilde q, \bar {\tilde q} ), 
\ee
where $\chi_i$ is an abuse of notation corresponding to $\chi_i(\tau) \chi_{\bar i }(\bar\tau)$ that follows from tracing the holomorphic and anti-holomorphic parts in $H_i$.
It is customary to denote
\be
 \tilde q = e^{2 i \pi \tilde \tau}, \,\, \tilde \tau = -1/\tau, \,\, q = e^{2i\pi \tau}.
\ee
The rhs of (\ref{eq:chiX1}) can be rewritten using its $S$ modular transformation property to yield
\be \label{eq:StransX}
\chi_X(\tilde q, \bar {\tilde q} ) = \sum_{i \in H^\mathcal{A},\alpha,J} \gamma_{x \,i_{\alpha,\alpha}}S_{ij} \chi_j( q, \bar q),
\ee
where the $S$ matrix here correspond precisely to that of the bulk phase. 
When the condensation multiplicities (i.e. elements of the W matrix) are either 0 or 1, one can readily show, using the identities (\ref{eq:defect_verlinde_e}, \ref{eq:Vgamma}), that (\ref{eq:StransX}) reduces to the following:
\be \label{eq:X_chi_decomp}
\chi_X(\tilde q, \bar{\tilde q}) = \sum_j W_{jx} \chi_j(q, \bar q).
\ee
We note that here $j$ runs also over sectors outside of $\mathcal{A}$. While we do not have a direct proof of this result for general $W_{i1}>1$ , from physical considerations -- that the edge excitation admits a decomposition as bulk anyons -- the result (\ref{eq:X_chi_decomp}) should remain true. 

Parts of these have been discussed in \cite{Lou:2019heg, Shen:2019rck} where $\mathcal{A}$ defines a bosonic modular invariant. 
These results readily apply to super modular invariants, with the fusion algebra of the topological defects again given by the defect Verlinde formula (\ref{eq:defect_verlinde_e}, \ref{eq:Vgamma}).

The novel structure that comes with a super modular invariant is the presence of fermion parity. In the following, we will discuss
also twisted characters with R-type boundary conditions in the time direction in the following. 

\subsection{Topological defects in the CFT and the fermion parity defect}

It is well known that in a CFT carrying global symmetries, one can define characters twisted by a generator $g$ of the global symmetry group $G$ in the time direction. 
\be
\chi^g_X(\tilde q,\bar{\tilde q}) = \textrm{tr}(g \hat X\tilde q^{L_0 - c/24} \bar{\tilde q}^{\bar{L}_0 -\bar c/24} ).
\ee
As a global symmetry $g$ would commute with $L_0$ and $\bar{L}_0$.
In the context of the fermion parity symmetry, $g = (-1)^F$, where $F$ is the fermion parity operator. 
We thus define
\be
\chi^R_X(\tilde q,\bar{\tilde q}) \equiv  \textrm{tr}( (-)^F \hat X \tilde q^{L_0 - c/24} \bar{\tilde q}^{\bar{L}_0 -\bar c/24} ).
\ee
By considering the fermion parity of the different sectors appearing in (\ref{eq:X_chi_decomp}), one concludes that
\be
\chi^R_X(\tilde q,\bar{\tilde q}) = \sum_i \Omega_{ix} \chi_i( q, \bar{q}).
\ee
Moreover, in a spin CFT, we should keep track of the spin structure in the spatial direction that follows from the insertion of $X$. 
For $X$ corresponding to an $NS\, (R)$-type object, it generates a $NS\, (R)$ type spin structure in the spatial cycle.  We note that
since $\mathcal{A}$ is a Lagrangian algebra, among the $\mathcal{A}$-modules only $\mathcal{A}$ is NS type. As expected-- none of the topological line operators are local wrt to $\mathcal{A}$. An important fact is that under an $S$ transformation, the spin structures of the two cycles are expected to swap. i.e.
\be \label{eq:S_trans}
\chi^s_{X(t)} (\tilde q, \bar{\tilde q}) = \sum_{Y(s)} S^{s, t}_{X(t) Y(s)} \chi^t_{Y(s)}(q, \bar q),
\ee
where $s,t \in \{NS,R\}$ denote the spin structures along the time and spatial cycle, and $S^{s,t}_{X(t) Y(s)}$ denotes the $S$ transformation matrix that would swap the spin structures, so that we only sum over defects $Y$ with spin structure $s$.  
The ``defect S-matrix'' is related to the W matrix and the bulk $S$ matrix. 
It is given by: 
\be
\sum_i W^t_{i X(s)} S_{ij} = \sum_{Y(t)} S^{s,t}_{X(s) Y(t)} W^s_{j Y(t)},
\ee
where we have introduced the shorthand $W^{1} \equiv \Omega$ and $W^0 \equiv W$.

Equation (\ref{eq:S_trans}) has a handy application. As discussed in (\ref{sec:twistedDVF}), there is a special defect $x_f$ that
generates an R-type spin structure. One can work out the the components $W_{ix_f}$ readily if we have $\Omega_{i 1}$ by applying (\ref{eq:S_trans}) -- noting that the trivial defect is the only $NS$ sector defect here:
\be
\chi^R_{0(NS)}( q, \bar{q}) =  \sum_i \Omega_{i1} \chi_i (\tilde q, \bar{\tilde q}) = \chi^{NS}_{x_f(R)}(\tilde q, \bar{\tilde q}) = \sum_i W_{ix_f} \chi_i(q,\bar q).
\ee
This finally gives
\be \label{eq:work_xf}
\sum_i \Omega_{i1} S_{ij} = W_{jx_f}.
\ee

Generically, given any other symmetries $g$, and corresponding defects generating $g$-twisted boundary conditions, the W matrix of the latter can be extracted using analogues of (\ref{eq:work_xf}).
For example, using this method, we identify analogues of RR, NSR, RNS, defects in a condensed theory involving multiple non-Abelian fermion condensation. These example concerning $SU(2)_{10}$ and $D(D_4)$
will be relegated to the appendix. 

\subsection{Revisiting the (twisted)- Verlinde formula}

The discussion above inspires a re-visit of the Verlinde formula for a spin CFT. The structure of the S-matrix of characters in a spin -CFT being decomposed into different sectors, namely $\{S^{NS,NS}, S^{NS, R}, S^{R,NS}, S^{R,R}\}$, have long been discussed in the literature.

The supersymmetric CFT literature has given a Verlinde formula (see for example \cite{Abdurrahman:1994ar,Eholzer_1994}), although that does not distinguish even and odd fusion channels.
On the other hand,  in a spin CFT that is graded by fermion parity, one should distinguish parity even and odd fusion channels. There is a separate identity isolating the difference between the even and odd channels, in the form of the twisted Verlinde formula. The derivation is based on consideration of dimensional reduction of the 3d spin TQFT to a 2d spin TQFT in \cite{Aasen:2017ubm}. We supplied an alternative derivation in the context of (non Abelian) fermion condensation in (\ref{eq:twisted_VLF}). 
Here, we will obtain a third derivation that depends solely on properties of the decomposition of the $S$ matrix recalled above. 

An extra key observation is that the fusion of two excitations whose worldlines cut across a common twist lines would fuse in a twisted way. The wave-function satisfies
i.e.
\be \label{eq:twistedfuse}
\chi^{NS}_{x\otimes y} = \sum_z {n}_{xy}^z \chi^{NS}_{z}, \qquad \chi^R_{x\otimes y} = \sum_z \tilde{n}_{xy}^z \chi^R_{z}
\ee

Now we can also evaluate the wavefunction via
\begin{align}\label{eq:SSS}
    \chi^{X}_{x(Z)\otimes y(Y)} & =\sum_{w(X)} S^{X, Y}_{y(Y) w(X)}   \hat{x}(Z) \chi^Y_{w(X)} = \sum_{w(X)}S^{X,Y}_{y(Y) w} \frac{S^{Z,X}_{x(Z) w(X)} }{S^{NS,X}_{0,w(X)}}  \chi^Y_{w(X)}  \nonumber \\
                                & = \sum_{w(X), u(Y)}S^{X,Y}_{y(Y) w(X)} \frac{S^{Z,X}_{x(Z) w(X)} }{S^{NS,X}_{0,w(X)}} (S^{-1})^{Y,Y.Z}_{w(X) u(Y.Z)} \chi^X_{u(Y.Z)}.
\end{align}
The expression $Y.Z$ corresponds to the aggregate spin structure after fusing two objects with spin structures $Y$ and $Z$ respectively. We note that they form a $\mathbb{Z}_2$ group structure,
with $R$ playing the role of a $\mathbb{Z}_2$ generator satisfying $R.R = NS$.

The second equality above is obtained by considering (\ref{eq:loop_identity}). It is a ``spin-structure enriched'' version  of a well-known identity. 

\begin{figure}[h]
\centering
\be \label{eq:loop_identity}
\begin{tikzpicture}
\draw [thick] (0,1.5) -- (0,0.1);
\draw [thick] (0,-1) -- (0,-0.1);
\draw [thick] (-0.1,0.4) to [out=180,in=30] (-0.8,0.2) to [out=330,in=180] (0,0) to [out=0,in=210] (0.8,0.2) to [out=150,in=0] (0.1,0.4);
\draw [thick] (2.5,1.5) -- (2.5,-1);
\node at (-1.3,0.25) {$\hat x(Z)$}
node at (0.7,1.5) {$\omega(X)$}
node at (3.2,1.5) {$\omega(X)$}
node at (1.7,0.25) {$=\lambda$}
node at (4.5,0.25) {$,\quad\lambda=\frac{S^{Z,X}_{x\omega}}{S^{NS,X}_{0\omega}}$};
\end{tikzpicture}
\ee
\end{figure}

Combining with (\ref{eq:twistedfuse}, \ref{eq:SSS}), we obtain
\be \label{eq:twistedverlinde}
({n^X)}_{x(Z) y(Y)}^{u(Z.Y)}  = \sum_{w(X), u(Y)}S^{X,Y}_{y(Y) w(X)} \frac{S^{Z,X}_{x(Z) w(X)} }{S^{NS,X}_{0w(X)}} (S^{-1})^{Y,X.Z}_{w(X) u(Y.Z)},
\ee
where we have again introduced the short-hand notation
\be
n^{NS} \equiv n, \qquad  n^R \equiv \tilde n.
\ee

Let us emphasize that the $S^{X,Y}$ matrix we are working with is not in a unitary basis. 
Let us take  the $C_2$ theory as an example which follows from the Ising theory (with three sectors $0, \psi, \sigma$ ) with the fermion $\psi$ condensed,
this theory only has 1 NS sector and 1 R sector.
i.e.
\be
0_{NS} = 0 \oplus \psi, \qquad \beta_R = \sigma. 
\ee
The corresponding characters in the $C_2$ theory can be written as
\be
\chi^{NS}_0(q,\bar{q}) = \chi_0(q,\bar{q})  + \chi_\psi(q,\bar{q}) , \qquad \chi^{R}_{0}= \chi_0 (q,\bar{q}) - \chi_\psi (q,\bar{q}) , \qquad \chi^{NS}_\beta (q,\bar{q}) = \chi_\sigma(q,\bar{q}) .  
\ee
In this basis, the $S^{X,Y}$-matrix is given by
\be
S^{NS NS}_{00} = 1, \qquad S^{NS R}_{0\beta} = \sqrt{2}, \qquad S^{R NS}_{\beta 0} = \frac{1}{\sqrt{2}}.
\ee
If instead we pick the normalization as 
\be
\tilde \chi_\beta^{NS} = \sqrt{2} \chi_\sigma, 
\ee
(corresponding to rescaling a $q$ type object by $\sqrt{2}$,
then $\tilde{S}^{NS R}_{0 \beta} = \tilde{S}^{R NS}_{\beta 0} =1$, and the Verlinde formula would require an extra factor of $\sqrt{e_x e_y e_u}$ on the rhs of (\ref{eq:twistedverlinde}), where $e_i$ is the dimension of the endormorphism of sector $i$. This is the version that appears in \cite{Aasen:2017ubm}. 

These are some general considerations that a priori are not connected to anyon condensation.

%#todo: searching algorithm

\section{Conclusion}

The main aim of the current paper is to study in detail gapped boundaries of 2+1 d topological orders characterized by an anyon condensate that contains fermions. 
The physics of these gapped boundaries include the different species of excitations, their fusion rules, and also the properties of junctions when two different gapped boundaries meet. 
In the case of bosonic condensates, these issues have been studied in detail by many, which we have mentioned in the introduction. It is realized that the underlying mathematical structure characterizing each boundary condensate is a commutative Frobenius algebra, and that the boundary excitations and junction excitations are modules and bi-modules of these Frobenius algebra, respectively.  In the current paper, we have generalized these considerations to cover gapped boundaries following from fermionic anyon condensation. The mathematical generalization is to replace  ``commutativity'' by ``super-commutativity''.  This has been discussed to some extent in \cite{Aasen:2017ubm} for simple current condensations. Here we generalize it to accommodate arbitrary fermionic anyon condensation. Moreover, we also extended the discussion to include junction excitations.  Particularly, we developed systematic ways to read off the endomorphism of a (bi)-module -- which describes whether the corresponding defect could host a Majorana mode.
Along the way, we clarify and generalize the defect Verlinde formula discussed in \cite{Shen_2019} to fermionic boundaries, as well as providing a systematic recipe to compute the half-linking numbers central to the formula. 

We also discussed the connection between these defects in a super-condensate and line operators in a super modular invariant CFTs -- as in the bosonic case, each fermionic condensate defines a ``super'' modular invariant, and these gapped boundary excitations are topological line operators.

There are some miscellaneous facts that we have omitted in the main text, but which maybe of interest.

Fermion condensation that preserves fermion parity at the end can be
reduced to an Abelian fermion condensation \cite{Wan:2016php}.
Consider a fermionic gapped boundary of a bosonic bulk topological order. If we adopt the strategy of a sequential condensation that condenses the bosons in the condensate first, the intermediate phase has only three different possible choices.
Namely, the toric code order (c=0), the Ising order  (c=1/2) , and the 3-fermion order $(D_4, 1)$ (c=4) \cite{rowell2007classification}.
These are the only bosonic modular tensor categories with at least one fermionic simple current and that they have quantum dimension 2 -- which would become fully confined by condensing just one more fermion. This simple fermion is responsible for carrying the odd fermion parity, and it is necessarily a simple current in the intermediate phase, as demonstrated in \cite{Wan:2016php}.
This fact gives a simple check that decides if a bulk order has a fermionic gapped boundary -- i.e. by staring at the topological central charge which is preserved under anyon condensation.

Another curious fact is the apparent scarcity of fermionic gapped boundaries, particularly those beyond simple fermion condensation. In the entire $SU(2)_k$ series of modular tensor categories, only $SU(2)_{10}$ contains a Lagrangian super-Frobenius algebra. We could not find any in $SU(3)_k$, by following the principle of sequential condensation. These conclusions are made by adopting the philosophy of ``sequential'' condensation. We first look up modular invariants of these models ($SU(2)_k$ series adopts an ADE classification, and $SU(3)_k$ has been classified for example in \cite{gannon1994classification}.) and among them look for candidates where one can condense a further simple fermionic current.  

To conclude, we note that there are various questions that are still pending.
We looked into super modular invariants in $D(D_4)$, and there, we found examples where the super-modular invariants do not appear to correspond to a super-commutative Frobenius algebra as we have defined it in the main text. It rather appeared as a non-commutative algebra where anyons in the condensate can be non-local wrt each other. It would also imply that the product of the tentative algebra must break fermion parity -- the product of total parity even anyons gives a parity odd anyon.  One could perhaps discard them as simply being unphysical. However, the mutual non-locality among the condensed anyons was only at worst a minus sign. It is a curiosity whether there might after all be an interpretation to these super modular invariants. 
 
The generalization to fermionic condensates suggests that it is possible to further extend the idea of anyon condensation to include anyons of arbitrary spin -- the new ingredient is to couple to the condensing anyon an appropriate gauge field -- the counter part of spin structures that would render the condensation consistent. It is believed that such condensates might be related to gapless boundary conditions. 

Finally, it is realized recently that these gapped boundaries are examples of spontaneous breaking of a categorical symmetry\cite{thorngren2019fusion,ji2019categorical,kong2020algebraic}. It is important to understand if there are other implications of the categorical symmetry, or how to systematically reverse the process of condensation (i.e. the analogue of gauging), and how these ideas could be generalized to higher dimensions.

\appendix

\section{Computing F-symbols of the gapped boundary/interface}

In the main text, we have reviewed that the representations of the super Frobenius algebra forms a super-fusion category.
We have discussed in detail how to recover the fusion rules of the resultant super-fusion category. Since fusion remains associative, there is a corresponding set of $F$ symbols for the representation category, which can be obtained using data of the parent phase. This has also be discussed in detail in the case of a bosonic condensates in \cite{Eliens:2013epa}, but only mentioned briefly in \cite{Aasen:2017ubm} in the case of a fermionic condensate.
Here we would like to present the details in a simple example for illustration purpose. The tricky part is to fix various sign conventions and keeping track of spin structures in constructing fusion basis in the representation category.
In the following, we will make the process completely explicit in a simple but non-trivial example. 

\subsection{Illustration via $SU(2)_6$}
%#done: detailed computation of su(2)_6 boundary F-symbols    
The basic topological data of $SU(2)_6$ is shown in table \ref{tab:su2_6}.
    The anyon $6$ in $su(2)_6$ is a simple current fermion with quantum dimension $d_6=1$. The monodromy \cite{bruillard2017fermionic} of each anyon with this fermion defines a $\Z_2$ grading on $su(2)_6$, thus dividing the bulk anyons into Neveu-Schwartz sector (anyons $0,2,4,6$) and Ramond sector (anyons $1,3,5$).
        \begin{table}[h]
            \centering
            \begin{tabular}{|C|C|C|}
                    \hline \text { anyons } a =2j & d_{a} & h_{a} \\
                    \hline 0 & 1 & 0 \\
                    1 & \sqrt{2+\sqrt{2}} & 3 / 32 \\
                    2 & 1+\sqrt{2} & 1 / 4 \\
                    3 & \sqrt{2(2+\sqrt{2})} & 15 / 32 \\
                    4 & 1+\sqrt{2} & 3 / 4 \\
                    5 & \sqrt{2+\sqrt{2}} & 3 / 32 \\
                    6 & 1 & 1/2 \\
                    \hline
            \end{tabular}
            \caption{$su(2)_{6}$ topological sectors, quantum dimensions, and topological spins. The label $2j$ gives the $SU(2)$ spins
            of the corresponding sector. } 
            \label{tab:su2_6}
        \end{table}
    The fusion of sector $a$ with the fermion is simply given by $a\otimes 6 = 6-a$.
    After condensing the fermion $6$, anyon $a$ and $6-a$ are identified as a single object $X_a = \{ a + (6-a) \}$ in the gapped interface. The spin structure of an interface object is directly inherited from the parent, since $a$ and $6-a$ have the same spin structure in this case. 
    The list of interface objects is given in table \ref{tab:su2_6_child}. Note that the anyon $3$ becomes a q-type object $X_3$ in the interface, because $3$ is invariant under the action of the fermion $6\otimes 3=3$. 
        \begin{table}[h]
            \centering
            \begin{tabular}{|c|c|c|c|c|}
                    \hline Interface object & $X_0= 0\oplus 6$  & $X_1 = 1\oplus 3$  & $X_2= 2\oplus 4$  & $X_3=3$ \\
                    \hline Object type & m-type & m-type & m-type & q-type\\
                    \hline Spin structure & NS & R & NS & R \\
                    \hline
            \end{tabular}
            \caption{Interface objects after condensing $6$ in $su(2)_6$} 
            \label{tab:su2_6_child} 
        \end{table}

        Their fusion rules are as follows:
        \begin{align}
        X_1 \otimes X_1 &=  X_0 \oplus X_1 \oplus X_2 \\
         X_2\otimes X_2 & = X_0 \oplus \mathbb{C}^{1|1} X_2\\
         X_3\otimes X_3 & =  \mathbb{C}^{1|1}X_0\oplus \mathbb{C}^{1|1} X_2 \\
        X_1 \otimes X_2 & =   X_1 \oplus X_3\\
        X_1\otimes X_3 & =  \mathbb{C}^{1|1} X_2 \\
        X_2\otimes X_3 & =  \mathbb{C}^{1|1} X_1 \oplus X_3 
        \end{align}

    The objects $X_0$ and $X_2$ are unconfined and can pass through the gapped interface, while $X_1$ and $X_3$ are confined at the interface. The unconfined phase $U=\{X_0, X_2\}$ corresponds to the super modular invariant $|\chi_0+\chi_6|^2 + |\chi_2+\chi_4|^2$.
    
    The folling notation is used to denote the interface excitations after condensing the simple current fermion $6$ in $su(2)_6$.
      
    {\bf Warning:  In this section, to make notations inline with related references, the time direction flows from the top to the bottom in each diagram.} i.e. We adopt an opposite convention for the flow of time in this sub-section, compared to other parts of the paper.

    \begin{figure}[H]
        \centering
        \begin{equation}
            \mathtikz{
                \draw[very thick, dashed] (0,0) -- (0,1.5cm);
                \node[below] at (0,0) {$X_0$};
                \begin{scope}[xshift=1cm]
                    \draw[very thick, green] (0,0) -- (0,1.5cm);
                    \node[below] at (0,0) {$X_1$};
                \end{scope}
                \begin{scope}[xshift=2cm]
                    \draw[very thick, orange] (0,0) -- (0,1.5cm);
                    \node[below] at (0,0) {$X_2$};
                \end{scope}
                \begin{scope}[xshift=3cm]
                    \draw[very thick, gray] (0,0) -- (0,1.5cm);
                    \node[below] at (0,0) {$X_3$};
                \end{scope}
            }
        \end{equation}
        \caption{The new vacuum $X_0$ is denoted by a dashed line, and the other three excitations are distinguished by different colors.}
        \label{fig:child_object_notation}
    \end{figure}
    The phase $\lambda\in U(1)$ which will appear later is defined as the fermion line with two end points as shown in figure \ref{fig:lambda}. For $su(2)_6$ the fermion $f$ is the sector $6$. The $\lambda$ can not be canceled from the solution of super Pentagon equations by a gauge transformation. Indeed, it can be shown that $\lambda$ is purely imaginary \cite{Aasen:2017ubm} and therefore $\lambda=\pm i$.
    \begin{figure}[H]
        \centering
        \begin{equation}
            \lambda = 
            \mathtikz{
                \draw[very thick, dotted, red] (0,0) -- (1cm,1cm);
                \draw[very thick, dotted, red] (0,0) -- (-1cm,1cm);
                \mysquare{1cm}{1cm}{gray};
                \mysquare{-1cm}{1cm}{gray};
                \node[below left] at (-0.5cm,0.5cm) {$f$};
            }
        \end{equation}
        \caption{The complex number $\lambda$ represents the vacuum splitting into two fermions with end points. Here the red dotted line represents simple current fermion (quantum dimension $=1$).}
        \label{fig:lambda}
    \end{figure}
    
    We use the following convention to solve the super Pentagon equation in child phase. If the fusion channel $X\otimes Y \rightarrow Z$ is even, then we identify the channel $X\otimes Y \rightarrow Z$ with the parent fusion channel $\tilde{X}\otimes \tilde{Y} \rightarrow \tilde{Z}$ where $\tilde{X}, \tilde{Y}, \tilde{Z}$ are bulk representatives of $X,Y,Z$ respectively.
    If the fusion channel $X\otimes Y \rightarrow Z$ is odd, then we identify this odd fusion channel with an even fusion channel where a fermion line segment ends at the $\tilde{Z}$ leg, the interval between the vertex and the ending point is changed to $f\tilde{Z}$.
    %\begin{figure}[H]
    %    \centering
    %    \includegraphics[width=0.5\textwidth]{pic/even_odd_channel_convention.jpg}
    %    \caption{The even fusion channel $X\otimes Y \rightarrow Z$ is identified with the corresponding %bulk fusion channel. The odd fusion channel $X\otimes Y \rightarrow Z$ is identified with the bulk %fusion channel decorated by a fermion line segment.}
    %    \label{fig:even_odd_channel_convention}
    %\end{figure}
    \begin{figure}[H]
        \centering
        \begin{equation*}
            \mathtikzS{0.3}{
                \antivertexIL{0}{0}{1.5cm}{$X$}{$Y$}{$Z$};
            }
            =
            \mathtikzS{0.3}{
                \antivertexIL{0}{0}{1.5cm}{$\tilde{X}$}{$\tilde{Y}$}{$\tilde{Z}$};
            }
            ,
            \mathtikzS{0.3}{
                \antivertexIL{0}{0}{1.5cm}{$X$}{$Y$}{$Z$};
                \fill [fill=red] (0,0) circle [radius=0.2cm];
            }
            =
            \mathtikzS{0.3}{
                \antivertexIL{0}{0}{1.5cm}{$\tilde{X}$}{$\tilde{Y}$}{$\tilde{Z}$};
                \draw[very thick, dotted, red] (0,-1cm) -- (2.5cm,1.5cm);
                \mysquare{2.5cm}{1.5cm}{gray};
                \node[below right] at (1cm,0) {$f$};
                \node[left] at (0,-0.5cm) {$f\tilde{Z}$};
            }
        \end{equation*}
        \caption{The even fusion channel $X\otimes Y \rightarrow Z$ is identified with the corresponding bulk fusion channel. The odd fusion channel $X\otimes Y \rightarrow Z$ is identified with the bulk fusion channel decorated by a fermion line segment.}
        \label{fig:even_odd_channel_convention}
    \end{figure}
    We choose $0,1,2,3$ as the bulk representatives of $X_0,X_1,X_2,X_3$ respectively, therefore we have a ``vertex lifting rule'' with respect to the above convention: 
    \begin{itemize}
        \item For any even channel vertex, replace leg label $X_i$ by bulk anyon $i$. As an example we look at the even channel in $X_2\otimes X_2 \rightarrow \mathbb{C}^{1|1} X_2$, this vertex is replaced by $2\otimes 2 \rightarrow 2$ vertex in the bulk.
            \begin{equation*}
                \mathtikzS{0.4}{
                \antivertexILC{0}{0}{1.5cm}{$X_2$}{$X_2$}{$X_2$}{orange};
            }
            =
            \mathtikzS{0.4}{
                \antivertexIL{0}{0}{1.5cm}{$2$}{$2$}{$2$};
            }
            \end{equation*}
        \item For any odd channel vertex, remove the oddness by attaching a fermion line and move downward according to figure \ref{fig:even_odd_channel_convention}. Here we list all the odd channel lifting rules:
        \newcommand{\tempScale}{0.35}
            \begin{align*}
                \mathtikzS{\tempScale}{
                    \antivertexILC{0}{0}{1.5cm}{$X_2$}{$X_2$}{$X_2$}{orange};
                    \fill [fill=red] (0,0) circle [radius=0.2cm];
                }
                =
                \mathtikzS{\tempScale}{
                    \antivertexIL{0}{0}{1.5cm}{$2$}{$2$}{$2$};
                    \draw[very thick, dotted, red] (0,-1cm) -- (2.5cm,1.5cm);
                    \mysquare{2.5cm}{1.5cm}{gray};
                    \node[below right] at (1cm,0) {$6$};
                    \node[left] at (0,-0.5cm) {$4$};
                }
                \quad,&\quad
                \mathtikzS{\tempScale}{
                    %\antivertexIL{0}{0}{1.5cm}{$X_3$}{$X_3$}{$X_0$};
                    \draw[thick,gray] (-1.5cm,1.5cm) -- (0,0) -- (1.5cm,1.5cm);
                    \draw[thick,dashed] (0,0) -- (0,-1.5cm);
                    \node[below] at (0,-1.5cm) {$X_0$};
                    \node[above] at (-1.5cm,1.5cm) {$X_3$};
                    \node[above] at (1.5cm,1.5cm) {$X_3$};
                    \fill [fill=red] (0,0) circle [radius=0.2cm];
                }
                =
                \mathtikzS{\tempScale}{
                    \antivertexIL{0}{0}{1.5cm}{$3$}{$3$}{$0$};
                    \draw[very thick, dotted, red] (0,-1cm) -- (2.5cm,1.5cm);
                    \mysquare{2.5cm}{1.5cm}{gray};
                    \node[below right] at (1cm,0) {$6$};
                    \node[left] at (0,-0.5cm) {$6$};
                }
                \\
                \mathtikzS{\tempScale}{
                    %\antivertexIL{0}{0}{1.5cm}{$X_3$}{$X_3$}{$X_0$};
                    \draw[thick,gray] (-1.5cm,1.5cm) -- (0,0) -- (1.5cm,1.5cm);
                    \draw[thick,orange] (0,0) -- (0,-1.5cm);
                    \node[below] at (0,-1.5cm) {$X_2$};
                    \node[above] at (-1.5cm,1.5cm) {$X_3$};
                    \node[above] at (1.5cm,1.5cm) {$X_3$};
                    \fill [fill=red] (0,0) circle [radius=0.2cm];
                }
                =
                \mathtikzS{\tempScale}{
                    \antivertexIL{0}{0}{1.5cm}{$3$}{$3$}{$2$};
                    \draw[very thick, dotted, red] (0,-1cm) -- (2.5cm,1.5cm);
                    \mysquare{2.5cm}{1.5cm}{gray};
                    \node[below right] at (1cm,0) {$6$};
                    \node[left] at (0,-0.5cm) {$4$};
                }
                \quad,&\quad
                \mathtikzS{\tempScale}{
                    %\antivertexIL{0}{0}{1.5cm}{$X_3$}{$X_3$}{$X_0$};
                    \draw[thick,gray] (0,0) -- (1.5cm,1.5cm);
                    \draw[thick,green] (0,0) -- (-1.5cm,1.5cm);
                    \draw[thick,orange] (0,0) -- (0,-1.5cm);
                    \node[below] at (0,-1.5cm) {$X_2$};
                    \node[above] at (-1.5cm,1.5cm) {$X_1$};
                    \node[above] at (1.5cm,1.5cm) {$X_3$};
                    \fill [fill=red] (0,0) circle [radius=0.2cm];
                }
                =
                \mathtikzS{\tempScale}{
                    \antivertexIL{0}{0}{1.5cm}{$1$}{$3$}{$2$};
                    \draw[very thick, dotted, red] (0,-1cm) -- (2.5cm,1.5cm);
                    \mysquare{2.5cm}{1.5cm}{gray};
                    \node[below right] at (1cm,0) {$6$};
                    \node[left] at (0,-0.5cm) {$4$};
                }
                \\
                \mathtikzS{\tempScale}{
                    %\antivertexIL{0}{0}{1.5cm}{$X_3$}{$X_3$}{$X_0$};
                    \draw[thick,green] (0,0) -- (1.5cm,1.5cm);
                    \draw[thick,gray] (0,0) -- (-1.5cm,1.5cm);
                    \draw[thick,orange] (0,0) -- (0,-1.5cm);
                    \node[below] at (0,-1.5cm) {$X_2$};
                    \node[above] at (-1.5cm,1.5cm) {$X_3$};
                    \node[above] at (1.5cm,1.5cm) {$X_1$};
                    \fill [fill=red] (0,0) circle [radius=0.2cm];
                }
                =
                \mathtikzS{\tempScale}{
                    \antivertexIL{0}{0}{1.5cm}{$3$}{$1$}{$2$};
                    \draw[very thick, dotted, red] (0,-1cm) -- (2.5cm,1.5cm);
                    \mysquare{2.5cm}{1.5cm}{gray};
                    \node[below right] at (1cm,0) {$6$};
                    \node[left] at (0,-0.5cm) {$4$};
                }
                \quad,&\quad
                \mathtikzS{\tempScale}{
                    %\antivertexIL{0}{0}{1.5cm}{$X_3$}{$X_3$}{$X_0$};
                    \draw[thick,gray] (0,0) -- (1.5cm,1.5cm);
                    \draw[thick,orange] (0,0) -- (-1.5cm,1.5cm);
                    \draw[thick,green] (0,0) -- (0,-1.5cm);
                    \node[below] at (0,-1.5cm) {$X_1$};
                    \node[above] at (-1.5cm,1.5cm) {$X_2$};
                    \node[above] at (1.5cm,1.5cm) {$X_3$};
                    \fill [fill=red] (0,0) circle [radius=0.2cm];
                }
                =
                \mathtikzS{\tempScale}{
                    \antivertexIL{0}{0}{1.5cm}{$2$}{$3$}{$1$};
                    \draw[very thick, dotted, red] (0,-1cm) -- (2.5cm,1.5cm);
                    \mysquare{2.5cm}{1.5cm}{gray};
                    \node[below right] at (1cm,0) {$6$};
                    \node[left] at (0,-0.5cm) {$5$};
                }
                \\
                \mathtikzS{\tempScale}{
                    %\antivertexIL{0}{0}{1.5cm}{$X_3$}{$X_3$}{$X_0$};
                    \draw[thick,orange] (0,0) -- (1.5cm,1.5cm);
                    \draw[thick,gray] (0,0) -- (-1.5cm,1.5cm);
                    \draw[thick,green] (0,0) -- (0,-1.5cm);
                    \node[below] at (0,-1.5cm) {$X_1$};
                    \node[above] at (-1.5cm,1.5cm) {$X_3$};
                    \node[above] at (1.5cm,1.5cm) {$X_2$};
                    \fill [fill=red] (0,0) circle [radius=0.2cm];
                }
                =
                \mathtikzS{\tempScale}{
                    \antivertexIL{0}{0}{1.5cm}{$3$}{$2$}{$1$};
                    \draw[very thick, dotted, red] (0,-1cm) -- (2.5cm,1.5cm);
                    \mysquare{2.5cm}{1.5cm}{gray};
                    \node[below right] at (1cm,0) {$6$};
                    \node[left] at (0,-0.5cm) {$5$};
                }
                \quad .&
            \end{align*}
    \end{itemize}
    
To calculate all the $F$-symbols in the child phase, we first lift all the vertices to bulk vertices according to figure \ref{fig:even_odd_channel_convention}, then we do $F$-moves and $R$-moves in the bulk, and finally we translate the bulk vertices back to child theory using figure \ref{fig:even_odd_channel_convention} again.
Consider the following child phase $F$-move, the legs $\alpha,\beta,\gamma,\delta,\xi,\eta \in \{X_0,X_1,X_2,X_3\}$ label an interface excitation, and the vertices $\mu,\nu,\mu^{\prime},\nu^{\prime}\in\mathbb{Z}_2$ label an even/odd channel.
%#done: F move figure
\begin{equation}
    \mathtikzS{0.4}{
        \coordinate["$\alpha$" above] (A) at (-3cm,3cm);
        \coordinate["$\beta$" above] (B) at (-1cm,3cm);
        \coordinate["$\gamma$" above] (C) at (1cm,3cm);
        \coordinate["$\delta$" below right] (O) at (0,0);
        \coordinate (D) at (-2cm,2cm);
        \coordinate (E) at (-1cm,1cm);
        \coordinate["$\xi$" left] (M) at (-1.2cm,1.3cm);
        \node[left] at (-0.7cm,0.8cm) {$\nu$};
        \node[left] at (-1.7cm,1.8cm) {$\mu$};
        %\fill (-1cm,1cm) circle [radius=0.1cm];
        %\fill (-2cm,2cm) circle [radius=0.1cm];
        \draw[thick] (O) -- (A);
        \draw[thick] (D) -- (B);
        \draw[thick] (E) -- (C);
    }
    \mathtikzS{0.4}{
        \draw[-{Triangle[width=18pt,length=12pt]}, line width=3pt] (-1.5cm, 0) -- (1.5cm, 0cm);
        \node[above left] at (0.2cm,0.2cm) {$F$};
    }
    \mathtikzS{0.4}{
        \coordinate["$\alpha$" above] (A) at (-3cm,3cm);
        \coordinate["$\beta$" above] (B) at (-1cm,3cm);
        \coordinate["$\gamma$" above] (C) at (1cm,3cm);
        \coordinate["$\delta$" below right] (O) at (0,0);
        \coordinate (E) at (-1cm,1cm);
        \coordinate (F) at (0,2cm);
        \coordinate["$\eta$" right] (N) at (-0.5cm,1.2cm);
        \node[left] at (-1cm,1cm) {$\nu^{\prime}$};
        \node[right] at (0cm,2cm) {$\mu^{\prime}$};
        %\fill (-1cm,1cm) circle [radius=0.1cm];
        %\fill (0cm,2cm) circle [radius=0.1cm];
        \draw[thick] (O) -- (A);
        \draw[thick] (C) -- (E);
        \draw[thick] (B) -- (F);
    }
    \label{eq:child_F}
\end{equation}
 Without loss of generality, we study the case where $\mu=\nu=0$, namely both vertices in the LHS of \ref{eq:child_F} are even. We provide a step-by-step calculation of the $F$-symbols in this case, calculation in other cases can be carried out with similar steps, and are therefore omitted in this paper. To perform an interface $F$-move, vertices are first lift to the bulk, followed by a bulk $F$-move.
\begin{align*}
    \mathtikzS{0.3}{
        %skeleton
        \draw[thick] (0,0) -- (2cm,2cm);
        \draw[thick] (0,0) -- (-2cm,2cm);
        \draw[thick] (0,0) -- (0,-1cm);
        \draw[thick] (-1cm,1cm) -- (0,2cm);
        \node[below] at (0,-1cm) {$\delta$};
        \node[above] at (-2cm,2cm) {$\alpha$};
        \node[above] at (0cm,2cm) {$\beta$};
        \node[above] at (2cm,2cm) {$\gamma$};
        \node[left] at (-0.2cm,0.2cm) {$\xi$};
    }
    =
    \mathtikzS{0.3}{
        %skeleton
        \draw[thick] (0,0) -- (2cm,2cm);
        \draw[thick] (0,0) -- (-2cm,2cm);
        \draw[thick] (0,0) -- (0,-1cm);
        \draw[thick] (-1cm,1cm) -- (0,2cm);
        \node[below] at (0,-1cm) {$\delta^{\prime}$};
        \node[above] at (-2cm,2cm) {$\alpha^{\prime}$};
        \node[above] at (0cm,2cm) {$\beta^{\prime}$};
        \node[above] at (2cm,2cm) {$\gamma^{\prime}$};
        \node[left] at (-0.2cm,0.2cm) {$\xi^{\prime}$};
    }
    %\qquad\text{lift vertices to the bulk}
    =\quad
    \sum_{\eta^{\prime}} \left(F^{\alpha^{\prime}\beta^{\prime}\gamma^{\prime}}_{\delta^{\prime}}\right)_{\xi^{\prime}\eta^{\prime}}
    \mathtikzS{0.3}{
        \draw[thick] (0,0) -- (2cm,2cm);
        \draw[thick] (0,0) -- (-2cm,2cm);
        \draw[thick] (0,0) -- (0,-1cm);
        \draw[thick] (1cm,1cm) -- (0,2cm);
        \node[below] at (0,-1cm) {$\delta^{\prime}$};
        \node[above] at (-2cm,2cm) {$\alpha^{\prime}$};
        \node[above] at (0cm,2cm) {$\beta^{\prime}$};
        \node[above] at (2cm,2cm) {$\gamma^{\prime}$};
        \node[right] at (0.2cm,0.2cm) {$\eta^{\prime}$};
    }
    %\qquad\text{bulk $F$-moves}
\end{align*}
When $\eta^{\prime}\in\{0,1,2,3\}$, the above fusion tree is directly transformed to an interface fusion tree with two even vertices through the first rule in figure \ref{fig:even_odd_channel_convention}.
\begin{equation*}
    \mathtikzS{0.3}{
        \draw[thick] (0,0) -- (2cm,2cm);
        \draw[thick] (0,0) -- (-2cm,2cm);
        \draw[thick] (0,0) -- (0,-1cm);
        \draw[thick] (1cm,1cm) -- (0,2cm);
        \node[below] at (0,-1cm) {$\delta^{\prime}$};
        \node[above] at (-2cm,2cm) {$\alpha^{\prime}$};
        \node[above] at (0cm,2cm) {$\beta^{\prime}$};
        \node[above] at (2cm,2cm) {$\gamma^{\prime}$};
        \node[right] at (0.3cm,0.3cm) {$\eta^{\prime}$};
    }
    =
    \mathtikzS{0.3}{
        \draw[thick] (0,0) -- (2cm,2cm);
        \draw[thick] (0,0) -- (-2cm,2cm);
        \draw[thick] (0,0) -- (0,-1cm);
        \draw[thick] (1cm,1cm) -- (0,2cm);
        \node[below] at (0,-1cm) {$\delta$};
        \node[above] at (-2cm,2cm) {$\alpha$};
        \node[above] at (0cm,2cm) {$\beta$};
        \node[above] at (2cm,2cm) {$\gamma$};
        \node[right] at (0.3cm,0.3cm) {$\eta$};
    }
    \quad \text{if } \eta^{\prime}\in\{0,1,2,3\}.
\end{equation*}
When $\eta^{\prime}\in\{4,5,6\}$, the bulk fusion tree can not be transformed to an interface fusion tree at first sight, because neither the first or the second rule in figure \ref{fig:even_odd_channel_convention} can be directly applied to the two vertices. However, the following diagram reveals the relation between the bulk and interface fusion trees.
\begin{align*}
    \hspace*{-1cm}
    \mathtikzS{0.3}{
        \draw[thick] (0,0) -- (4cm,4cm);
        \draw[thick] (0,0) -- (-4cm,4cm);
        \draw[thick] (0,0) -- (0,-2cm);
        \draw[thick] (3cm,3cm) -- (2cm,4cm);
        \node[below] at (0,-2cm) {$\delta$};
        \node[above] at (-4cm,4cm) {$\alpha$};
        \node[above] at (2cm,4cm) {$\beta$};
        \node[above] at (4cm,4cm) {$\gamma$};
        %%%%
        \fill [fill=red] (0,0) circle [radius=0.2cm];
        \fill [fill=red] (3cm,3cm) circle [radius=0.2cm];
        \node[left] at (1.9cm,1.7cm) {\tiny $6-\eta$};
    }
    &=
    \mathtikzS{0.3}{
        \draw[thick] (0,0) -- (4cm,4cm);
        \draw[thick] (0,0) -- (-4cm,4cm);
        \draw[thick] (0,0) -- (0,-2cm);
        \draw[thick] (3cm,3cm) -- (2cm,4cm);
        \node[below] at (0,-2cm) {$\delta^{\prime}$};
        \node[above] at (-4cm,4cm) {$\alpha^{\prime}$};
        \node[above] at (2cm,4cm) {$\beta^{\prime}$};
        \node[above] at (4cm,4cm) {$\gamma^{\prime}$};
        %%%%%
        \draw[thick,dotted,red] (2cm,2cm) -- (6.5cm,3.5cm);
        \mysquare{6.5cm}{3.5cm}{gray};
        \draw[thick,dotted,red] (0,-1cm) -- (10.5cm,2.5cm);
        \mysquare{10.5cm}{2.5cm}{gray};
        %%%
        \node[left] at (1.9cm,1.7cm) {\tiny $6-\eta^{\prime}$};
        \node[left] at (2.9cm,2.7cm) {\tiny $\eta^{\prime}$};
        \node[left] at (0.2,-0.5cm) {\tiny $6-\delta^{\prime}$};
    }
    %\qquad \text{apply the second rule in figure \ref{fig:even_odd_channel_convention}}
    =\left(F^{\alpha^{\prime}(6-\eta^{\prime})6}_{\delta^{\prime}}\right)_{(6-\delta^{\prime})\eta^{\prime}}
    \hspace*{-1cm}
    \mathtikzS{0.3}{
        \draw[thick] (0,0) -- (4cm,4cm);
        \draw[thick] (0,0) -- (-4cm,4cm);
        \draw[thick] (0,0) -- (0,-2cm);
        \draw[thick] (3cm,3cm) -- (2cm,4cm);
        \node[below] at (0,-2cm) {$\delta^{\prime}$};
        \node[above] at (-4cm,4cm) {$\alpha^{\prime}$};
        \node[above] at (2cm,4cm) {$\beta^{\prime}$};
        \node[above] at (4cm,4cm) {$\gamma^{\prime}$};
        %%%%%
        \draw[thick,dotted,red] (2cm,2cm) -- (6.5cm,3.5cm);
        \mysquare{6.5cm}{3.5cm}{gray};
        \draw[thick,dotted,red] (1cm,1cm) -- (8.5cm,3.5cm);
        \mysquare{8.5cm}{3.5cm}{gray};
        %%%
        \node[left] at (1.9cm,1.7cm) {\tiny $6-\eta^{\prime}$};
        \node[left] at (2.9cm,2.7cm) {\tiny $\eta^{\prime}$};
        \node[left] at (0.9cm,0.7cm) {\tiny $\eta^{\prime}$};
    }
    %\qquad \text{bulk $F$-moves}
    \\
    &=\left(F^{\alpha^{\prime}(6-\eta^{\prime})6}_{\delta^{\prime}}\right)_{(6-\delta^{\prime})\eta^{\prime}}
    \left( F^{\eta^{\prime}66}_{\eta^{\prime}} \right)_{(6-\eta^{\prime})0}
    \mathtikzS{0.2}{
        \draw[thick] (0,0) -- (4cm,4cm);
        \draw[thick] (0,0) -- (-4cm,4cm);
        \draw[thick] (0,0) -- (0,-2cm);
        \draw[thick] (3cm,3cm) -- (2cm,4cm);
        \node[below] at (0,-2cm) {$\delta^{\prime}$};
        \node[above] at (-4cm,4cm) {$\alpha^{\prime}$};
        \node[above] at (2cm,4cm) {$\beta^{\prime}$};
        \node[above] at (4cm,4cm) {$\gamma^{\prime}$};
        %%%%%
        \node[left] at (1.9cm,1.7cm) {\tiny $\eta^{\prime}$};
        \begin{scope}[xshift=7cm,scale=1.5]
        \draw[thick, dotted, red] (0,0) -- (1cm,1cm);
        \draw[thick, dotted, red] (0,0) -- (-1cm,1cm);
        \mysquare{1cm}{1cm}{gray};
        \mysquare{-1cm}{1cm}{gray};
        \end{scope}
    }
    %\quad\text{bulk $F$-moves}
    \\
    &=\lambda
    \left(F^{\alpha^{\prime}(6-\eta^{\prime})6}_{\delta^{\prime}}\right)_{(6-\delta^{\prime})\eta^{\prime}}
    \left( F^{\eta^{\prime}66}_{\eta^{\prime}} \right)_{(6-\eta^{\prime})0}
    \mathtikzS{0.2}{
        \draw[thick] (0,0) -- (4cm,4cm);
        \draw[thick] (0,0) -- (-4cm,4cm);
        \draw[thick] (0,0) -- (0,-2cm);
        \draw[thick] (3cm,3cm) -- (2cm,4cm);
        \node[below] at (0,-2cm) {$\delta^{\prime}$};
        \node[above] at (-4cm,4cm) {$\alpha^{\prime}$};
        \node[above] at (2cm,4cm) {$\beta^{\prime}$};
        \node[above] at (4cm,4cm) {$\gamma^{\prime}$};
        %%%%%
        \node[right] at (1.7cm,1.5cm) {\tiny $\eta^{\prime}$};
    }
    %\quad\text{replace the fermion lines by $\lambda$ according to figure \ref{fig:lambda}}
\end{align*}
where in the first step we apply the vertex lifting rule in figure \ref{fig:even_odd_channel_convention}, in the second and third step we perform a bulk $F$-move, and in the last step we replace the fermion lines by $\lambda$ according to figure \ref{fig:lambda}. For simplicity we sometimes label $X_{6-\eta}$ by $6-\eta$, this notation is unambiguous.
It follows that 
\begin{equation*}
    \mathtikzS{0.2}{
        \draw[thick] (0,0) -- (4cm,4cm);
        \draw[thick] (0,0) -- (-4cm,4cm);
        \draw[thick] (0,0) -- (0,-2cm);
        \draw[thick] (3cm,3cm) -- (2cm,4cm);
        \node[below] at (0,-2cm) {$\delta^{\prime}$};
        \node[above] at (-4cm,4cm) {$\alpha^{\prime}$};
        \node[above] at (2cm,4cm) {$\beta^{\prime}$};
        \node[above] at (4cm,4cm) {$\gamma^{\prime}$};
        %%%%%
        \node[left] at (1.9cm,1.7cm) {\tiny $\eta^{\prime}$};
    }
    =
    \lambda^{-1}
    \left(F^{\alpha^{\prime}(6-\eta^{\prime})6}_{\delta^{\prime}}\right)_{(6-\delta^{\prime})\eta^{\prime}}^{-1}
    \left( F^{\eta^{\prime}66}_{\eta^{\prime}} \right)_{(6-\eta^{\prime})0}^{-1}
    \mathtikzS{0.2}{
        \draw[thick] (0,0) -- (4cm,4cm);
        \draw[thick] (0,0) -- (-4cm,4cm);
        \draw[thick] (0,0) -- (0,-2cm);
        \draw[thick] (3cm,3cm) -- (2cm,4cm);
        \node[below] at (0,-2cm) {$\delta$};
        \node[above] at (-4cm,4cm) {$\alpha$};
        \node[above] at (2cm,4cm) {$\beta$};
        \node[above] at (4cm,4cm) {$\gamma$};
        %%%%
        \fill [fill=red] (0,0) circle [radius=0.2cm];
        \fill [fill=red] (3cm,3cm) circle [radius=0.2cm];
        \node[right] at (1.7cm,1.5cm) {\tiny $6-\eta$};
    }
\end{equation*}
Now we have found all the interface $F$-symbols in the case $\mu=\nu=0$.
\begin{equation*}
    \hspace*{-1cm}
    \mathtikzS{0.3}{
        %skeleton
        \draw[thick] (0,0) -- (2cm,2cm);
        \draw[thick] (0,0) -- (-2cm,2cm);
        \draw[thick] (0,0) -- (0,-1cm);
        \draw[thick] (-1cm,1cm) -- (0,2cm);
        \node[below] at (0,-1cm) {$\delta$};
        \node[above] at (-2cm,2cm) {$\alpha$};
        \node[above] at (0cm,2cm) {$\beta$};
        \node[above] at (2cm,2cm) {$\gamma$};
        \node[left] at (-0.3cm,0.3cm) {$\xi$};
    }
    \hspace*{-0.5cm}
    =
    \sum_{\eta^{\prime}} \left(F^{\alpha^{\prime}\beta^{\prime}\gamma^{\prime}}_{\delta^{\prime}}\right)_{\xi^{\prime}\eta^{\prime}} 
    \begin{cases}
        \mathtikzS{0.3}{
        \draw[thick] (0,0) -- (2cm,2cm);
        \draw[thick] (0,0) -- (-2cm,2cm);
        \draw[thick] (0,0) -- (0,-1cm);
        \draw[thick] (1cm,1cm) -- (0,2cm);
        \node[below] at (0,-1cm) {$\delta$};
        \node[above] at (-2cm,2cm) {$\alpha$};
        \node[above] at (0cm,2cm) {$\beta$};
        \node[above] at (2cm,2cm) {$\gamma$};
        \node[right] at (0.3cm,0.3cm) {$\eta$};
    }
    &\text{if } \eta^{\prime}\in\{0,1,2,3\} 
    \\
    \lambda^{-1}
    \left(F^{\alpha^{\prime}(6-\eta^{\prime})6}_{\delta^{\prime}}\right)_{(6-\delta^{\prime})\eta^{\prime}}^{-1}
    \left( F^{\eta^{\prime}66}_{\eta^{\prime}} \right)_{(6-\eta^{\prime})0}^{-1}
    \hspace*{-1cm}
    \mathtikzS{0.15}{
        \draw[thick] (0,0) -- (4cm,4cm);
        \draw[thick] (0,0) -- (-4cm,4cm);
        \draw[thick] (0,0) -- (0,-2cm);
        \draw[thick] (3cm,3cm) -- (2cm,4cm);
        \node[below] at (0,-2cm) {$\delta$};
        \node[above] at (-4cm,4cm) {$\alpha$};
        \node[above] at (2cm,4cm) {$\beta$};
        \node[above] at (4cm,4cm) {$\gamma$};
        %%%%
        \fill [fill=red] (0,0) circle [radius=0.2cm];
        \fill [fill=red] (3cm,3cm) circle [radius=0.2cm];
        \node[right] at (1.5cm,1.3cm) {\tiny $6-\eta$};
    }
    &\text{if } \eta^{\prime}\in\{4,5,6\} 
    \end{cases}
\end{equation*}

 With similar steps we can solve for all the $F$-symbols in the interface, the detailed calculation is omitted. A part of the solutions are shown in figure \ref{fig:F_moves_part}, and the complete list can be found in the attached file.
 \begin{figure}[H]
     \centering
     \includegraphics[width=\textwidth]{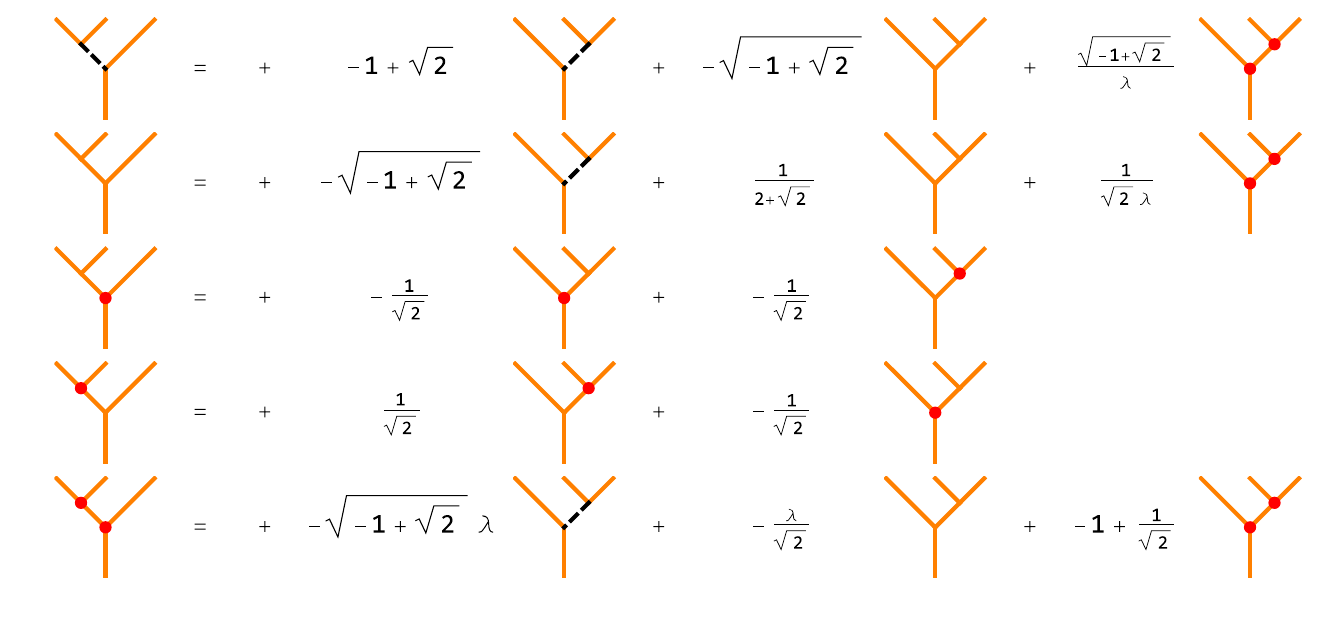}
     \caption{The nontrivial F moves in the child phase}
     \label{fig:F_moves_part}
 \end{figure}

\section{Revisiting the gapped boundaries of the Toric code in Abelian Chern-Simons theory}

In this appendix, we will look at both the bosonic -fermionic junctions and the entanglement on a cylinder with fermionic boundaries in the toric code model, through the lens of Abelian Chern Simons theory. 

\subsection{The bosonic-fermionc junctions}

In \ref{sec:toric} we have illustrated that there must be an even number of Majorana modes on the $e-m-f$ boundary if we put it on a disk using the lattice theory. Now in this sub-section we'll demonstrate the same story using the Abelian Chern-Simons theory representation of the same phase. 

The action of a generic Abelian Chern Simons theory is characterized by a $K$ matrix
\be \label{eq:CSaction}
S = \frac{1}{4\pi} \int K_{ij} a^i \wedge d a^j,
\ee
where $K$ is an $N\times N$ matrix if there are $N$ different $U(1)$ gauge fields $a^i$, $i \in \{1, \cdots, N\}$.
It is well known that the Chern-Simons theory with $K$ matrix 
\be
K_{\mathbb{Z}_2}= 
\begin{pmatrix}
    0 & 2 \\ 
    2 & 0
    \end{pmatrix}
  \ee 
    is the low energy effective field theory of the Toric code topological order. Excitations of the topological order is characterzed by their charge vectors $l_i \in \mathbb{Z}$, $i \in \{1, \cdots, N\}$  \cite{Wang:2012am}. 
The fermionic excitation $f$, is  characterized by the class of charge vectors 
\be
\{ l_f \}= 
\begin{matrix}\begin{pmatrix}
    2n+1 \\ 
    2m+1
    \end{pmatrix}&n,m\in\mathbb{Z} \end{matrix}
    \ee
    To form a gapped boundary, where $f$ is condensed, we have to pair up the fermionic excitation with a free fermion. This can be achieved by introducing extra fields by extending the $K$ matrix as follows
    \begin{equation} \label{eq:extendedK}
    \begin{matrix}
    K_{\mathbb{Z}_2 \otimes \mathbb{Z}^f_0}=\begin{pmatrix}
    0 & 2 & 0 & 0 \\ 
    2 & 0 & 0 & 0 \\
    0 & 0 & 1 & 0 \\
    0 & 0 & 0 & -1 
    \end{pmatrix}
    \end{matrix}
    \end{equation}
   
   The extended block in the K-matrix carries odd integers in the diagonal components. It is well known that the resultant Chern-Simons theory is a ``spin-TQFT'' \cite{Belov:2005ze}, that is sensitive to the spin-structure of the manifold. Indeed physically, it corresponds to introducing free fermions, whose charge vectors are given by the classes of vectors as follows:
   \be
\{ l_\psi \}= 
\begin{matrix}\begin{pmatrix}
    2n \\ 
    2m \\
    2p+1\\
    2q+1
    \end{pmatrix}&n,m,p,q\in\mathbb{Z} \end{matrix}
    \ee
   One could easily check that the exchange statistics of two charge vectors $l_{1,2} \in \{l_\psi\}$ obtained by $\pi l_1 K^{-1} l_2$ gives $\pi$.
    
  In the presence of boundaries, one could obtain a boundary action from e.g. gauge fixing $a^i_0 = 0$ and solving for the flatness constraint that follows from the eom of $a^i_0$ which gives
  \be
  a_\mu^i = \partial_\mu \phi^i.
  \ee
  Substituting into the action (\ref{eq:CSaction}) produces a 1+1 d effective action on the bosons $\phi^i$. 
  For completeness, it takes the following form
  \be
  S_{\partial \mathcal{M}} = \frac{1}{4\pi} \int_{\partial \mathcal{M}} dt dx \, (K_{ij} \partial_t \phi^i \partial_x \phi^j - V_{ij} \partial_x \phi^i \partial_x \phi^j ),
  \ee
    where it is discussed in many places \cite{Wen:1995qn} that $V_{ij}$ determines the velocity of the edge modes that is not determined universally by the topological bulk, but it is controlled by the specific materials that realize the theory. 
 Regardless of the precise values of $V_{ij}$, as long as the matrix is not singular,  the $\phi^i$ and $\phi^j$ can be related to free left and right moving modes in the edge. In the theory defined by the $K$-matrix (\ref{eq:extendedK})
    \begin{equation}
    \left\{\begin{matrix}
    \phi_{1,2}=\frac{1}{2}(\phi_L\pm\phi_R)\\
    \phi_{3,4}=\tilde\phi_{L,R}
    \end{matrix}\right.
    \end{equation}
    i.e. There are two pairs of free (anti)-chiral bosons in this edge theory. 
    
    Now we would like to consider a disk geometry as in figure \ref{fig:disk2}, and we would like to understand the kind of excitations that
    are localized at the junction. One way to go about this problem is to consider regulating the junction by considering a small segment ending at the two gapped boundaries. This is also illustrated in figure \ref{fig:disk2}. 
    
     One then quantizes the edge theory on the little segment that ends on the gapped boundary.

 It is already noted in \cite{Shen:2019rck,dong_topological_2008} that a segment ending on a gapped boundary satisfies boundary
 conditions depending on the charge vectors of anyons condensing at the boundary \cite{Shen:2019rck,dong_topological_2008}. 
 To be precise, the boundary conditions at a given boundary is
 \be \label{eq:bcgap}
 l_i  \partial_t \phi^i \vert_{x=x_0} =0,  \qquad l_i \in L_{\mathcal{A}},
 \ee
 where $x_0$ is the position of the end point of the segment, $L_{\mathcal{A}}$ is a collection of charge vectors that condense at the gapped boundary, with mutual statistics $l_i. K^{-1}. l_j =0$ for all $l_{i,j} \in L_{\mathcal{A}}$.
 
   The charge vectors that define the condensate for the electric, magnetic and fermionic gapped boundaries are given respectively by \cite{levin_protected_2013,Wang:2012am,kapustin_topological_2011}
    \begin{equation} \label{eq:3condensates}
    \begin{matrix}
    L_e=\begin{Bmatrix}
    \begin{pmatrix}
    1\\0\\0\\0
    \end{pmatrix} &
    \begin{pmatrix}
    0\\0\\1\\a
    \end{pmatrix} 
    \end{Bmatrix}
    &
    L_m=\begin{Bmatrix}
    \begin{pmatrix}
    0\\1\\0\\0
    \end{pmatrix} &
    \begin{pmatrix}
    0\\0\\1\\b
    \end{pmatrix} 
    \end{Bmatrix}
    &
    L_f=\begin{Bmatrix}
    \begin{pmatrix}
    1\\1\\0\\c
    \end{pmatrix} &
    \begin{pmatrix}
    1\\-1\\d\\0
    \end{pmatrix} 
    \end{Bmatrix}
    \quad a,b,c,d=\pm 1
    \end{matrix}
    \end{equation}
    We note that the choice of these charge vectors are subjected to some ambiguities. i.e. $a,b,c,d$ can take either $\pm 1$ and they would still describe the condensation of the same topological charges at the boundary. This lead to freedom in determining boundary conditions in (\ref{eq:bcgap}). As we are going to see -- this ambiguity is precisely the freedom to insert Majorana modes at a junction. 
    
    Now we put three distinct boundaries on a disk. As mentioned above, we analyse each junction by considering the open segment ending on two different adjacent boundaries. This is illustrated in figure \ref{fig:disk2}.

        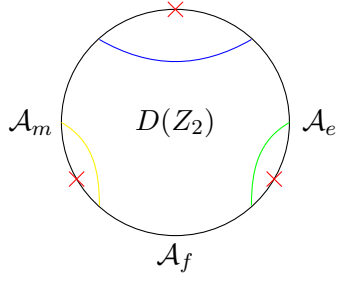
\begin{figure}[h]
    \centering
    \begin{tikzpicture}
    \draw (-4,0) circle [radius=1.5];
    \draw [red] (-2.6,-0.85) -- (-2.8,-0.65);
    \draw [red] (-2.6,-0.65) -- (-2.8,-0.85);
    \draw [red] (-3.9,1.6) -- (-4.1,1.4);
    \draw [red] (-3.9,1.4) -- (-4.1,1.6);
    \draw [red] (-5.2,-0.85) -- (-5.4,-0.65);
    \draw [red] (-5.2,-0.65) -- (-5.4,-0.85);
    \node at (-5.9,0) {$\mathcal{A}_m$};
    \node at (-2.1,0) {$\mathcal{A}_e$};
    \node at (-4,-1.8) {$\mathcal{A}_f$};
    \node at (-4,0) {$D(Z_2)$};
    \draw [blue] (-5,1.1) to [out=-30,in=-150] (-3,1.1);
    \draw [yellow] (-5.5,0) to [out=-30,in=90] (-5,-1.1);
    \draw [green] (-2.5,0) to [out=-150,in=90] (-3,-1.1);
    %\draw [->] (-5,1.5) to [out=30,in=150] (-3,1.5);
    %\node at (-4,2) {$t$};
    \end{tikzpicture}
    \caption{Three different boundaries separated by three junctions on a disk. The bulk phase is the toric code order. }
    \label{fig:disk2}
    \end{figure}
    Let $x$ be the parameter along the strings and takes value between $1$ and $l$. Then the boundary conditions for three strings are given by
    \begin{equation}
    \begin{matrix}
    \left\{
    \begin{matrix}
    \partial_t\phi^{[Y]}_L|_{x=l}=-c\partial_t\tilde\phi^{[Y]}_R|_{x=l}\\
    \partial_t\phi^{[Y]}_R|_{x=l}=-d\partial_t\tilde\phi^{[Y]}_L|_{x=l}\\
    \partial_t\phi^{[Y]}_L|_{x=0}=\partial_t\phi^{[Y]}_R|_{x=0}\\
    \partial_t\tilde\phi^{[Y]}_L|_{x=0}=-b\partial_t\tilde\phi^{[Y]}_R|_{x=0}
    \end{matrix}
    \right.
    &
    \left\{
    \begin{matrix}
    \partial_t\phi^{[B]}_L|_{x=l}=\partial_t\phi^{[B]}_R|_{x=l}\\
    \partial_t\tilde\phi^{[B]}_L|_{x=l}=-b\partial_t\tilde\phi^{[B]}_R|_{x=l}\\
    \partial_t\phi^{[B]}_L|_{x=0}=-\partial_t\phi^{[B]}_R|_{x=0}\\
    \partial_t\tilde\phi^{[B]}_L|_{x=0}=-a\partial_t\tilde\phi^{[B]}_R|_{x=0}
    \end{matrix}
    \right.
    &
    \left\{
    \begin{matrix}
    \partial_t\phi^{[G]}_L|_{x=l}=-\partial_t\phi^{[G]}_R|_{x=l}\\
    \partial_t\tilde\phi^{[G]}_L|_{x=l}=-a\partial_t\tilde\phi^{[G]}_R|_{x=l}\\
    \partial_t\phi^{[G]}_L|_{x=0}=-c\partial_t\tilde\phi^{[G]}_R|_{x=0}\\
    \partial_t\phi^{[G]}_R|_{x=0}=-d\partial_t\tilde\phi^{[G]}_L|_{x=0}
    \end{matrix}
    \right.
    \end{matrix}
    \end{equation}
    Where the upper index $\{Y,B,G\}$ indicating the strings with color yellow, blue and green. 
    It is also convenient to present these boundary conditions by ``unfolding'' the gluing conditions connecting
    a left and a right moving mode above and combine them into a ``closed string'' with either
    periodic or anti-periodic boundary conditions.

        \begin{equation}
    \begin{matrix}
    \Phi^{[Y]}(\tilde x)=
    \left\{
    \begin{matrix}
    \phi_L^{[Y]}(-x)\\
    -\phi_R^{[Y]}(x)\\
    -d\tilde\phi_L^{[Y]}(-x)\\
    -bd\tilde\phi_R^{[Y]}(x)
    \end{matrix}
    \right.
    &
    \Phi^{[G]}(\tilde x)=
    \left\{
    \begin{matrix}
    \phi_L^{[G]}(-x)&\ \tilde x\in(-2l,-l)\\
    c\tilde\phi_R^{[G]}(x)&\ \tilde x\in(-l,0)\\
    ac\tilde\phi_L^{[G]}(-x)&\ \tilde x\in(0,l)\\
    acd\phi_R^{[G]}(x)&\ \tilde x\in(l,2l)
    \end{matrix}
    \right.
    \\&
    \left\{
    \begin{matrix}
    \Phi_1^{[B]}(\tilde{\tilde x})=
    \left\{
    \begin{matrix}
    \phi_L^{[B]}(-x)&\ \tilde{\tilde x}\in(-l,0)\\
    \phi_R^{[B]}(x)&\ \tilde{\tilde x}\in(0,l)
    \end{matrix}
    \right.\\
    \Phi_2^{[B]}(\tilde{\tilde x})=
    \left\{
    \begin{matrix}
    \phi_L^{[B]}(-x)&\ \tilde{\tilde x}\in(-l,0)\\
    a\phi_R^{[B]}(x)&\ \tilde{\tilde x}\in(0,l)\\
    \end{matrix}
    \right.
    \end{matrix}
    \right.
    \end{matrix}
    \end{equation}
    with boundary conditions
    \be
    \begin{matrix}
    \Phi^{[Y]}(-2l)=-bcd\Phi^{[Y]}(2l)
    &
    \Phi^{[G]}(-2l)=acd\Phi^{[G]}(2l)
    &
    \left\{
    \begin{matrix}
    \Phi_1^{[B]}(-l)=-\Phi_1^{[B]}(l)\\
    \tilde\Phi_2^{[B]}(-l)=ab\tilde\Phi_2^{[B]}(l)
    \end{matrix}
    \right.
    \end{matrix}
    \ee
  The mode expansion of a chiral boson depend on the boundary conditions. 
    \be
    \Phi^{osci}(x)=i\sum_{n}\frac{\alpha_n}{n}e^{-i\frac{2n\pi}{4l}}
    \ee
 Here we are focusing on the oscillator parts of the expansion. In the case of a periodic boundary condition, the sum over $n$ runs over integers. On the other hand, for anti-periodic boundary conditions, $n \in \mathbb{Z} + 1/2$. 
 These half-integer modes are related to the majorana mode (and its Virasoro descendent).     
 The bottom line is therefore, that by making different choice of $a,b,c,d$ in (\ref{eq:3condensates}), one could change the boundary conditions of the bosons describing the junctions altering their moding, which is  equivalent to adding or deleting Majorana modes at the given junction.    
     However, we can only have altogether an even number of chiral bosons with anti-periodic boundary conditions as we take different combinations of these variables $a,b,c,d=\pm1$. This is consistent with the fact that any physically realizable situation has an even number of Majorana modes.

     In principle we can have more junctions on a disk and it is straight forward to generalize the result above to those cases.\\ 
    This reproduces the observation made in section \ref{sec:bfjunction} on the lattice model. 
\subsection{Entanglement of a strip region in a cylinder}

In the previous subsection, we reviewed the Chern-Simons formulation of these gapped boundaries. 
In this section we take up the formulation to compute the entanglement of a strip region in a cylinder with $e-f$ boundaries.
These are direct generaliztions of the computations in \cite{Lou:2019heg,Shen:2019rck}. 

We will first consider a strip region on a cylinder. The top and bottom boundaries are each characterized by some condensate, or Lagrangian algebra. This is illustrated in figure \ref{fig:striponcylinder}.

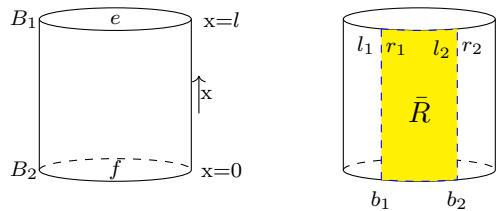
\begin{figure}[h]
\centering
\begin{tikzpicture}
\draw [dashed] (-1,0) arc (180:0:1cm and 0.15cm);
\draw (-1,0) arc (180:360:1cm and 0.15cm);
\draw (0,2) ellipse [x radius=1cm, y radius=0.15cm];
\draw (-1,0) -- (-1,2);
\draw (1,0) -- (1,2);
\draw [->] (1.1,0.75) -- (1.1,1.25);
\node at (1.2,1) {\scriptsize x};
\node at (1.4,0) {\scriptsize x=0}
node at (1.4,2) {\scriptsize x=$l$}
node at (-1.2,2) {\scriptsize $B_1$}
node at (-1.2,0) {\scriptsize $B_2$};
\node at (0,2) {\scriptsize$e$}
node at (0,0) {\scriptsize$f$};

\draw [dashed] (3,0) arc (180:0:1cm and 0.15cm);
\draw (3,0) arc (180:360:1cm and 0.15cm);
\draw (4,2) ellipse [x radius=1cm, y radius=0.15cm];
\draw (3,0) -- (3,2);
\draw (5,0) -- (5,2);
\draw [blue, dashed, fill=yellow] (3.5,1.87) -- (3.5,-0.13) to [out=-4, in=184] (4.5,-0.13) -- (4.5,1.87) to [out=184, in=-4] (3.5,1.87);

\node at (4,0.8) {$\bar R$}
node at (3.5,-0.4) {\scriptsize$b_1$}
node at (4.5,-0.4) {\scriptsize$b_2$}
node at (3.3,1.65) {\scriptsize$l_1$}
node at (3.7,1.6) {\scriptsize$r_1$}
node at (4.3,1.6) {\scriptsize$l_2$}
node at (4.7,1.62) {\scriptsize$r_2$};
\end{tikzpicture}
\caption{Entanglement entropy of a strip $R$ on a cylinder, cutting through the top and bottom boundaries. Each of these boundaries are gapped, and is characterized by some condensate. }
\label{fig:striponcylinder}
\end{figure}

\subsubsection{Anti-periodic boundary condition}
First we consider the Lagrangian algebras $L_e$ and $L_f$ characterizing the top and bottom boundaries respectively. For concreteness, we first consider taking $a=c=d=-1$. 
The boundary conditions for edge modes living at the entanglement cuts are thus given by
\begin{equation}
\begin{matrix}
\left\{
\begin{matrix}
\partial_t\phi_L|_{x=l}=-\partial_t\phi_R|_{x=l}\\
\partial_t\tilde\phi_L|_{x=l}=\partial_t\tilde\phi_R|_{x=l}
\end{matrix}
\right.
&
\left\{
\begin{matrix}
\partial_t\phi_L|_{x=0}=\partial_t\tilde\phi_R|_{x=0}\\
\partial_t\tilde\phi_L|_{x=0}=\partial_t\phi_R|_{x=0}
\end{matrix}
\right.
\end{matrix}
\end{equation}
Again it is convenient to combine the left and right moving modes and express it in terms of  a chiral boson with anti-periodic boundary condition.
\begin{equation}
\Phi(\tilde x)=
\left\{
\begin{matrix}
\phi_L(-x)&\ \tilde x\in(-2l,-l)\\
-\tilde\phi_R(x)&\ \tilde x\in(-l,0)\\
\tilde\phi_L(-x)&\ \tilde x\in(0,l)\\
-\phi_R(x)&\ \tilde x\in(l,2l)
\end{matrix}
\right.
\end{equation}

\begin{equation}
\Phi(-2l)=-\Phi(2l)
\end{equation}
Hence the mode expansion is
\begin{equation}
\Phi(\tilde x)=i\sum_{n\in\mathbb{Z}+\frac{1}{2}}\frac{\alpha_n}{n}e^{-i\frac{2n\pi\tilde x}{4l}}
\end{equation}
Exactly as in \cite{Shen:2019rck}, one can construct the Ishibashi state describing the gluing across the entanglement cut:
\begin{equation}
\begin{aligned}
&|0\rangle\rangle_{b_i}=e^{-\frac{2\pi\epsilon H_i}{4l}}exp(\sum_{n\in\mathbb{N}+\frac{1}{2}}\frac{1}{n}\alpha_{i,-n}\bar\alpha_{i,-n})|0\rangle_{b_i}\\
&|0\rangle\rangle=|0\rangle\rangle_{b_1}\otimes|0\rangle\rangle_{b_2}
\end{aligned}
\end{equation}
where the Hamiltonian $H_i=L^i_0+\bar L^i_0-1/12$ is inserted as a UV regularization with the cutoff scale $\epsilon$. The normalization constant of this state is 
\be
    \begin{aligned}
    \mathcal{N}^{-1}&=\langle\langle0|0\rangle\rangle\\
    &=q^{-\frac{1}{16}}\sqrt{\frac{\eta(q)}{\theta_4(q)}}
    \end{aligned}\quad q=e^{-\frac{8\pi\epsilon}{4l}}
    \ee
where $\eta(q)$ $\theta_4(q)$ and the $\theta_2(q)$ below are the Dedekind $\eta$-function and Jacobi $\theta$-functions respectively. Then the reduced density matrix $\rho$ is obtained by tracing out the chiral or anti-chiral part of the density matrix $\mathcal{N}|0\rangle\rangle\langle\langle0|$. Therefore the trace of the n-th-power of the reduced density matrix is given by
\be
\begin{aligned}
tr\rho^n&=\left(\left(\sqrt{\frac{\theta_4(q)}{\eta(q)}}\right)^n\sqrt{\frac{\eta(q^n)}{\theta_4(q^n)}}\right)^2\\
&=\left(\left(\sqrt{\frac{\theta_2(\tilde q)}{\eta(\tilde q)}}\right)^n\sqrt{\frac{\eta(\tilde q^{\frac{1}{n}})}{\theta_2(\tilde q^{\frac{1}{n}})}}\right)^2\\
&\overset{l/\epsilon\rightarrow\infty}{\rightarrow}2^{n-1}\tilde q^{\frac{1}{12}(n-\frac{1}{n})}, 	\quad \tilde q=e^{-\frac{2\pi l}{\epsilon}}
\end{aligned}
\ee
Hence the entanglement entropy of this state is given by
\begin{equation}
S=\lim_{n\rightarrow1}\frac{1}{n}tr\rho^n=2(\frac{\pi l}{6\epsilon}-log\sqrt{2})
\end{equation}
Here the $-\ln \sqrt{2}$ is the contributoin of the two ground states, indicating the trapped Majorana mode in the junction of $e-f$ boundary.

\subsubsection{Periodic boundary condition}
Then we consider the same Lagrangian algebra with $a=-c=-d=1$. The only difference from the previous calculation is that here we have periodic boundary condition and non trivial zero modes.
\begin{equation}
\begin{matrix}
\left\{
\begin{matrix}
\partial_t\phi_L|_{x=l}=-\partial_t\phi_R|_{x=l}\\
\partial_t\tilde\phi_L|_{x=l}=-\partial_t\tilde\phi_R|_{x=l}
\end{matrix}
\right.
&
\left\{
\begin{matrix}
\partial_t\phi_L|_{x=0}=\partial_t\tilde\phi_R|_{x=0}\\
\partial_t\tilde\phi_L|_{x=0}=\partial_t\phi_R|_{x=0}
\end{matrix}
\right.
\end{matrix}
\end{equation}

\begin{equation}
p_L=p_R=-\tilde p_R=-\tilde p_L=C\qquad C\in\mathbb{Z}
\end{equation}

\begin{equation}
\Phi(\tilde x)=
\left\{
\begin{matrix}
\phi_L(-x)&\ \tilde x\in(-2l,-l)\\
-\tilde\phi_R(x)&\ \tilde x\in(-l,0)\\
-\tilde\phi_L(-x)&\ \tilde x\in(0,l)\\
\phi_R(x)&\ \tilde x\in(l,2l)
\end{matrix}
\right.
\end{equation}

\begin{equation}
\Phi(-2l)=\Phi(2l)
\end{equation}
So we have the following mode exopansion 
\begin{equation}
\Phi(\tilde x)=\Phi_0+P\tilde x+i\sum_{n\neq0}\frac{\alpha_n}{n}e^{-i\frac{2n\pi\tilde x}{4l}}\qquad P=C
\end{equation}
The Ishibashi state across the entanglement cut is given by
\begin{equation}
\begin{aligned}
&|B\rangle\rangle_{b_i}=\sum_{C_i\in\mathbb{Z}}e^{-\frac{2\pi\epsilon H_i}{4l}}exp(\sum_{n=1}\frac{1}{n}\alpha_{i,-n}\bar\alpha_{i,-n})|P_{i}=\bar P_{i}=C_i\rangle_{b_i}\\
&|B\rangle\rangle=|B\rangle\rangle_{b_1}\otimes|B\rangle\rangle_{b_2}
\end{aligned}
\end{equation}
Following the same method, the entanglement entropy is
\begin{equation}
S=2\cdot\frac{\pi l}{6\epsilon}
\end{equation}
As shown before the junction between $e-f$ boundaries can be chosen to trap an unpaired Majorana mode or not. 
The ground states on a cylinder with different boundary conditions at the top and bottom could be understood as a blown up of the junction. Such an ambiguity in the trapped Majorana mode can be revealed by studying the entanglement entropy. 

\subsection{Entanglement of a cylindrical region in a cylinder}
In this section we consider the entanglement of a cylindrical region embedded in the cylinder. This is illustrated in figure \ref{fig:cylinder_in_cylinder}. 
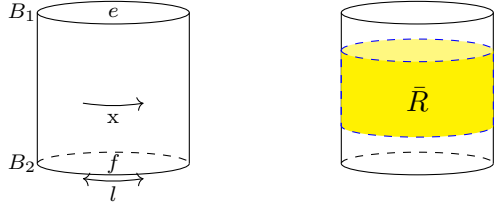
\begin{figure}[h]
\centering
\begin{tikzpicture}
\draw [dashed] (-1,0) arc (180:0:1cm and 0.15cm);
\draw (-1,0) arc (180:360:1cm and 0.15cm);
\draw (0,2) ellipse [x radius=1cm, y radius=0.15cm];
\draw (-1,0) -- (-1,2);
\draw (1,0) -- (1,2);
\draw [->] (-0.4,0.8) to [out=-8, in=188] (0.4,0.8);
\node at (0,0.6) {\scriptsize x};
\draw [<->] (-0.4,-0.2) to [out=-8, in=188] (0.4,-0.2);
\node at (0,-0.4) {\scriptsize $l$};
\node at (-1.2,2) {\scriptsize $B_1$}
node at (-1.2,0) {\scriptsize $B_2$};
\node at (0,2) {\scriptsize$e$}
node at (0,0) {\scriptsize$f$};
\draw [dashed] (3,0) arc (180:0:1cm and 0.15cm);
\draw (3,0) arc (180:360:1cm and 0.15cm);
\draw (4,2) ellipse [x radius=1cm, y radius=0.15cm];
\draw (3,0) -- (3,2);
\draw (5,0) -- (5,2);
\draw [dashed, blue, fill=yellow, fill opacity=0.5] (3,1.5) arc (180:0:1cm and 0.15cm) -- (5,0.5) arc (0:180:1cm and 0.15cm) -- (3,1.5);
\draw [dashed, blue, fill=yellow] (3,1.5) arc (180:360:1cm and 0.15cm) -- (5,0.5) arc (0:-180:1cm and 0.15cm) -- (3,1.5);
\node at (4,0.85) {$\bar R$};
\end{tikzpicture}
\caption{Entanglement entropy of a cylindrical region embedded in a cylinder.}
\label{fig:cylinder_in_cylinder}
\end{figure}

 Recall in \cite{Lou:2019heg} that in caseas as such, the gluing condition defining the Ishibashi states is determined by the allowed anyon lines crossing the cut. In the configuration of the entanglement we have chosen, the anyon line allowed has to be a common condensed anyon shared by the top and bottom boundary. The only such anyon is the trivial anyon, and therefore, the Ishiashi state contains only the trivial sector. 
 The entanglement entropy can be similarly computed, which is given by
\begin{equation}
S=\frac{\pi l}{6\epsilon}-2\ln 2.
\end{equation}    
Here $\ln 2$ is again the quantum dimension of the bulk toric code order, and there is a factor of two, following from independent contributions from the two entanglement cuts. 

\section{Some topological data of $D(S_3)$ and its gapped boundaries}
We would like to review here some basic data of the $D(S_3)$ model.
There're $8$ anyons in $D(S_3)$, labeled by letters $A$ through $H$. Of them $A,\ B,\ C,\ D,\ F$ are bosons and $E$ is a fermion. 
A summary of quandum dimension and twist of all the anyons are listed below. (The trivial object is conventionally denoted ``$A$''.)
    
    \scalebox{0.88}{
        \begin{tabular}{|c|c|c|c|c|c|c|c|c|}
            \hline
            anyon $a$ & $A$ & $B$ & $C$ & $D$ & $E$ & $F$ & $G$ & $H$ \\
            \hline
            quantum dimension $d_a$ & 1 & 1 & 2 & 3 & 3 & 2 & 2 & 2\\
            \hline
            twist $\theta_a$ & 1 & 1 & 1 & 1 & -1 & 1 & $e^{2\pi i/3}$ & $e^{-2\pi i/3}$\\
            \hline
        \end{tabular} 
    }

    Their fusion rules are given in Table \ref{tab:s3_fusion}
    \begin{table}
        \centering
        \scalebox{0.7}{
        \begin{tabular}{|c|c|c|c|c|c|c|c|c|}
            \hline
            $\otimes$ & $A$ & $B$ & $C$ & $D$ & $E$ & $F$ & $G$ & $H$ \\ \hline
            $A$ & $A$ & $B$ & $C$ & $D$ & $E$ & $F$ & $G$ & $H$ \\ \hline
            $B$ & $B$ & $A$ & $C$ & $E$ & $D$ & $F$ & $G$ & $H$ \\ \hline
            $C$ & $C$ & $C$ & $A\oplus B\oplus C$ & $D\oplus E$ & $D\oplus E$ & $G\oplus H$ & $F\oplus H$ & $F\oplus G$ \\ \hline
            $D$ & $D$ & $E$ & $D\oplus E$ & \makecell{$A\oplus C\oplus F$ \\ $\oplus G\oplus H$} & \makecell{$B\oplus C\oplus F$ \\ $\oplus G\oplus H$} & $D\oplus E$ & $D\oplus E$ & $D\oplus E$ \\ \hline
            $E$ & $E$ & $D$ & $D\oplus E$ & \makecell{$B\oplus C$ \\$\oplus F\oplus G\oplus H$} & \makecell{$A\oplus C$ \\ $\oplus F\oplus G\oplus H$} & $D\oplus E$ & $D\oplus E$ & $D\oplus E$ \\ \hline
            $F$ & $F$ & $F$ & $G\oplus H$ & $D\oplus E$ & $D\oplus E$ & $A\oplus B\oplus F$ & $C\oplus H$ & $C\oplus G$ \\ \hline
            $G$ & $G$ & $G$ & $F\oplus H$ & $D\oplus E$ & $D\oplus E$ & $C\oplus H$ & $A\oplus B\oplus G$ & $C\oplus F$ \\\hline
            $H$ & $H$ & $H$ & $F\oplus G$ & $D\oplus E$ & $D\oplus E$ & $C\oplus G$ & $C\oplus F$ & $A\oplus B\oplus H$ \\
            \hline
        \end{tabular}
        }
        \caption{Fusion table of $D(S_3)$}
        \label{tab:s3_fusion}
    \end{table}

    The $S$-matrix of $D(S_3)$ is given by
    \begin{eqnarray}
        \label{eq:S3_S_matrix}
        S=\frac{1}{6}\left(
        \begin{array}{cccccccc}
           1 & 1 & 2 & 3 & 3 & 2 & 2 & 2 \\
           1 & 1 & 2 & -3 & -3 & 2 & 2 & 2 \\
           2 & 2 & 4 & 0 & 0 & -2 & -2 & -2 \\
           3 & -3 & 0 & 3 & -3 & 0 & 0 & 0 \\
           3 & -3 & 0 & -3 & 3 & 0 & 0 & 0 \\
           2 & 2 & -2 & 0 & 0 & 4 & -2 & -2 \\
           2 & 2 & -2 & 0 & 0 & -2 & -2 & 4 \\
           2 & 2 & -2 & 0 & 0 & -2 & 4 & -2 \\
        \end{array}    
        \right).
    \end{eqnarray} 

    There're four types of gapped boundaries for the $D(S_3)$ bulk, labeled by the four subgroups of $S_3$, or equivalently, by the four Lagrangian algebras $\mA_{1,2,3,4}$ shown in Table \ref{tab:s3_multifusion_cat}. The boundary excitations at one particular boundary form a fusion category, as represented by the diagonal cells of Table \ref{tab:s3_multifusion_cat}. The off-diagonal cells give the number and quantum dimension of distinct defects located between two different boundaries.
    \begin{table}
        \centering
        \scalebox{0.78}{
        \begin{tabular}{|C|C|C|C|C|} 
            \hline
            & \mA_1 & \mA_2 & \mA_3 & \mA_4 \\ \hline
            \makecell{\mA_1=A\oplus B\oplus 2C\\ K_1=\{1\}} & \Vect_{S_3} & \{\sqrt{3},\sqrt{3}\} & \{\sqrt{2},\sqrt{2},\sqrt{2}\} & \{\sqrt{6}\} \\ \hline
            \makecell{\mA_2=A\oplus B\oplus 2F\\ K_2=\Z_3} & \{\sqrt{3},\sqrt{3} \} & \Vect_{S_3} & \{\sqrt{6}\} & \{\sqrt{2},\sqrt{2},\sqrt{2}\} \\ \hline
            \makecell{\mA_3=A\oplus C\oplus D\\ K_3=\Z_2} & \{\sqrt{2},\sqrt{2},\sqrt{2}\} & \{\sqrt{6}\} & \Rep(S_3) & \{\sqrt{3},\sqrt{3} \} \\ \hline
            \makecell{\mA_4=A\oplus F\oplus D\\ K_4=S_3} & \{\sqrt{6}\} & \{\sqrt{2},\sqrt{2},\sqrt{2}\} & \{\sqrt{3},\sqrt{3} \} & \Rep(S_3) \\ \hline
        \end{tabular}
        }
        \caption{Summary of the distinct boundaries labeled by four different condensates $\mA_{1,2,3,4}$, and the quantum dimension of defects/excitations localized between them. This is reproduced from \cite{cong_defects_2017}. The diagonal cells give the fusion category describing the boundary excitations of each type of boundary. }
        \label{tab:s3_multifusion_cat}
    \end{table}

\section{Advanced level -- multiple non-Abelian fermionic condensation  }
In the discussion in the main text, we have focused on situations where the condensate only contains at most one fermionic anyon. 
We consider here more complicated examples where the condensate could contain multiple fermions.
One reason for considering these examples is that (modular invariant) supersymmetric CFT with multiple super-charges can also be understood in terms of anyon condensates -- with each super charge related to a species of condensed fermion. In these supersymmetric CFT's, one could allow each fermionic sector to carry independent spin structures. 
One would have naively expect that this structure should carry through. It does -- when the condensed fermions are all simple currents with quantum dimension equals 1. 
We study multiple fermion condensations in  $SU(2)_{10}$, and also in $D(D_4)$. We found that not all the cases have
a well defined independent spin structures. 
The independent spin structures may appear ``deformed'' -- as in the case of $SU(2)_{10}$. 
It is known that Lagrangian algebra in the modular tensor category is in 1-1 correspondence with modular invariants.
In $D(D_4)$, we find that for every super Frobenius algebra we studied it corresponds to a ``super'' modular invariant (i.e. combinations of anyons that are invariant under $S$ and $T^2$).
 However, the converse is no longer true. Some super modular invariants do not appear to correspond to super Frobenius algebra.
 There are also super modular invariants that appear to describe condensates that break fermion parity.
 It is not clear whether such modular invariants or strange condensates are physically relevant or realizable. 
 Junctions between boundaries following from these strange condensates are also mysterious, if physical at all.
 
 For completeness, we will give the list of these strange super modular invariants below.

\subsection{$SU(2)_{10}$}

The distinct topological sectors in the $SU(2)_{10}$ topological order is reviewed in the table below. 

\begin{center}
\begin{table}[h]
    \begin{tabular}{|C|C|C|}
        %\begin{array}{|c|c|c|}
            \hline \text { sectors } 2j & d_{a} & h_{a} \\
            \hline 0 & 1 & 0 \\
            1 & \sqrt{2+\sqrt{3}} & 1 / 16 \\
            2 & 1+\sqrt{3} & 1 / 6 \\
            3 & \sqrt{2}+\sqrt{2+\sqrt{3}} & 5 / 16 \\
            4 & 2+\sqrt{3} & 1 / 2 \\
            5 & 2 \sqrt{2+\sqrt{3}} & 35 / 48 \\
            6 & 2+\sqrt{3} & 1 \\
            7 & \sqrt{2}+\sqrt{2+\sqrt{3}} & 21 / 16 \\
            8 & 1+\sqrt{3} & 5 / 3 \\
            9 & \sqrt{2+\sqrt{3}} & 33 / 16 \\
            10 & 1 & 5 / 2 \\
            \hline
        %\end{array}
    \end{tabular}
    \caption{$su(2)_{10}$ topological sectors, quantum dimensions, and topological spins. The label $j$ gives the $SU(2)$ spins
    of these sectors.} 
    \label{tab:su2_10}
\end{table}
\end{center}
In earlier works \cite{Aasen:2017ubm}, the theory is part of the series $SU(2)_{4k +2}$ for all integers $k>0$ which contains a fermionic simple current (i.e with $SU(2)$ spin $2j= 4k+2$ that can be condensed. In the case of $SU(2)_{10}$, the fermionic simple current is
$2j =10$. 

 The condensate of interest to us however, is $\mathcal{A}= 0\oplus 4\oplus 6\oplus 10$, which has also been discussed in \cite{Wan:2016php}.
 There, it is noted that this forms a Lagrangian algebra, and therefore all the topological sectors except $\mathcal{A}$ itself
 is confined, giving rise to a gapped boundary. 
 Here, we look into the confined sectors more closely. Using the methods discussed in the main text, it is possible
 to work out all the modules of $\mathcal{A}$. These results are summarized in the table below. 
 Since $2j = 4$ and $2j=10$ are both fermionic, one might wonder whether one can define independent spin structures.
 
 \begin{table}[h]
     \begin{tabular}{|c|c|c|c|c|}
    \hline
    X& $ \mathcal{A}= 0\oplus 4\oplus  6 \oplus 10$ & $x_1 = 1\oplus 3 \oplus 2 \times 5 \oplus 7 \oplus 9$ & $x_2= 2\oplus 4 \oplus 6 \oplus 8$  & $x_3= 3 \oplus 7$ \\
    \hline
  fp & NS, n-type & R, n-type & NS, n-type & R, q-type\\
  \hline
 $(\chi_4,\chi_{10})_{fp} $ &  (NS,NS) & (``NS,R'') & (``R,NS'') & (R,R)\\
    \hline
    
   \end{tabular}
   \caption{Defects (modules) at the gapped boundary where $\mathcal{A} $ condenses. }
   \end{table}
   
   The labels $(NS,R)$ and $(R,NS)$ are {\it in quotes} because these spin structures carry some caveats.
   
    The way we determine spin structures is based on the trick described in (\ref{eq:work_xf}). 
We assign the following fermion parity to the condensates 

\begin{tabular}{|c|c|c|c|c|}
\hline
$2j$ & 0 & 4 & 6& 10 \\
\hline
$(\sigma_4^{2j}, \sigma_{10}^{2j} )$ & (1,1) & (-1,1) & (1,-1) & (-1,-1)  \\
\hline
\end{tabular}

We obtain
\be
W_{(2j) \,x_{(\chi^x_4, \chi^x_{10})}} \equiv \sum_{2k = \{0,4,6,10\}} S_{(2j) (2k)} (-1)^{\chi^x_4 \sigma^{2k}_4 + \chi^x_{10} \sigma^{2k}_{10}/2 }.
\ee
Here, we have substituted $\chi^x_{4\,, 10} = 1$ if $x$ is in the R-sector wrt 4 and 10 respectively, and 0 in the corresponding NS-sector. 
While taking the (NS,NS) sector simply recovers the W matrix of $\mathcal{A}$, and the  (R,R) sector above produces the defect $3\oplus 7$, the $(NS,R)$ , and the $(NS,NS)$and $(R,NS)$ produces a {\it linear combinations of defects}!
In particular, the (R,NS) defect is a combination of $\mathcal{A}$ and $2\oplus 4 \oplus 6 \oplus 8$, while (NS,R) a combination of 
$3\oplus 7$ and $1\oplus 3\oplus 2 \times 5 \oplus 7 \oplus 9$. This explains why the spin-structure labels of these two defects are ``in quotes'', since they appear non-standard. 

This appears to suggest that the defects $2\oplus 4 \oplus 6 \oplus 8$ and  $1\oplus 3\oplus 2 \times 5 \oplus 7 \oplus 9$ generate some symmetry transformation that does not simply form a $\mathbb{Z}_2$ group. This should be an example of the algebraic symmetry discussed recently in \cite{Kong:2020cie}.

\subsection{$D(D_4)$} \label{sec:dd4}
\subsubsection*{Basic data}
We review here some basic data of the quantum double $D(D_4)$.
There're $22$ anyons in $D(S_3)$, labeled by letters $a$ through $v$. Of them $a,b,c,d,e,f,g,h,i,j,k,l,m,n$ are bosons and $p,q,r,s,t,u$ are fermions.
The $S$ matrix and $T$ matrix are given by \cite{lan_gapped_2015}
\begin{equation}
    S=\frac{{1}}{{8}}\resizebox{0.9\hsize}{!}{$
    \left(
    \begin{array}{rrrrrrrrrrrrrrrrrrrrrr}
        1 & 1  & 1  & 1  & 1  & 1  & 1  & 1  & 2  & 2  & 2  & 2  & 2  & 2  & 2  & 2  & 2  & 2  & 2  & 2  & 2  & 2  \\
        1 & 1  & 1  & 1  & 1  & 1  & 1  & 1  & -2 & 2  & 2  & -2 & -2 & 2  & -2 & -2 & 2  & 2  & -2 & -2 & 2  & -2 \\
        1 & 1  & 1  & 1  & 1  & 1  & 1  & 1  & 2  & -2 & 2  & -2 & 2  & -2 & -2 & 2  & -2 & 2  & -2 & 2  & -2 & -2 \\
        1 & 1  & 1  & 1  & 1  & 1  & 1  & 1  & 2  & 2  & -2 & 2  & -2 & -2 & -2 & 2  & 2  & -2 & 2  & -2 & -2 & -2 \\
        1 & 1  & 1  & 1  & 1  & 1  & 1  & 1  & -2 & -2 & 2  & 2  & -2 & -2 & 2  & -2 & -2 & 2  & 2  & -2 & -2 & 2  \\
        1 & 1  & 1  & 1  & 1  & 1  & 1  & 1  & -2 & 2  & -2 & -2 & 2  & -2 & 2  & -2 & 2  & -2 & -2 & 2  & -2 & 2  \\
        1 & 1  & 1  & 1  & 1  & 1  & 1  & 1  & 2  & -2 & -2 & -2 & -2 & 2  & 2  & 2  & -2 & -2 & -2 & -2 & 2  & 2  \\
        1 & 1  & 1  & 1  & 1  & 1  & 1  & 1  & -2 & -2 & -2 & 2  & 2  & 2  & -2 & -2 & -2 & -2 & 2  & 2  & 2  & -2 \\
        2 & -2 & 2  & 2  & -2 & -2 & 2  & -2 & 4  & 0  & 0  & 0  & 0  & 0  & 0  & -4 & 0  & 0  & 0  & 0  & 0  & 0  \\
        2 & 2  & -2 & 2  & -2 & 2  & -2 & -2 & 0  & 4  & 0  & 0  & 0  & 0  & 0  & 0  & -4 & 0  & 0  & 0  & 0  & 0  \\
        2 & 2  & 2  & -2 & 2  & -2 & -2 & -2 & 0  & 0  & 4  & 0  & 0  & 0  & 0  & 0  & 0  & -4 & 0  & 0  & 0  & 0  \\
        2 & -2 & -2 & 2  & 2  & -2 & -2 & 2  & 0  & 0  & 0  & 4  & 0  & 0  & 0  & 0  & 0  & 0  & -4 & 0  & 0  & 0  \\
        2 & -2 & 2  & -2 & -2 & 2  & -2 & 2  & 0  & 0  & 0  & 0  & 4  & 0  & 0  & 0  & 0  & 0  & 0  & -4 & 0  & 0  \\
        2 & 2  & -2 & -2 & -2 & -2 & 2  & 2  & 0  & 0  & 0  & 0  & 0  & 4  & 0  & 0  & 0  & 0  & 0  & 0  & -4 & 0  \\
        2 & -2 & -2 & -2 & 2  & 2  & 2  & -2 & 0  & 0  & 0  & 0  & 0  & 0  & -4 & 0  & 0  & 0  & 0  & 0  & 0  & 4  \\
        2 & -2 & 2  & 2  & -2 & -2 & 2  & -2 & -4 & 0  & 0  & 0  & 0  & 0  & 0  & 4  & 0  & 0  & 0  & 0  & 0  & 0  \\
        2 & 2  & -2 & 2  & -2 & 2  & -2 & -2 & 0  & -4 & 0  & 0  & 0  & 0  & 0  & 0  & 4  & 0  & 0  & 0  & 0  & 0  \\
        2 & 2  & 2  & -2 & 2  & -2 & -2 & -2 & 0  & 0  & -4 & 0  & 0  & 0  & 0  & 0  & 0  & 4  & 0  & 0  & 0  & 0  \\
        2 & -2 & -2 & 2  & 2  & -2 & -2 & 2  & 0  & 0  & 0  & -4 & 0  & 0  & 0  & 0  & 0  & 0  & 4  & 0  & 0  & 0  \\
        2 & -2 & 2  & -2 & -2 & 2  & -2 & 2  & 0  & 0  & 0  & 0  & -4 & 0  & 0  & 0  & 0  & 0  & 0  & 4  & 0  & 0  \\
        2 & 2  & -2 & -2 & -2 & -2 & 2  & 2  & 0  & 0  & 0  & 0  & 0  & -4 & 0  & 0  & 0  & 0  & 0  & 0  & 4  & 0  \\
        2 & -2 & -2 & -2 & 2  & 2  & 2  & -2 & 0  & 0  & 0  & 0  & 0  & 0  & 4  & 0  & 0  & 0  & 0  & 0  & 0  & -4 
    \end{array}
    \right)
    $}
\end{equation}
\begin{equation}
    T=\Diag(1,1,1,1,1,1,1,1,1,1,1,1,1,1,i,-1,-1,-1,-1,-1,-1,-i)
\end{equation}
The total quantum dimension of $D(D_4)$ is
\begin{equation}
    D=\sqrt{\sum_{i}d_i^2}=8
\end{equation}

\subsubsection*{Super modular invariant and super Lagrangian algebra}
We use a Mathematica program to search for all possible super modular invariants and super Lagrangian algebras of $D(D_4)$. In order to have a condensation from $D(D_4)$ to the vacuum, it's necessary that the condensate ${A}$ has dimension $\text{dim}(A):=\sum_{i\in A}d_i=D=8$. 
\noindent The algorithm used to search for super modular invariants is:
\begin{enumerate}
    \item Generate all candidates of condensate with $\dim(A)=8$.
    \item In the above results, find all $T^2$-invariant candidates (only bosons and fermions can be present).
    \item In the above results, find all $S$-invariant candidates.
\end{enumerate}
\noindent The algorithm used to search for super Lagrangian algebra is:
\begin{enumerate}
    \item Generate all candidates of condensate with $\dim(A)=8$
    \item In the above results, find all candidates composed only of bosons and fermions. ($T^2$-invariance)
    \item In the above results, find all algebras. This requirement is equivalent to the existence of fusion channel: $\forall x,y\in \mA, \exists z\in \mA \ s.t \  N_{xy}^z >0$.
    \item In the above results, find all connected algebras (the vacuum $\mathbf{1}=a$ appears once).
    \item In the above results, find all separable algebras \footnote{Here we apply the corollary 3.9 of \cite{cong_topological_2016} without proving it in the fermionic case}
\end{enumerate}

\begin{figure}[h]
    \includegraphics{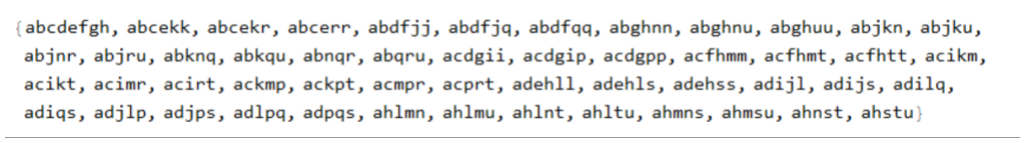}
    \caption{51 super modular invariants}
    \label{fig:super_mod_inv_51}
\end{figure}
\begin{figure}[h]
    \includegraphics{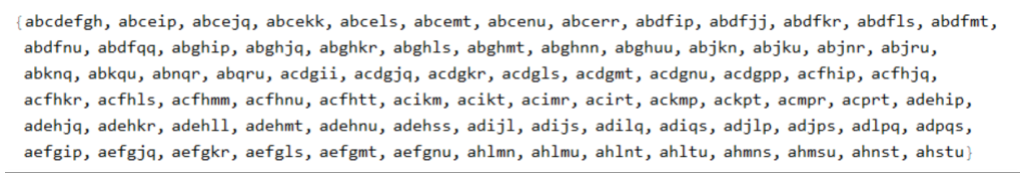}
    \caption{81 super Lagrangian algebras}
    \label{fig:super_lag_alg_81}
\end{figure}
Using the above algorithms we have found 51 super modular invariants and 81 super Lagrangian algebras shown in figure \ref{fig:super_mod_inv_51} and figure \ref{fig:super_lag_alg_81}, a detailed classification of them will be given later. To simplify notation, a set of anyons together with multiplicity information is represented by a string, e.g. ``abcekk'' can mean both the condensate $\mA=a+b+c+e+k+k$ in the list of super Lagrangian algebras, and the character $\chi=\chi_a+\chi_b+\chi_c+\chi_e+2\chi_k$ in the list of super modular invariants.

The super modular invariants and super Lagrangian algebras thus found actually contain usual bosonic condensates such as $\mA=abcekk$. To this end we introduce the notion of \textit{proper fermion condensation}, by \textit{proper} we mean there's at least one fermion in the condensate.

\subsubsection*{Double Toric Code $TC\boxtimes\overline{TC}$}
The above super Lagrangian algebras and super modular invariants can be understood from sequential condensation of $D(D_4)$. First we perform the simple boson condensation $\mA=a+b$ in $D(D_4)$, where $b$ is another boson with quantum dimension $1$. It turns out that the unconfined child theory of this boson condensation is the double toric code $TC\boxtimes\overline{TC}$. There's a gapped interface between the bulk $D(D_4)$ phase and child $TC\boxtimes\overline{TC}$ phase. Under the bulk-to-boundary condensation functor $F:i\to i\otimes \mA$, the bulk anyons are mapped to interface excitations as shown in the table(\ref{tab:dd4_ab}). In the first part of the table, we see that the $8$ simple current bosons in the bulk become $4$ simple current bosons in the interface, these $4$ bosons are unconfined and can enter the child phase. In the third part of the table, we see that the interface excitations ``$ip$", ``$ls$", ``$mt$" and ``$ov$" are confined in the sense of bosonic anyon condensation, because each has a lift to bulk anyons with different spin. 
\begin{table}[h]
    \centering
\begin{tabular}{|c|c|c|}
    \hline
    $i$ & $i\otimes A$ & $TC\boxtimes\overline{TC}$ objects\\
    \hline
    $a,\ b$ & $ab$ & $1\bar{1}$\\
    $c,\ e$ & $ce$ & $m\bar{m}$\\
    $d,\ f$ & $df$ & $e\bar{e}$\\
    $g,\ h$ & $gh$ & $f\bar{f}$\\
    \hline
    $j$ & $j^+ + j^-$ & $1\bar{e}+e\bar{1}$\\
    $k$ & $k^+ + k^-$ & $1\bar{m}+m\bar{1}$\\
    $n$ & $n^+ + n^-$ & $e\bar{m}+m\bar{e}$\\
    $q$ & $q^+ + q^-$ & $m\bar{f}+f\bar{m}$\\
    $r$ & $r^+ + r^-$ & $e\bar{f}+f\bar{e}$\\
    $u$ & $u^+ + u^-$ & $1\bar{f}+f\bar{1}$\\
    \hline
    $i,\ p$ & $ip$ &\\
    $l,\ s$ & $ls$ & Confined\\
    $m,\ t$ & $mt$ &\\
    $o,\ v$ & $ov$ &\\
    \hline
\end{tabular}
\caption{Boson condensation with condensate $\mA=a+b$.}
\label{tab:dd4_ab}
\end{table}

What's interesting is the second part of the table, in which some quantum dimension $2$ bosons/fermions split into two different particles in the interface, and all of them are unconfined and can enter the child phase. The fact that they split into two different particles rather than one particle with multiplicity $2$ can be easily checked from the modular invariant. If they had split into two different particles, the corresponding modular invariant would be $|\chi_a+\chi_b|^2 + |\chi_c+\chi_e|^2 + |\chi_d+\chi_f|^2 + |\chi_g+\chi_h|^2 + 2|\chi_j|^2 + 2|\chi_k|^2 + 2|\chi_n|^2 + 2|\chi_q|^2 + 2|\chi_r|^2 + 2|\chi_u|^2$, which is indeed a modular invariant. However, if they split into one particle with multiplicity=2, then the expected modular invariant would be $|\chi_a+\chi_b|^2 + |\chi_c+\chi_e|^2 + |\chi_d+\chi_f|^2 + |\chi_g+\chi_h|^2 + |2\chi_j|^2 + |2\chi_k|^2 + |2\chi_n|^2 + |2\chi_q|^2 + |2\chi_r|^2 + |2\chi_u|^2$, which is not a modular invariant by directly checking the $S$-transformation.

After the condensation of $\mA=a+b$, the unconfined U theory now has 16 anyons and each one has quantum dimension $1$. There're 10 bosons and 6 fermions in the unconfined theory. The $10$ bosons are grouped into two collections, namely $\{ ``ab", ``ce", ``df", ``gh"\}$ and $\{ ``j^+", ``j^-", ``k^+", ``k^-", ``n^+", ``n^-" \}$. The first $4$ bosons come from identifying two bulk bosons, while the latter $6$ bosons come from splitting of a quantum dimension $2$ bulk boson. The $6$ fermions are $\{ ``q^+", ``q^-", ``r^+", ``r^-", ``u^+", ``u^-"\}$. Here we have used $``j^+"$ and $``j^-"$ to denote the two different particles after splitting anyon $j$ in the bulk.

Another abelian topological order with $10$ simple current bosons and $6$ simple current fermions is the double Toric Code $TC\boxtimes\overline{TC}$. In the $10$ bosons, $4$ are different from others and are known as the diagonal bosons, namely $1\bar{1}$, $e\bar{e}$, $m\bar{m}$ and $f\bar{f}$. 

Indeed, the $S$-matrix of the child phase can be calculated by means of \cite{Eliens:2013epa} and identified with the $S$-matrix of the double Toric Code $TC\boxtimes\overline{TC}$. We omit the calculation and only show the object identification of the child phase and $TC\boxtimes\overline{TC}$. First, the 4 special bosons in the two theories are identified: $ab=1\bar{1}$, $df=e\bar{e}$, $ce=m\bar{m}$ and $gh=f\bar{f}$. Second, the remaining 6 bosons and 6 fermions in the two theories are identified through fusion rules. The full dictionary is listed in the last column of table(\ref{tab:dd4_ab}). However, this identification is not unique due to the topological symmetries in $TC\boxtimes\overline{TC}$. 
Similar analysis and identification with $TC\boxtimes\overline{TC}$ can be carried out in three other boson condensation $\mA=a+c$, $\mA=a+d$ and $\mA=a+h$. The dictionaries are shown respectively in table(\ref{tab:dic_ac}), table(\ref{tab:dic_ad}) and table(\ref{tab:dic_ah}).
\begin{table}[h]
    \centering
\begin{tabular}{|c|c|c|}
    \hline
    $i$ & $i\otimes A$ & $TC\boxtimes\overline{TC}$ particle \\
    \hline
    a,\ c & $ac$ & $1\bar{1}$ \\
    b,\ e & $be$ & $m\bar{m}$ \\
    d,\ g & $dg$ & $e\bar{e}$ \\
    f,\ h & $fh$ & $f\bar{f}$ \\
    \hline
    i& $i^+ + i^-$ & $1\bar{e}+e\bar{1}$\\
    k& $k^+ + k^-$ & $1\bar{m}+m\bar{1}$\\
    m& $m^+ + m^-$ & $e\bar{m}+m\bar{e}$\\
    p& $p^+ + p^-$ & $m\bar{f}+f\bar{m}$\\
    r& $r^+ + r^-$ & $e\bar{f}+f\bar{e}$\\
    t& $t^+ + t^-$ & $1\bar{f}+f\bar{1}$\\
    \hline
\end{tabular}
\caption{Identify child theory of $A=a+c$ condensation with $TC\boxtimes\overline{TC}$.}
\label{tab:dic_ac}
\end{table}

\begin{table}[h]
    \centering
\begin{tabular}{|c|c|c|}
    \hline
    $i$ & $i\otimes A$ & $TC\boxtimes\overline{TC}$ particle \\
    \hline
    a,\ d & $ad$ & $1\bar{1}$ \\
    b,\ f & $bf$ & $m\bar{m}$ \\
    c,\ g & $cg$ & $e\bar{e}$ \\
    e,\ h & $eh$ & $f\bar{f}$ \\
    \hline
    i& $i^+ + i^-$ & $1\bar{e}+e\bar{1}$\\
    j& $j^+ + j^-$ & $1\bar{m}+m\bar{1}$\\
    l& $l^+ + l^-$ & $e\bar{m}+m\bar{e}$\\
    p& $p^+ + p^-$ & $m\bar{f}+f\bar{m}$\\
    q& $q^+ + q^-$ & $e\bar{f}+f\bar{e}$\\
    s& $s^+ + s^-$ & $1\bar{f}+f\bar{1}$\\
    \hline
\end{tabular}
\caption{Identify child theory of $A=a+d$ condensation with $TC\boxtimes\overline{TC}$.}
\label{tab:dic_ad}
\end{table}

\begin{table}[h]
    \centering
\begin{tabular}{|c|c|c|}
    \hline
    $i$ & $i\otimes A$ & $TC\boxtimes\overline{TC}$ particle \\
    \hline
    a,\ h & $ah$ & $1\bar{1}$ \\
    b,\ g & $bg$ & $m\bar{m}$ \\
    c,\ f & $cf$ & $e\bar{e}$ \\
    d,\ d & $de$ & $f\bar{f}$ \\
    \hline
    l& $l^+ + l^-$ & $1\bar{e}+e\bar{1}$\\
    m& $m^+ + m^-$ & $1\bar{m}+m\bar{1}$\\
    n& $n^+ + n^-$ & $e\bar{m}+m\bar{e}$\\
    s& $s^+ + s^-$ & $m\bar{f}+f\bar{m}$\\
    t& $t^+ + t^-$ & $e\bar{f}+f\bar{e}$\\
    u& $u^+ + u^-$ & $1\bar{f}+f\bar{1}$\\
    \hline
\end{tabular}
\caption{Identify child theory of $A=a+h$ condensation with $TC\boxtimes\overline{TC}$.}
\label{tab:dic_ah}
\end{table}

\subsubsection*{Classification of fermion condensation in $D(D4)$}
The above found super modular invariants and super Lagrangian algebras of $D(D_4)$ are the results of sequential condensation, and can be understood from anyon condensation in the child phase $TC\boxtimes\overline{TC}$.
With our knowledge of $TC\boxtimes\overline{TC}$, the proper fermion condensation in $D(D_4)$ can be regrouped according to fermion condensation in $TC\boxtimes\overline{TC}$. In each case we can do the sequential condensation, namely we first arrive at stage $TC\boxtimes\overline{TC}$ and try condense the fermions in $TC\boxtimes\overline{TC}$. Since $TC\boxtimes\overline{TC}$ is abelian, fermion condensation in $TC\boxtimes\overline{TC}$ is much easier than in the original $D(D_4)$.

Fermion condensation in $TC\boxtimes\overline{TC}$ is classified according to the number of fermions in the condensate, and are divided into the following three collections.

\textbf{b+b+b+f}\\
There's only one fermion in the condensate in this case. For example, we can condense $\mA=1\bar{1}+e\bar{e}+m\bar{m}+f\bar{f}$ in $TC\boxtimes\overline{TC}$ according to table(\ref{tab:dd4_ab}). In terms of the $D(D_4)$ anyons this condensate of $TC\boxtimes\overline{TC}$ can be rewritten as $\mA=ab + j^+ + k^+ + u^+$, and hence equivalent to a direct condensation of $\mA=a+b+j+k+u$ in $D(D_4)$.
There're $12$ such condensates, namely \{ $abjku$, $abjnr$, $abknq$, $acikt$, $acimr$, $ackmp$, $adijs$, $adilq$, $adjlp$, $ahlmu$, $ahlnt$, $ahmns$ \}.

\textbf{b+b+f+f}\\
There're two fermions in the condensate in this case. There're $18$ such condensates, when rewritten using $D(D_4)$ labels, they are \{ $abjru$, $abkqu$, $abnqr$, $acirt$, $ackpt$, $acmpr$, $adiqs$, $adjps$, $adlpq$, $ahltu$, $ahmsu$, $ahnst$, $abcerr$, $abdfqq$, $abghuu$, $acdgpp$, $acfhtt$, $adehss$ \}.

\textbf{b+f+f+f}\\
There're three fermions in the condensate in this case. There're $4$ such condensates,  when rewritten using $D(D_4)$ labels, they are \{ $abqru$, $acprt$, $adpqs$, $ahstu$ \}.

\subsubsection*{Super modular invariants in $TC\boxtimes\overline{TC}$}
We use a similar algorithm to generate all super modular invariants of gapped boundary in $TC\boxtimes\overline{TC}$. 

The super modular invariants composed of purely bosonic anyons are:
\begin{eqnarray*}
    &1\bar{1}+1\bar{e}+e\bar{1}+e\bar{e}&\circlearrowleft 
    \\
    &1\bar{1}+1\bar{m}+m\bar{1}+m\bar{m}&\circlearrowleft 
    \\
    &1\bar{1}+e\bar{m}+m\bar{e}+f\bar{f}&\circlearrowleft 
    \\
    &1\bar{1}+e\bar{e}+m\bar{m}+f\bar{f}&\circlearrowleft 
    \\
    &1\bar{1}+1\bar{e}+m\bar{1}+m\bar{e}&\leftrightarrow 1\bar{1}+e\bar{1}+1\bar{m}+e\bar{m}
\end{eqnarray*}
where the symbol $\leftrightarrow$ and $\circlearrowleft$ denote the $Z_2$ orbit under the action of $i\bar{j}\rightarrow j\bar{i}$ in $TC\boxtimes\overline{TC}$.

The super modular invariants with at least one fermion are:
\begin{eqnarray*}
    &1\bar{1}+e\bar{e}+m\bar{f}+f\bar{m}&\circlearrowleft 
\\
&1\bar{1}+e\bar{f}+m\bar{m}+f\bar{e}&\circlearrowleft 
\\
&1\bar{1}+1\bar{f}+f\bar{1}+f\bar{f}&\circlearrowleft 
\\
&1\bar{1}+1\bar{e}+f\bar{1}+f\bar{e}&\leftrightarrow 1\bar{1}+e\bar{1}+1\bar{f}+e\bar{f}
\\
&1\bar{1}+1\bar{f}+m\bar{1}+m\bar{f}&\leftrightarrow 1\bar{1}+f\bar{1}+1\bar{m}+f\bar{m}
\\
&1\bar{1}+e\bar{f}+m\bar{e}+f\bar{m}&\leftrightarrow 1\bar{1}+f\bar{e}+e\bar{m}+m\bar{f}
\end{eqnarray*}
It is found that all super modular invariants of $TC\boxtimes\overline{TC}$ with fermion present are of the form \textbf{b+b+f+f}. As the second step of sequential condensation in $D(D_4)$, it is weird that the above listed \textbf{b+b+b+f} and \textbf{b+f+f+f} cases are NOT super modular invariants in $TC\boxtimes\overline{TC}$, although they're super modular invariants in the parent theory $D(D_4)$. 

Strange as it may seem, we observe that a special linear combination of super modular invariants of form \textbf{b+b+b+f} (or \textbf{b+f+f+f}) is indeed a super modular invariant of $TC\boxtimes\overline{TC}$. For example $\chi = \chi_a + \chi_b + \chi_j + \chi_k + \chi_u$ is a super modular invariant of $D(D_4)$, after condensing $\mA=a+b$ these anyons become $a,b\rightarrow 1\bar{1}$, $j\rightarrow 1\bar{e}$, $k\rightarrow 1\bar{m}$, $u\rightarrow 1\bar{f}$ as shown in table \ref{tab:dd4_ab}. It can be easily checked that $\chi_1 = \chi_{1\bar{1}} + \chi_{1\bar{e}} + \chi_{1\bar{m}} + \chi_{1\bar{f}}$ is not a super modular invariant of $TC\boxtimes\overline{TC}$. However, due to the topological symmetry in $TC\boxtimes\overline{TC}$ there's another identification $a,b\rightarrow 1\bar{1}$, $j\rightarrow e\bar{1}$, $k\rightarrow m\bar{1}$, $u\rightarrow f\bar{1}$. Under this identification the super modular invariant $\chi$ descends to $\chi_2 = \chi_{1\bar{1}} + \chi_{e\bar{1}} + \chi_{m\bar{1}} + \chi_{f\bar{1}}$. Although neither $\chi_1$ nor $\chi_2$ is a super modular invariant in the child phase $TC\boxtimes\overline{TC}$, the arithmetic average of the two if a super modular invariant in $TC\boxtimes\overline{TC}$: $\frac{1}{2}(\chi_1+\chi_2) = \chi_{1\bar{1}} + \frac{1}{2}(\chi_{1\bar{e}}+\chi_{e\bar{1}}) + \frac{1}{2}(\chi_{1\bar{m}}+\chi_{m\bar{1}}) + \frac{1}{2}(\chi_{1\bar{f}}+\chi_{f\bar{1}})$ is invariant under $S$-transformation.

For the above reason, we have in total $16$ weird condensates in $D(D_4)$, which have form \textbf{b+b+b+f} or \textbf{b+f+f+f} in the child pahse $TC\boxtimes\overline{TC}$ from the point of view of sequential condensation, shown in the following table.
\begin{longtable}{|l|l|l|l|}\hline
    \multicolumn{4}{|l|}{Weird condensates}\\\hline
    \{$a,b,j,k,u$\} &\{$a,c,i,k,t$\} &\{$a,d,i,j,s$\} &\{$a,h,l,m,u$\} \\\hline
    \{$a,b,j,n,r$\} &\{$a,c,i,m,r$\} &\{$a,d,i,l,q$\} &\{$a,h,l,n,t$\} \\\hline
    \{$a,b,k,n,q$\} &\{$a,c,k,m,p$\} &\{$a,d,j,l,p$\} &\{$a,h,m,n,s$\} \\\hline
    \{$a,b,q,r,u$\} &\{$a,c,p,r,t$\} &\{$a,d,p,q,s$\} &\{$a,h,s,t,u$\} \\\hline
    \caption{16 weird condensates in $D(D_4)$}
    \label{tab:weird_examples}
\end{longtable}

If we take these super modular invariants as fermion condensations, the resulting phases are all trivial with fermion parity broken. However there is still a $Z_2$-symmetry generated by a combination of fermion and boson or two fermions in the condensate.
All these examples have similiar structures so in the following table we select one example to illustrate. 
\begin{longtable}{l|l|l|l}
Condensate & Modules & $Z_2$ parity odd (1) & $Z_2$ parity odd (2) \\\hline
\multirow{5}*{$\{a,b,k,n,q\}$} & \multirow{1}{*}{$A=a\oplus b\oplus k\oplus n\oplus q$} & $\{k,q\}$ & $\{n,q\}$ \\\cline{2-4}
                                                                                    & \multirow{1}{*}{$X_1=g\oplus h\oplus j\oplus n\oplus r$} & $\{j,r\}$ & $\{j,n\}$
\\\cline{2-4}
                                                                                    & \multirow{1}{*}{$X_2=d\oplus f\oplus q\oplus r\oplus u$} & $\{q,r\}$ & $\{q,u\}$\\\cline{2-4}
                                                                                    & \multirow{1}{*}{$ X_3=c\oplus e\oplus j\oplus k\oplus u$} & $\{j,k\}$ & $\{j,u\}$ \\\cline{2-4}
&                                                                                    
$X_4=i\oplus l\oplus m\oplus o\oplus p\oplus s\oplus t\oplus v$
& $\{l,m,s,t\}$ & $\{l,o,s,v\}$ \\\hline
\end{longtable}
In the above table we list the anyons with odd parity for two $Z_2$ symmetries in the $3^{rd}$ and $4^{th}$ columun. These $Z_2$ symmetries are generated by the combination of $\{k,q\}$ and $\{n,q\}$ respectively.
\\
The fusion rule of these modules is given by
\begin{longtable}{|l|l|l|l|l|l|}
\hline
$\bigotimes$&$\mathbf{A}$&$\mathbf{X_1}$&$\mathbf{X_2}$&$\mathbf{X_3}$&$\mathbf{X_4}$\\\hline
$\mathbf{A}$&$A$&$X_1$&$X_2$&$X_3$&$X_4$\\\hline
$\mathbf{X_1}$&$X_1$&$A$&$X_3$&$X_2$&$X_4$\\\hline
$\mathbf{X_2}$&$X_2$&$X_3$&$A$&$X_1$&$X_4$\\\hline
$\mathbf{X_3}$&$X_3$&$X_2$&$X_1$&$A$&$X_4$\\\hline
$\mathbf{X_4}$&$X_4$&$X_4$&$X_4$&$X_4$&$A\oplus X_1\oplus X_2\oplus X_3$\\\hline
\end{longtable}

%This observation suggests that \textbf{b+b+b+f} and \textbf{b+f+f+f} cases (the 16 weird examples in Jiaqi's note) are not well defined as algebras??? or that the failure of sequential condensation is admissible at fermionic boundary???

%\begin{figure}
%    \centering
%    %\includegraphics[width=0.4\textwidth]{pic/%psu2_6_all_F22Symbols_lambda.pdf}
%    \caption{}
%    \label{fig:su2_6_VLC_all_F22}
%\end{figure}

%\begin{figure}
%    \centering
%    %\includegraphics[width=0.4\textwidth]{pic/%psu2_6_all_F22Symbols_lambda.pdf}
%    \caption{}
%    \label{fig:su2_6_VLC_all_F22}
%\end{figure}

  \section*{Acknowledgements}
We would like to thank Davide Gaiotto for encouraging us to look into fermionic condensations and topological defects in spin CFT, and for helpful discussions.   We would like to thank David Aasen for patiently explaining to us details of constructing fusion basis in the condensed theory with careful treatment of the spin structures in \cite{Aasen:2017ubm}. 
We are also grateful to Zheng-Cheng Gu, Liang Kong and Tian Lan for discussions and suggestions.
We thank Wei Li for initial collaborations. 
We would also like to thank Yuting Hu, Yidun Wan and Juven Wang for many past and on-going conversations on the subject.
 LYH acknowledges the support of NSFC (Grant No. 11922502, 11875111) and the
Shanghai Municipal Science and Technology Major Project (Shanghai Grant No.2019SHZDZX01). We would also like to thank the support of the Emmy Noether Fellowship, and the hospitality of Perimeter Institute, where part of the work was completed.

 \bibliography{ref}
    \bibliographystyle{alpha}
    \end{document}